\documentclass[sn-mathphys,Numbered,iicol,pdflatex]{sn-jnl}

\usepackage{graphicx}%
\usepackage{multirow}%
\usepackage{amsmath,amssymb,amsfonts}%
\usepackage{amsthm}%
\usepackage{mathrsfs}%
\usepackage[title]{appendix}%
\usepackage{xcolor}%
\usepackage{textcomp}%
\usepackage{manyfoot}%
\usepackage{booktabs}%
\usepackage{algorithm}%
\usepackage{algorithmicx}%
\usepackage{algpseudocode}%
\usepackage{listings}%
\usepackage{hyperref}

\newcommand{\tcut}{\ensuremath{t_{\text{cut}}}}
\newcommand{\Qmax}{\ensuremath{Q^2_{\text{max}}}}
\newcommand{\Qmin}{\ensuremath{Q^2_{\text{min}}}}
\newcommand{\GEp}{\ensuremath{G_E^p}}
\newcommand{\ofQ}{\ensuremath{\left(Q^2\right)}}
\newcommand{\ofZ}{\ensuremath{\left(Z\right)}}
\newcommand{\redchi}{\ensuremath{\chi^2_r}}

\begin{document}

\title{Direct Comparison of using a \texorpdfstring{$Z$}{Z}-Transformation instead of the traditional \texorpdfstring{$Q^2$}{Q2} for Extraction of the Proton Radius from \texorpdfstring{$e-p$}{e-p} Scattering Data}

\author*[1,2]{Tyler J. Hague}
\email{TJHague@gmail.com}

\author[3]{Douglas W. Higinbotham}
\email{doug@jlab.org}

\author[3,4]{Spencer Portuese}
\nomail

\affil[1]{%
 \orgname{North Carolina A\&T State University}, \orgaddress{\city{Greensboro}, \state{North Carolina} \postcode{27411}, \country{USA}}
}
\affil[2]{
 \orgname{Lawrence Berkeley National Laboratory}, \orgaddress{\city{Berkeley}, \state{California} \postcode{94720}, \country{USA}}
}%
\affil[3]{
 \orgname{Thomas Jefferson National Accelerator Facility}, \orgaddress{\city{Newport News}, \state{Virginia} \postcode{23606}, \country{USA}}
}%
\affil[4]{%
 \orgname{The Governor's School for Science and Technology}, \orgaddress{\city{Hampton}, \state{Virginia} \postcode{23666}, \country{USA}}
 }

\date{\today}

\abstract{
A discrepancy in the determination of the proton's charge radius, $r_p$, between muonic hydrogen spectroscopy versus classic atomic spectroscopy and electron scattering data has become known as the proton radius puzzle.
Extractions of $r_p$ from electron scattering data require determination of the slope of the proton's charge form factor, \GEp{}, in the limit of $Q^2\rightarrow0$ through fitting and extrapolation.
Some works have presented the $Z$-transformation fitting technique as the best choice for this type of extraction due to the true functional form of \GEp{} being mathematically guaranteed to exist within the parameter-space of the fit function.
In this work, we test this claim by examining the mathematical bias and variances introduced by this technique as compared to the more traditional $Q^2$ fits using statistically sampled \GEp{} parameterizations with known input radii.
Our tests conclude that the quality of the $Z$-transformation technique depends on the range of data used.
In the case of new experiments, the fit function and technique should be selected in advance by generating realistic pseudodata and assessing the power of different techniques.
}

\maketitle

\section{Motivation}

The \textit{Proton Radius Puzzle} and the recent increase in available elastic electron-proton scattering data has invigorated discussions into the best methodology for analyzing elastic scattering data to extract the proton electric form factor, \GEp{}~\cite{Lorce:2020onh}. 
The \textit{Proton Radius Puzzle} refers to a $\sim7\sigma$ disagreement in the proton rms charge radius, $\sqrt{\left<r_p^2\right>}$ (written as $r_p$ henceforth), when measured with a new muonic hydrogen spectroscopy technique versus conventional measurement techniques (atomic hydrogen spectroscopy and $e-p$ scattering).

Of particular note is the discussion surrounding the technique for extracting $r_p$ from $e-p$ scattering data, as it often requires one to make model-dependent analysis choices that affect the final result~\cite{Barcus:2019skg}.
In the case where a theoretical model, such as in Ref.~\cite{Alarcon:2018zbz}, is not used, the most impactful choice is that of the \GEp{} fit function and any constraints on that function.

As has been shown~\cite{Eides:2000xc,Miller:2018ybm} that the consistent definition of the proton's charge radius for all types of radius measurements is
\begin{equation}
    r_p \equiv \sqrt{-6\left.\frac{\partial\GEp\ofQ}{\partial Q^2}\right\rvert_{Q^2=0}}
    \label{eq:rp}
\end{equation} and thus simply determined by the slope of $\GEp$ at a $Q^2=0$.
For electron scattering, which cannot be measured to $Q^2=0$, the implication of this is that \GEp{} must be fit with some functional form and then extrapolated to $Q^2=0$ to determine the slope as it approaches the limit.
This has proven to be a challenge, as the true functional form of \GEp{} is not known.
As such, the form chosen must be sufficiently flexible so as to contain the true value of $r_p$ within its parameter-space as well as robust enough to not diverge rapidly outside of the region of fitted data, particularly as $Q^2\rightarrow0$.

One such method that has seen increased use is the so-called $Z$-transformation~\cite{Hill:2010yb}.
In this scheme, the negative four-momentum transfer, $Q^2$, range is conformally mapped to a unit circle.
This mapping is done with the formalism:
\begin{equation}
    Z\left(t,\tcut,t_0\right) = \frac{\sqrt{\tcut-t}-\sqrt{\tcut-t_0}}{\sqrt{\tcut-t}+\sqrt{\tcut-t_0}}
    \label{eq:ztransform}
\end{equation}
where $t=q^2=-Q^2$, \tcut{} is the highest mapped $Q^2$ value, and $t_0$ is a free parameter.

The use of the $Z$-transformation is well-motivated.
This method is the standard tool for meson transition form factor studies~\cite{Hill:2010yb, Boyd:1995sq, Arnesen:2005ez, Becher:2005bg, Hill:2006ub, Hill:2006bq, Bourrely:2008za}.
By applying this mapping with a cut below the two-pion production threshold, the form factor is restricted to the region of analyticity (i.e. it can be represented by a convergent power series).
This implies that with a sufficient number of parameters the uncertainty from truncating a power series fit can be minimized, while still arriving at a function with sufficient predictive power.
In all, it is an enticing choice that has been used in a large number of analyses~\cite{Paz:2011qr,Epstein:2014zua,Lee:2015jqa,Ye:2017gyb,Sick:2018fzn}.

When fitting $Z$-transformed data, Eq.~\ref{eq:rp} must be adjusted to accommodate this coordinate transform:
\begin{equation}
    r_p \equiv \sqrt{-6 \left.\frac{\partial \GEp\ofZ}{\partial Z}\cdot\frac{\partial Z\left(-Q^2,\tcut,t_0\right)}{\partial Q^2}\right\rvert_{Q^2=0}}
    \label{eq:rp_Z}
\end{equation}

However, this method is not without it's drawbacks~\cite{Bernauer:2016ziz}.
Namely, by nature of design, while the true form factor exists within this parameter-space, so do many incorrect form factor parameterizations that can create local $\chi^2$ minima in the fitting routine.
One posited solution to this is the use of physically motivated constraints on the power series coefficients (Ref.~\cite{Hill:2010yb} suggests using $\left|p_i\right|\leq10$ as the coefficients for a $i$-th order power series).
The use of these constraints, while conservative, introduces a model-dependence to the fit.
Model-dependence, in and of itself, is not problematic; constraints placed on typical form factor polynomial fit (or any other commonly used functions) are model assumptions on the shape and behavior of the form factor.
This is only to say that one must take care and be aware of the ramifications of model inputs.

Studies similar to this one have been performed previously, as in Refs.~\cite{Kraus:2014qua,Yan:2018bez}, which we have used as a framework for our study.
Particularly, we aim to address claims made in Ref.~\cite{Paz:2020prs} that reanalyzes the PRad~\cite{Xiong:2019umf} proton electric form factor data using the $Z$-transformation technique.
A claim is made that the PRad proton radius uncertainties are underestimated due to the use of the ``Rational(1,1)'' fit function instead of the $Z$-transformation technique.
This claim, if justified, naturally leads to a conclusion that the results from the $Z$-transformation technique are a lower limit on precision and accuracy in the extraction of the proton radius when using realistic data.
This work aims to test that claim.

Here, we aim study the power of this technique to accurately measure the proton charge radius relative to that of other techniques.
We acknowledge that the Ref.~\cite{Hill:2010yb} posit that trial fits of model data are problematic, as models have to make assumptions on the behavior of the form factor.
However, we argue that by studying the technique with several models the affects of these assumptions can be mitigated and a better understanding of the technique can be realized.

\section{Method}
\label{sec:method}

In this study we closely follow the methodology of Ref.~\cite{Kraus:2014qua}.
To test the robustness of the $Z$-transformation, we generate \GEp{} pseudodata from parameterizations with a known radius.
These pseudodata are then analyzed with several fit functions both in $Q^2$ and transformed to $Z$.
This technique allow us to assess the sensitivity of the $Z$ conformal mapping to the input and statistical fluctuations of data.
For this study, we use 6 different parameterizations of \GEp{} as generators:

\begin{itemize}
    \item Alarc\'{o}n, Higinbotham, Weiss, and Ye (AW) fit~\cite{Alarcon:2018zbz}, a parameterization in terms of radius-independent and -dependent parts ($r_p=0.844$fm)
    \item Arrington, Melnitchouk, and Tjon (AMT) fit~\cite{Arrington:2007ux}, a rational(3,5) parameterization ($r_p=0.878$fm)
    \item Arrington fit~\cite{Arrington:2003qk}, an inverse-polynomial fit ($r_p=0.829$fm)
    \item Bernauer fit~\cite{Bernauer:2010,A1:2010nsl,A1:2013fsc}, a 10th-order polynomial fit ($r_p=0.887$fm)
    \item Standard dipole fit~\cite{Hand:1963zz} ($r_p=0.811$fm)
    \item Kelly fit~\cite{Kelly:2004hm}, a rational(1,3) fit ($r_p=0.863$fm)
\end{itemize}

Each of these parameterizations are used to generate equally spaced (in $Q^2$) values of \GEp{} and apply an uncertainty of $0.2\%$ to each point, corresponding to a $0.4\%$ uncertainty on the measured cross section.
These pseudodata begin at $\Qmin=0.004$ GeV$^2$ and are spaced every $0.001$ GeV$^2$ up to a variable cutoff \Qmax{}.
\Qmax{} is varied by adding a single data point at a time up to a total of 500 points (that is, $0.005$ to $0.504$ GeV$^2$).
At each \Qmax{} value, 5000 data sets are generated for each parameterization where each data point is smeared by a Gaussian distribution with a standard deviation equal to the $0.2\%$ uncertainty on the point.

A keen eye may note that some of these parameterizations incorporate corrections for two-photon exchange (TPE), while others do not.
We make no attempt to change this as, particularly at high $\epsilon$, the corrections for TPE are quite small often at the sub-percent level~\cite{Arrington:2012dq}.
While TPE has been suggested as a contribution to the Proton Radius and Proton Form Factor Puzzles, there is no clear consensus as to whether it can account for these differences~\cite{Schmidt:2019vpr}.
While the inclusion or exclusion of TPE will change the underlying form of the data, our work here assumes no knowledge of the underlying form and is testing different fit functions ability to extrapolate the slope to $Q^2=0$ without this knowledge.
We do not need the fit function to explain any other physical quantities or to be a true representation of the underlying function of the form factor.

Each data set is fit with polynomials of increasing order, $n=1\ldots4$, as defined by
\begin{equation}
    f\ofQ = p_0 \times \left[ 1 + \sum\limits_{i=1}^{n}p_i\left(Q^2\right)^i \right]
    \label{eq:poly}
\end{equation}
where $p_0$ is a normalization term.
By examining Eqs.~\ref{eq:rp} and \ref{eq:poly}, the radius is then extracted as $r_p = \sqrt{-6 p_1}$.
We also explore applying bounds to the fitting parameters and the use of Rational(N,M)-type functions as used in the recent PRad analysis~\cite{Xiong:2019umf}.
For each of the 5000 fits done with each polynomial at each \Qmax{} point, the extracted radius, $r_p$, and $\chi^2$ are recorded.
The final radius from each polynomial at each \Qmax{} point is treated to be the radius extracted from the mean slope of all fits and the uncertainty of the slope is taken to be the standard deviation of the slopes fitted and then propagated to the mean extracted radius.
Additionally, to assess the power of the fitting techniques, at each \Qmax{} point we calculate the Mean Squared Error as 
\begin{equation}
    \text{MSE} = \text{bias}^2 + \sigma^2,
    \label{eq:MSE}
\end{equation}
where $\text{bias}$ is the difference between the mean extracted radius and the input radius, and $\sigma$ is the rms-spread of extracted radii.

Each of these data sets is then transformed from $Q^2$ to $Z$ using Eq.~\ref{eq:ztransform}.
For this work, we set $t_{\text{cut}}=4m_\pi^2$, where $m_\pi$ is the charged pion mass, to restrict data below the two-pion production threshold and $t_0=0$.
These values were chosen to match the values that were used in the proof-of-concept performed in Ref.~\cite{Hill:2010yb}.
The $Z$-transformed data is then fit with polynomials of increasing order, $n=1\ldots4$ (analogous to Eq.~\ref{eq:poly} with $Q^2$ replaced by $Z$).
This functional form, along with this choice of $t_0$, allows us to write Eq.~\ref{eq:rp_Z} as $r_p = \sqrt{-1.5 p_1 / t_{\text{cut}}}$.

These $Q^2$ and $Z$-transformed data are then refit with a polynomial but with added bounds to constrain the fits.
In each of these, prescribed bounds from literature are followed for the $Q^2$~\cite{Barcus:2019skg} and $Z$~\cite{Hill:2010yb} data.
As the $Z$-transformation dramatically changes the shape of the $G_E^p$ data so that bounds for data in $Q^2$ are not applicable to $Z$-transformed data and vice versa.
This means that the bounds applied to fits in $Q^2$ and fits in $Z$ are \textit{not} equivalent to each other.
However, given that the techniques allow for different formulation of bounds on the fit parameters, the authors assert that the most fair technique to compare (un)bounded fits in $Q^2$ to (un)bounded fits in $Z$.

When assessing the quality of these fits in this study, there are two qualities to keep in mind:
\begin{enumerate}
    \item How well does the fit reflect the data?
    \item How well does the fit extract the input radius?
\end{enumerate}
The first question is easily addressed with a simple `goodness of fit' test.
In this work, we report the mean and standard deviation of the reduced chi-squared, \redchi{}, calculated for each fit at each \Qmax{}.
A \redchi{} of approximately 1 indicates that the fit represents the data well, whereas $\redchi\gg1$ indicates underfitting and $\redchi\ll1$ indicates overfitting.
An important note is that while a satisfactory \redchi{} value is critical for extracting the proton radius, as the data must be properly fit, it does not indicate that the fit will accurately extrapolate to guarantee a correct radius extraction~\cite{andrae2010dos}.
This statement is even more pertinent as additional terms are added to the fit.
Additional terms bring in additional moments that can cause the fit function to vary wildly and unpredictably when extended beyond the range of the measured (or simulated in this case) data.

The second question is addressed by studying the bias and the variance of the extracted radii.
The bias is defined as the difference of the mean radius extraction at each \Qmax{} and the input model radius.
The variance is the standard deviation of the mean extracted radius.
In this context, both of these quantities are best assessed when compared to the magnitude of the proton radius puzzle; an unreasonably large bias could yield a result that suggests the incorrect solution to the puzzle, whereas an unreasonably large variance would be unable to discern between the purported solutions.
To aid in this assessment, lines representing the magnitude of the proton radius puzzle are added to the bias-variance plots in this work.
As these two values can often be inversely correlated when optimizing a fit, we choose to define the error using the Mean Squared Error as in Eq.~\ref{eq:MSE}.
A smaller MSE indicates an overall improvement in the radius extraction.
It should be noted that just as having $\redchi\approx1$ does not indicate that the extracted radius will be correct, having a low MSE is also insufficient if the fit does not well represent the data.
The two properties must be taken together to show that the fit can both reproduce the data and extract the correct radius.

All figures in this article show the fitting techniques applied to the AMT parameterization as an illustrative set. See the Supplemental Information for this work for the full set of figures for all parameterizations as well as textual descriptions of all figures.

\section{Results with unbounded polynomials}

The first test we perform on using this methodology is with unbounded polynomials.
The authors note that while Ref.~\cite{Hill:2010yb} proposes that bounds of $\left|p_i\right|\leq10$ be used, these bounds, while conservative, are based on a vector-dominance model.
In order to initially assess the robustness in the least model-dependent way, we opt to start by omitting any bounds.

\begin{figure}
    \centering
    \includegraphics[width=\linewidth]{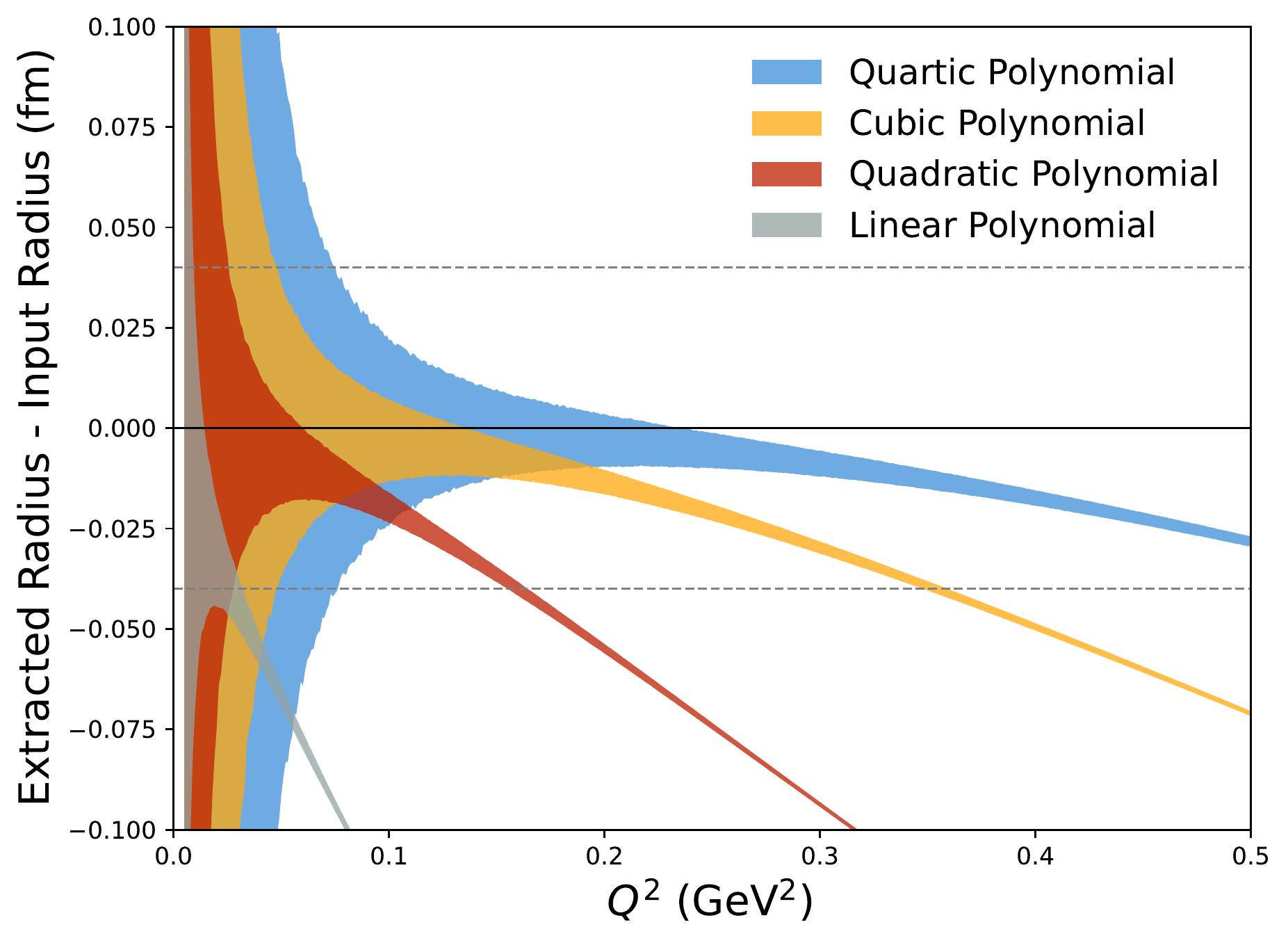}
    \includegraphics[width=\linewidth]{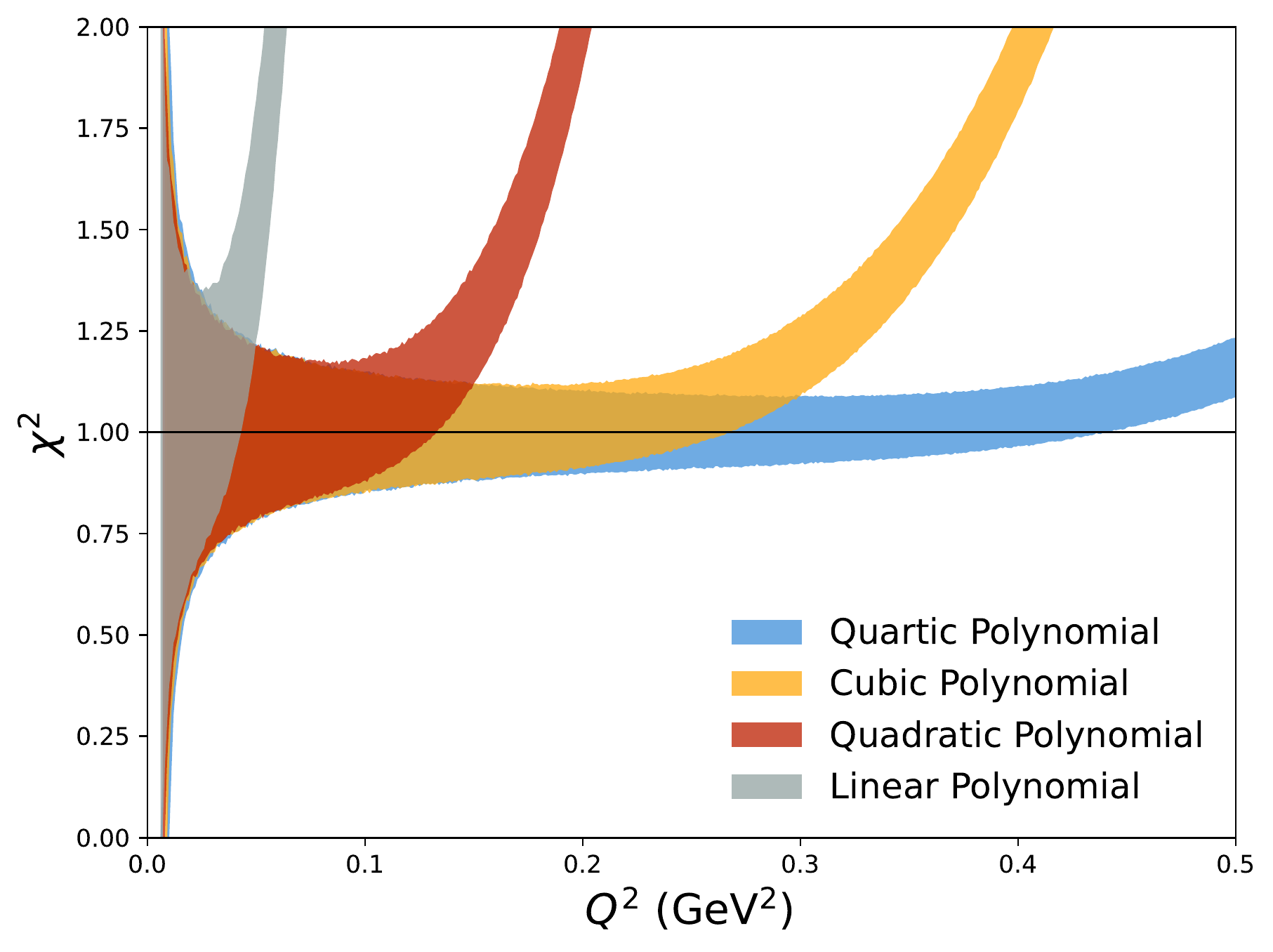}
    \includegraphics[width=\linewidth]{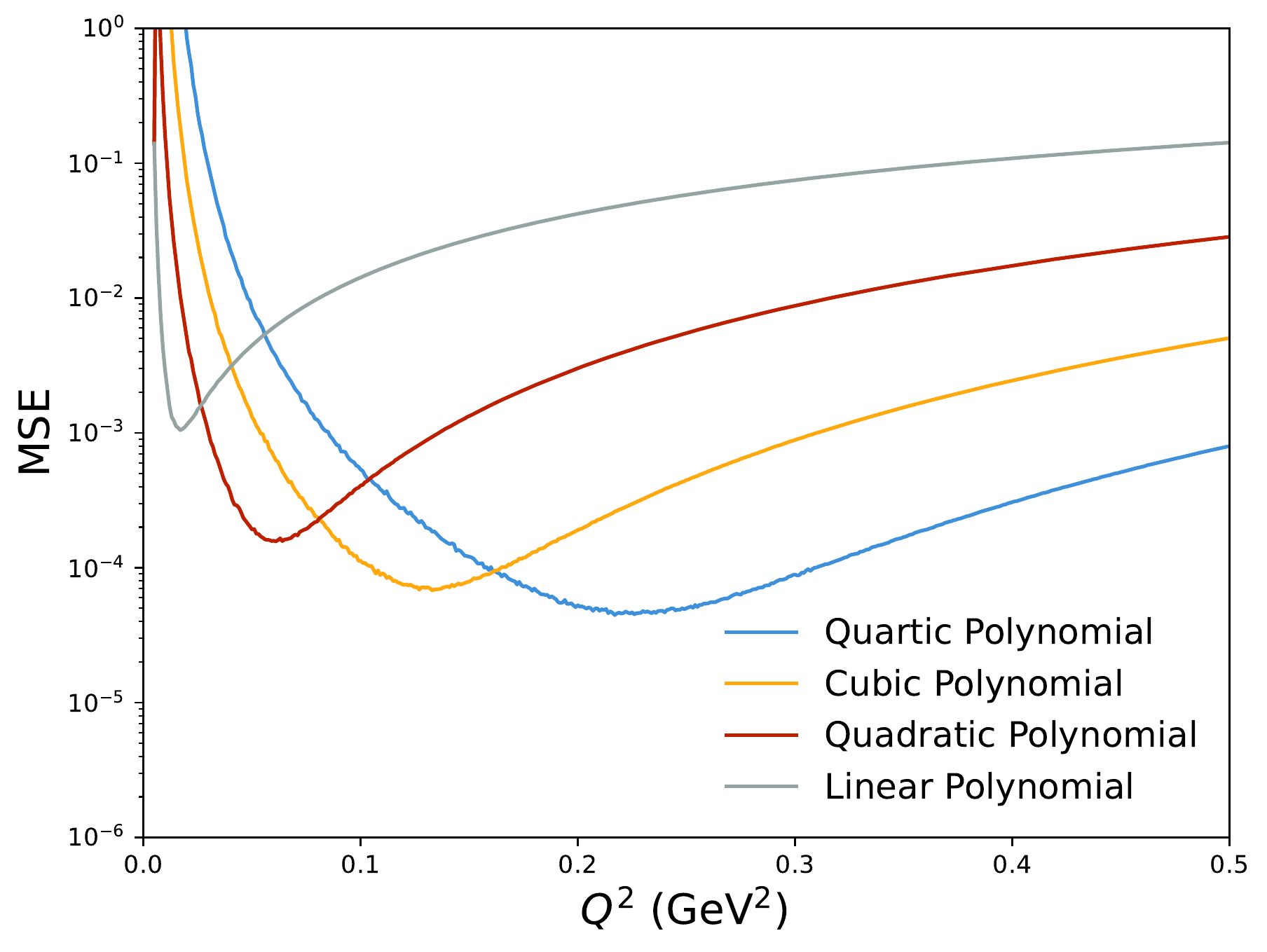}
    \caption{[Top] Results from unbounded polynomial fits in $Q^2$ for the AMT parameterization of \GEp{}. The vertical axis is the bias, the difference between the extracted radius and the input radius. Deviations from $0$ represent an ``incorrect'' extraction. The spread at each \Qmax{} is the rms spread of all 5000 test fits. A larger spread implies higher uncertainty. The grey dashed lines represent the magnitude of the \textit{Proton Radius Puzzle} in either direction from 0. [Middle] $\chi^2$ results from unbounded polynomial fits in $Q^2$ for the AMT parameterization of \GEp{}. [Bottom] The Mean Squared Error from unbounded polynomial fits in $Q^2$ for the AMT parameterization of \GEp{}.}
    \label{fig:Q2_unbound}
\end{figure}

\begin{figure}
    \centering
    \includegraphics[width=\linewidth]{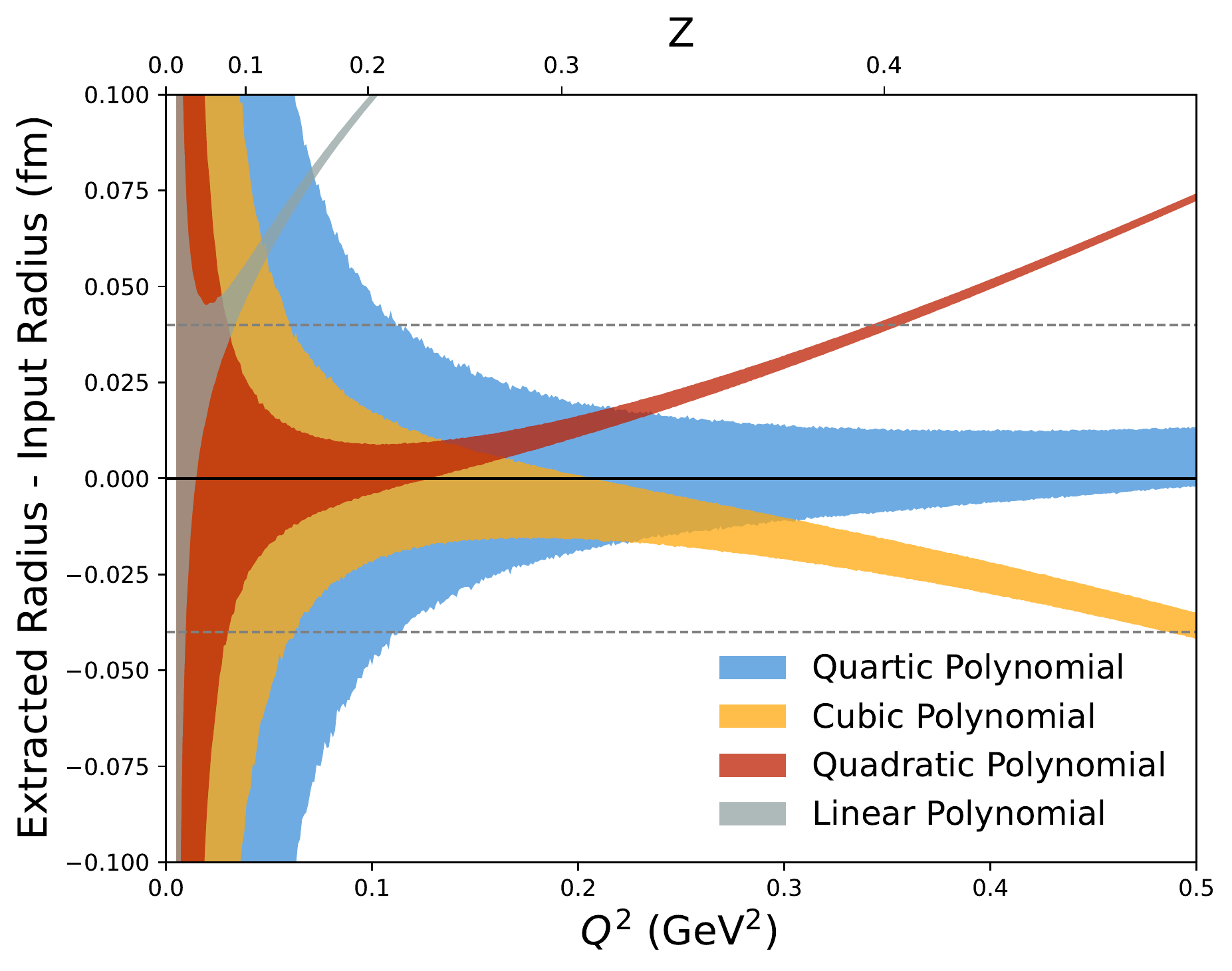}
    \includegraphics[width=\linewidth]{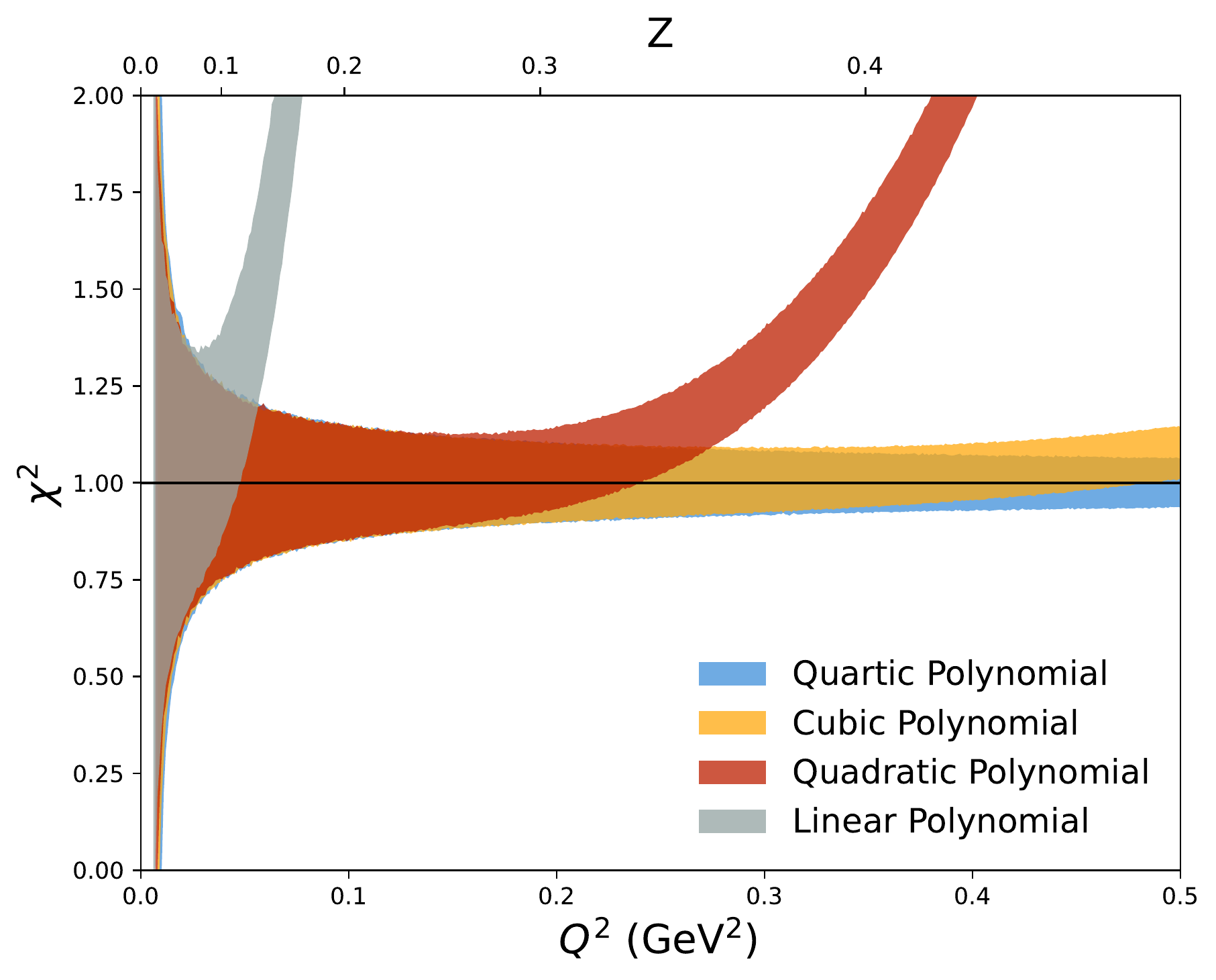}
    \includegraphics[width=\linewidth]{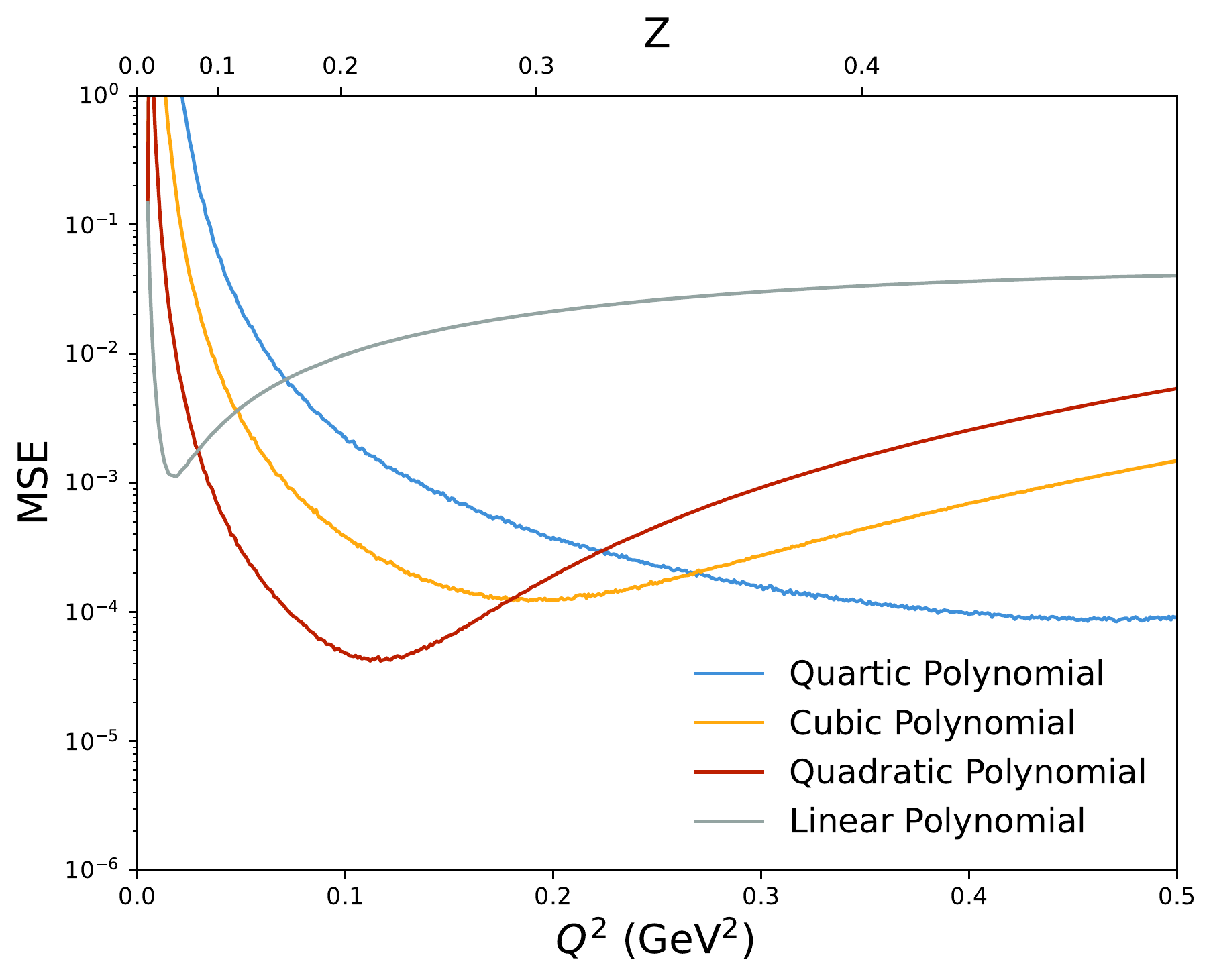}
    \caption{[Top] Results from unbounded polynomial fits in $Z$ for the AMT parameterization of \GEp{}. See Fig.~\ref{fig:Q2_unbound} for information on interpretation. [Middle] $\chi^2$ results from unbounded polynomial fits in $Z$ for the AMT parameterization of \GEp{}. [Bottom] The Mean Squared Error from unbounded polynomial fits in $Z$ for the AMT parameterization of \GEp{}.}
    \label{fig:Z_unbound}
\end{figure}

Comparing the outputs of the unbounded polynomial fits in Figs.~\ref{fig:Q2_unbound} and \ref{fig:Z_unbound}, we can see that there is a trade-off between choosing to fit in $Q^2$ or $Z$.
The unbounded $Q^2$ fit has a larger mean bias, that decreases with increasing polynomial order.
However, this fit has very little variance in the fit value once a moderate $Q^2$ cutoff is reached.
The unbounded $Z$ fit, on the other hand, has a much smaller mean bias for most models checked (though for some models it diverges rapidly once past a moderate $Q^2$ cutoff).
The trade-off is that the variance in results from the $Z$ fit is quite large.

Given these results, it is ill-advised to use either of these methods blindly.
There may exist some data set where one of these is the optimal choice; however that data set does not exist within this exercise.
When comparing the fits in $Q^2$ to $Z$ the choice is, in essence, between a highly repeatable incorrect answer or a potentially correct answer that is not repeatable.

\section{Results from bounded polynomial fits}

The next test involves placing bounds on the fit parameters.
As discussed in the previous section, Ref.~\cite{Hill:2010yb} proposes that, for the fit of $Z$-transformed data, bounds of $\left|p_i\right|\leq10$ be used.
The bounds are determined by beginning with a model-independent dispersion relation analysis, but then applies a model-dependent vector-dominance ansatz.
The calculations could prescribe a more rigid set of bounds than this, however Ref.~\cite{Hill:2010yb} decided that it was most appropriate to suggest a conservative implementation to minimize the model dependence.

For the $Q^2$ fit we follow the prescription of Ref.~\cite{Barcus:2019skg} and require that the fit parameters alternate signs.
This bounded regression forces the result to approximate a monotonic function (in this case, non-increasing).
This technique reduces the risk that the fit parameters will have dramatic fluctuations when extrapolated beyond the fit region.

\begin{figure}
    \centering
    \includegraphics[width=\linewidth]{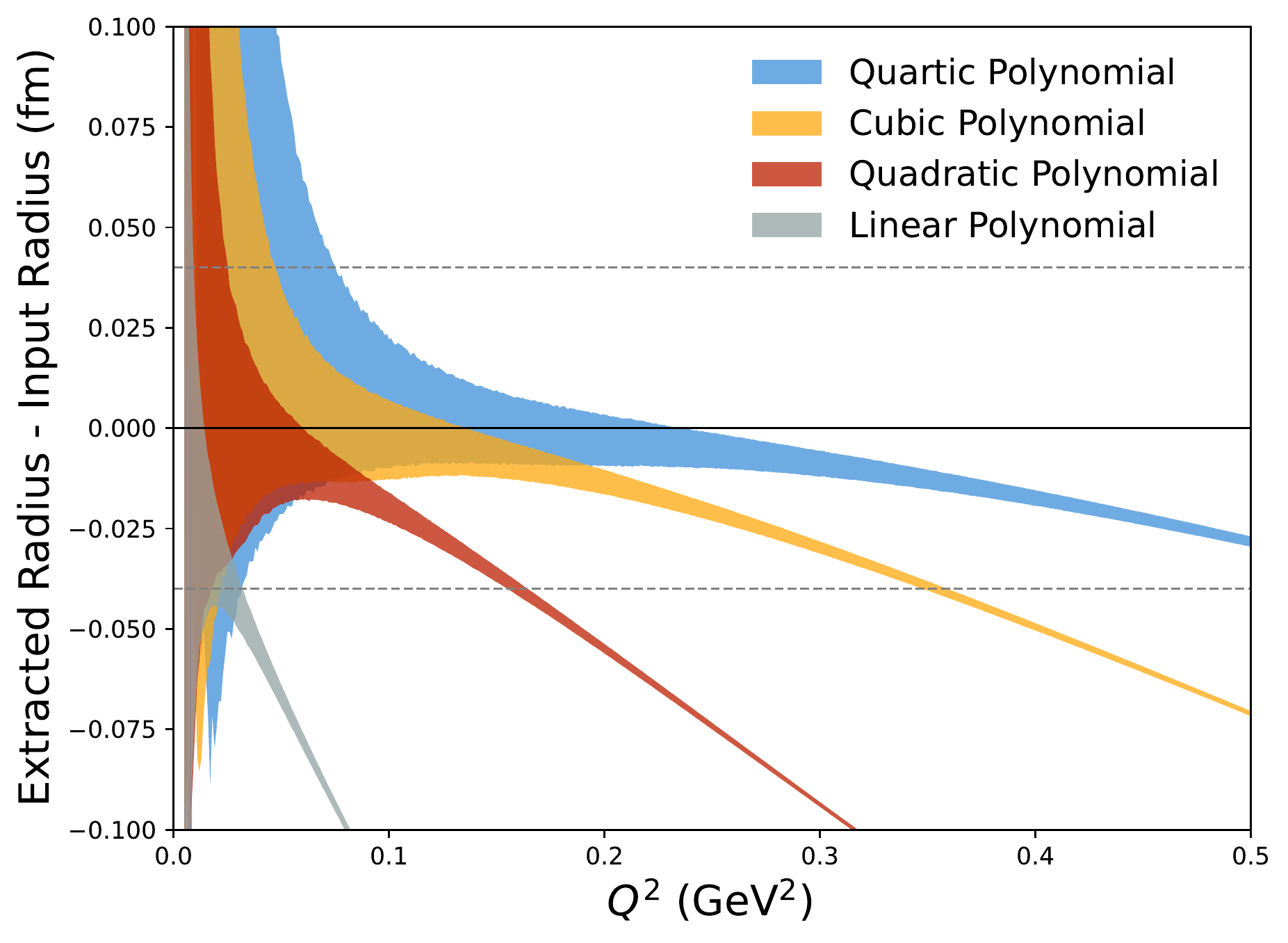}
    \includegraphics[width=\linewidth]{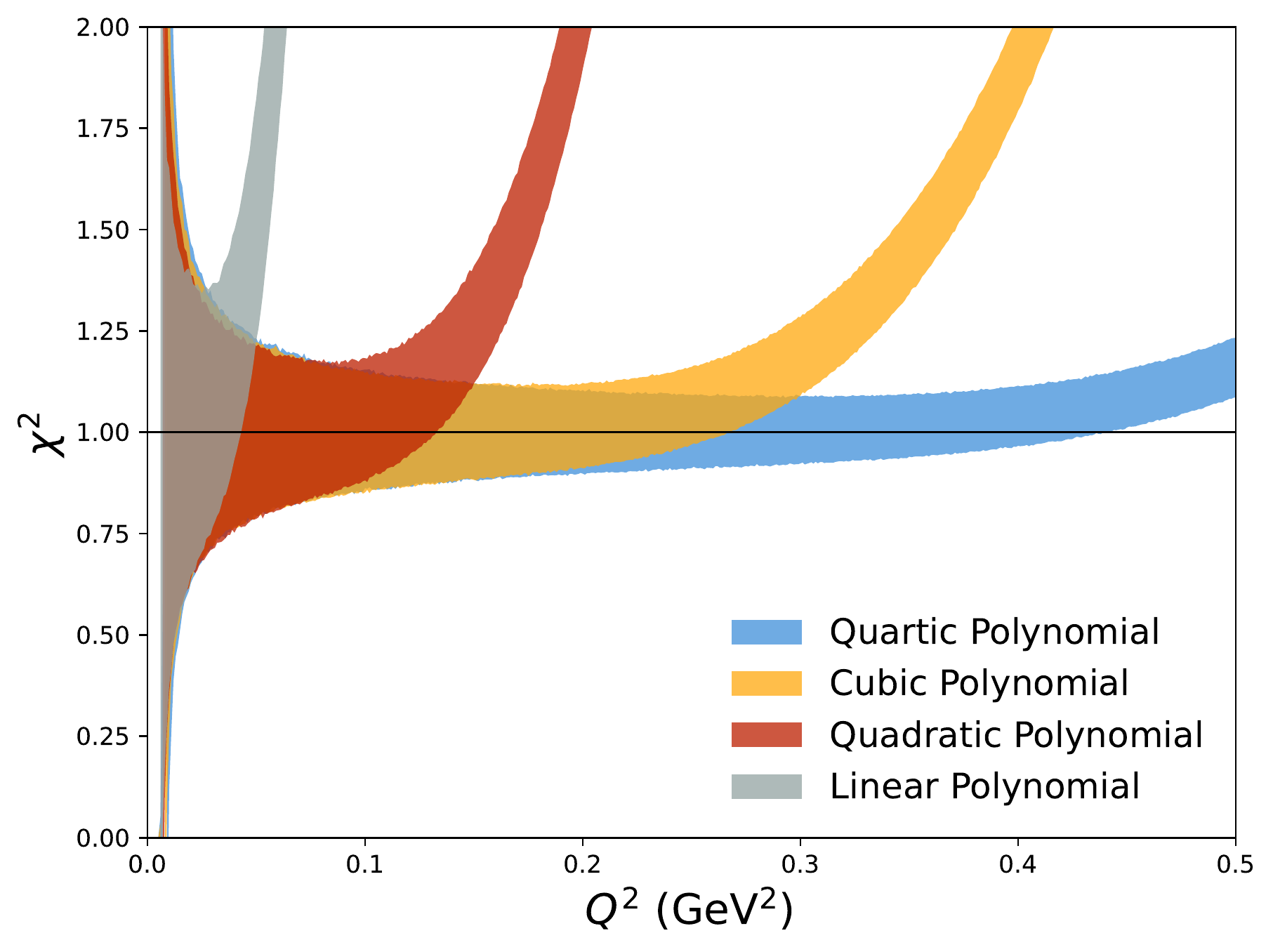}
    \includegraphics[width=\linewidth]{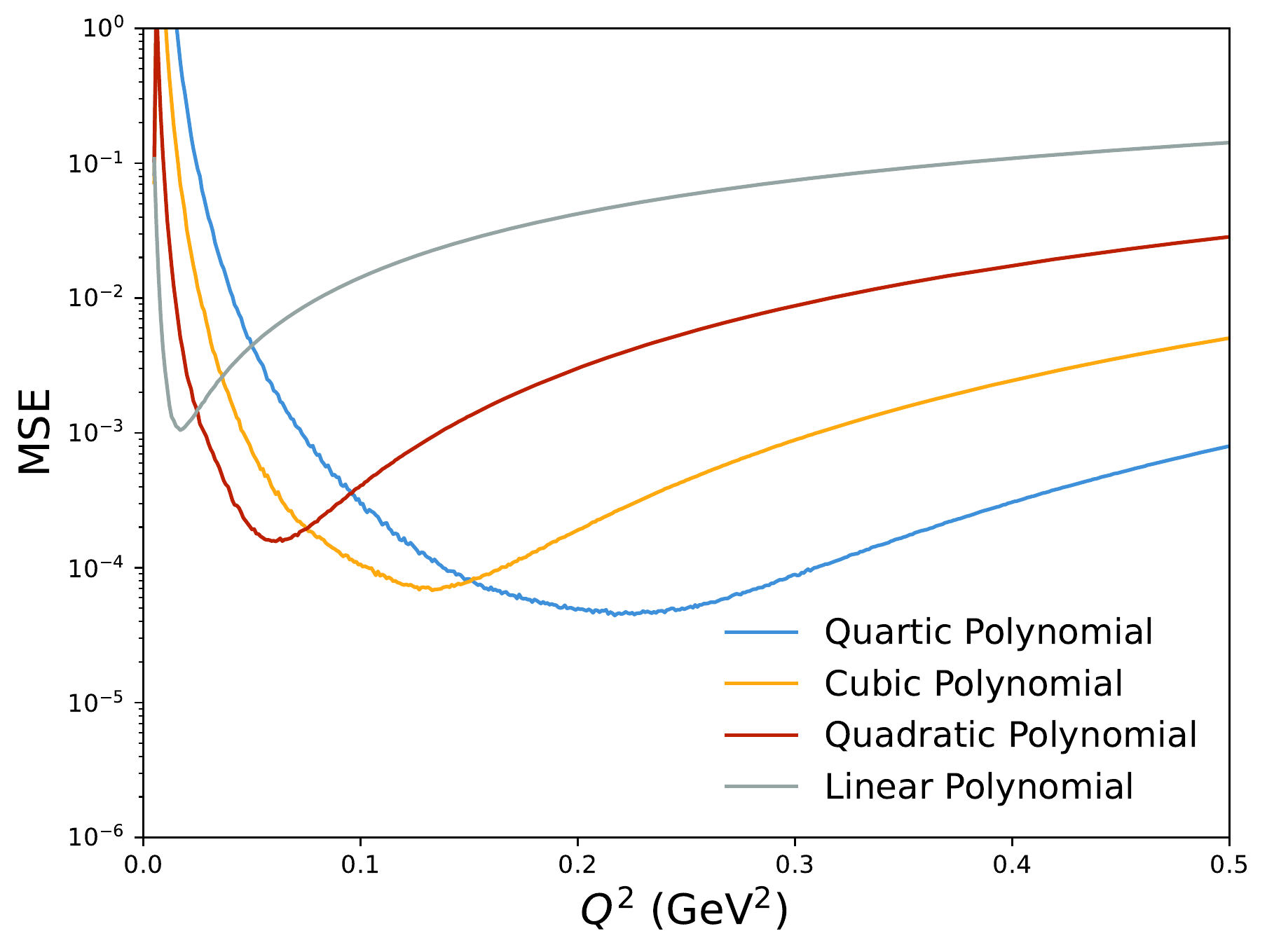}
    \caption{[Top] Results from bounded polynomial fits in $Q^2$ for the AMT parameterization of \GEp{}. See Fig.~\ref{fig:Q2_unbound} for information on interpretation. [Middle] $\chi^2$ results from bounded polynomial fits in $Q^2$ for the AMT parameterization of \GEp{}. [Bottom] The Mean Squared Error from bounded polynomial fits in $Q^2$ for the AMT parameterization of \GEp{}.}
    \label{fig:Q2_bound}
\end{figure}

\begin{figure}
    \centering
    \includegraphics[width=\linewidth]{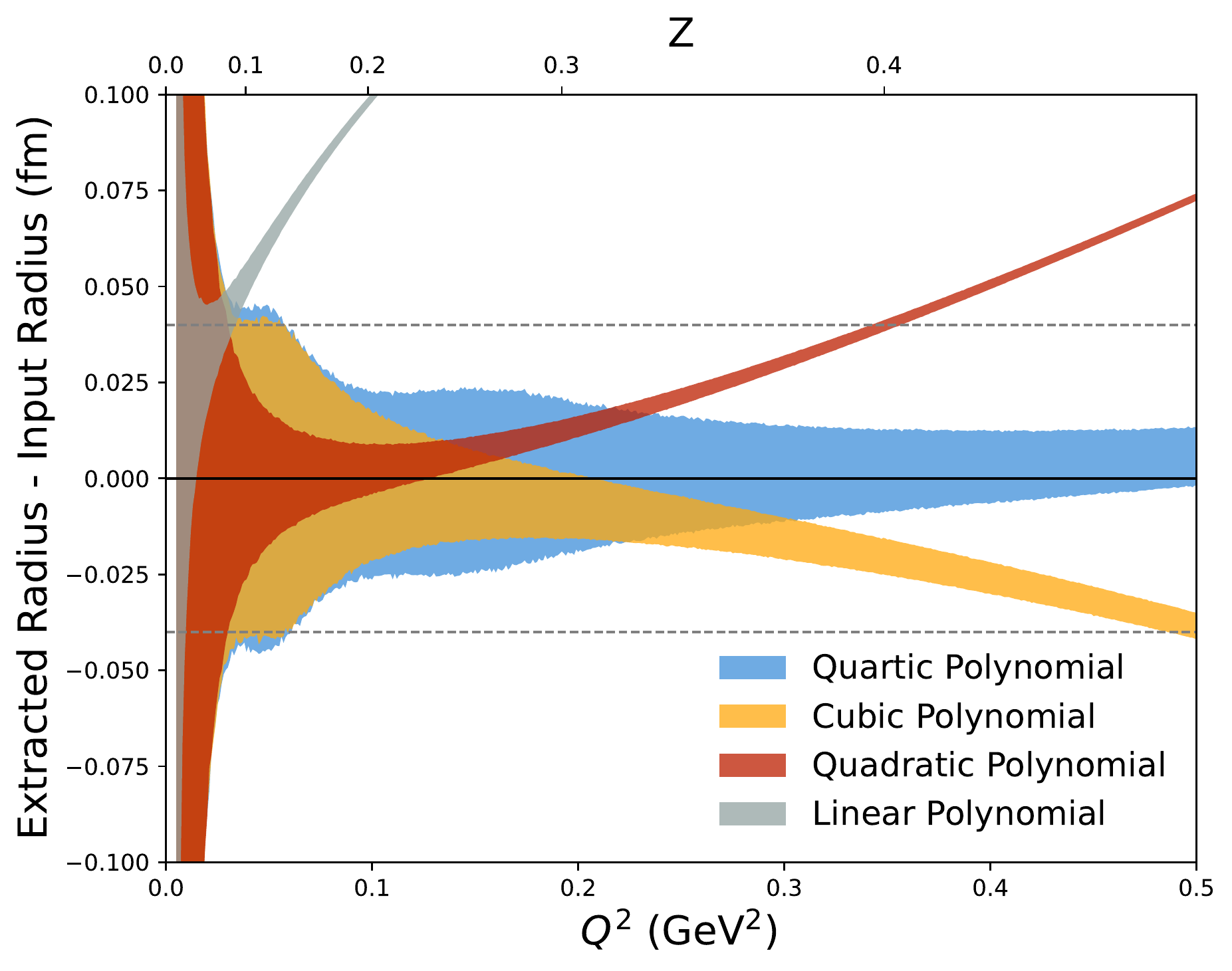}
    \includegraphics[width=\linewidth]{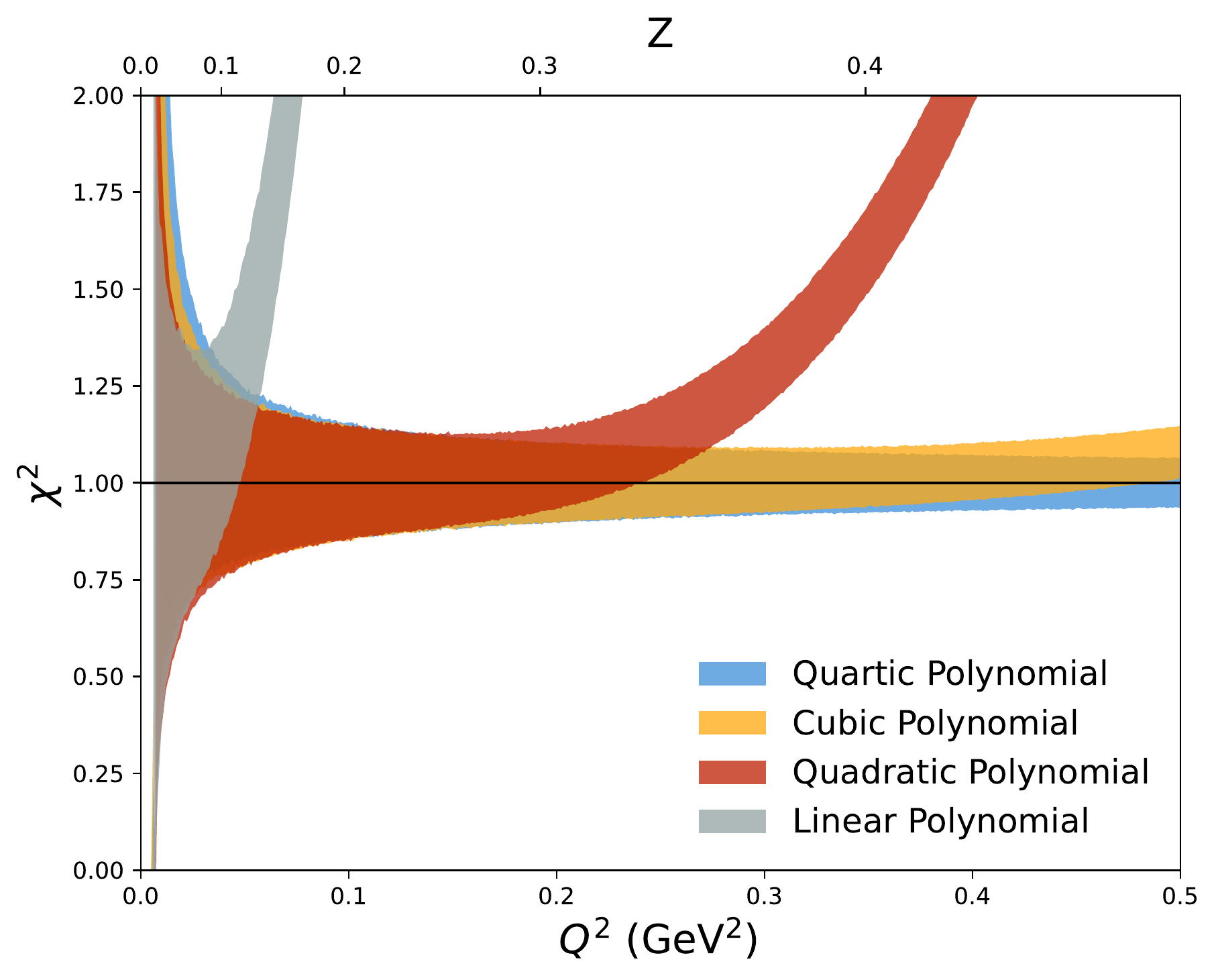}
    \includegraphics[width=\linewidth]{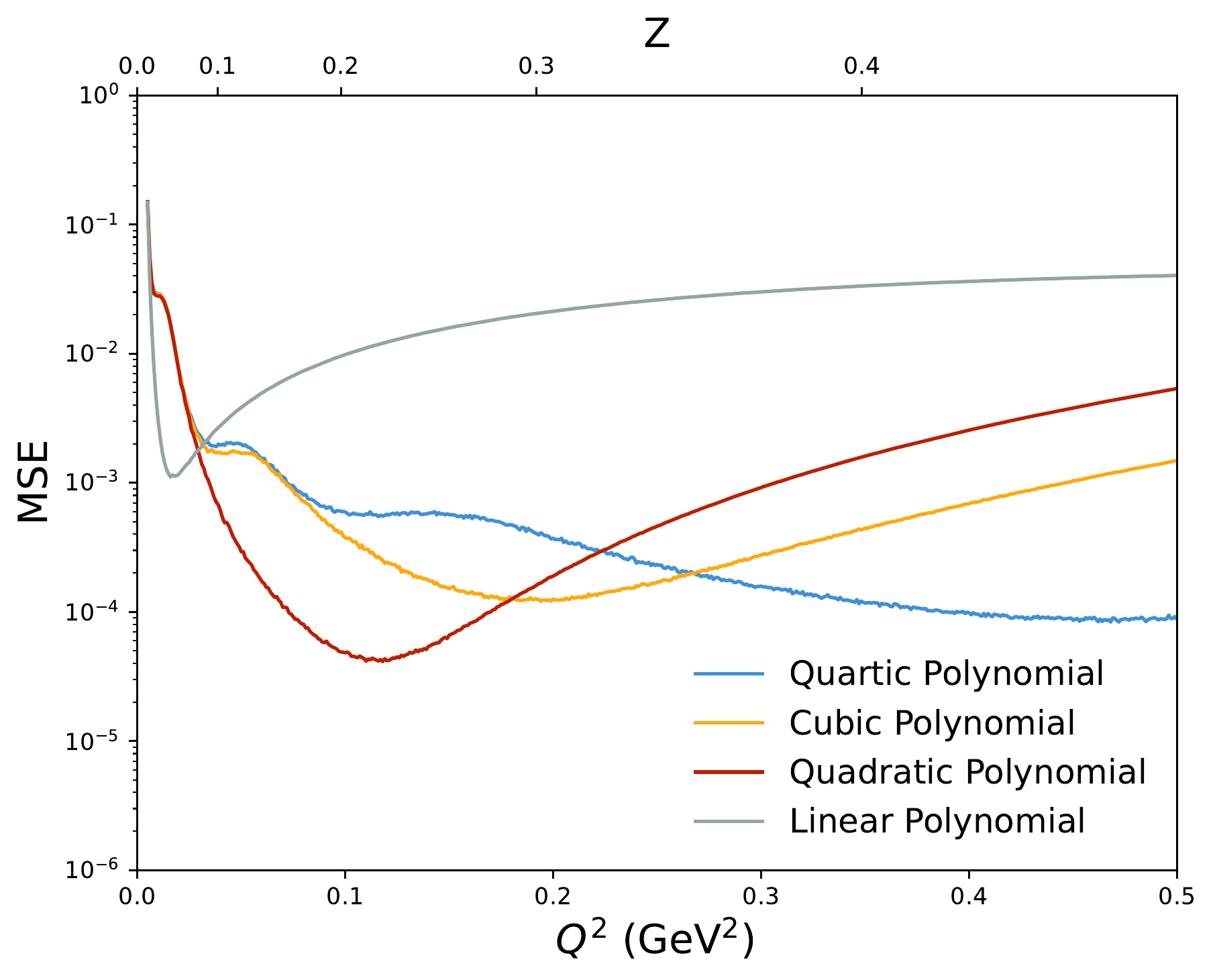}
    \caption{[Top] Results from bounded polynomial fits in $Z$ for the AMT parameterization of \GEp{}. See Fig.~\ref{fig:Q2_unbound} for information on interpretation. [Middle] $\chi^2$ results from bounded polynomial fits in $Z$ for the AMT parameterization of \GEp{}. [Bottom] The Mean Squared Error from bounded polynomial fits in $Z$ for the AMT parameterization of \GEp{}.}
    \label{fig:Z_bound}
\end{figure}

The results of these fits, seen in Figs.~\ref{fig:Q2_bound} and \ref{fig:Z_bound}, fall largely in line with the results of the unbounded polynomial fits.
The main effect is a reduction of the variance, most notably seen below $\Qmax{}\approx0.1$ GeV$^2$.
In this region, the variance is still too large to effectively resolve the proton radius puzzle.
At \Qmax{} above this, the difference is negligible.
The fit in $Q^2$ still has a very large bias and the fit in $Z$ with the vector-dominance bounds has very large variance.

\section{Results from fitting with a Rational(N,M) function}

The final functional form used is a Rational(N,M) function.
A Rational(N,M) function is defined as the form
\begin{equation}
    f\ofQ=p_0\frac{1+\sum\limits_{i=1}^Nn_iQ^2}{1+\sum\limits_{j=1}^Mm_jQ^2}
\end{equation}
with $p_0$ as a normalization factor.
As an example, the Rational(1,1) function then parameterizes the data as
\begin{equation}
    f\ofQ=p_0\GEp\ofQ=p_0\frac{1+n_1Q^2}{1+m_1Q^2}
\end{equation}
where the extracted radius is $\sqrt{-6\left(n_1-m_1\right)}$.
This functional form was used for the extraction of the proton radius from the PRad experiment~\cite{Xiong:2019umf}.
The decision was made by studying various fitters on generated pseudodata (much like this study) and selecting the most robust fitter for the $Q^2$ range of the measurement~\cite{Yan:2018bez}.

\begin{figure}[ht]
    \centering
    \includegraphics[width=\linewidth]{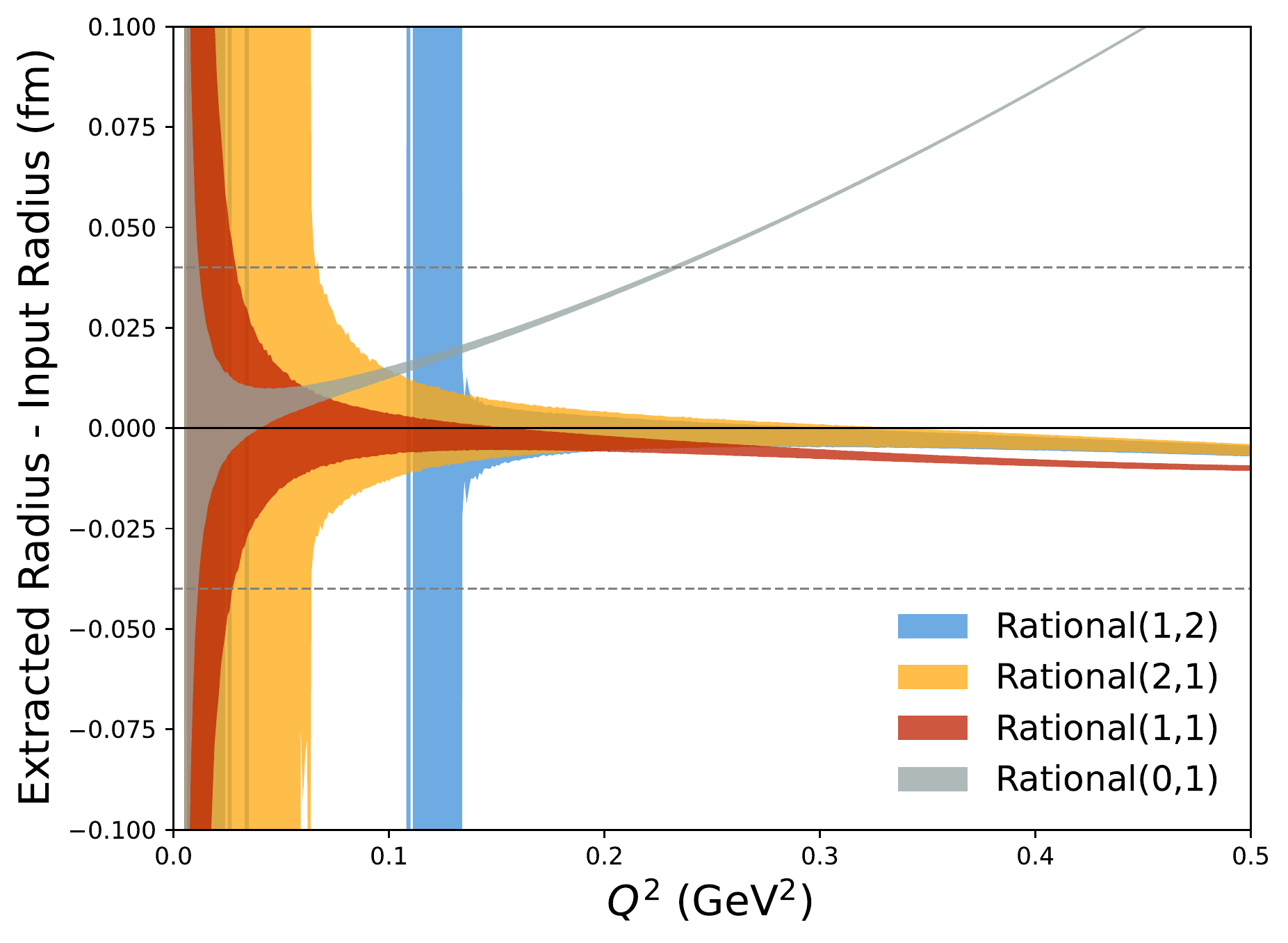}
    \includegraphics[width=\linewidth]{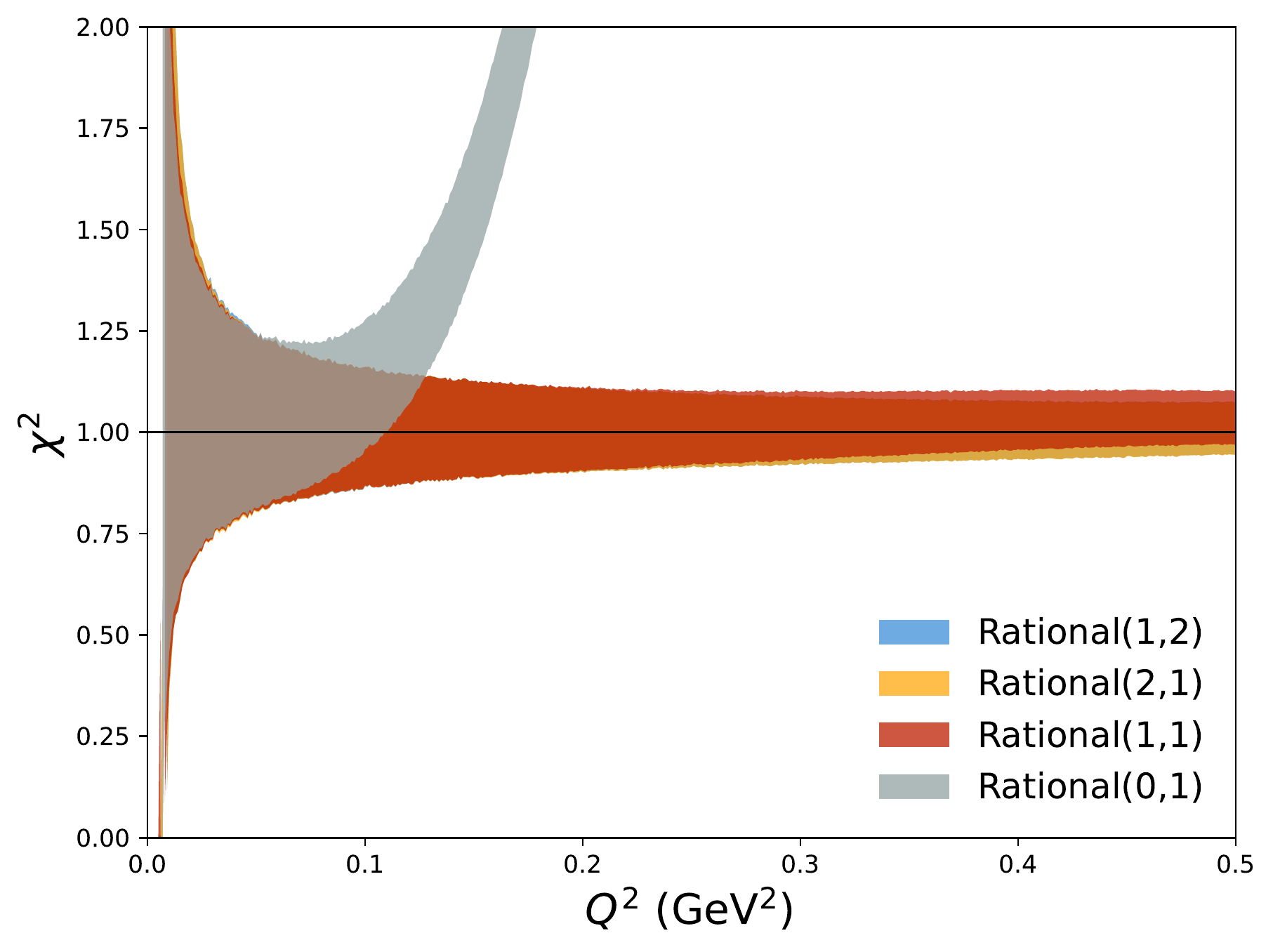}
    \includegraphics[width=\linewidth]{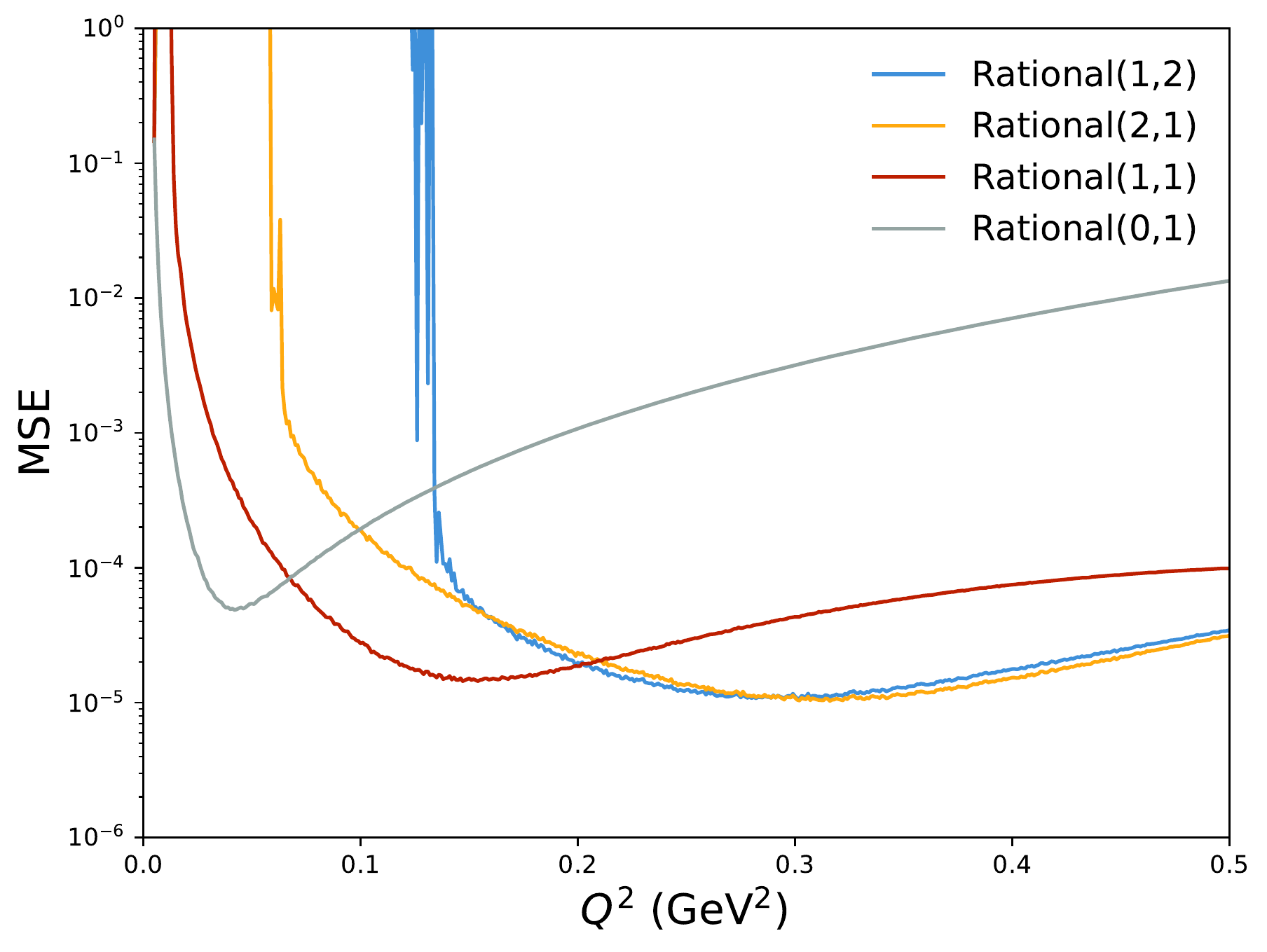}
    \caption{[Top] Results from Rational(N,M) fits in $Q^2$ for the AMT parameterization of \GEp{}. See Fig.~\ref{fig:Q2_unbound} for information on interpretation. [Middle] $\chi^2$ results from Rational(N,M) fits in $Q^2$ for the AMT parameterization of \GEp{}. [Bottom] The Mean Squared Error from Rational(N,M) fits in $Q^2$ for the AMT parameterization of \GEp{}.}
    \label{fig:RationalNM}
\end{figure}

In this study, we vary N and M within the range of $0-2$ while restricting them to be within $\pm1$ of each other.
The results of these fits are seen in Fig.~\ref{fig:RationalNM}.
An important note here is that the Rational(1,0) function is simply a linear polynomial which yields identical behavior to the linear fit in Fig.~\ref{fig:Q2_unbound}.
Increasing the orders N and M simultaneously increases the \Qmax{} at which the bias is small compared to the proton radius puzzle, the minimum \Qmax{} at which the variance is small compared to the proton radius puzzle, and the time taken for the fits to converge.
The reason for the increase in the time to convergence is that Rational(N,M) functions cover a very large parameter-space, such that increasing the order introduces additional local minima that slow down the fit.
It is noted that it was attempted to include the Rational(2,2) function here, but convergence of the illustrative fits took a prohibitively lengthy amount of time.

For the fits in $Q^2$, the functional form of a Rational(N,M) function (provided both N and M at least 1) is by far the most precise and has the least bias.
Unsurprisingly, given the study done for the PRad experiment, the Rational(1,1) function has minimal bias around $\Qmax{}\approx0.07\text{GeV}^2$ for nearly all models (though perfect agreement with the PRad study should not be expected as this analysis uses a higher \Qmin{}, different point density, and different point-to-point uncertainties).

This functional form, when applied to $Q^2$ data, shows by far the most robust fitting of the parameterizations considered.
However, as with all of the fitting forms studied, it will not always be the best choice in all scenarios.

\section{Conclusion}

From these illustrative fits, there are a number of conclusions that can be drawn:

\begin{itemize}
    \item When comparing fits in $Q^2$ and $Z$, both unbound and bound, no technique is clearly superior for extracting the proton radius. The technique that yields lower MSE depends not only on the order of polynomial and the \Qmax{} range, but also on the \GEp{} parameterization fit.
    \item In contrast, the Rational(N,M) fitting procedure, when N and M are both at least 1 and a minimum \Qmax{} is reached, consistently has an MSE value of nearly an order of magnitude lower than the other fits, with the exception of when fitting the Bernauer parameterization. It should also be noted that the Rational(N,M) fits behave very poorly for low \Qmax{} values, even giving unphysical results for higher (N,M) orders. A caveat to the smaller MSE values is that it is partially driven by the substantially smaller variance of the fits as there appear to be fewer local minima in the parameter-space.
    \item Applying bounds to the fitting procedure decreases the variance of the extracted radius. This is expected behavior, as bounds limit the phase space of possible solutions. However, bounds appear to do very little to improve the extracted bias. While physically motivated bounds (e.g. the slope must be negative to ensure a positive charge radius) are a useful tool, additional bounds may induce a false belief in the accuracy of the extraction.
    \item As the true functional form of \GEp{} is not known, no single fitting function will be ideal for every possible data set. As such, it is imperative that any analysis that seeks to extract the proton charge radius does their due diligence to determine the best fitting function for their data. Ref.~\cite{Yan:2018bez} details a procedure for an analysis of the most robust fitter for a given data range. Highlighting the care that must be taken, this reference also found that Rational(N,M) functions were inadequate for particularly low $Q^2$ data.
    \item The true form factor is guaranteed to lie within the phase space of a polynomial fit in $Z$-transformed data. This technique aims not only to accurately extract the proton radius, but to gain insight into the true functional form of the proton form factor. This additional insight comes with a trade-off of increased uncertainties on the extracted proton radius. As thoroughly described in Ref.~\cite{explain_predict}, modeling to explain data and modeling to predict data are two different goals that are not always achieved with identical techniques. The $Z$-transformation technique aims to achieve a deeper explanation of the data than the other techniques investigated, which in turn can negatively impact its predictive power.
    \item As discussed in Sec.~\ref{sec:method}, it must be stressed that $\chi^2$ or MSE alone cannot be used to assess the extracted radius from a fit. The $\chi^2$ of a fit is \textit{only} related to the ability of the fit to reproduce the fitted data. The MSE of a fit is \textit{only} related to the extracted and input radius, with no regard to data points that were fit. Both values must be taken together to assess the strength of a particular fit.
\end{itemize}

In our work, there was no single data set where applying a $Z$ conformal mapping to the $Q^2$ range improved the extraction of the proton charge radius.
This result should not imply that this is not a useful technique.
Rather, the $Z$ transformation should be considered a useful tool in the physicists toolbox after they have assessed the correct tool for the job such as Ref.~\cite{Yan:2018bez} finds that a 2\textsuperscript{nd} order polynomial fit in $Z$ is a robust fit for the PRad data points.

\section*{Acknowledgments}
This material is based upon work supported by the U.S. Department of Energy, Office of Science, Office of Nuclear Physics under contracts DE-AC05-06OR23177 and DE-AC02-05CH11231. This material is based upon work supported in part by NSF award PHY-1812421.

\newpage

\clearpage
\backmatter

\bibliography{ZTransform.bib}

\end{document}


\title{Supplemental Information for:\\\textit{Direct Comparison of using a \texorpdfstring{$Z$}{Z}-Transformation instead of the traditional \texorpdfstring{$Q^2$}{Q2} for for Extraction of the Proton Radius from \texorpdfstring{$e-p$}{e-p} Scattering Data}}

\author*[1,2]{Tyler J. Hague}
\email{TJHague@gmail.com}

\author[3]{Douglas W. Higinbotham}
\email{doug@jlab.org}

\author[3,4]{Spencer Portuese}
\nomail

\affil[1]{%
 \orgname{North Carolina A\&T State University}, \orgaddress{\city{Greensboro}, \state{North Carolina} \postcode{27411}, \country{USA}}
}
\affil[2]{
 \orgname{Lawrence Berkeley National Laboratory}, \orgaddress{\city{Berkeley}, \state{California} \postcode{94720}, \country{USA}}
}%
\affil[3]{
 \orgname{Thomas Jefferson National Accelerator Facility}, \orgaddress{\city{Newport News}, \state{Virginia} \postcode{23606}, \country{USA}}
}%
\affil[4]{%
 \orgname{The Governor's School for Science and Technology}, \orgaddress{\city{Hampton}, \state{Virginia} \postcode{23666}, \country{USA}}
 }

\date{\today}

\maketitle

\section{Introduction}

In our paper \textit{Direct Comparison of using a $Z$-Transformation instead of the traditional \texorpdfstring{$Q^2$}{Q2} for for Extraction of the Proton Radius from \texorpdfstring{$e-p$}{e-p} Scattering Data}, we discuss the work we performed to compare the robustness of proton electric form factor, \GEp{}, fitting techniques for extracting the proton radius.
Included in the paper are several illustrative figures to support the discussion that are non-exhaustive of the work performed.
This supplemental materials is provided as a complete set of the figures produced from this work.

For these tests, six \GEp{} parameterizations were used:
\begin{itemize}
    \item Alarc\'{o}n, Higinbotham, Weiss, and Ye (AW) fit~\cite{Alarcon:2018zbz}, a parameterization in terms of radius-independent and -dependent parts ($r_p=0.844$fm)
    \item Arrington, Melnitchouk, and Tjon (AMT) fit~\cite{Arrington:2007ux}, a rational(3,5) parameterization ($r_p=0.878$fm)
    \item Arrington fit~\cite{Arrington:2003qk}, an inverse-polynomial fit ($r_p=0.829$fm)
    \item Bernauer fit~\cite{Bernauer:2010}, a 10th-order polynomial fit ($r_p=0.887$fm)
    \item Standard dipole fit ($r_p=0.811$fm)
    \item Kelly fit~\cite{Kelly:2004hm}, a rational(1,3) fit ($r_p=0.863$fm)
\end{itemize}
Each of these parameterizations are used to generate equally spaced (in $Q^2$) values of \GEp{} and apply an uncertainty of $0.2\%$ to each point, corresponding to a $0.4\%$ uncertainty on the measured cross section.
These psuedodata begin at $\Qmin=0.004$ GeV$^2$ and are spaced every $0.001$ GeV$^2$ up to a variable cutoff \Qmax{}.
\Qmax{} is varied by adding a single data point at a time up to a total of 500 points (that is, $0.005$ to $0.504$ GeV$^2$).
At each \Qmax{} value, 5000 data sets are generated for each parameterization where each data point is smeared by a normal distribution with a standard deviation equal to the $0.2\%$ uncertainty on the point.

These figures are organized first by the \GEp{} parameterization used, and then by the fitting technique used.
Each \GEp{} parameterization and fitting technique pair has a set of 3 figures:
\begin{itemize}
    \item Bias-Variance
    \begin{itemize}
        \item This plot shows the difference between the mean extracted radius and the input radius, as well as the standard deviation of the mean extracted radius for each \Qmax{}
    \end{itemize}
    \item $\chi^2$
    \begin{itemize}
        \item This plot shows the mean and standard deviation of the $\chi^2$ of the fits at each \Qmax{}
    \end{itemize}
    \item Mean Squared Error
    \begin{itemize}
        \item The Mean Squared Error, defined as
        \begin{equation}
            \text{MSE}=\text{bias}^2+\sigma^2,
            \label{eq:MSE}
        \end{equation}
        is used as a measure of the robustness of a fitting technique for the given data set
    \end{itemize}
\end{itemize}

\newpage
\clearpage

\section{Alarc\'{o}n, Higinbotham, Weiss, and Ye (AW)}

\subsection{Polynomial Fits in $Q^2$}

\begin{figure}[hb]
    \centering
    \includegraphics[height=0.2\textheight]{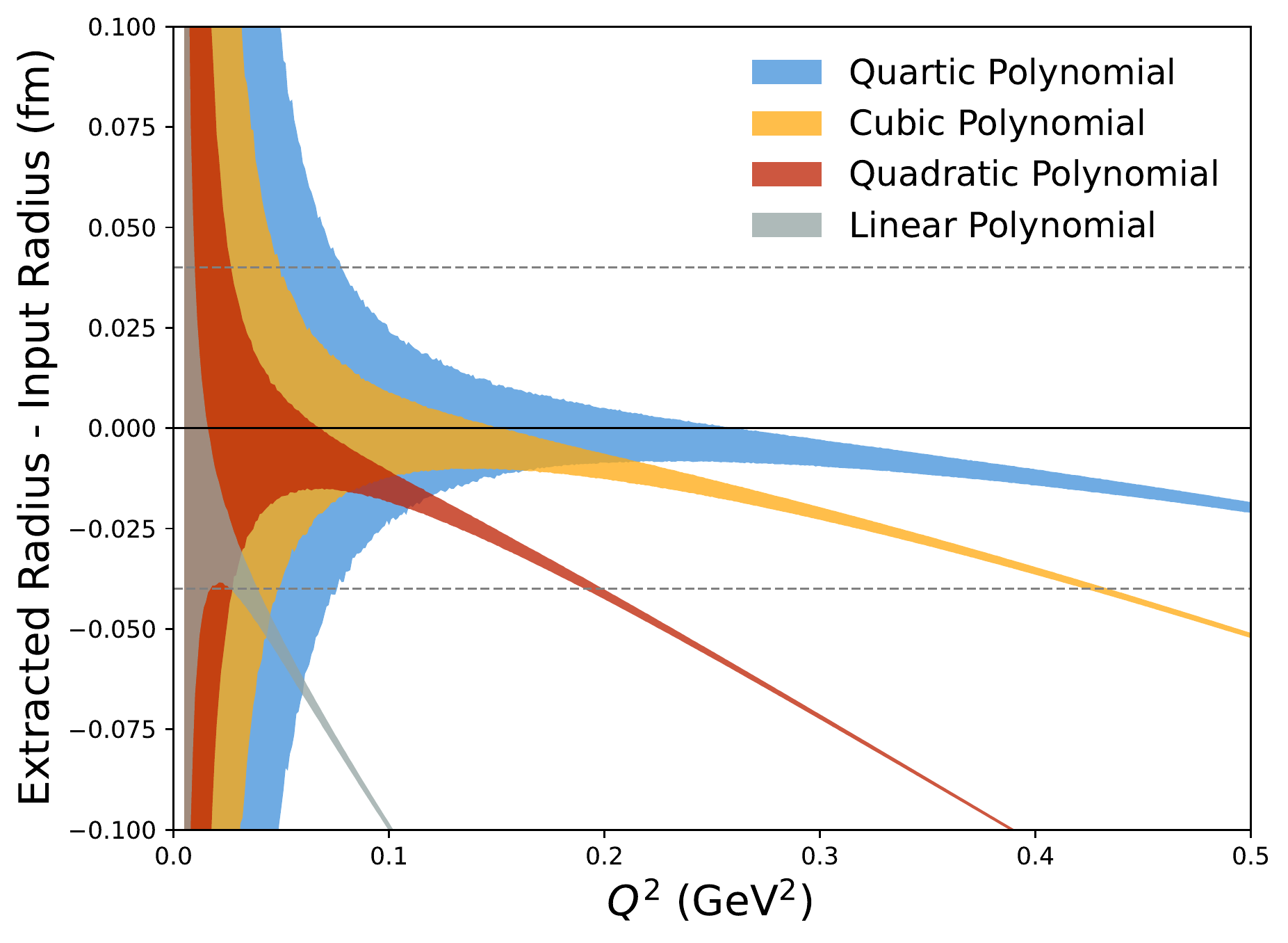}
    \caption{Bias-Variance plot for polynomial fits in $Q^2$ of the AW \GEp{} parameterization}
\end{figure}

\begin{figure}[hb]
    \centering
    \includegraphics[height=0.2\textheight]{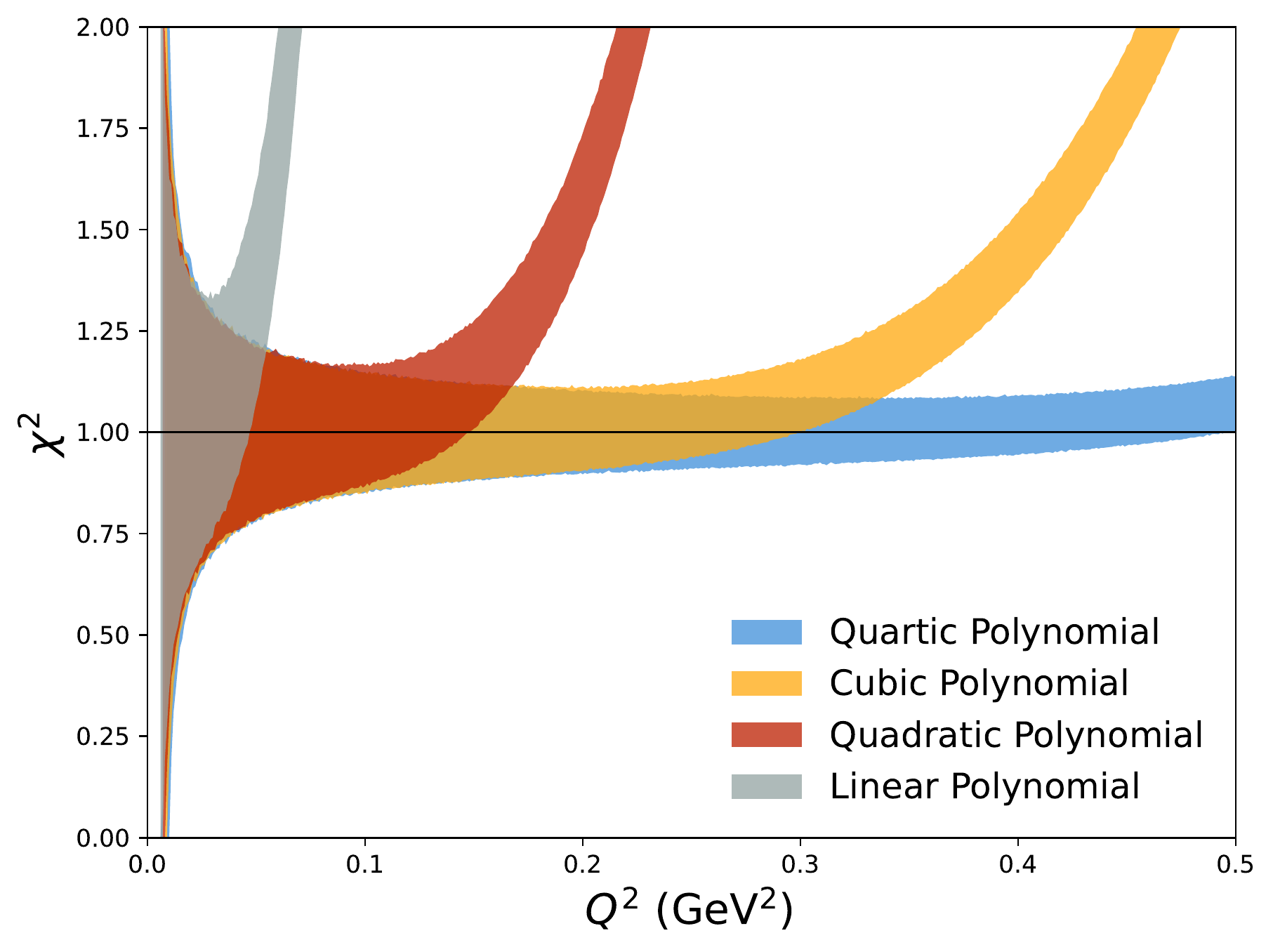}
    \caption{$\chi^2$ plot for polynomial fits in $Q^2$ of the AW \GEp{} parameterization}
\end{figure}

\begin{figure}[hb]
    \centering
    \includegraphics[height=0.2\textheight]{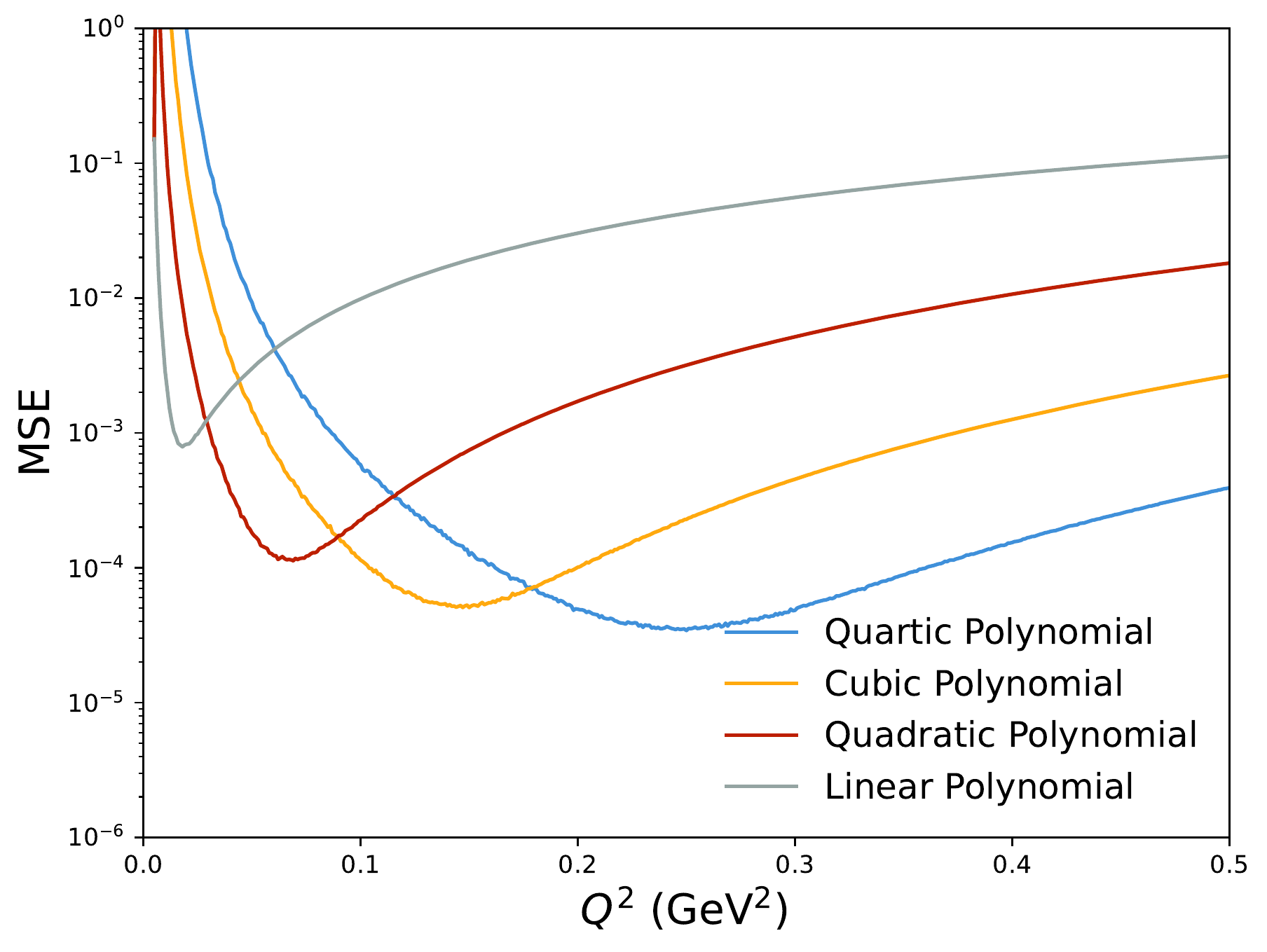}
    \caption{Mean Squared Error plot for polynomial fits in $Q^2$ of the AW \GEp{} parameterization}
\end{figure}

\newpage

\subsubsection{Figure Descriptions}

Three plots, each with “$Q^2$ (GeV$^2$)” from 0 to 0.5 on the x axis. Each plot has four sets of data shown, one for each power of polynomial used, labeled as “Linear Polynomial”, “Quadratic Polynomial”, “Cubic Polynomial”, and “Quartic Polynomial”.

The top plot has “Extracted Radius - Input Radius (fm)” from -0.1 to 0.1 on the y axis. Two horizontal lines drawn at $\pm$0.04 fm show the magnitude of the proton radius puzzle. Each data set is shown as a band. All bands have variance higher than the axis scale at small $Q^2$ and systematically return a small radius that decreases with increasing $Q^2$ range. Higher order polynomials decrease slower.

The middle plot has $\chi^2$ from 0 to 2 on the y axis. Each data set is shown as a band. All bands begin centered at 1 and then at some $Q^2$ have a rapid upturn. Higher order polynomials delay the upturn until higher $Q^2$. The deviation from $\chi^2\approx1$ occurs at significantly higher $Q^2$ than when the extracted radius deviates from the input radius.

The bottom plot has MSE, Mean Squared Error as described in Eq.~\ref{eq:MSE}, from $10^{-6}$ to $10^0$ with a logarithmic scale on the y axis. Each set of data is represented as a curve. Each curve rapidly drops to the minimum MSE for the fit and then begins to increase with $Q^2$ and start to level off about two orders of magnitude higher than the minima. Higher order polynomials have lower minima that occur at higher $Q^2$ and then increase slower. The lowest minima is about $4\cdot10^{-5}$.

\newpage

\subsection{Bounded Polynomial Fits in $Q^2$}

\begin{figure}[hb]
    \centering
    \includegraphics[height=0.2\textheight]{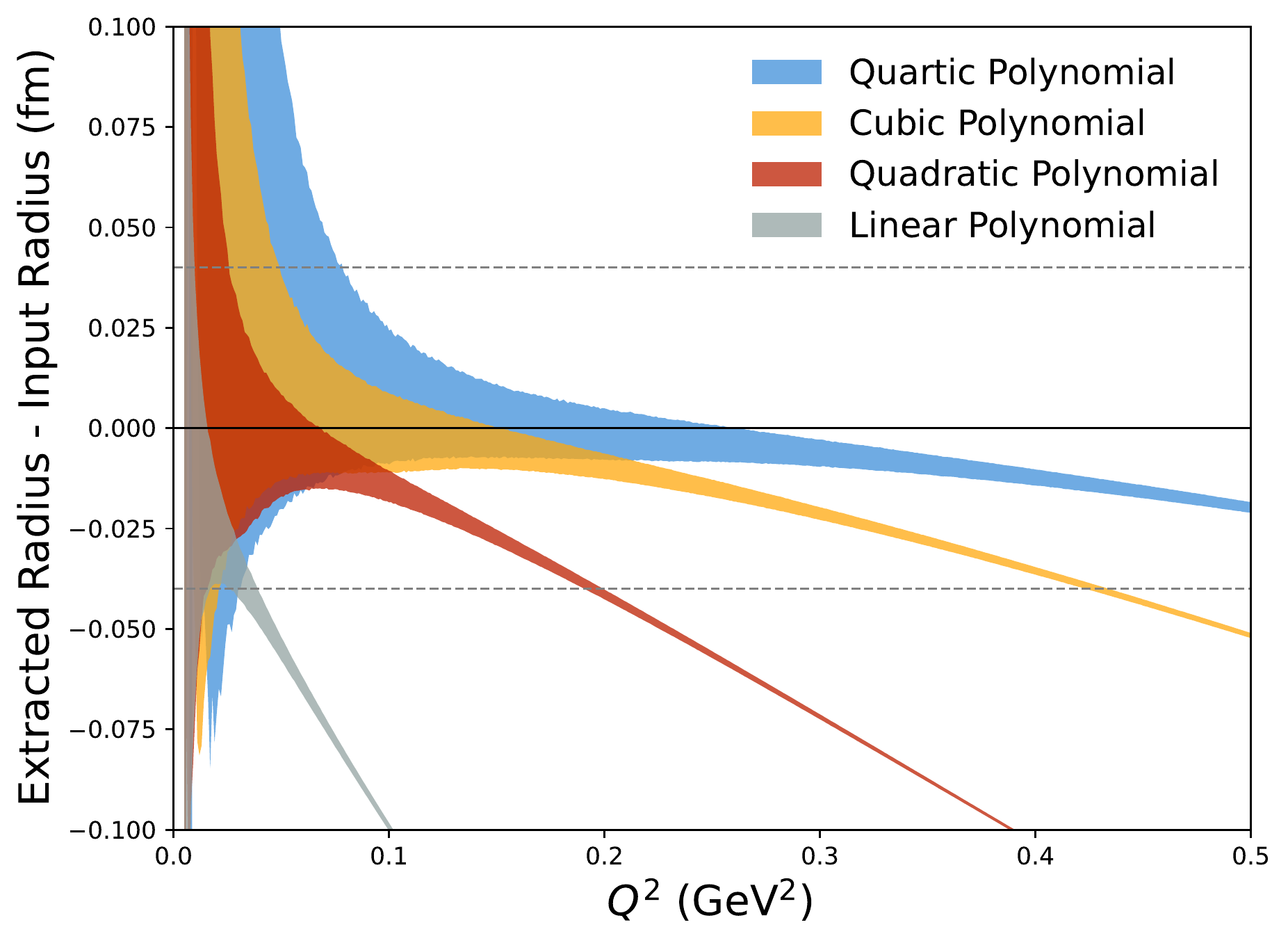}
    \caption{Bias-Variance plot for bounded polynomial fits in $Q^2$ of the AW \GEp{} parameterization}
\end{figure}

\begin{figure}[hb]
    \centering
    \includegraphics[height=0.2\textheight]{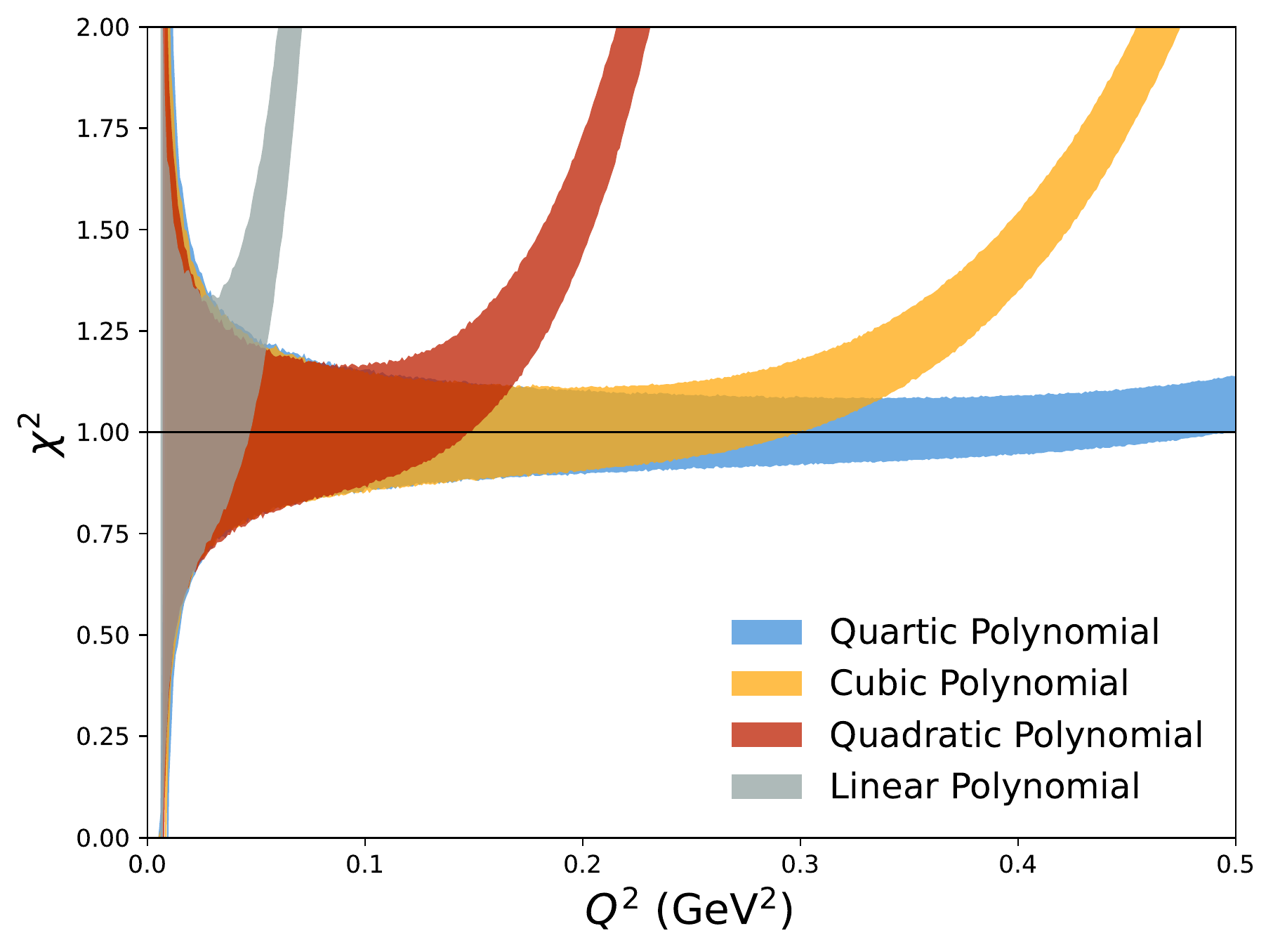}
    \caption{$\chi^2$ plot for bounded polynomial fits in $Q^2$ of the AW \GEp{} parameterization}
\end{figure}

\begin{figure}[hb]
    \centering
    \includegraphics[height=0.2\textheight]{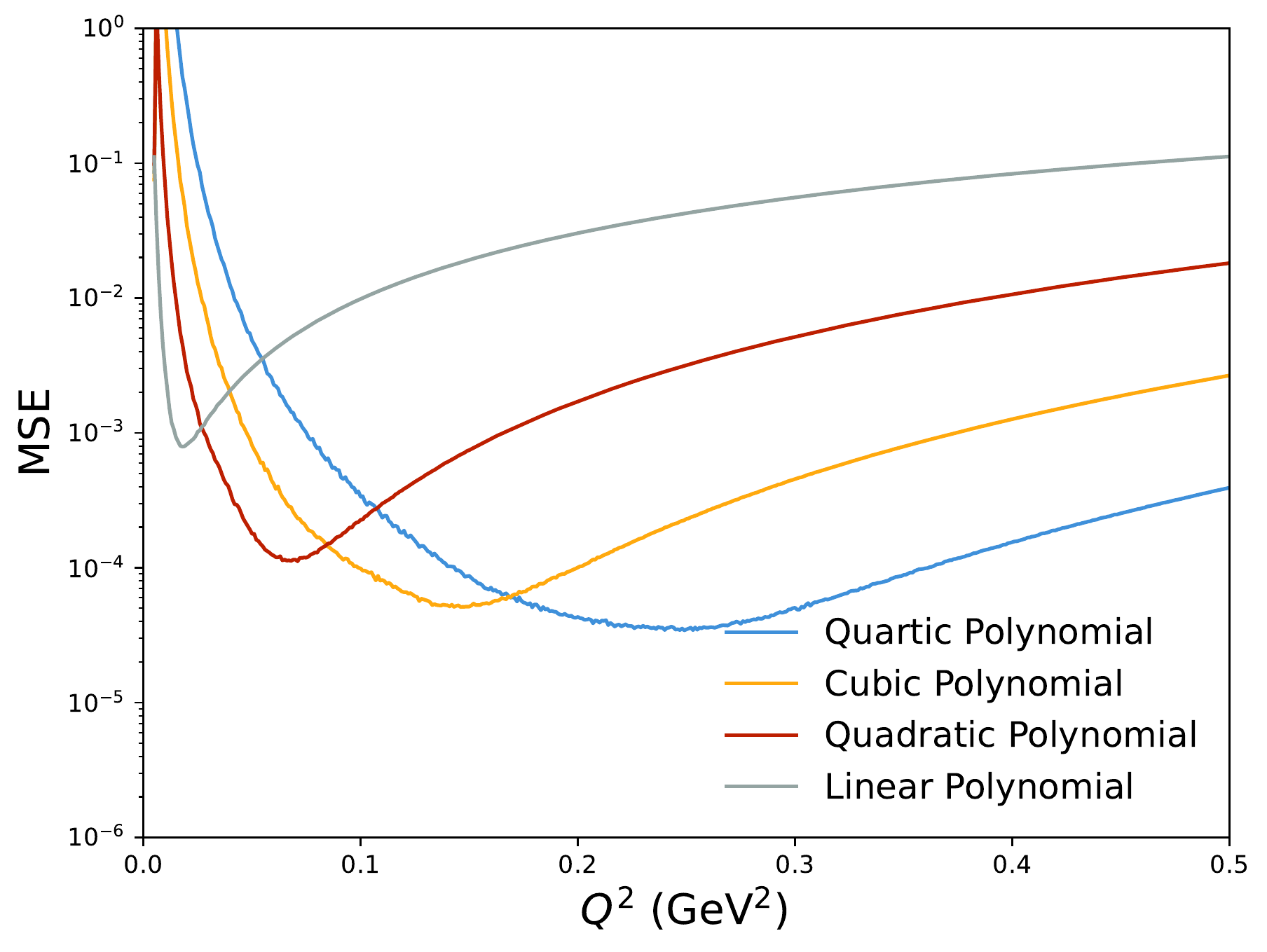}
    \caption{Mean Squared Error plot for bounded polynomial fits in $Q^2$ of the AW \GEp{} parameterization}
\end{figure}

\newpage

\subsubsection{Figure Descriptions}

Three plots, each with “$Q^2$ (GeV$^2$)” from 0 to 0.5 on the x axis. Each plot has four sets of data shown, one for each power of polynomial used, labeled as “Linear Polynomial”, “Quadratic Polynomial”, “Cubic Polynomial”, and “Quartic Polynomial”.

The top plot has “Extracted Radius - Input Radius (fm)” from -0.1 to 0.1 on the y axis. Two horizontal lines drawn at $\pm$0.04 fm show the magnitude of the proton radius puzzle. Each data set is shown as a band. The low $Q^2$ variance is smaller than the unbound fits, but is not overall qualitatively different.

The middle plot has $\chi^2$ from 0 to 2 on the y axis. Each data set is shown as a band. The data is not qualitatively different than the unbound fits.

The bottom plot has MSE, Mean Squared Error as described in Eq.~\ref{eq:MSE}, from $10^{-6}$ to $10^0$ with a logarithmic scale on the y axis. Each set of data is represented as a curve. The data is not qualitatively different than the unbound fits.

\newpage

\subsection{Polynomial Fits in $Z$}

\begin{figure}[hb]
    \centering
    \includegraphics[height=0.2\textheight]{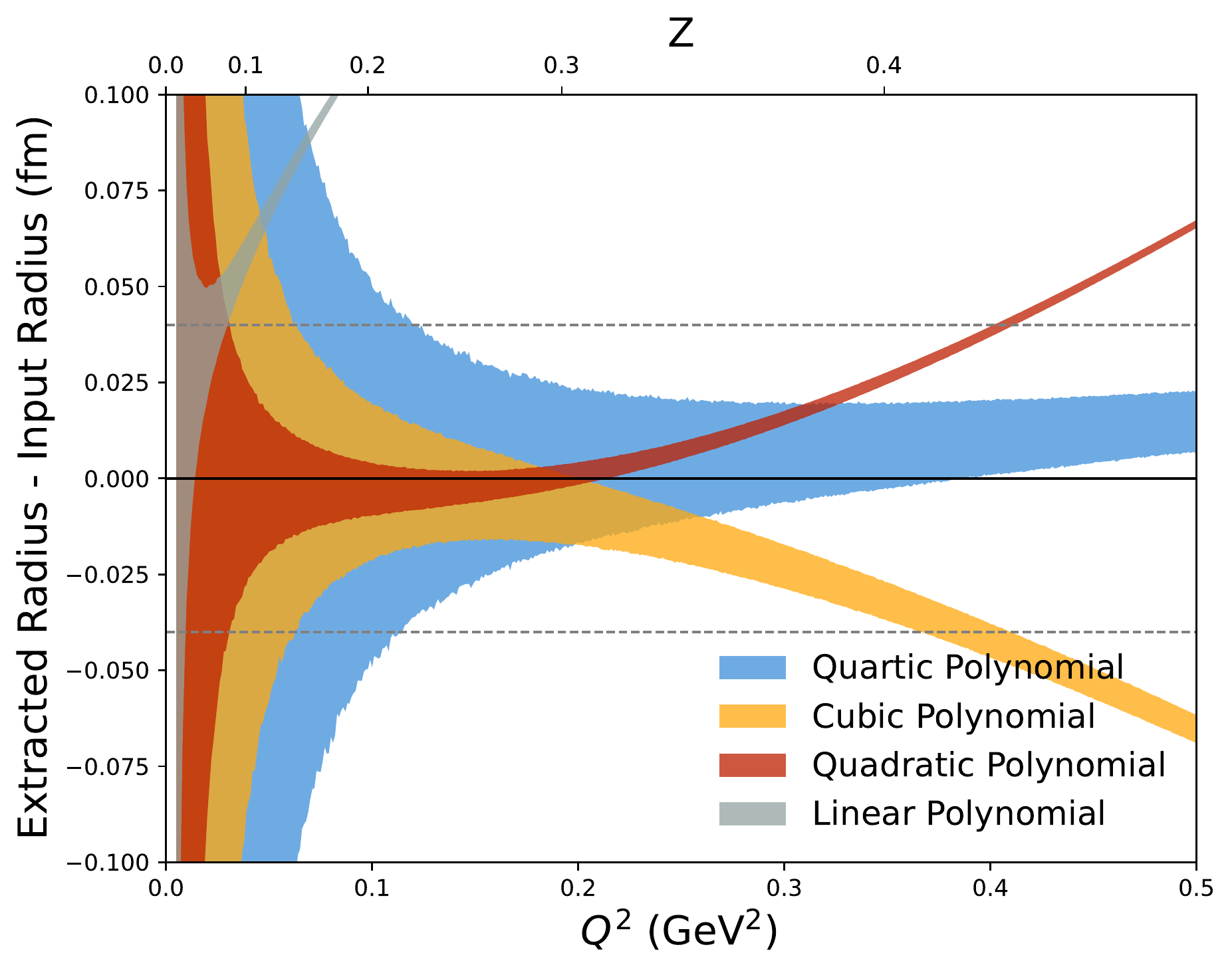}
    \caption{Bias-Variance plot for polynomial fits in $Z$ of the AW \GEp{} parameterization}
\end{figure}

\begin{figure}[hb]
    \centering
    \includegraphics[height=0.2\textheight]{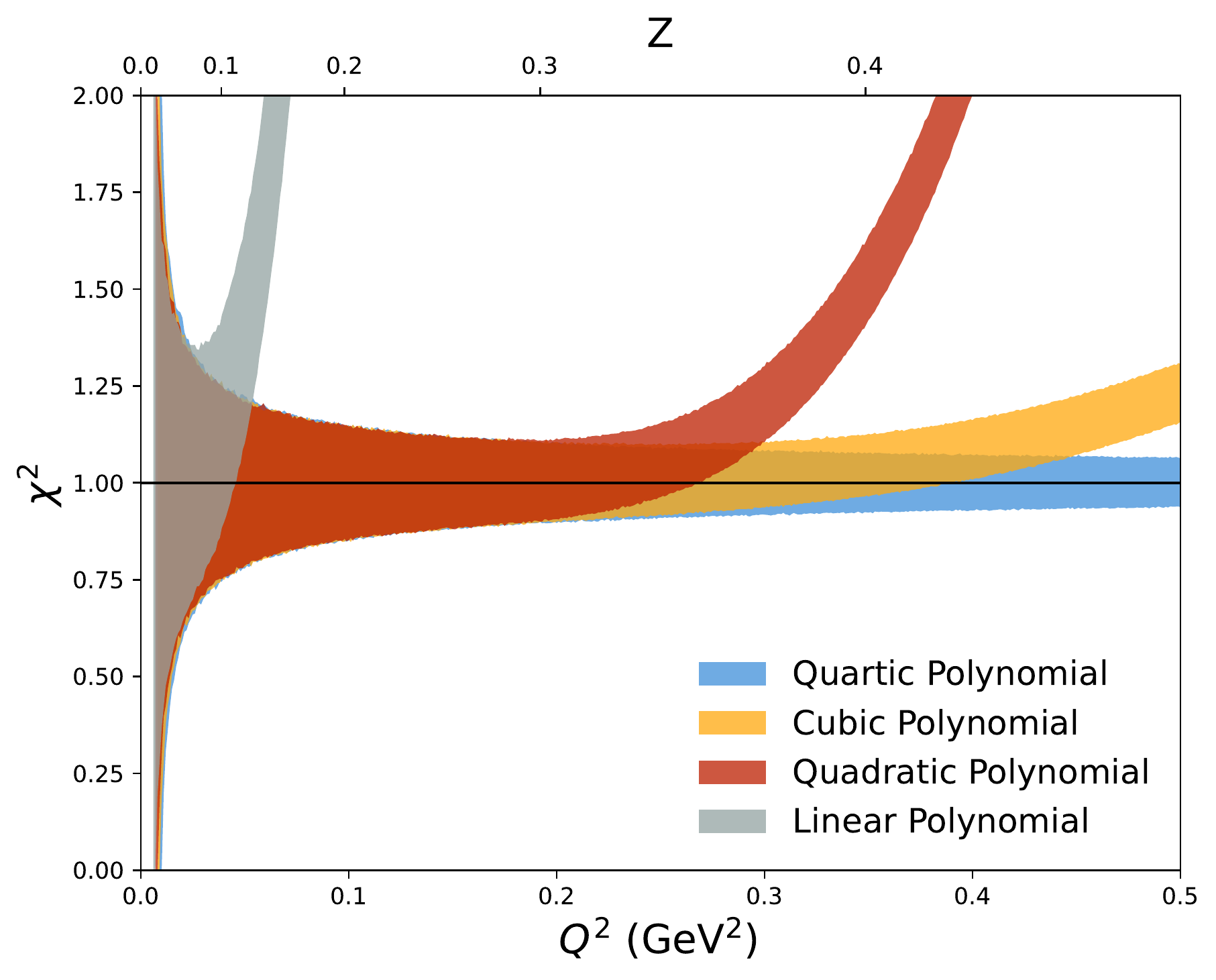}
    \caption{$\chi^2$ plot for polynomial fits in $Z$ of the AW \GEp{} parameterization}
\end{figure}

\begin{figure}[hb]
    \centering
    \includegraphics[height=0.2\textheight]{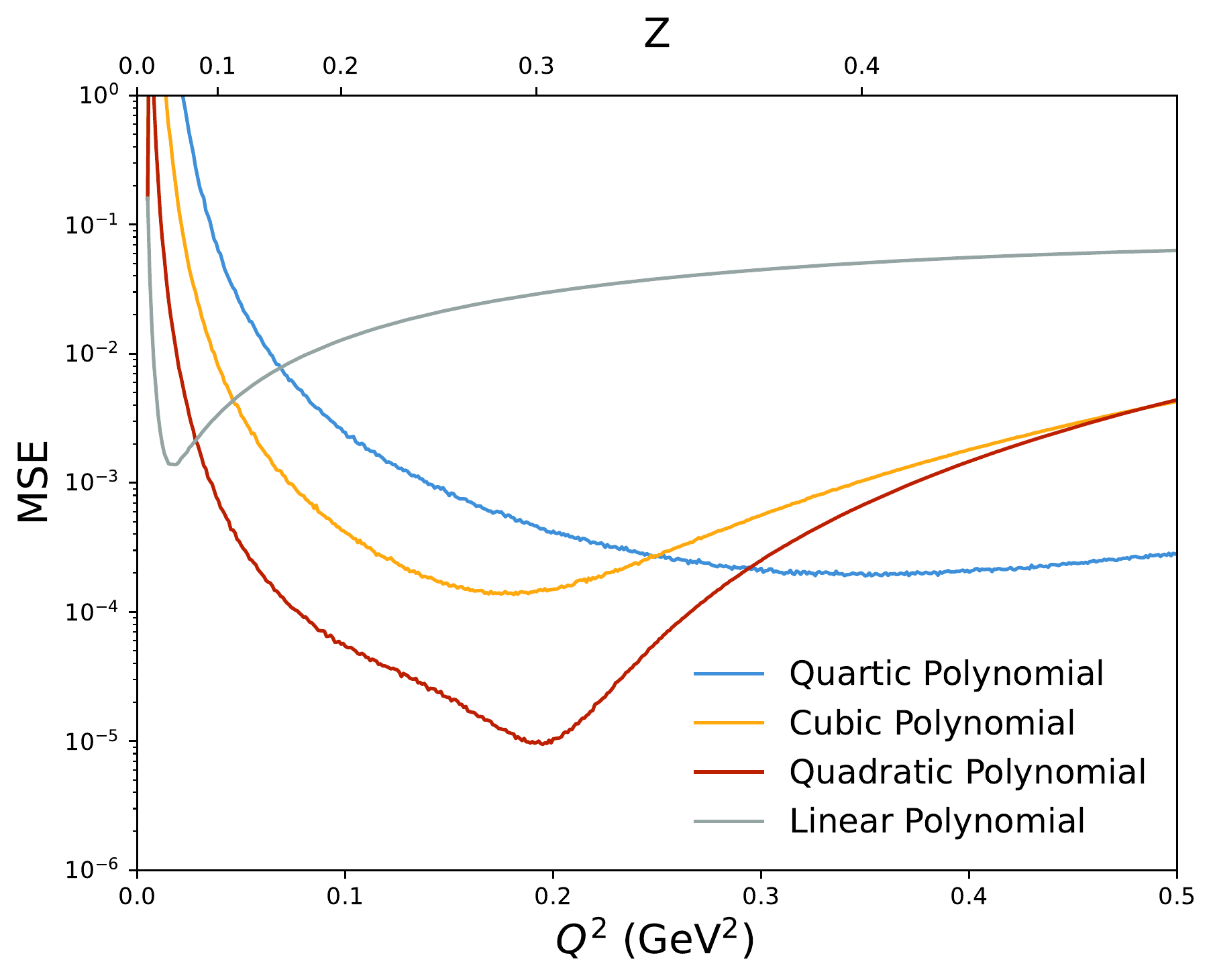}
    \caption{Mean Squared Error plot for polynomial fits in $Z$ of the AW \GEp{} parameterization}
\end{figure}

\newpage

\subsubsection{Figure Descriptions}

Three plots, each with “$Q^2$ (GeV$^2$)” from 0 to 0.5 on the x axis. A second x axis on top shows the mapping of $Q^2$ to $Z$, which spreads out low $Q^2$ points and bunches up high $Q^2$ points. Each plot has four sets of data shown, one for each power of polynomial used, labeled as “Linear Polynomial”, “Quadratic Polynomial”, “Cubic Polynomial”, and “Quartic Polynomial”.

The top plot has “Extracted Radius - Input Radius (fm)” from -0.1 to 0.1 on the y axis. Two horizontal lines drawn at $\pm$0.04 fm show the magnitude of the proton radius puzzle. Each data set is shown as a band. The high variance region extends to higher $Q^2$ than fits in $Q^2$ and the variance is higher than the $Q^2$ fits for the full range of the plot. The curves rapidly diverge from the input radius at approximately the same $Q^2$ as the fits in $Q^2$, but are better centered at the correct radius at lower $Q^2$. Odd-order polynomials diverge to a large radius and even-order polynomials diverge to a small radius.

The middle plot has $\chi^2$ from 0 to 2 on the y axis. Each data set is shown as a band. All bands begin centered at 1 and then at some $Q^2$ have a rapid upturn. Higher order polynomials delay the upturn until higher $Q^2$. Much like the fits in $Q^2$, the deviation from $\chi^2\approx1$ occurs at significantly higher $Q^2$ than when the extracted radius deviates from the input radius.

The bottom plot has MSE, Mean Squared Error as described in Eq.~\ref{eq:MSE}, from $10^{-6}$ to $10^0$ with a logarithmic scale on the y axis. Each set of data is represented as a curve. Each curve rapidly drops to the minimum MSE for the fit and then begins to increase with $Q^2$ and start to level off about two orders of magnitude higher than the minima. The quadratic polynomial has the lowest minimum at about $1\cdot10^{-5}$. The minima are spread over a larger $Q^2$ range and the range grows with polynomial order.

\newpage

\subsection{Bounded Polynomial Fits in $Z$}

\begin{figure}[hb]
    \centering
    \includegraphics[height=0.2\textheight]{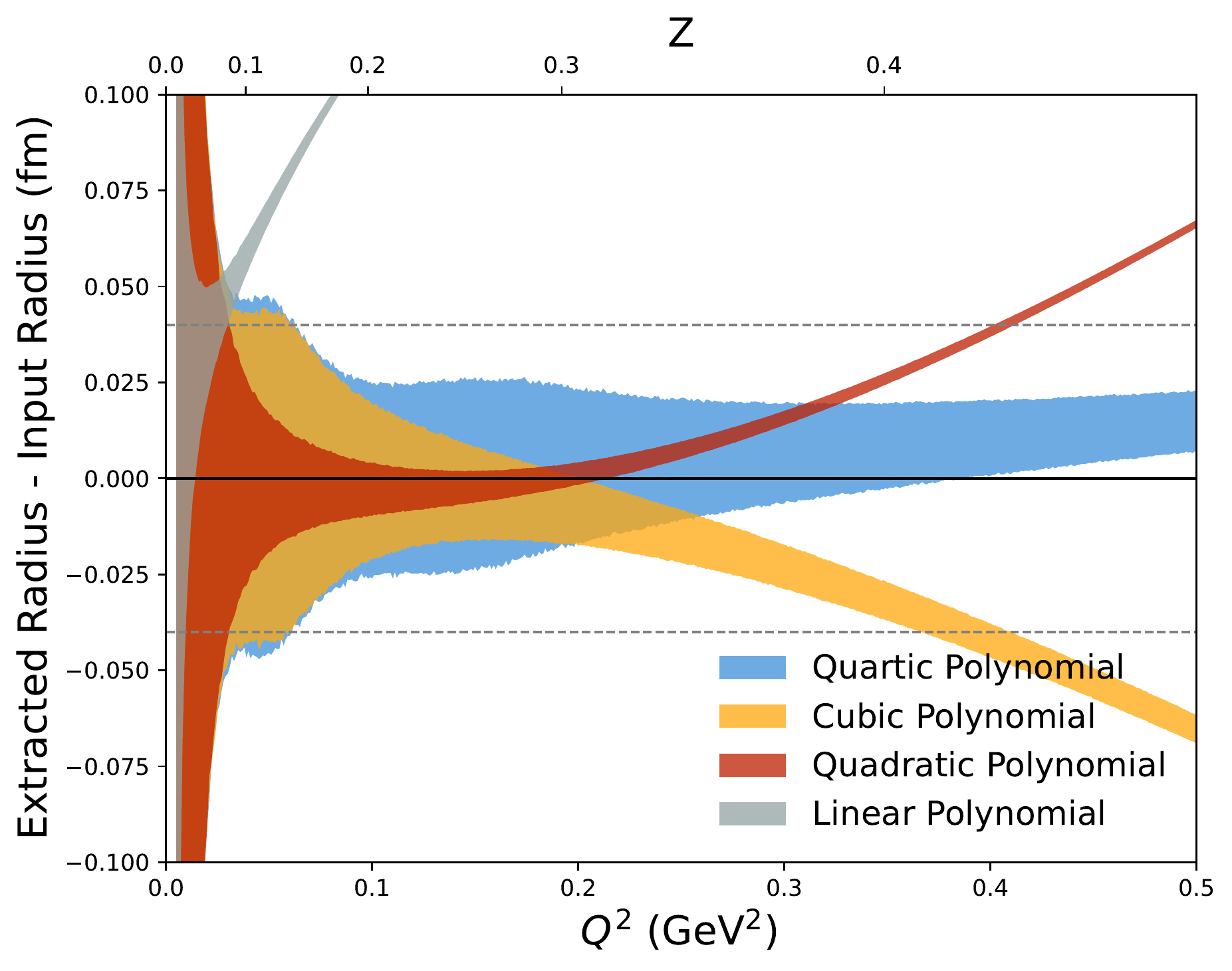}
    \caption{Bias-Variance plot for bounded polynomial fits in $Z$ of the AW \GEp{} parameterization}
\end{figure}

\begin{figure}[hb]
    \centering
    \includegraphics[height=0.2\textheight]{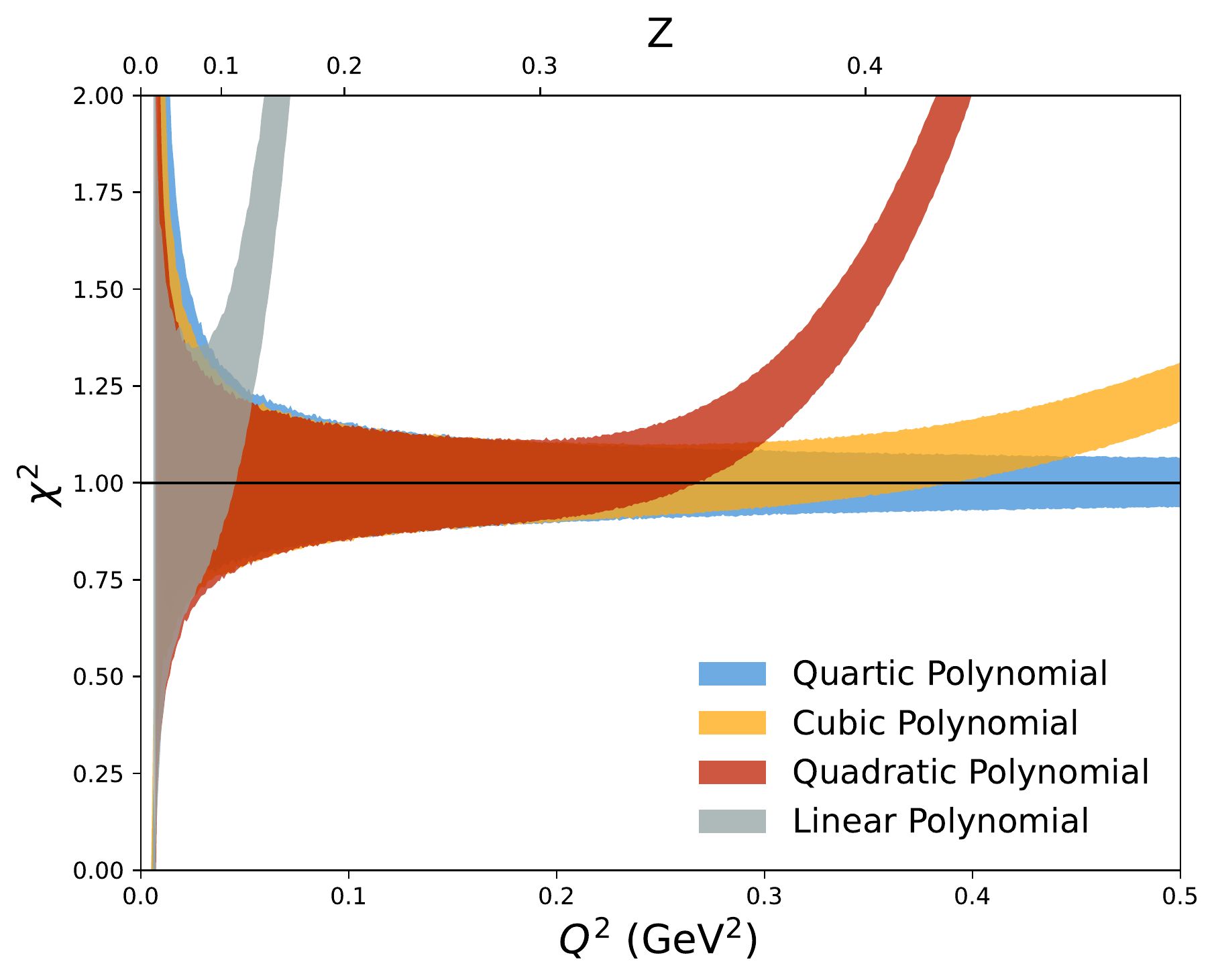}
    \caption{$\chi^2$ plot for bounded polynomial fits in $Z$ of the AW \GEp{} parameterization}
\end{figure}

\begin{figure}[hb]
    \centering
    \includegraphics[height=0.2\textheight]{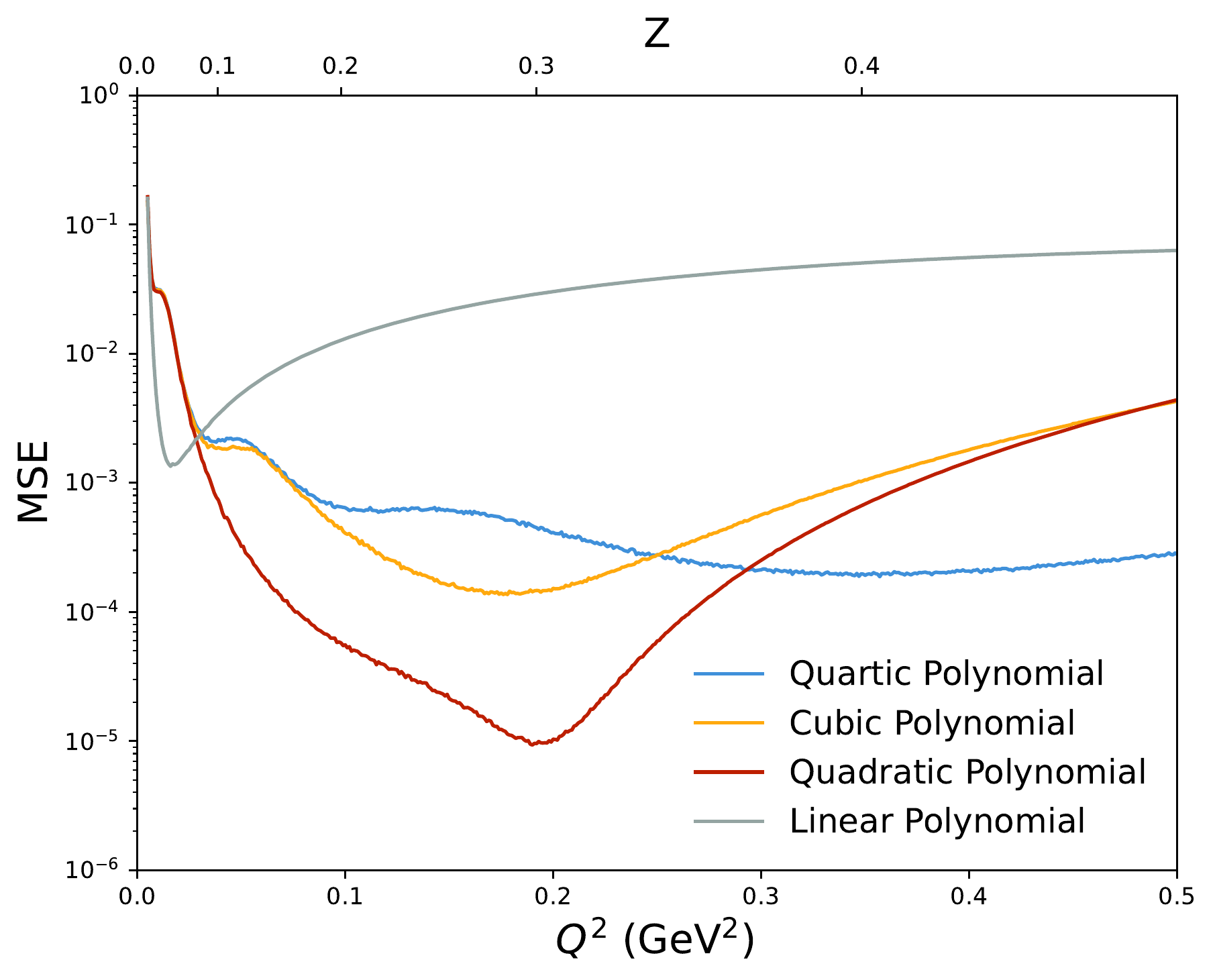}
    \caption{Mean Squared Error plot for bounded polynomial fits in $Z$ of the AW \GEp{} parameterization}
\end{figure}

\newpage

\subsubsection{Figure Descriptions}

Three plots, each with “$Q^2$ (GeV$^2$)” from 0 to 0.5 on the x axis. A second x axis on top shows the mapping of $Q^2$ to $Z$, which spreads out low $Q^2$ points and bunches up high $Q^2$ points. Each plot has four sets of data shown, one for each power of polynomial used, labeled as “Linear Polynomial”, “Quadratic Polynomial”, “Cubic Polynomial”, and “Quartic Polynomial”.

The top plot has “Extracted Radius - Input Radius (fm)” from -0.1 to 0.1 on the y axis. Two horizontal lines drawn at $\pm$0.04 fm show the magnitude of the proton radius puzzle. Each data set is shown as a band. The low $Q^2$ variance is significantly tempered from the unbound fits in $Z$. However, as $Q^2$ increases the curves are not qualitatively different from the unbound fits.

The middle plot has $\chi^2$ from 0 to 2 on the y axis. Each data set is shown as a band. The data is not qualitatively different than the unbound fits.

The bottom plot has MSE, Mean Squared Error as described in Eq.~\ref{eq:MSE}, from $10^{-6}$ to $10^0$ with a logarithmic scale on the y axis. Each set of data is represented as a curve. The data approaches their minima more rapidly than the unbound fits, but as $Q^2$ increases they are not qualitatively different from the unbound fits.

\newpage

\subsection{Rational(N,M) Fits in $Q^2$}

\begin{figure}[hb]
    \centering
    \includegraphics[height=0.2\textheight]{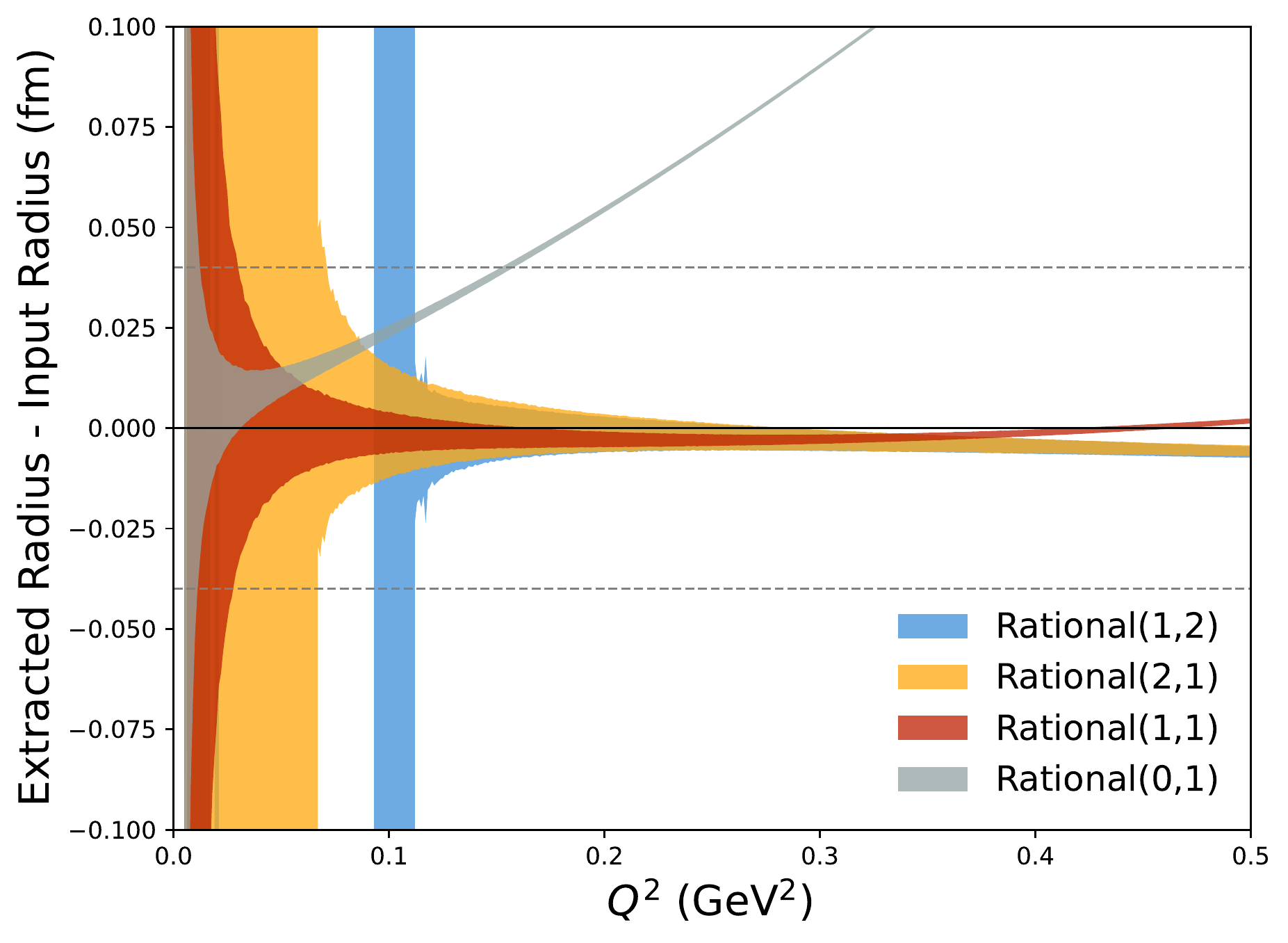}
    \caption{Bias-Variance plot for Rational(N,M) fits in $Q^2$ of the AW \GEp{} parameterization}
\end{figure}

\begin{figure}[hb]
    \centering
    \includegraphics[height=0.2\textheight]{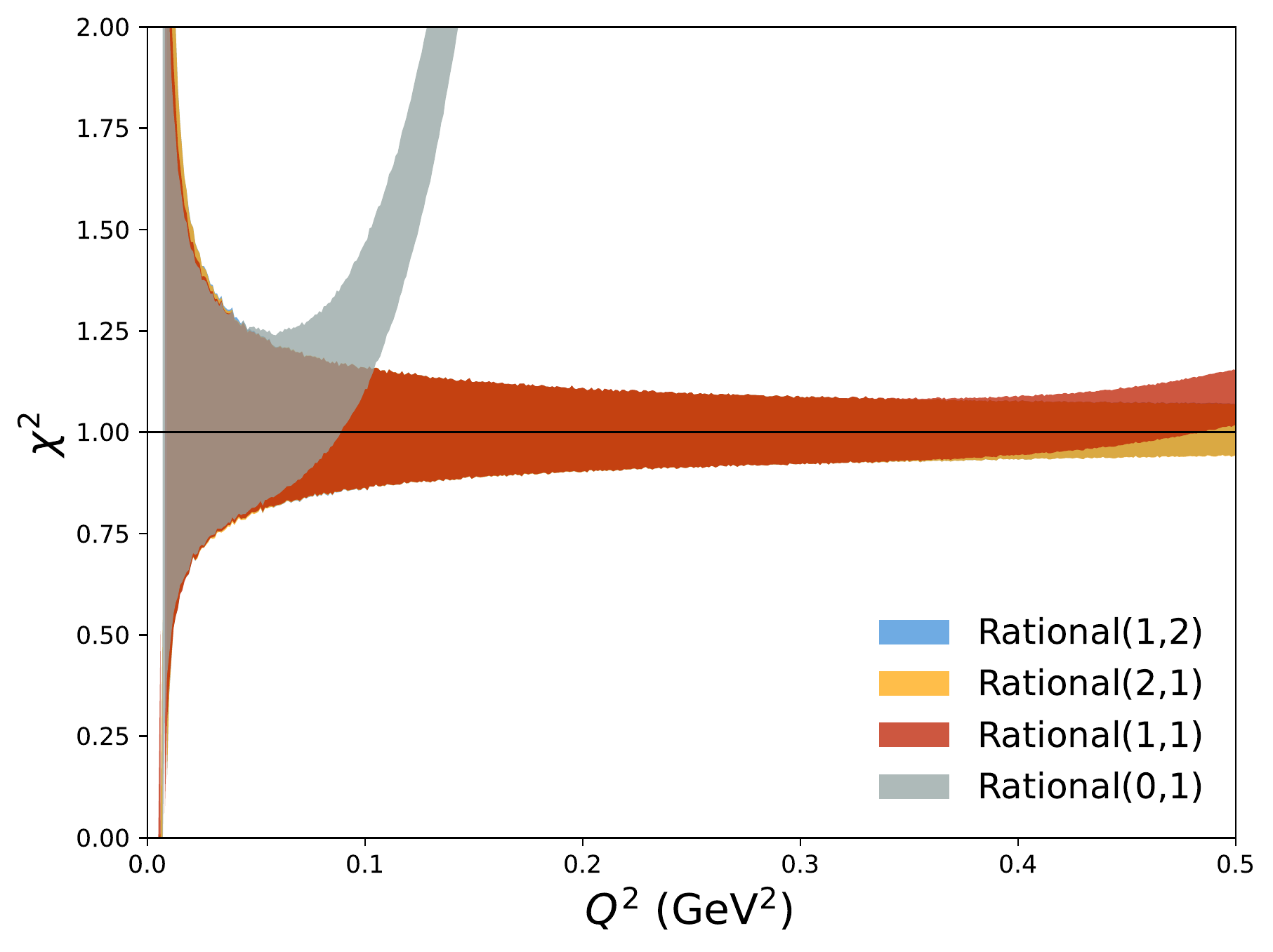}
    \caption{$\chi^2$ plot for Rational(N,M) fits in $Q^2$ of the AW \GEp{} parameterization}
\end{figure}

\begin{figure}[hb]
    \centering
    \includegraphics[height=0.2\textheight]{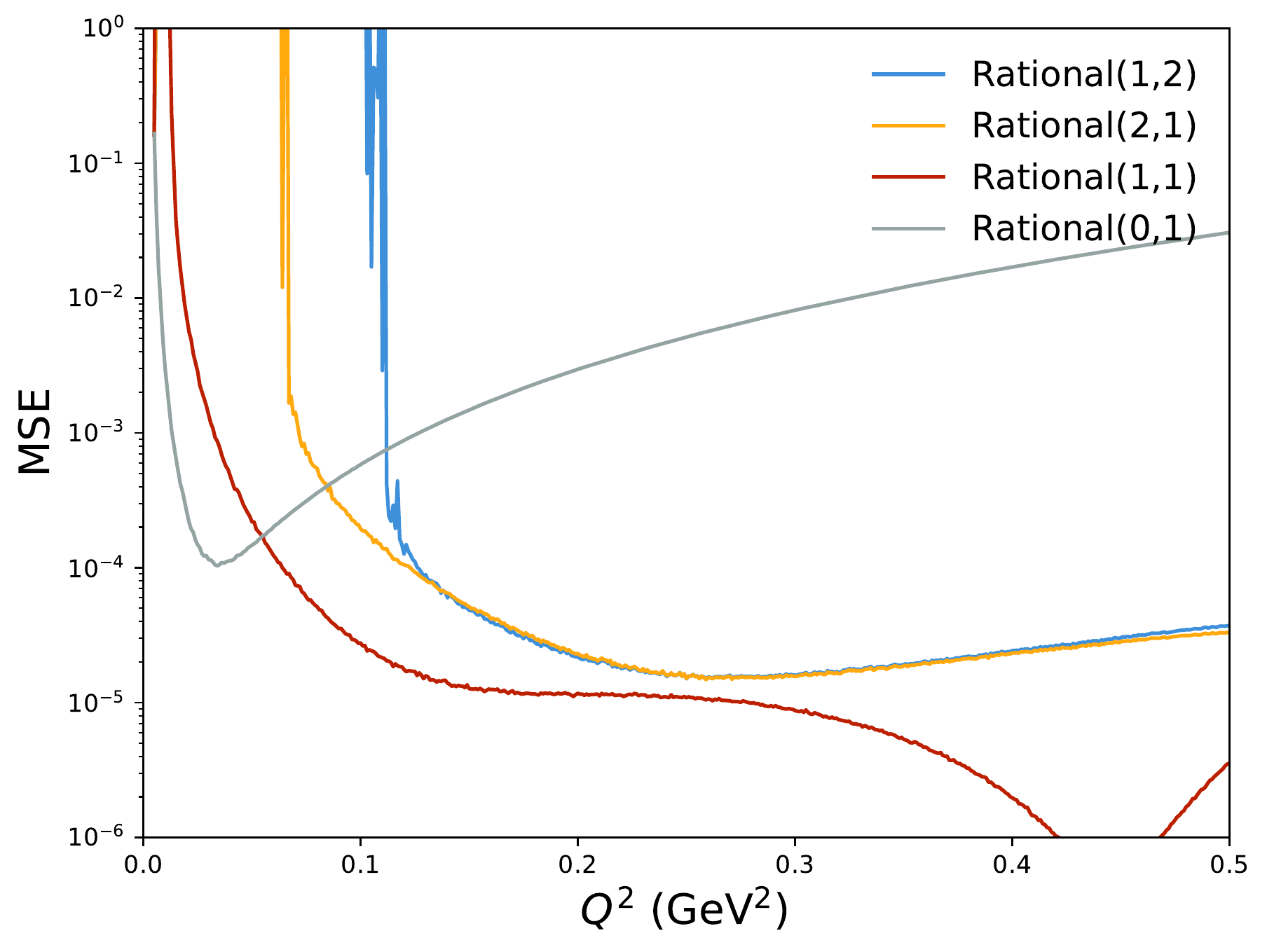}
    \caption{Mean Squared Error plot for Rational(N,M) fits in $Q^2$ of the AW \GEp{} parameterization}
\end{figure}

\newpage

\subsubsection{Figure Descriptions}

Three plots, each with “$Q^2$ (GeV$^2$)” from 0 to 0.5 on the x axis. Each plot has four sets of data shown, one for each power of Rational(N,M) function used, labeled as “Rational(0,1)”, “Rational(1,1)”, “Rational(2,1)”, and “Rational(1,2)”.

The top plot has “Extracted Radius - Input Radius (fm)” from -0.1 to 0.1 on the y axis. Two horizontal lines drawn at $\pm$0.04 fm show the magnitude of the proton radius puzzle. Each data set is shown as a band. This data shows larger variance at low $Q^2$ than any other fit and the variance increases with increasing N and M, with M having a larger influence on the variance. The Rational(0,1) curve rapidly diverges from the input radius. The Rational(1,2) curve shows no data below $Q^2\approx0.1$ GeV$^2$ as the mean radius was unphysical. Curves with both N and M of at least 1 stay very close to the input radius for the full range with Rational(1,2) and Rational(2,1) slightly trending towards a small radius at high $Q^2$. The variances of all Rational(N,M) fits are less than half that of any of the other fits.

The middle plot has $\chi^2$ from 0 to 2 on the y axis. Each data set is shown as a band. All begin centered at 1, with Rational(0,1) having a rapid upturn. The others stay centered at 1 for the full range, with Rational(1,1) slightly increasing at the highest $Q^2$.

The bottom plot has MSE, Mean Squared Error as described in Eq.~\ref{eq:MSE}, from $10^{-6}$ to $10^0$ with a logarithmic scale on the y axis. Each set of data is represented as a curve. The Rational(0,1) fit reaches a minima of $10^{-4}$ and then quickly rises. The others approach their minima and then remain near there for the full range. These are the lowest MSE values of all fit types.

\newpage

\section{Arrington, Melnitchouk, and Tjon (AMT)}

\subsection{Polynomial Fits in $Q^2$}

\begin{figure}[hb]
    \centering
    \includegraphics[height=0.2\textheight]{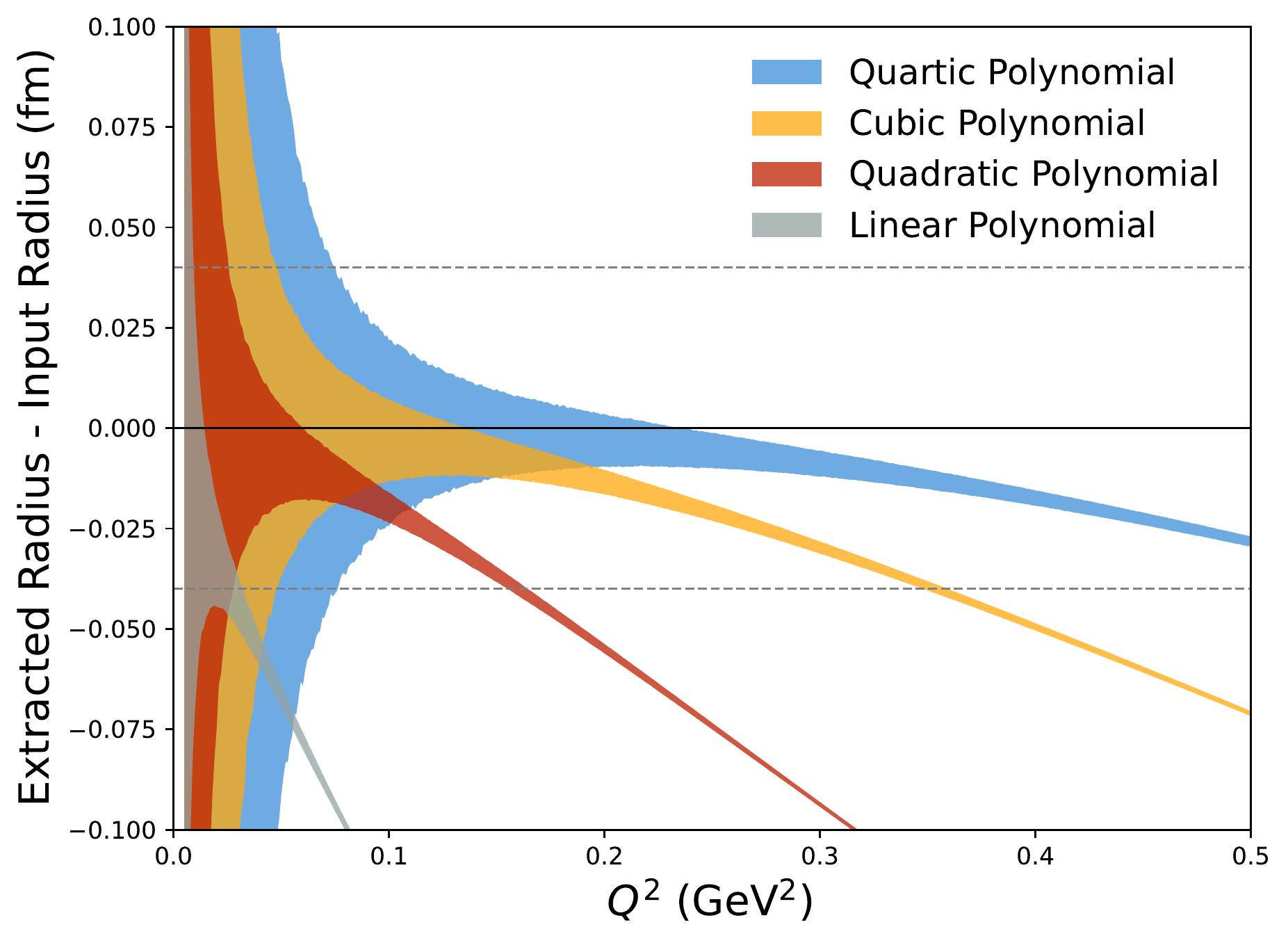}
    \caption{Bias-Variance plot for polynomial fits in $Q^2$ of the AMT \GEp{} parameterization}
\end{figure}

\begin{figure}[hb]
    \centering
    \includegraphics[height=0.2\textheight]{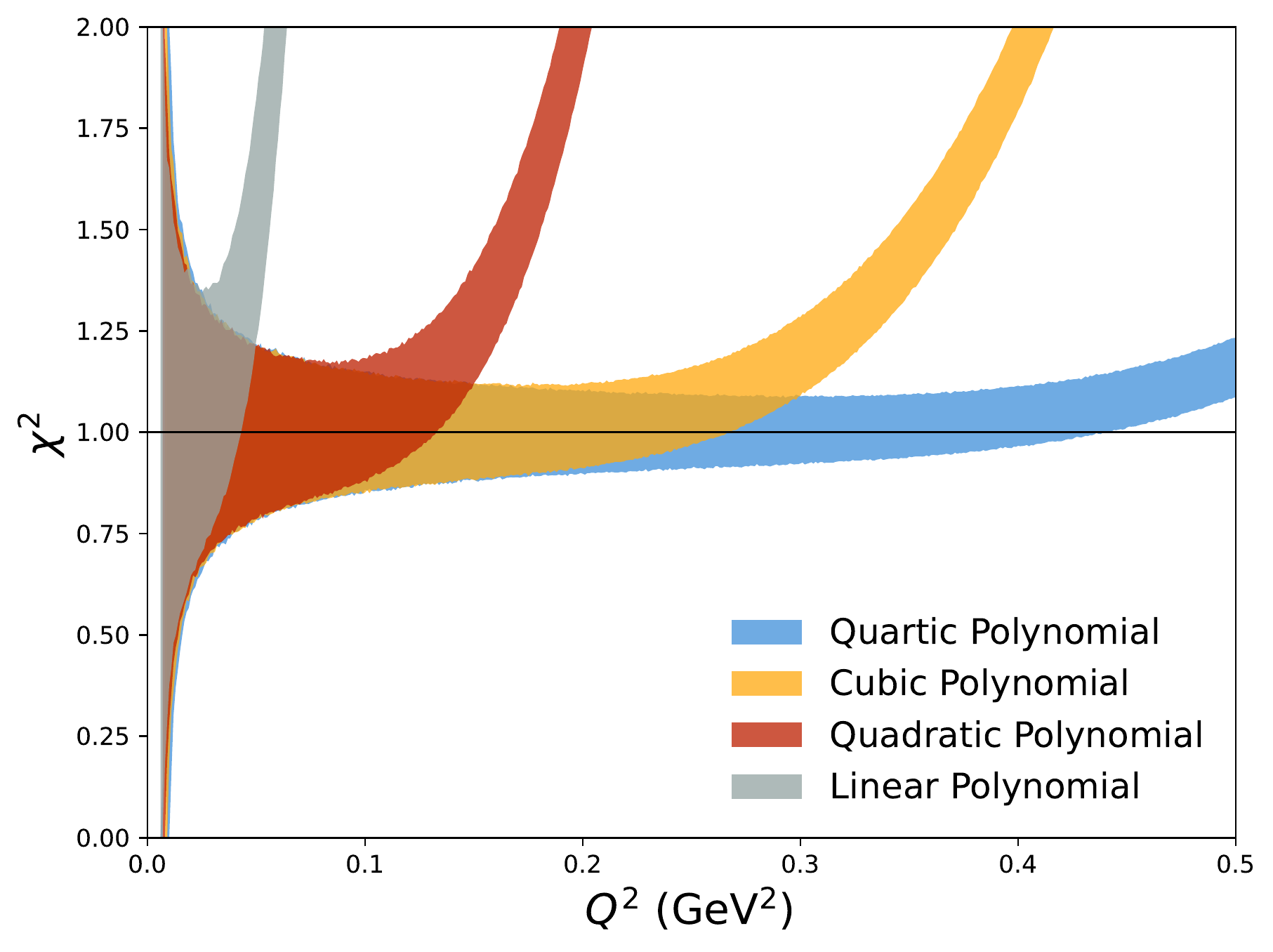}
    \caption{$\chi^2$ plot for polynomial fits in $Q^2$ of the AMT \GEp{} parameterization}
\end{figure}

\begin{figure}[hb]
    \centering
    \includegraphics[height=0.2\textheight]{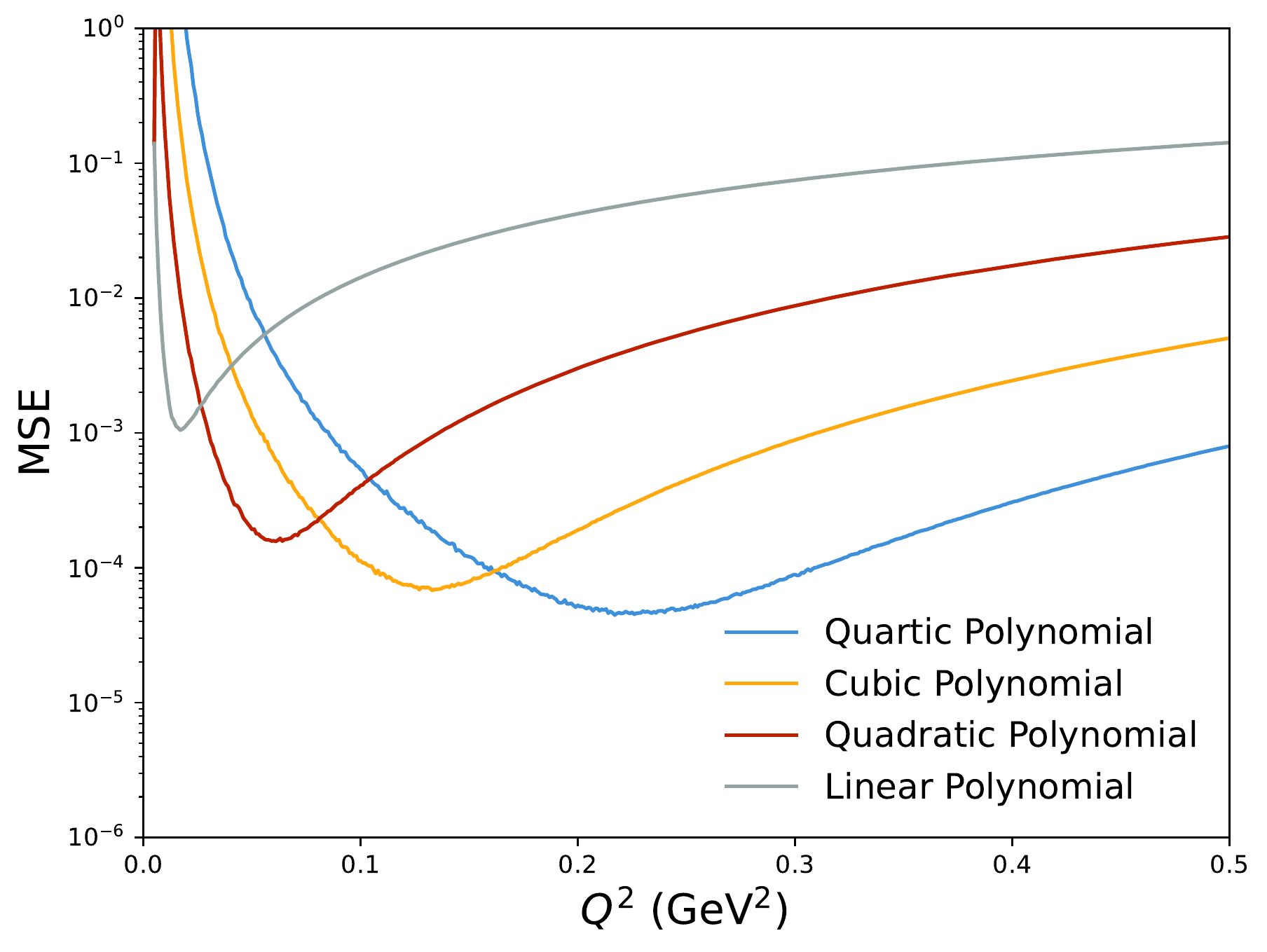}
    \caption{Mean Squared Error plot for polynomial fits in $Q^2$ of the AMT \GEp{} parameterization}
\end{figure}

\newpage

\subsubsection{Figure Descriptions}

Three plots, each with “$Q^2$ (GeV$^2$)” from 0 to 0.5 on the x axis. Each plot has four sets of data shown, one for each power of polynomial used, labeled as “Linear Polynomial”, “Quadratic Polynomial”, “Cubic Polynomial”, and “Quartic Polynomial”.

The top plot has “Extracted Radius - Input Radius (fm)” from -0.1 to 0.1 on the y axis. Two horizontal lines drawn at $\pm$0.04 fm show the magnitude of the proton radius puzzle. Each data set is shown as a band. All bands have variance higher than the axis scale at small $Q^2$ and systematically return a small radius that decreases with increasing $Q^2$ range. Higher order polynomials decrease slower.

The middle plot has $\chi^2$ from 0 to 2 on the y axis. Each data set is shown as a band. All bands begin centered at 1 and then at some $Q^2$ have a rapid upturn. Higher order polynomials delay the upturn until higher $Q^2$. The deviation from $\chi^2\approx1$ occurs at significantly higher $Q^2$ than when the extracted radius deviates from the input radius.

The bottom plot has MSE, Mean Squared Error as described in Eq.~\ref{eq:MSE}, from $10^{-6}$ to $10^0$ with a logarithmic scale on the y axis. Each set of data is represented as a curve. Each curve rapidly drops to the minimum MSE for the fit and then begins to increase with $Q^2$ and start to level off about two orders of magnitude higher than the minima. Higher order polynomials have lower minima that occur at higher $Q^2$ and then increase slower. The lowest minima is about $4\cdot10^{-5}$.

\newpage

\subsection{Bounded Polynomial Fits in $Q^2$}

\begin{figure}[hb]
    \centering
    \includegraphics[height=0.2\textheight]{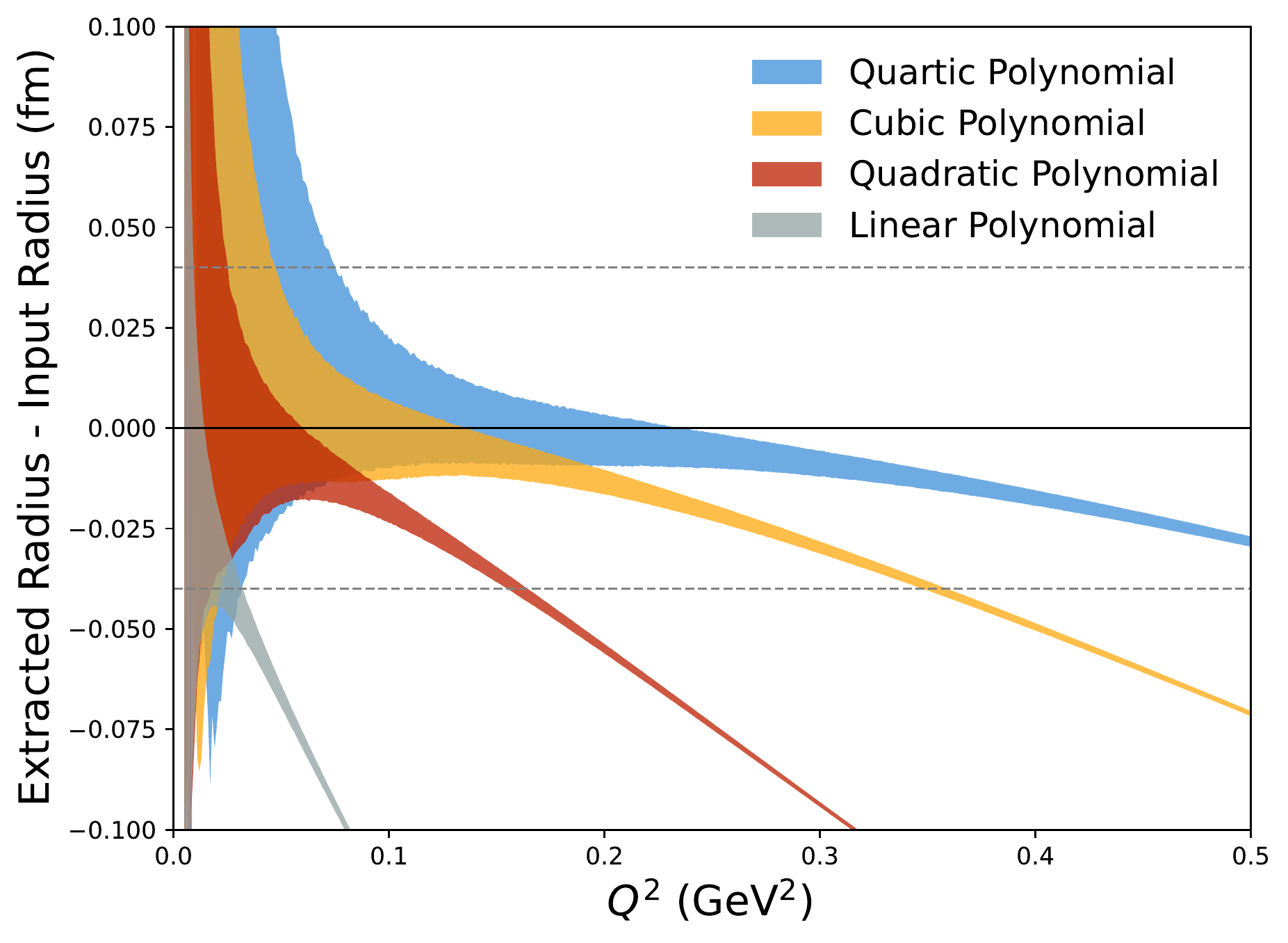}
    \caption{Bias-Variance plot for bounded polynomial fits in $Q^2$ of the AMT \GEp{} parameterization}
\end{figure}

\begin{figure}[hb]
    \centering
    \includegraphics[height=0.2\textheight]{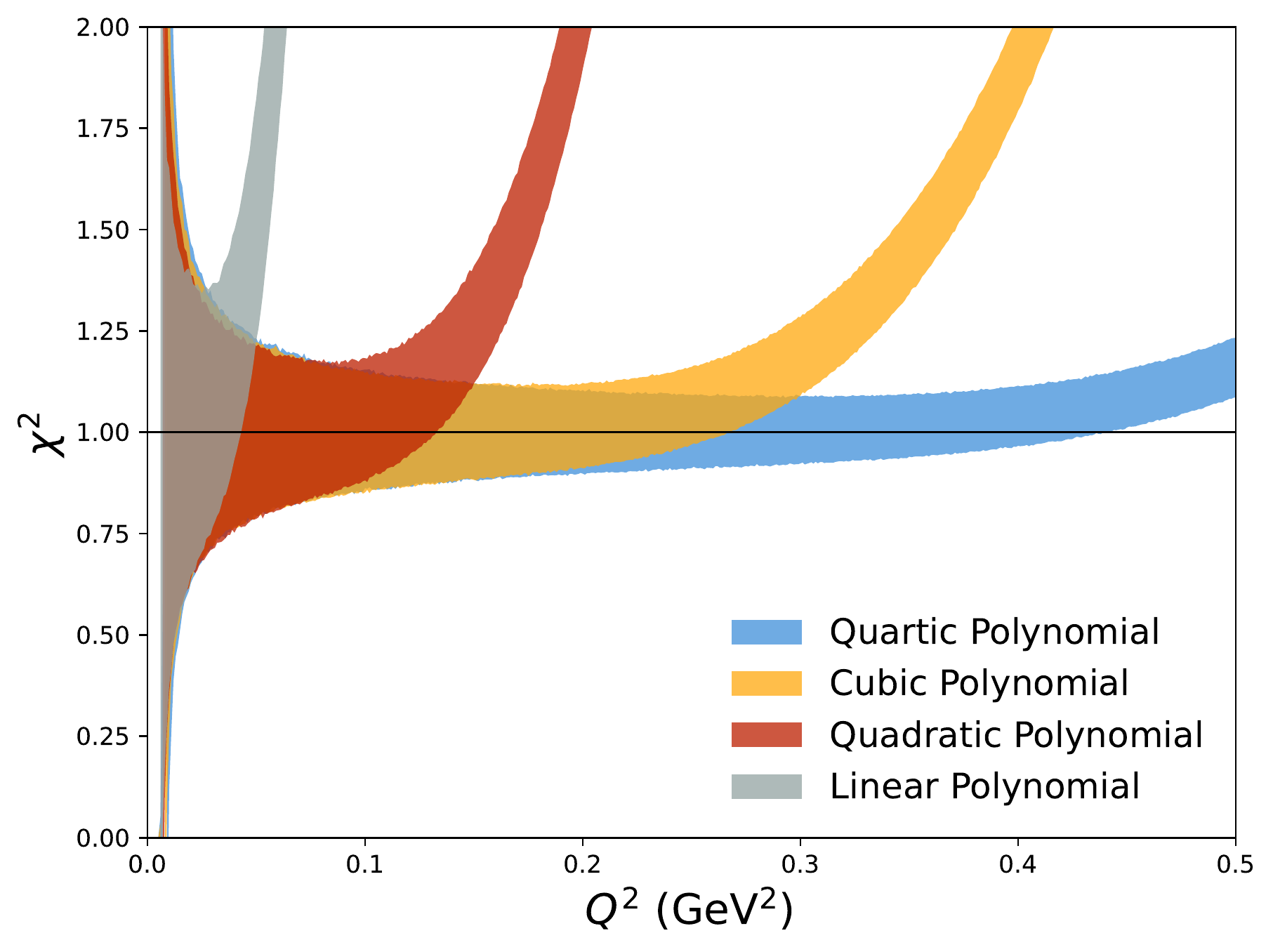}
    \caption{$\chi^2$ plot for bounded polynomial fits in $Q^2$ of the AMT \GEp{} parameterization}
\end{figure}

\begin{figure}[hb]
    \centering
    \includegraphics[height=0.2\textheight]{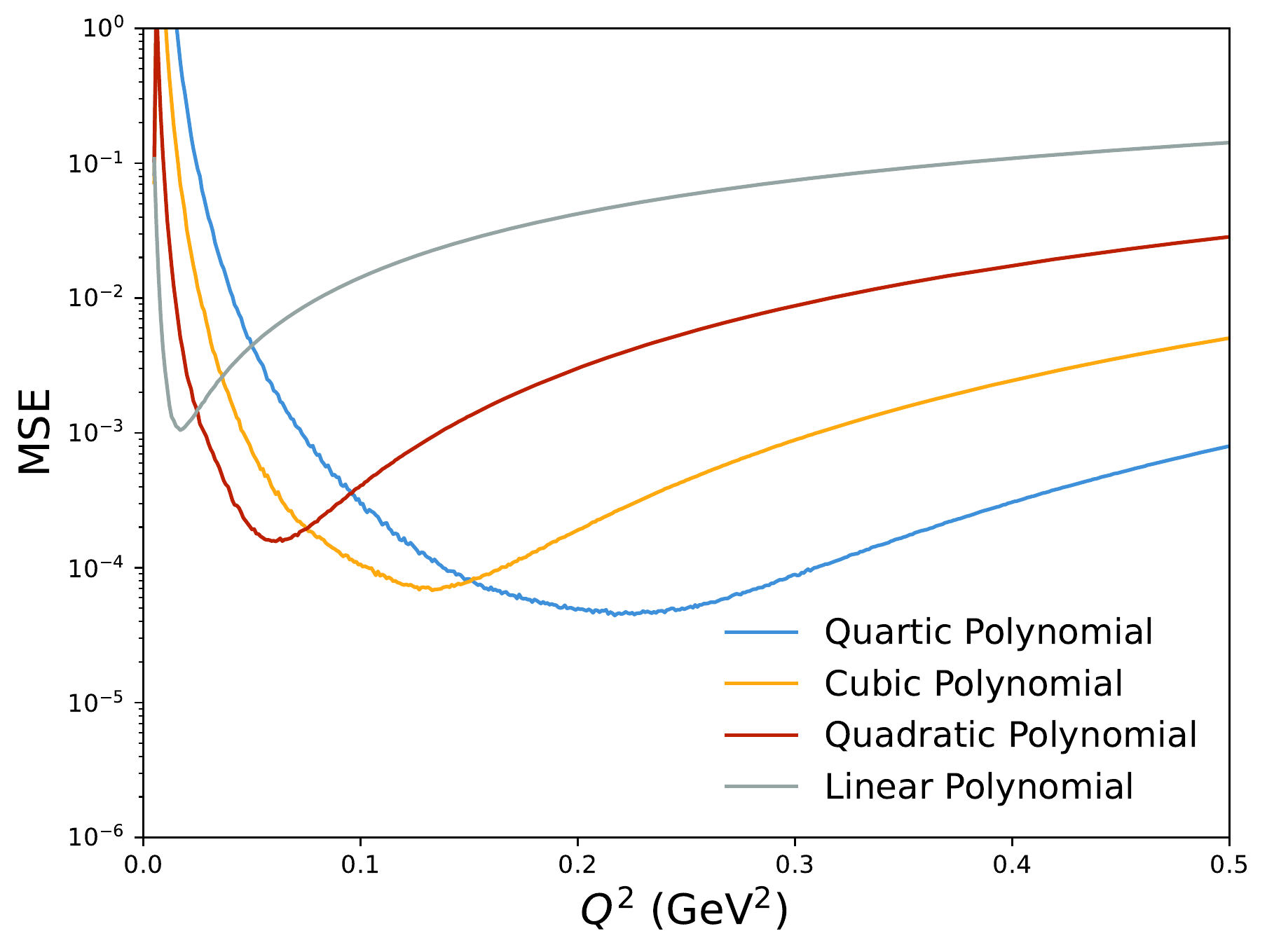}
    \caption{Mean Squared Error plot for bounded polynomial fits in $Q^2$ of the AMT \GEp{} parameterization}
\end{figure}

\newpage

\subsubsection{Figure Descriptions}

Three plots, each with “$Q^2$ (GeV$^2$)” from 0 to 0.5 on the x axis. Each plot has four sets of data shown, one for each power of polynomial used, labeled as “Linear Polynomial”, “Quadratic Polynomial”, “Cubic Polynomial”, and “Quartic Polynomial”.

The top plot has “Extracted Radius - Input Radius (fm)” from -0.1 to 0.1 on the y axis. Two horizontal lines drawn at $\pm$0.04 fm show the magnitude of the proton radius puzzle. Each data set is shown as a band. The low $Q^2$ variance is smaller than the unbound fits, but is not overall qualitatively different.

The middle plot has $\chi^2$ from 0 to 2 on the y axis. Each data set is shown as a band. The data is not qualitatively different than the unbound fits.

The bottom plot has MSE, Mean Squared Error as described in Eq.~\ref{eq:MSE}, from $10^{-6}$ to $10^0$ with a logarithmic scale on the y axis. Each set of data is represented as a curve. The data is not qualitatively different than the unbound fits.

\newpage

\subsection{Polynomial Fits in $Z$}

\begin{figure}[hb]
    \centering
    \includegraphics[height=0.2\textheight]{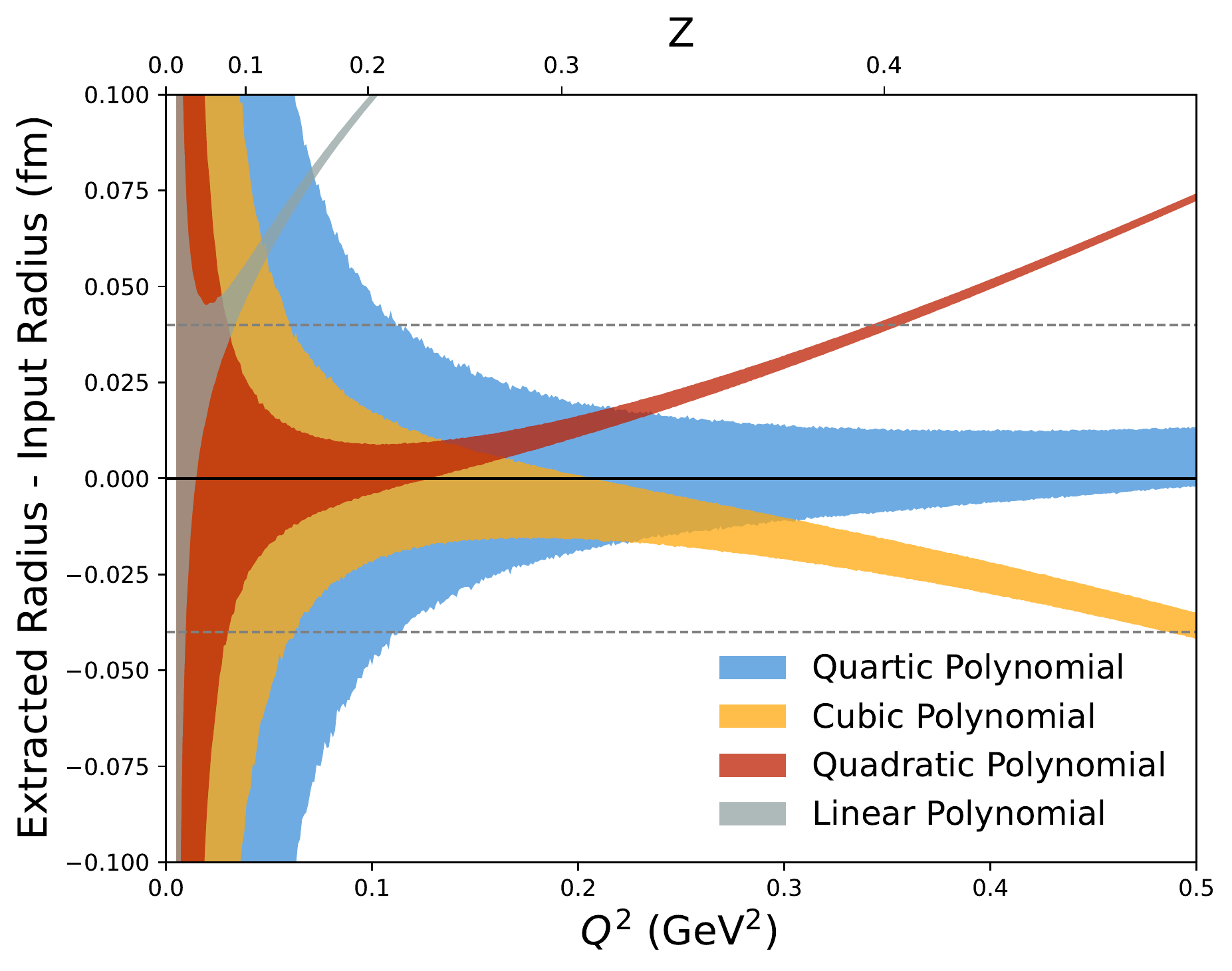}
    \caption{Bias-Variance plot for polynomial fits in $Z$ of the AMT \GEp{} parameterization}
\end{figure}

\begin{figure}[hb]
    \centering
    \includegraphics[height=0.2\textheight]{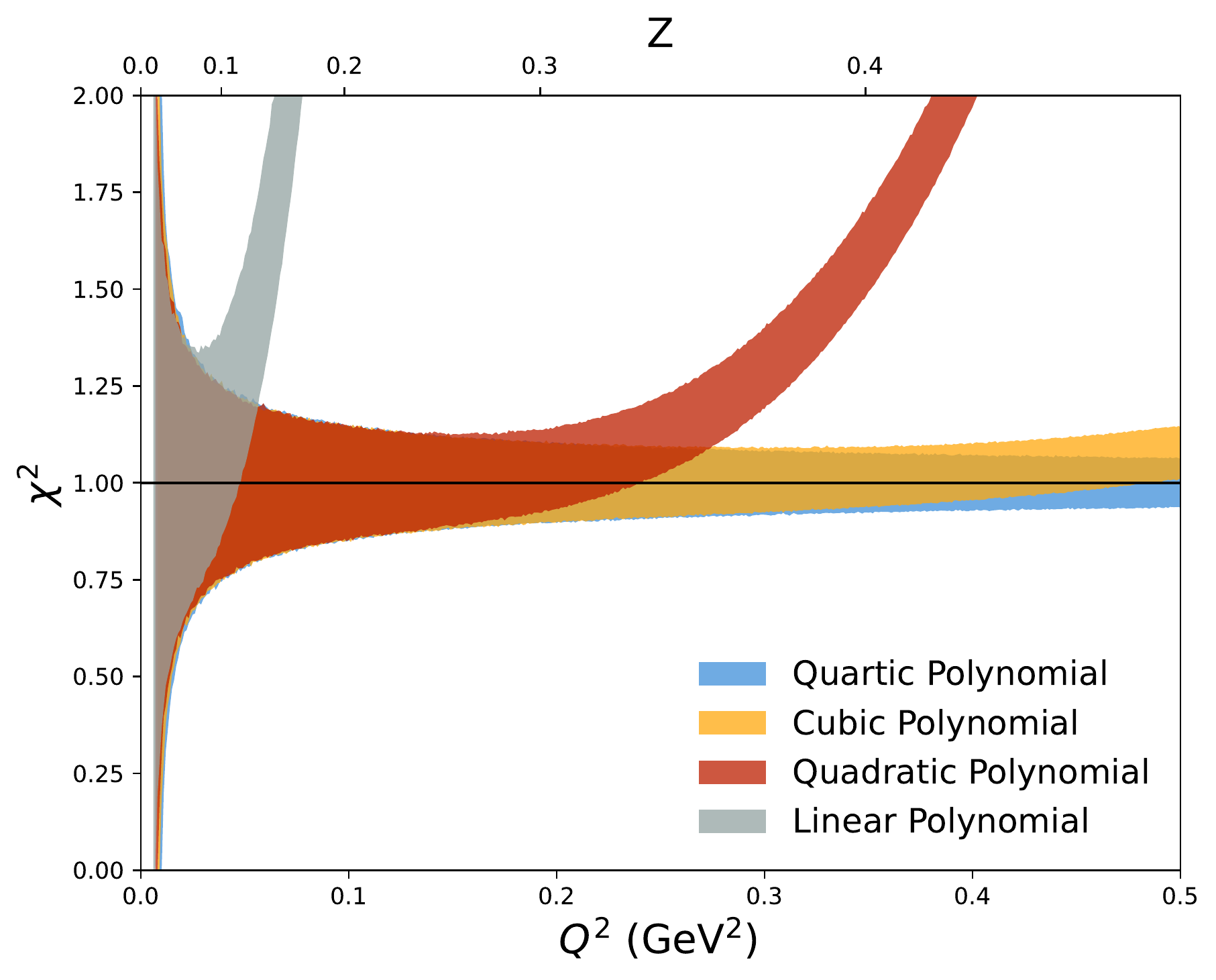}
    \caption{$\chi^2$ plot for polynomial fits in $Z$ of the AMT \GEp{} parameterization}
\end{figure}

\begin{figure}[hb]
    \centering
    \includegraphics[height=0.2\textheight]{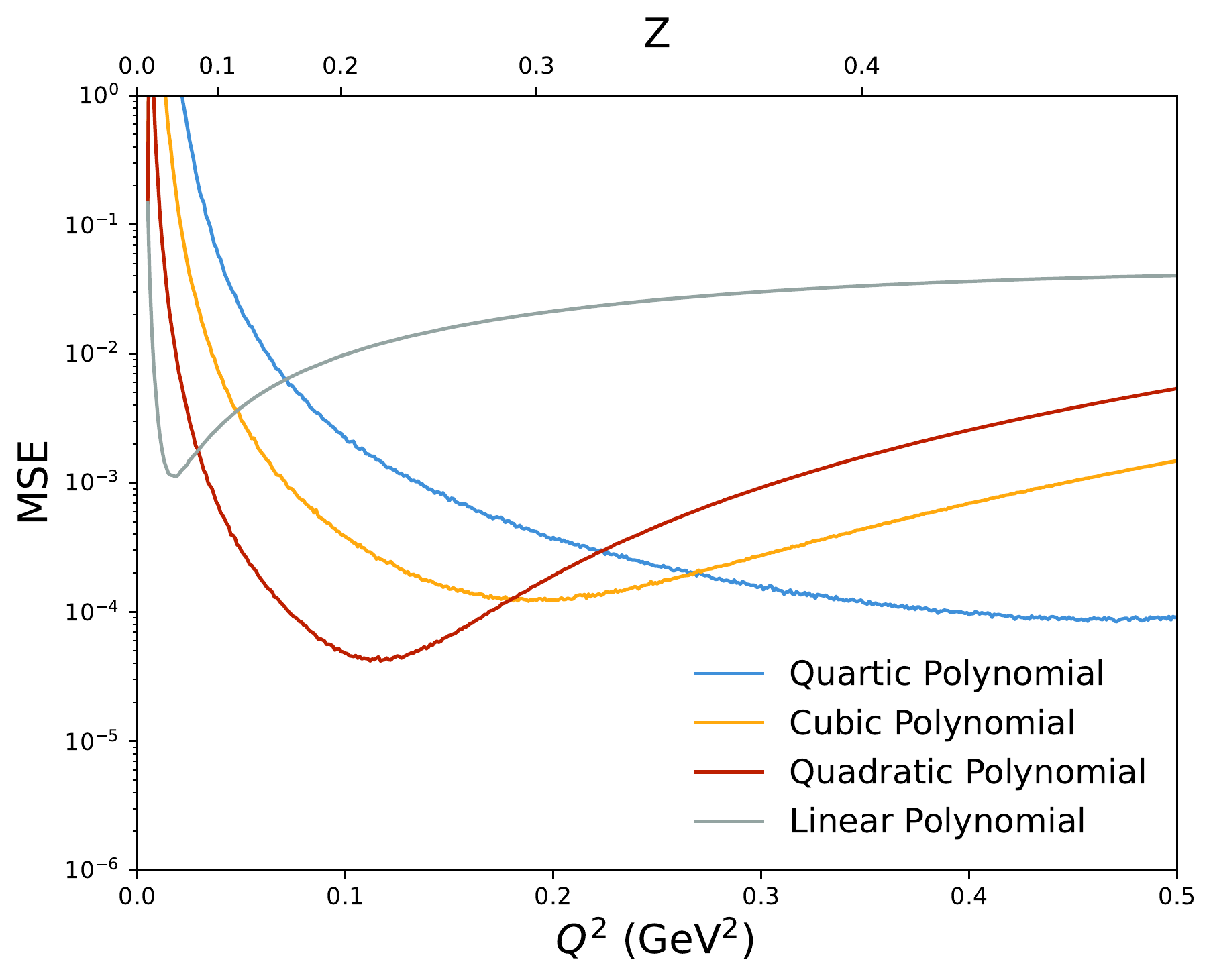}
    \caption{Mean Squared Error plot for polynomial fits in $Z$ of the AMT \GEp{} parameterization}
\end{figure}

\newpage

\subsubsection{Figure Descriptions}

Three plots, each with “$Q^2$ (GeV$^2$)” from 0 to 0.5 on the x axis. A second x axis on top shows the mapping of $Q^2$ to $Z$, which spreads out low $Q^2$ points and bunches up high $Q^2$ points. Each plot has four sets of data shown, one for each power of polynomial used, labeled as “Linear Polynomial”, “Quadratic Polynomial”, “Cubic Polynomial”, and “Quartic Polynomial”.

The top plot has “Extracted Radius - Input Radius (fm)” from -0.1 to 0.1 on the y axis. Two horizontal lines drawn at $\pm$0.04 fm show the magnitude of the proton radius puzzle. Each data set is shown as a band. The high variance region extends to higher $Q^2$ than fits in $Q^2$ and the variance is higher than the $Q^2$ fits for the full range of the plot. With the exception of the quartic polynomial, the curves rapidly diverge from the input radius at approximately the same $Q^2$ as the fits in $Q^2$, but are better centered at the correct radius at lower $Q^2$. The quartic polynomial remains well centered with a slight deviation to a large radius at high $Q^2$. The linear and quadratic polynomials diverge to large radii and the cubic polynomial diverges to a small radius.

The middle plot has $\chi^2$ from 0 to 2 on the y axis. Each data set is shown as a band. All bands begin centered at 1 and then at some $Q^2$ have a rapid upturn. Higher order polynomials delay the upturn until higher $Q^2$. Much like the fits in $Q^2$, the deviation from $\chi^2\approx1$ occurs at significantly higher $Q^2$ than when the extracted radius deviates from the input radius. The quartic polynomial remains well centered at 1 for the full range.

The bottom plot has MSE, Mean Squared Error as described in Eq.~\ref{eq:MSE}, from $10^{-6}$ to $10^0$ with a logarithmic scale on the y axis. Each set of data is represented as a curve. Each curve rapidly drops to the minimum MSE for the fit and then begins to increase with $Q^2$ and start to level off about two orders of magnitude higher than the minima. The quadratic polynomial has the lowest minimum at about $4\cdot10^{-5}$. The minima are spread over a larger $Q^2$ range and the range grows with polynomial order.

\newpage

\subsection{Bounded Polynomial Fits in $Z$}

\begin{figure}[hb]
    \centering
    \includegraphics[height=0.2\textheight]{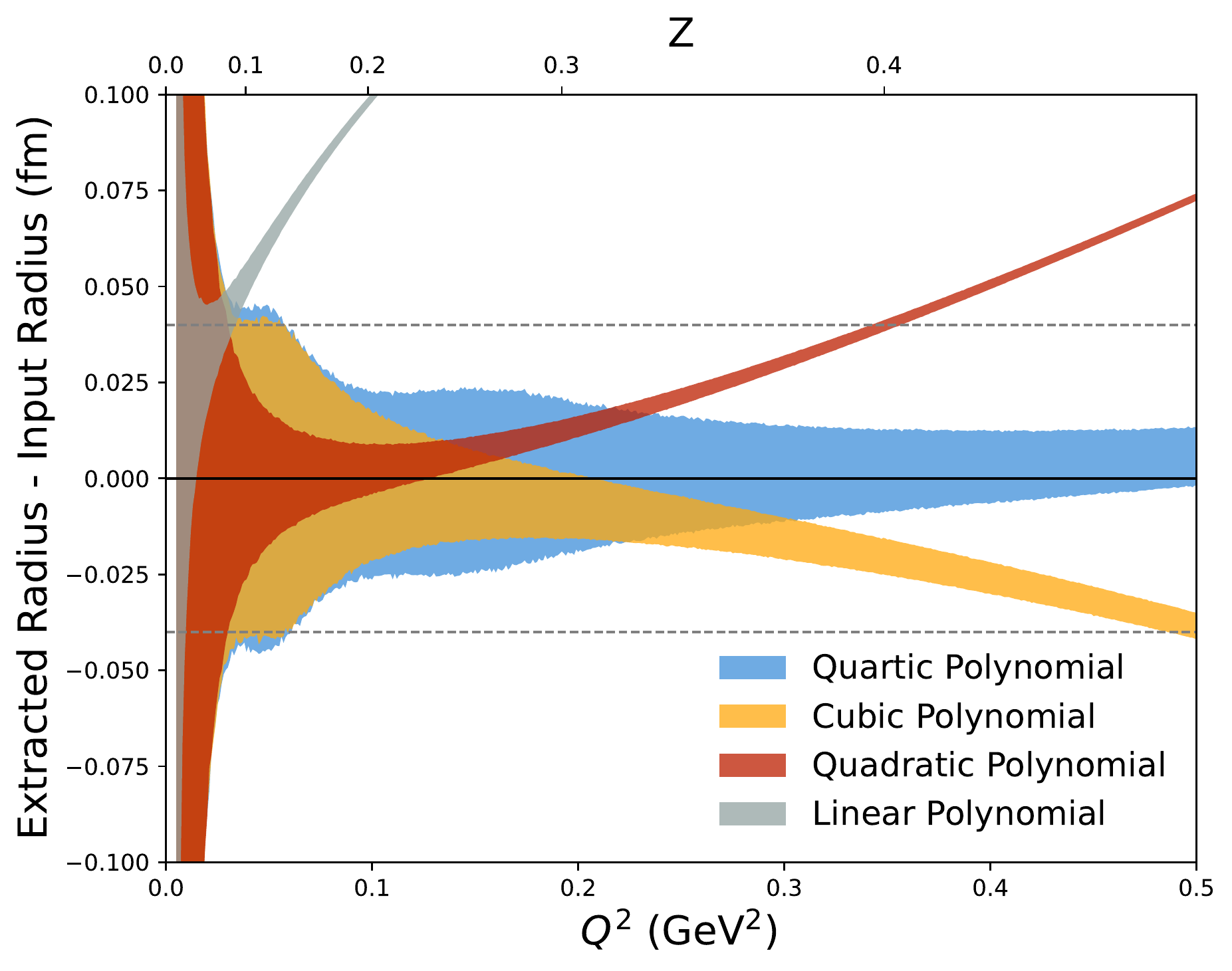}
    \caption{Bias-Variance plot for bounded polynomial fits in $Z$ of the AMT \GEp{} parameterization}
\end{figure}

\begin{figure}[hb]
    \centering
    \includegraphics[height=0.2\textheight]{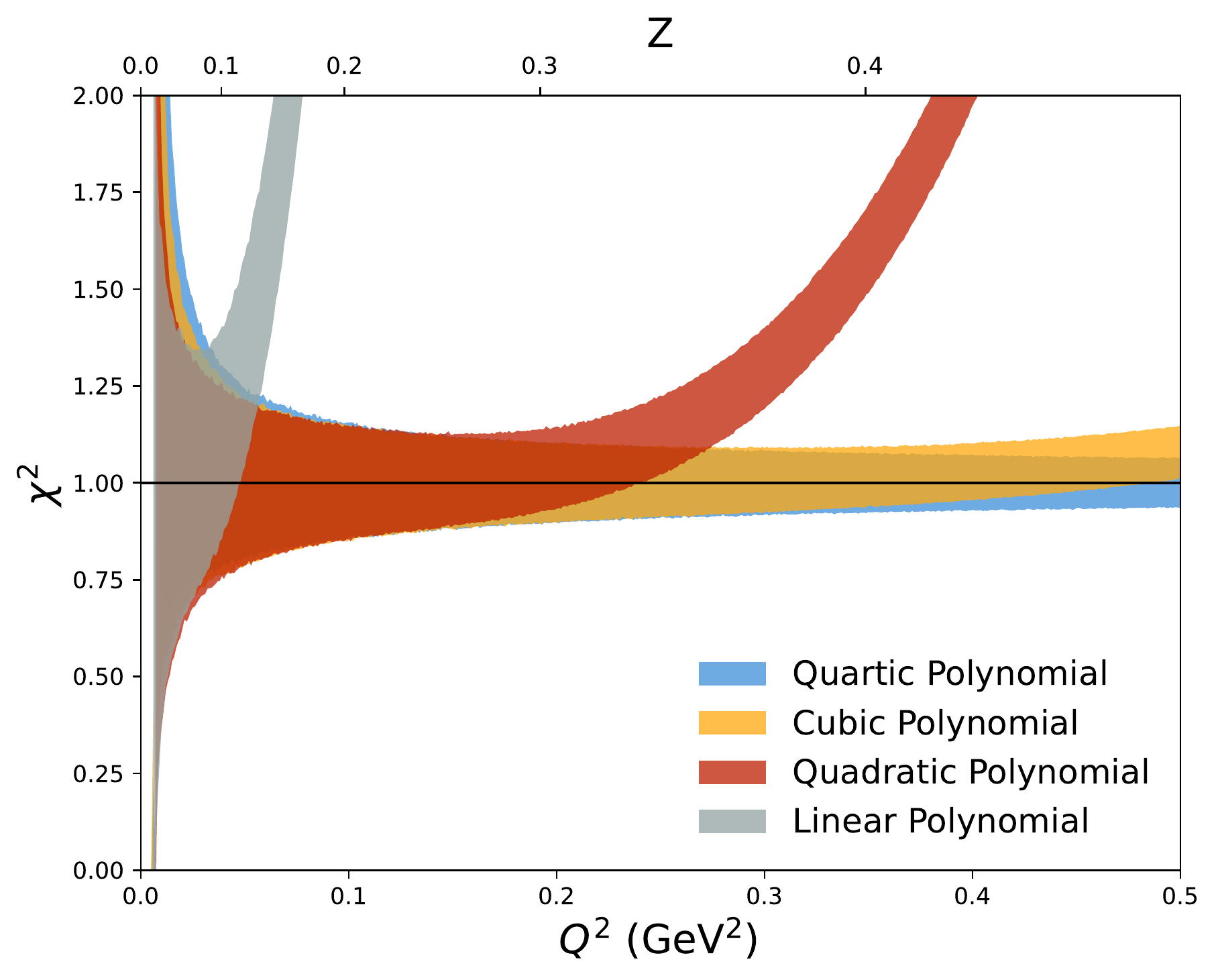}
    \caption{$\chi^2$ plot for bounded polynomial fits in $Z$ of the AMT \GEp{} parameterization}
\end{figure}

\begin{figure}[hb]
    \centering
    \includegraphics[height=0.2\textheight]{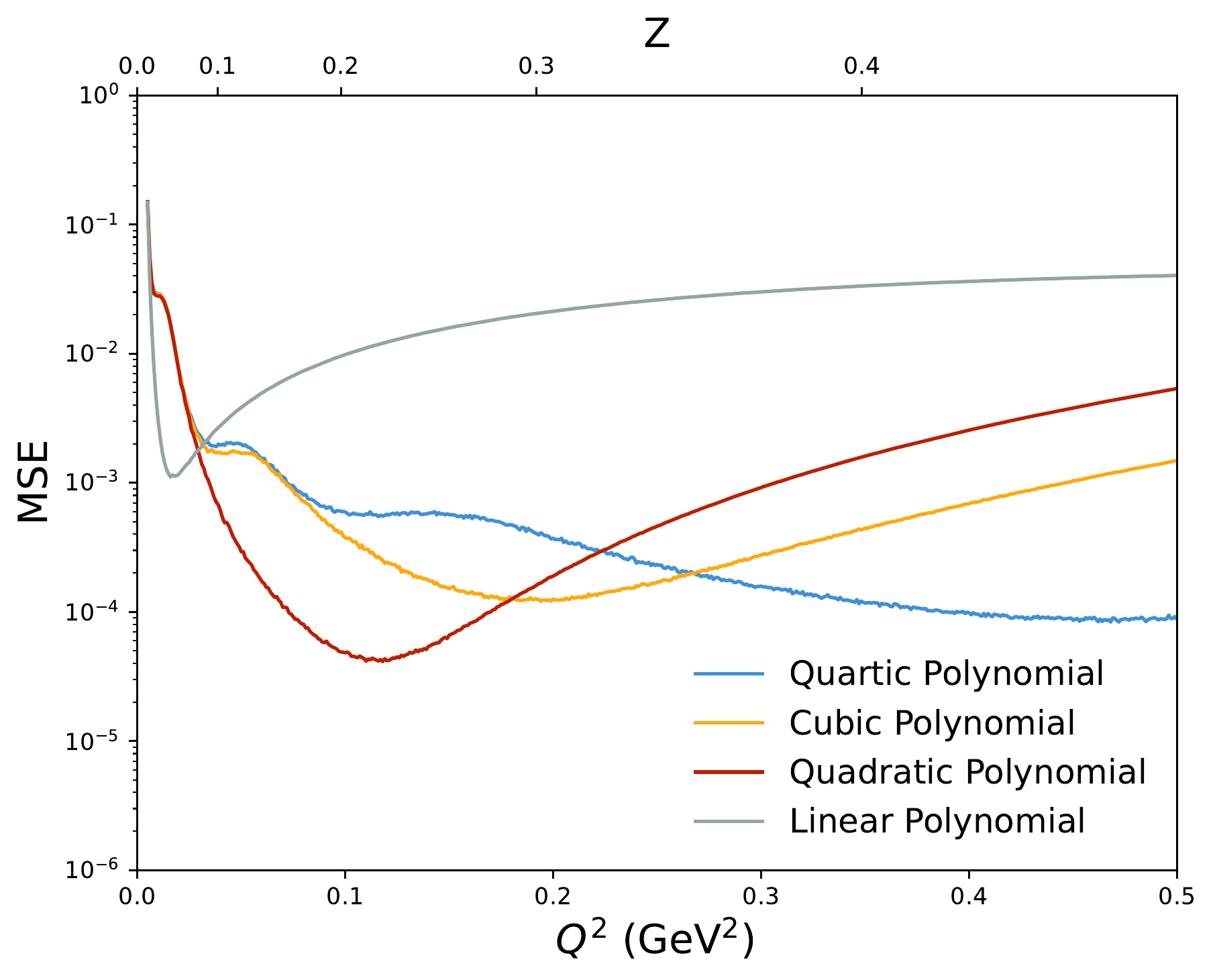}
    \caption{Mean Squared Error plot for bounded polynomial fits in $Z$ of the AMT \GEp{} parameterization}
\end{figure}

\newpage

\subsubsection{Figure Descriptions}

Three plots, each with “$Q^2$ (GeV$^2$)” from 0 to 0.5 on the x axis. A second x axis on top shows the mapping of $Q^2$ to $Z$, which spreads out low $Q^2$ points and bunches up high $Q^2$ points. Each plot has four sets of data shown, one for each power of polynomial used, labeled as “Linear Polynomial”, “Quadratic Polynomial”, “Cubic Polynomial”, and “Quartic Polynomial”.

The top plot has “Extracted Radius - Input Radius (fm)” from -0.1 to 0.1 on the y axis. Two horizontal lines drawn at $\pm$0.04 fm show the magnitude of the proton radius puzzle. Each data set is shown as a band. The low $Q^2$ variance is significantly tempered from the unbound fits in $Z$. However, as $Q^2$ increases the curves are not qualitatively different from the unbound fits.

The middle plot has $\chi^2$ from 0 to 2 on the y axis. Each data set is shown as a band. The data is not qualitatively different than the unbound fits.

The bottom plot has MSE, Mean Squared Error as described in Eq.~\ref{eq:MSE}, from $10^{-6}$ to $10^0$ with a logarithmic scale on the y axis. Each set of data is represented as a curve. The data approaches their minima more rapidly than the unbound fits, but as $Q^2$ increases they are not qualitatively different from the unbound fits.

\newpage

\subsection{Rational(N,M) Fits in $Q^2$}

\begin{figure}[hb]
    \centering
    \includegraphics[height=0.2\textheight]{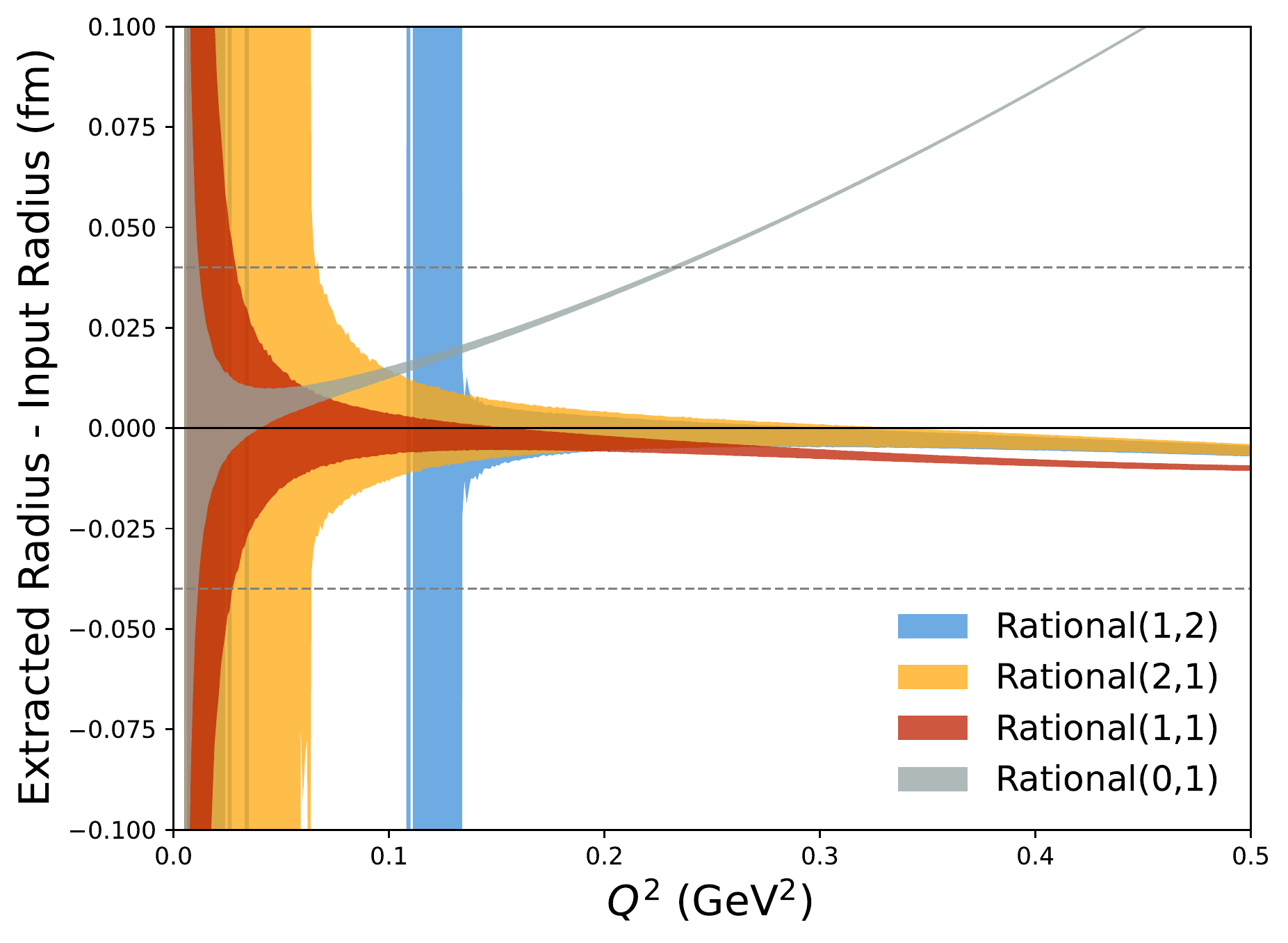}
    \caption{Bias-Variance plot for Rational(N,M) fits in $Q^2$ of the AMT \GEp{} parameterization}
\end{figure}

\begin{figure}[hb]
    \centering
    \includegraphics[height=0.2\textheight]{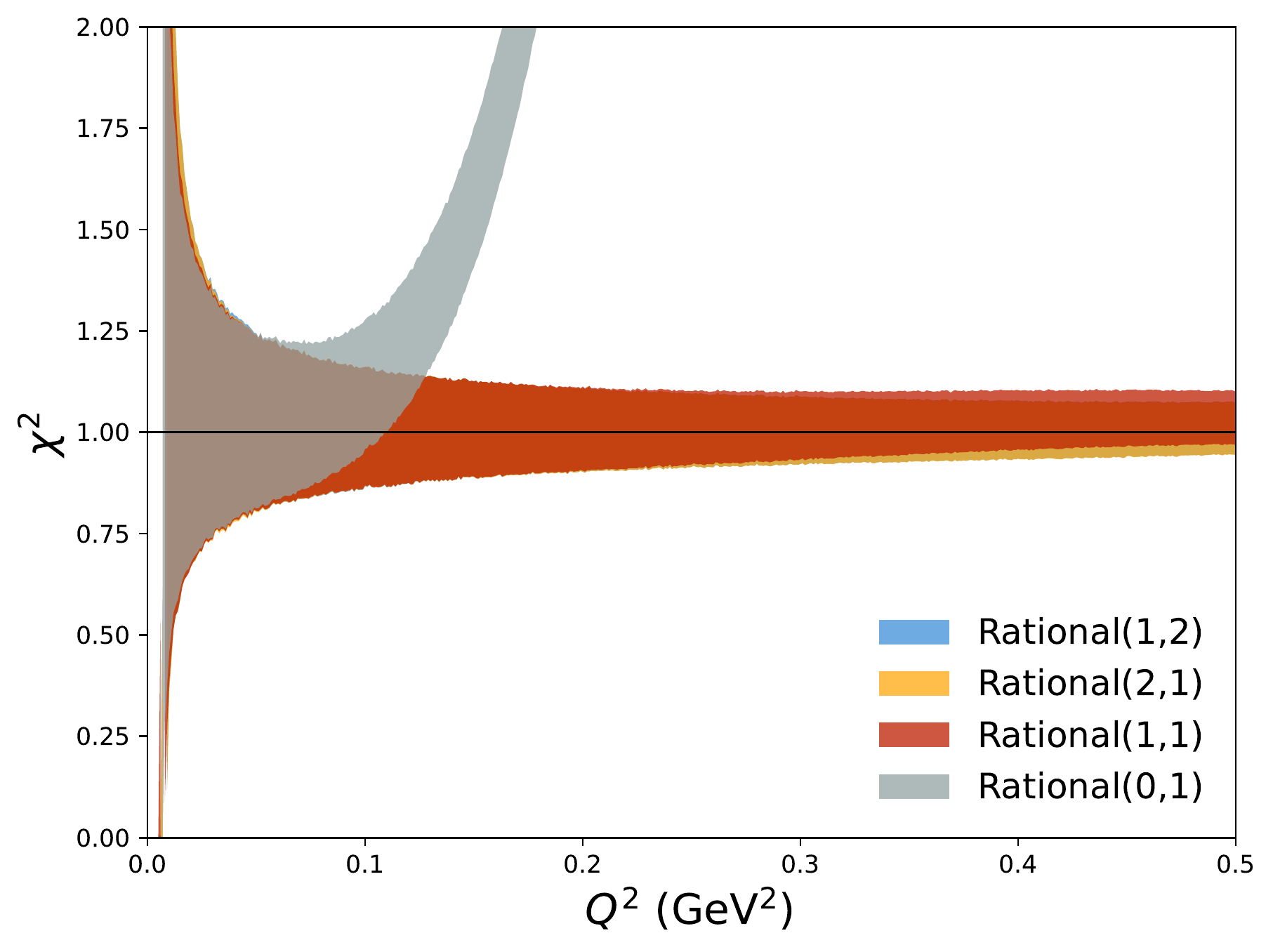}
    \caption{$\chi^2$ plot for Rational(N,M) fits in $Q^2$ of the AMT \GEp{} parameterization}
\end{figure}

\begin{figure}[hb]
    \centering
    \includegraphics[height=0.2\textheight]{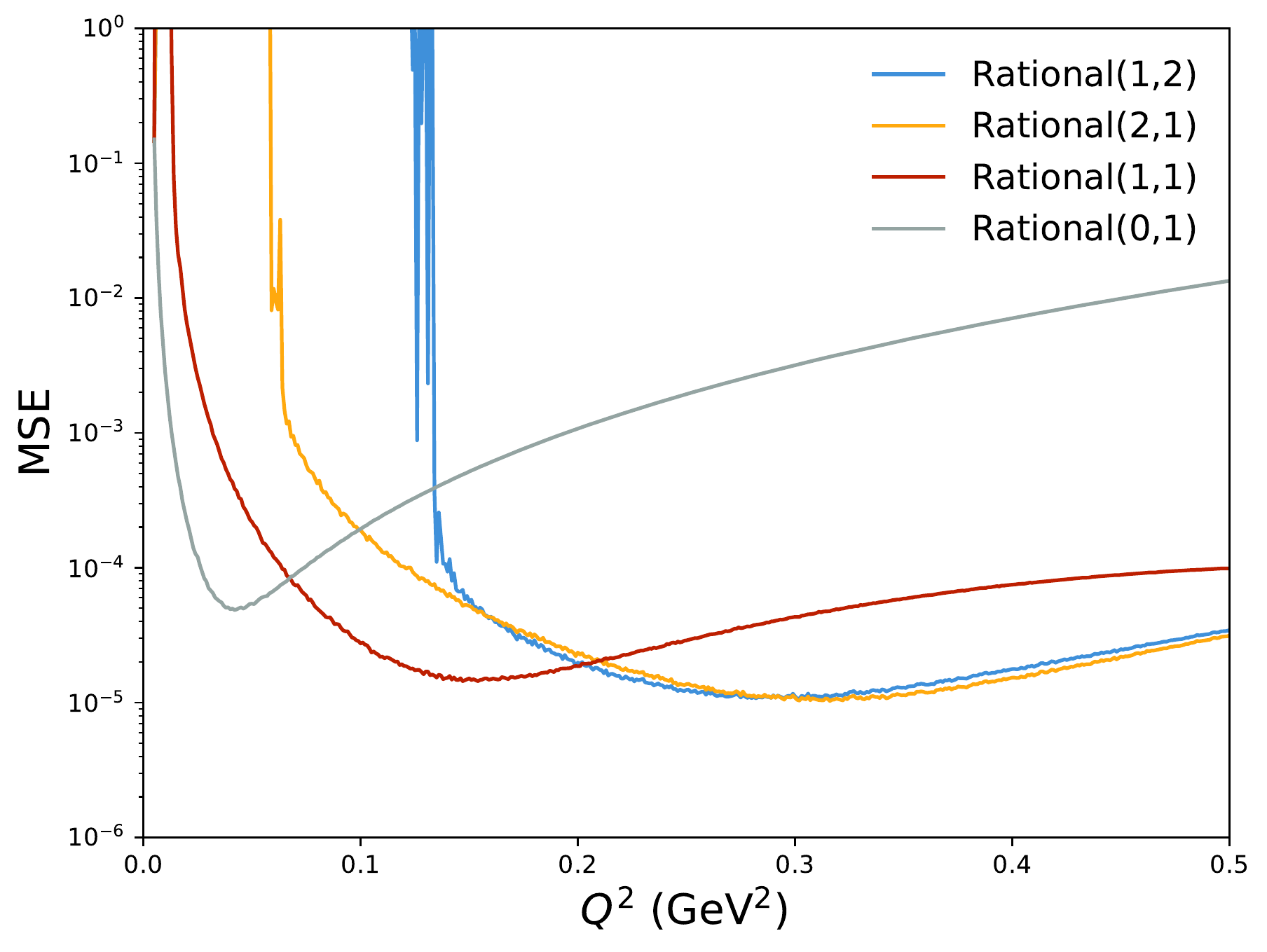}
    \caption{Mean Squared Error plot for Rational(N,M) fits in $Q^2$ of the AMT \GEp{} parameterization}
\end{figure}

\newpage

\subsubsection{Figure Descriptions}

Three plots, each with “$Q^2$ (GeV$^2$)” from 0 to 0.5 on the x axis. Each plot has four sets of data shown, one for each power of Rational(N,M) function used, labeled as “Rational(0,1)”, “Rational(1,1)”, “Rational(2,1)”, and “Rational(1,2)”.

The top plot has “Extracted Radius - Input Radius (fm)” from -0.1 to 0.1 on the y axis. Two horizontal lines drawn at $\pm$0.04 fm show the magnitude of the proton radius puzzle. Each data set is shown as a band. This data shows larger variance at low $Q^2$ than any other fit and the variance increases with increasing N and M, with M having a larger influence on the variance. The Rational(0,1) curve rapidly diverges from the input radius. The Rational(1,2) curve shows no data below $Q^2\approx0.1$ GeV$^2$ as the mean radius was unphysical. Curves with both N and M of at least 1 stay close to the input radius, but trend towards a small radius with increasing $Q^2$. The variances of all Rational(N,M) fits are less than half that of any of the other fits.

The middle plot has $\chi^2$ from 0 to 2 on the y axis. Each data set is shown as a band. All begin centered at 1, with Rational(0,1) having a rapid upturn. The others stay centered at 1 for the full range, with Rational(1,1) slightly increasing at the highest $Q^2$.

The bottom plot has MSE, Mean Squared Error as described in Eq.~\ref{eq:MSE}, from $10^{-6}$ to $10^0$ with a logarithmic scale on the y axis. Each set of data is represented as a curve. The Rational(0,1) fit reaches a minima of $4\cdot10^{-5}$ and then quickly rises. The others approach their minima and then remain near there for the full range. These are the lowest MSE values of all fit types.

\newpage

\section{Arrington}

\subsection{Polynomial Fits in $Q^2$}

\begin{figure}[hb]
    \centering
    \includegraphics[height=0.2\textheight]{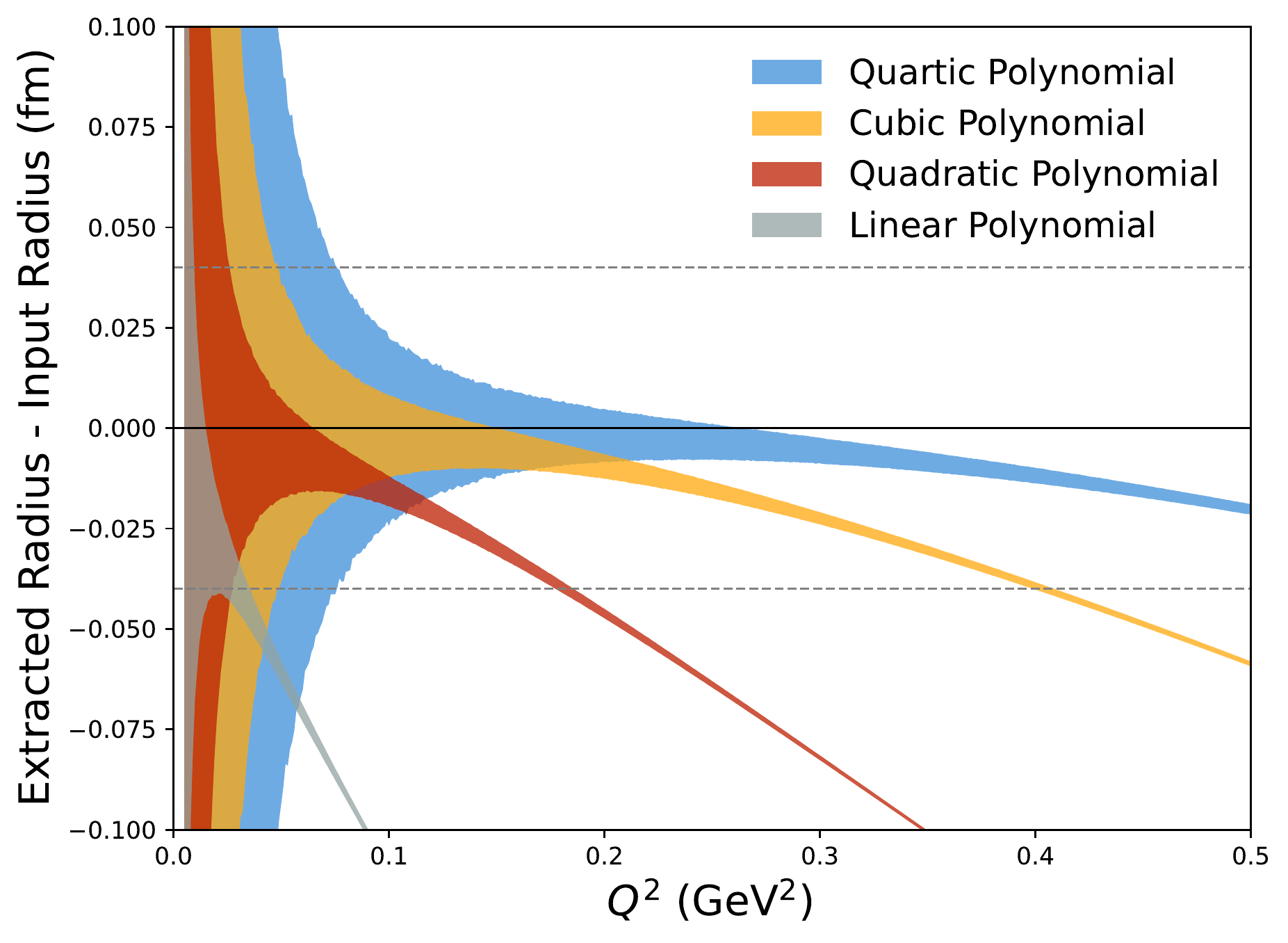}
    \caption{Bias-Variance plot for polynomial fits in $Q^2$ of the Arrington \GEp{} parameterization}
\end{figure}

\begin{figure}[hb]
    \centering
    \includegraphics[height=0.2\textheight]{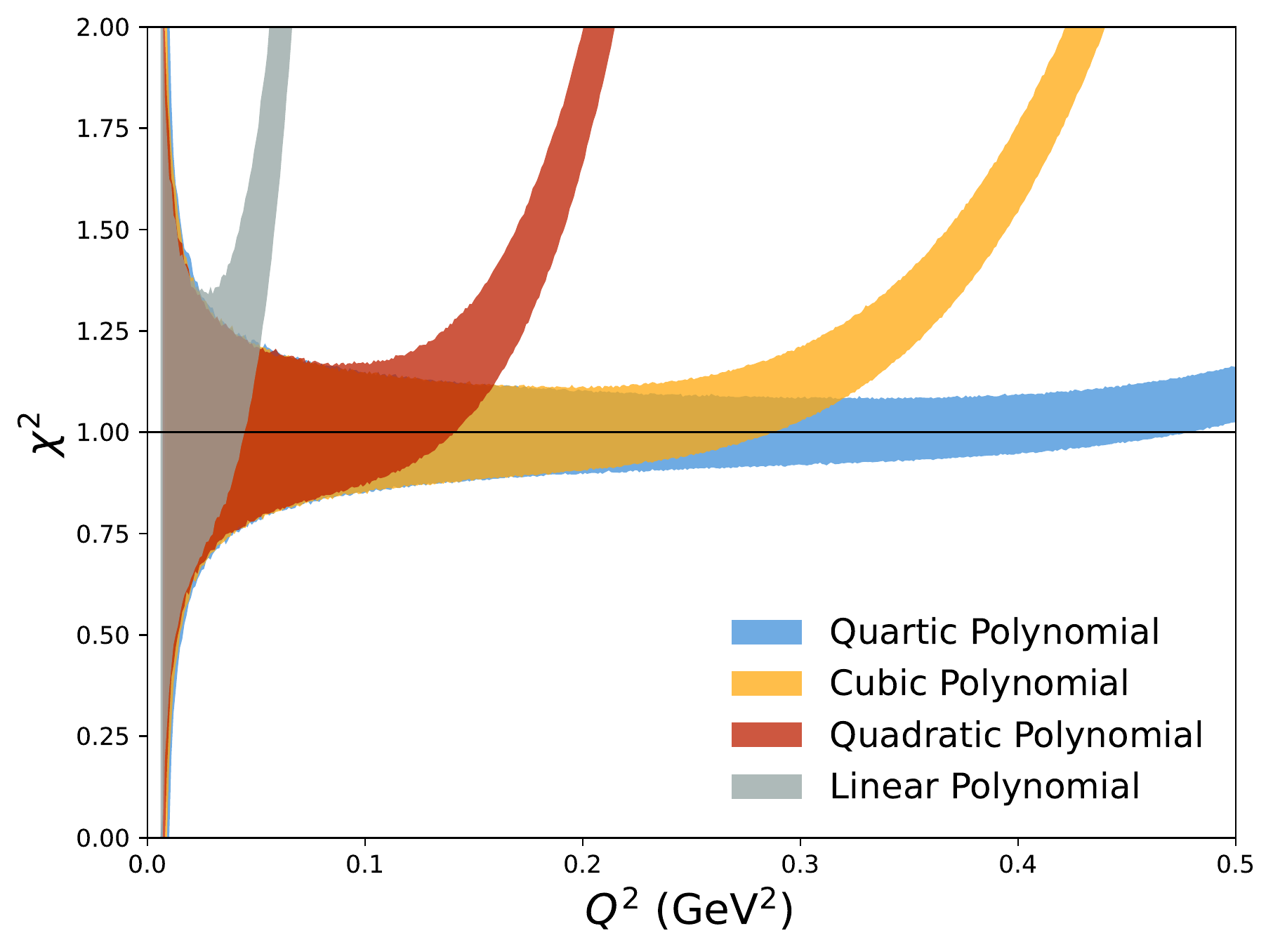}
    \caption{$\chi^2$ plot for polynomial fits in $Q^2$ of the Arrington \GEp{} parameterization}
\end{figure}

\begin{figure}[hb]
    \centering
    \includegraphics[height=0.2\textheight]{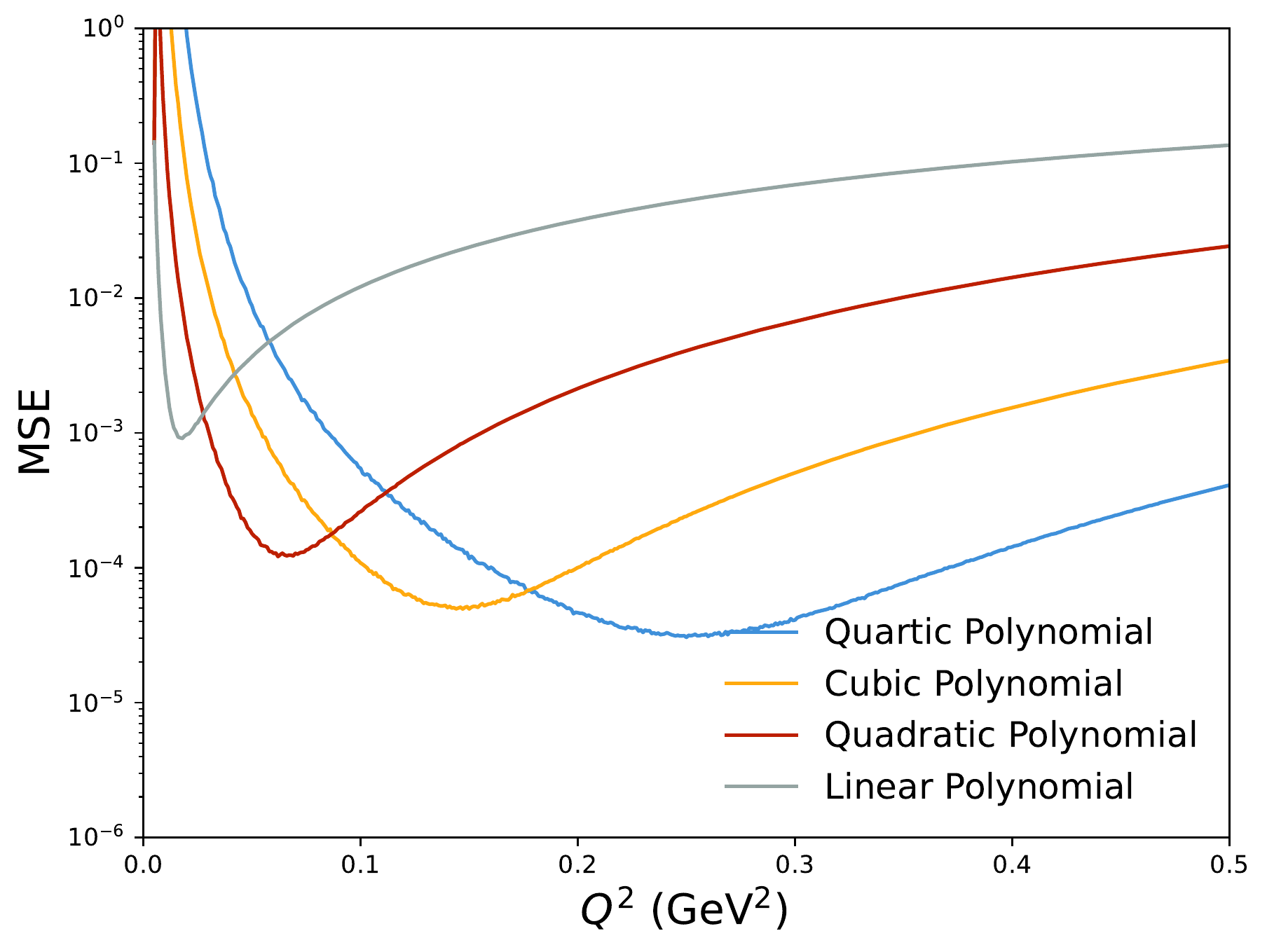}
    \caption{Mean Squared Error plot for polynomial fits in $Q^2$ of the Arrington \GEp{} parameterization}
\end{figure}

\newpage

\subsubsection{Figure Descriptions}

Three plots, each with “$Q^2$ (GeV$^2$)” from 0 to 0.5 on the x axis. Each plot has four sets of data shown, one for each power of polynomial used, labeled as “Linear Polynomial”, “Quadratic Polynomial”, “Cubic Polynomial”, and “Quartic Polynomial”.

The top plot has “Extracted Radius - Input Radius (fm)” from -0.1 to 0.1 on the y axis. Two horizontal lines drawn at $\pm$0.04 fm show the magnitude of the proton radius puzzle. Each data set is shown as a band. All bands have variance higher than the axis scale at small $Q^2$ and systematically return a small radius that decreases with increasing $Q^2$ range. Higher order polynomials decrease slower.

The middle plot has $\chi^2$ from 0 to 2 on the y axis. Each data set is shown as a band. All bands begin centered at 1 and then at some $Q^2$ have a rapid upturn. Higher order polynomials delay the upturn until higher $Q^2$. The deviation from $\chi^2\approx1$ occurs at significantly higher $Q^2$ than when the extracted radius deviates from the input radius.

The bottom plot has MSE, Mean Squared Error as described in Eq.~\ref{eq:MSE}, from $10^{-6}$ to $10^0$ with a logarithmic scale on the y axis. Each set of data is represented as a curve. Each curve rapidly drops to the minimum MSE for the fit and then begins to increase with $Q^2$ and start to level off about two orders of magnitude higher than the minima. Higher order polynomials have lower minima that occur at higher $Q^2$ and then increase slower. The lowest minima is about $3\cdot10^{-5}$.

\newpage

\subsection{Bounded Polynomial Fits in $Q^2$}

\begin{figure}[hb]
    \centering
    \includegraphics[height=0.2\textheight]{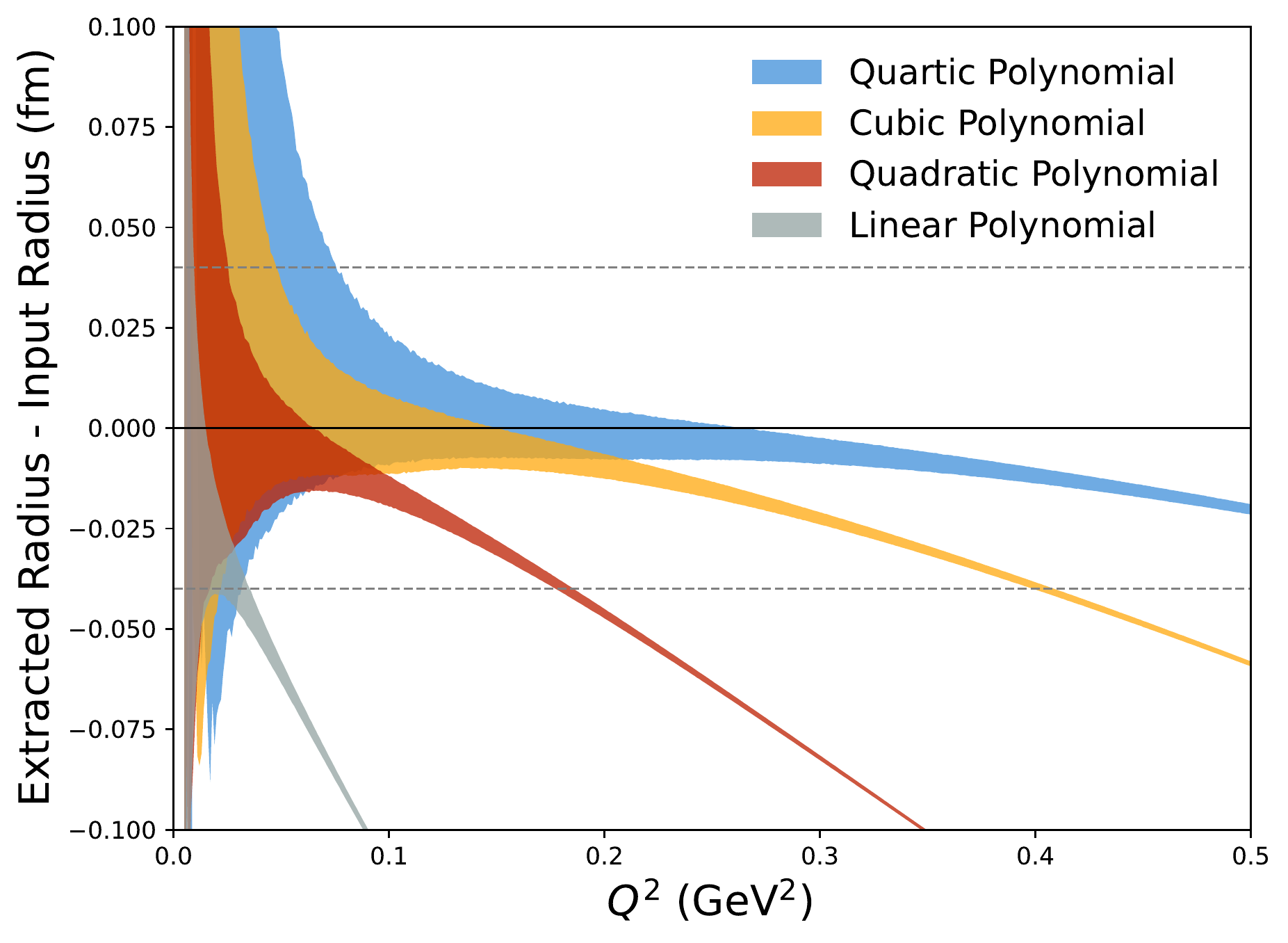}
    \caption{Bias-Variance plot for bounded polynomial fits in $Q^2$ of the Arrington \GEp{} parameterization}
\end{figure}

\begin{figure}[hb]
    \centering
    \includegraphics[height=0.2\textheight]{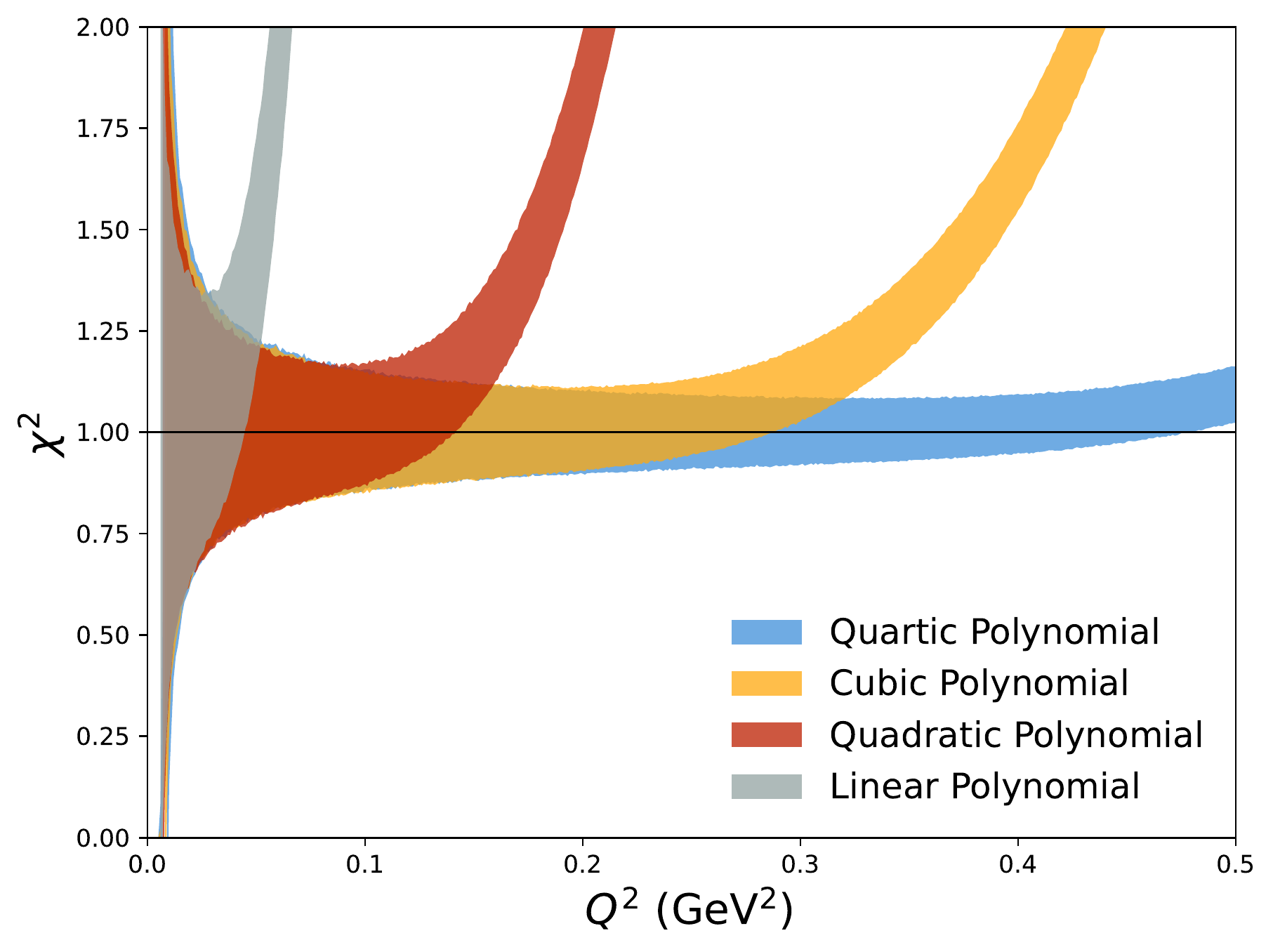}
    \caption{$\chi^2$ plot for bounded polynomial fits in $Q^2$ of the Arrington \GEp{} parameterization}
\end{figure}

\begin{figure}[hb]
    \centering
    \includegraphics[height=0.2\textheight]{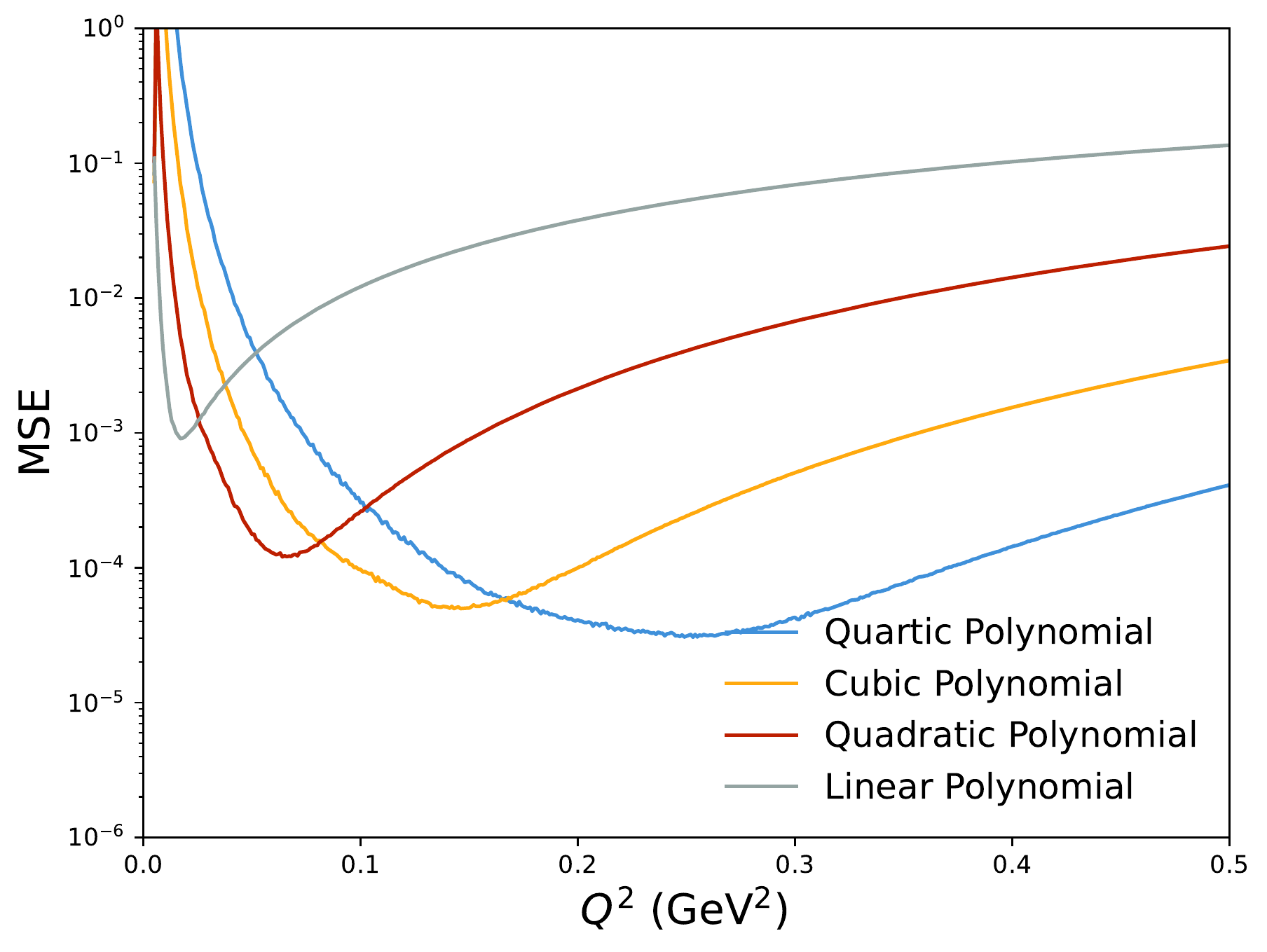}
    \caption{Mean Squared Error plot for bounded polynomial fits in $Q^2$ of the Arrington \GEp{} parameterization}
\end{figure}

\newpage

\subsubsection{Figure Descriptions}

Three plots, each with “$Q^2$ (GeV$^2$)” from 0 to 0.5 on the x axis. Each plot has four sets of data shown, one for each power of polynomial used, labeled as “Linear Polynomial”, “Quadratic Polynomial”, “Cubic Polynomial”, and “Quartic Polynomial”.

The top plot has “Extracted Radius - Input Radius (fm)” from -0.1 to 0.1 on the y axis. Two horizontal lines drawn at $\pm$0.04 fm show the magnitude of the proton radius puzzle. Each data set is shown as a band. The low $Q^2$ variance is smaller than the unbound fits, but is not overall qualitatively different.

The middle plot has $\chi^2$ from 0 to 2 on the y axis. Each data set is shown as a band. The data is not qualitatively different than the unbound fits.

The bottom plot has MSE, Mean Squared Error as described in Eq.~\ref{eq:MSE}, from $10^{-6}$ to $10^0$ with a logarithmic scale on the y axis. Each set of data is represented as a curve. The data is not qualitatively different than the unbound fits.

\newpage

\subsection{Polynomial Fits in $Z$}

\begin{figure}[hb]
    \centering
    \includegraphics[height=0.2\textheight]{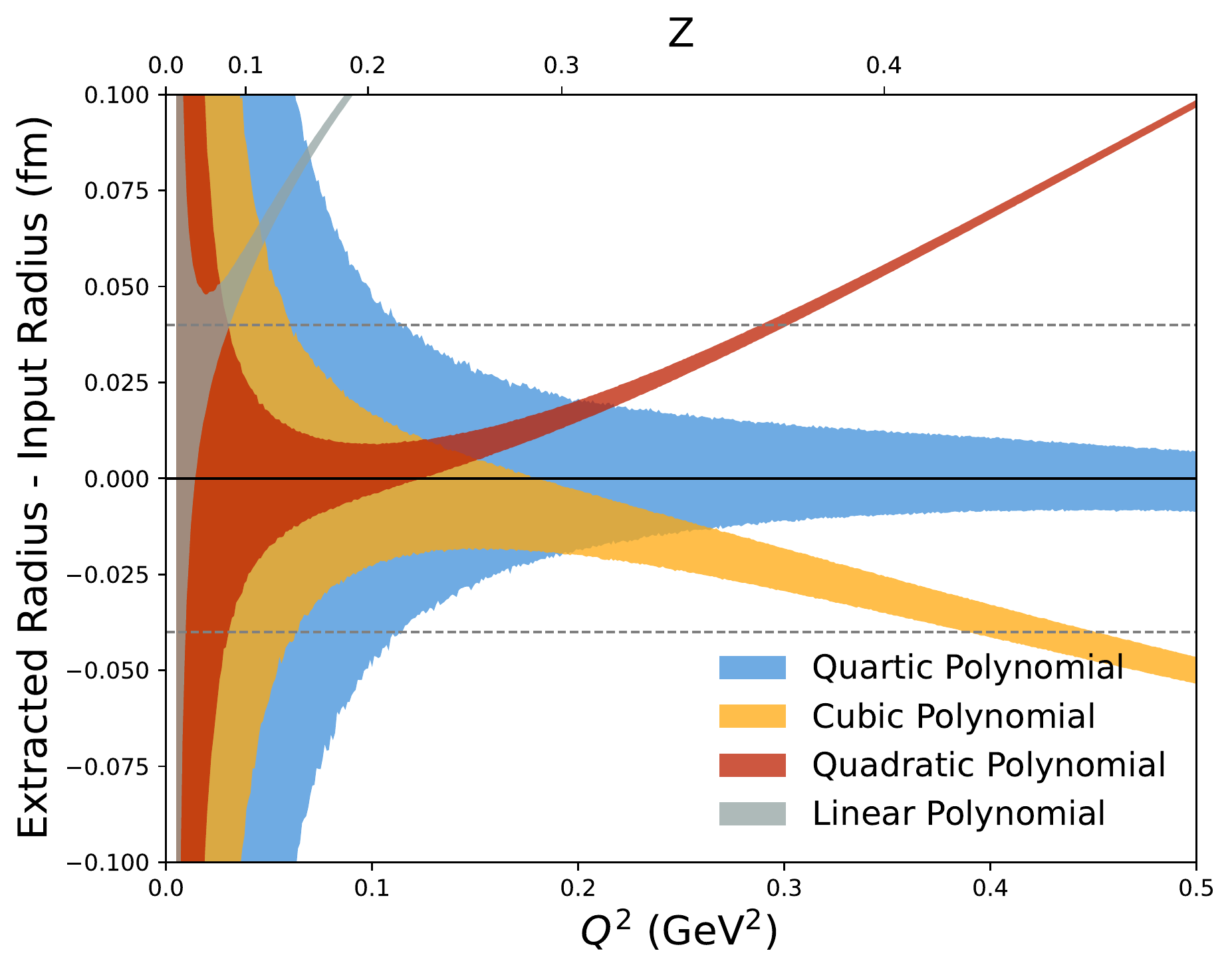}
    \caption{Bias-Variance plot for polynomial fits in $Z$ of the Arrington \GEp{} parameterization}
\end{figure}

\begin{figure}[hb]
    \centering
    \includegraphics[height=0.2\textheight]{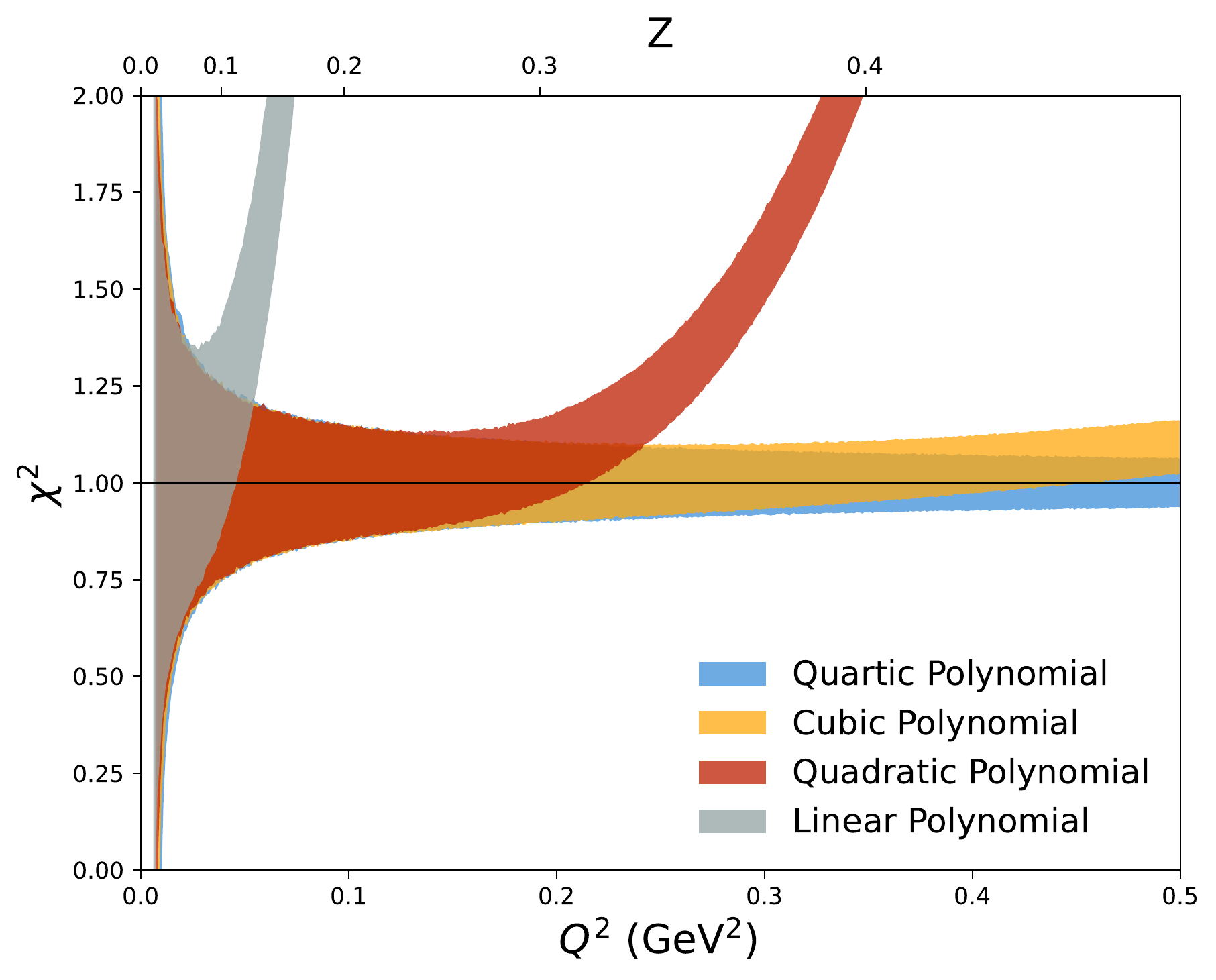}
    \caption{$\chi^2$ plot for polynomial fits in $Z$ of the Arrington \GEp{} parameterization}
\end{figure}

\begin{figure}[hb]
    \centering
    \includegraphics[height=0.2\textheight]{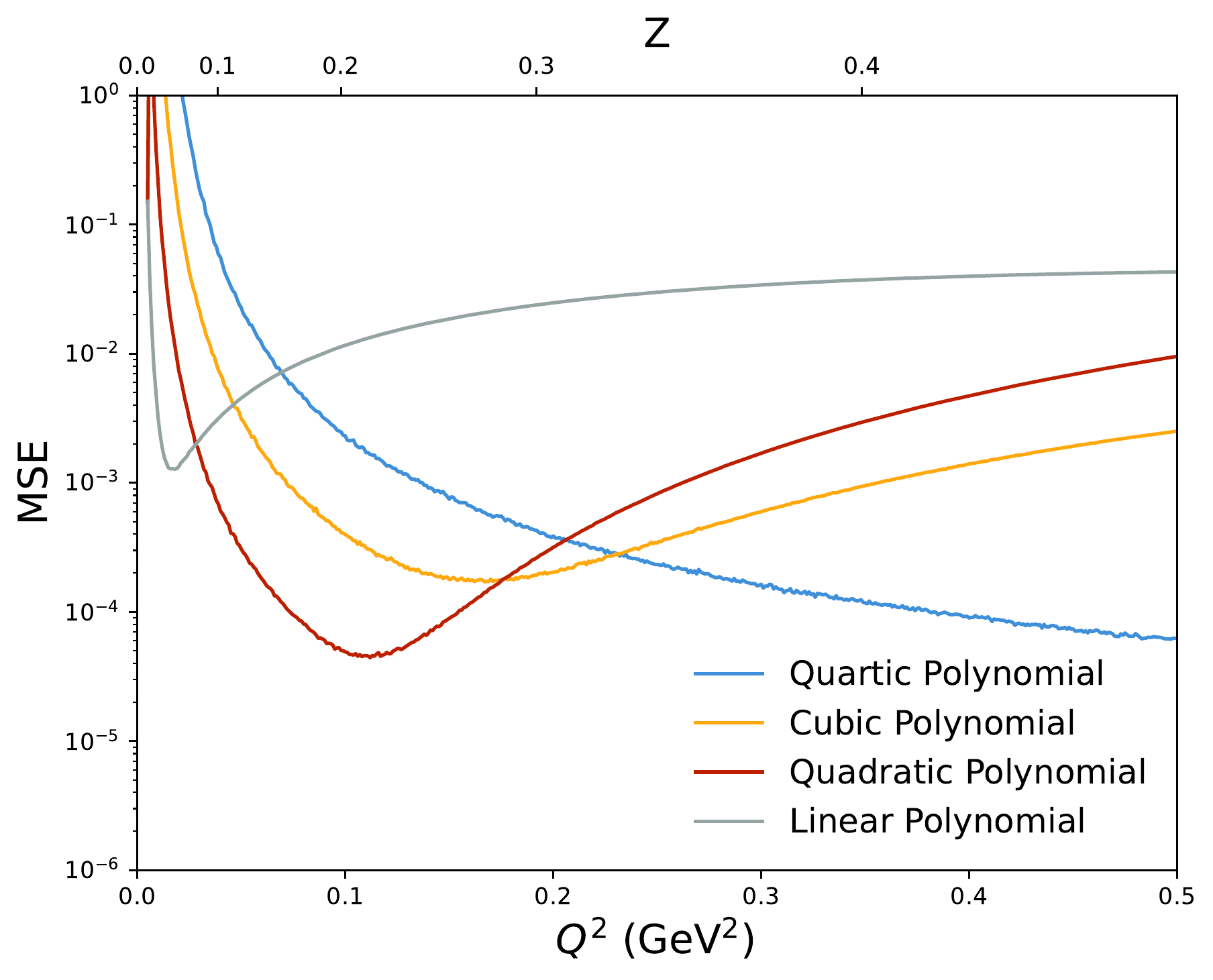}
    \caption{Mean Squared Error plot for polynomial fits in $Z$ of the Arrington \GEp{} parameterization}
\end{figure}

\newpage

\subsubsection{Figure Descriptions}

Three plots, each with “$Q^2$ (GeV$^2$)” from 0 to 0.5 on the x axis. A second x axis on top shows the mapping of $Q^2$ to $Z$, which spreads out low $Q^2$ points and bunches up high $Q^2$ points. Each plot has four sets of data shown, one for each power of polynomial used, labeled as “Linear Polynomial”, “Quadratic Polynomial”, “Cubic Polynomial”, and “Quartic Polynomial”.

The top plot has “Extracted Radius - Input Radius (fm)” from -0.1 to 0.1 on the y axis. Two horizontal lines drawn at $\pm$0.04 fm show the magnitude of the proton radius puzzle. Each data set is shown as a band. The high variance region extends to higher $Q^2$ than fits in $Q^2$ and the variance is higher than the $Q^2$ fits for the full range of the plot. With the exception of the quartic polynomial, the curves rapidly diverge from the input radius at approximately the same $Q^2$ as the fits in $Q^2$, but are better centered at the correct radius at lower $Q^2$. The quartic polynomial remains well centered with a slight indication of trending towards a small radius at high $Q^2$. The linear and quadratic polynomials diverge to large radii and the cubic polynomial diverges to a small radius.

The middle plot has $\chi^2$ from 0 to 2 on the y axis. Each data set is shown as a band. All bands begin centered at 1 and then at some $Q^2$ have a rapid upturn. Higher order polynomials delay the upturn until higher $Q^2$. Much like the fits in $Q^2$, the deviation from $\chi^2\approx1$ occurs at significantly higher $Q^2$ than when the extracted radius deviates from the input radius. The quartic polynomial remains well centered at 1 for the full range.

The bottom plot has MSE, Mean Squared Error as described in Eq.~\ref{eq:MSE}, from $10^{-6}$ to $10^0$ with a logarithmic scale on the y axis. Each set of data is represented as a curve. Each curve rapidly drops to the minimum MSE for the fit and then begins to increase with $Q^2$ and start to level off about two orders of magnitude higher than the minima. The quadratic polynomial has the lowest minimum at about $4\cdot10^{-5}$. The minima are spread over a larger $Q^2$ range and the range grows with polynomial order. The quartic polynomial curve is decreasing over the full range.

\newpage

\subsection{Bounded Polynomial Fits in $Z$}

\begin{figure}[hb]
    \centering
    \includegraphics[height=0.2\textheight]{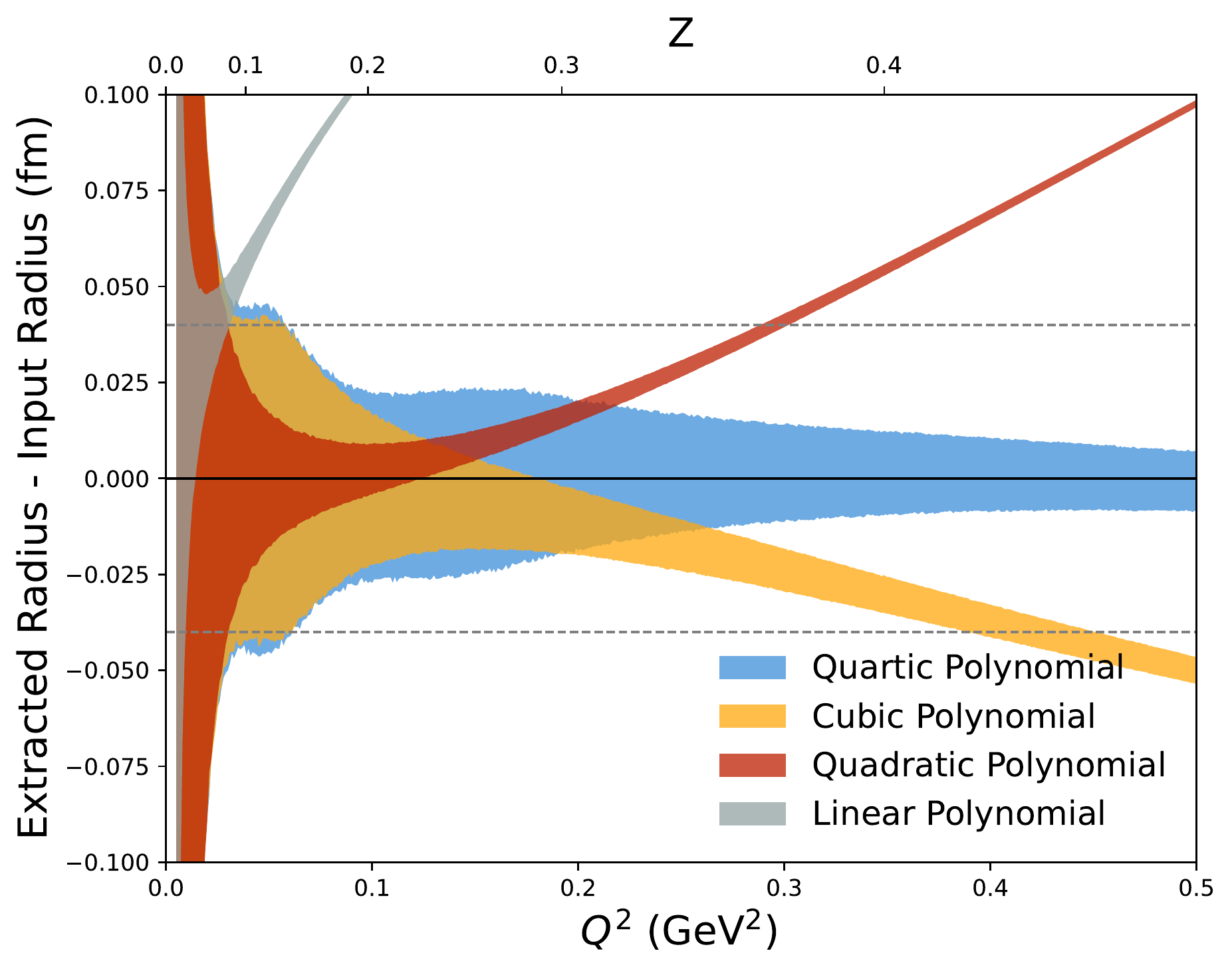}
    \caption{Bias-Variance plot for bounded polynomial fits in $Z$ of the Arrington \GEp{} parameterization}
\end{figure}

\begin{figure}[hb]
    \centering
    \includegraphics[height=0.2\textheight]{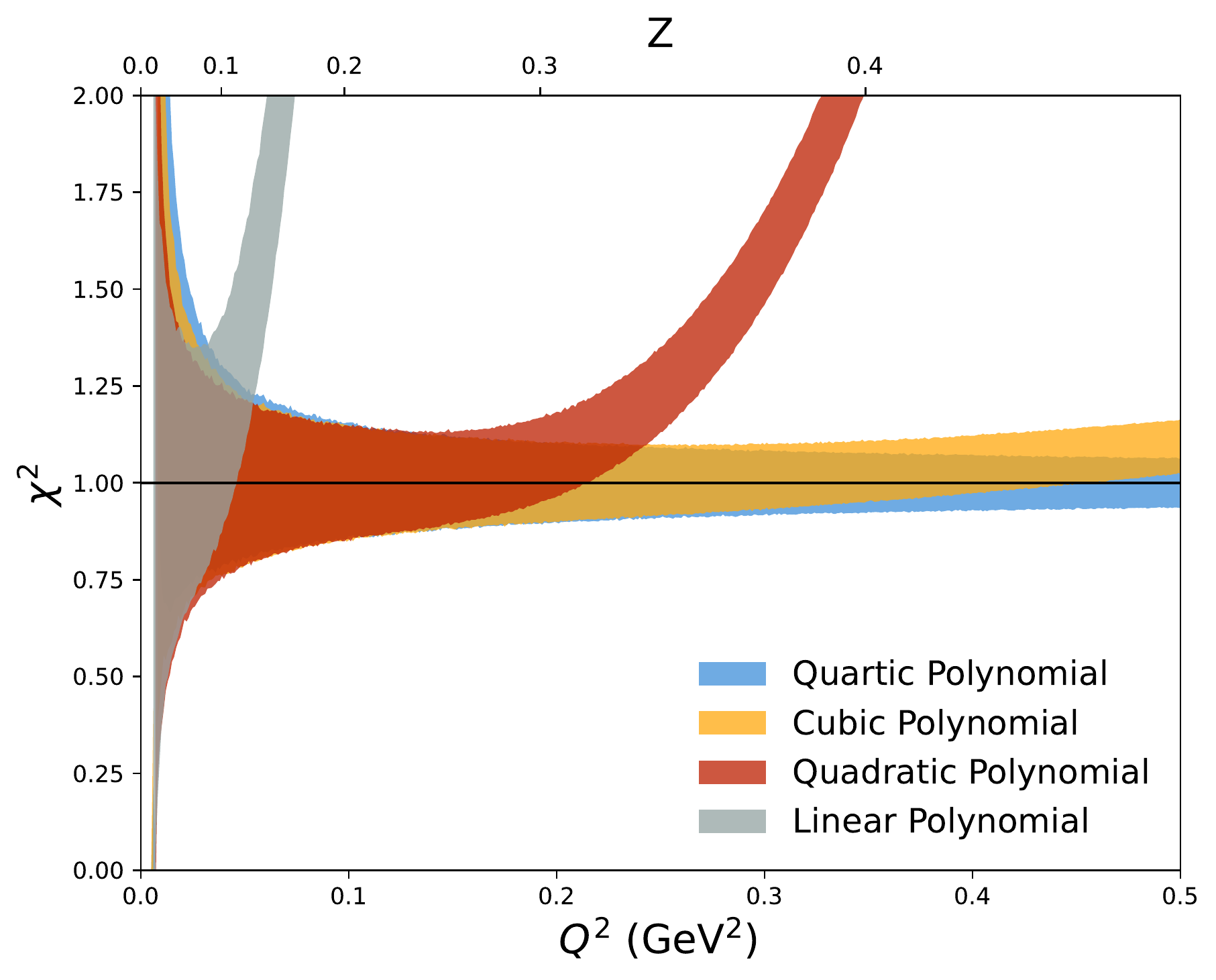}
    \caption{$\chi^2$ plot for bounded polynomial fits in $Z$ of the Arrington \GEp{} parameterization}
\end{figure}

\begin{figure}[hb]
    \centering
    \includegraphics[height=0.2\textheight]{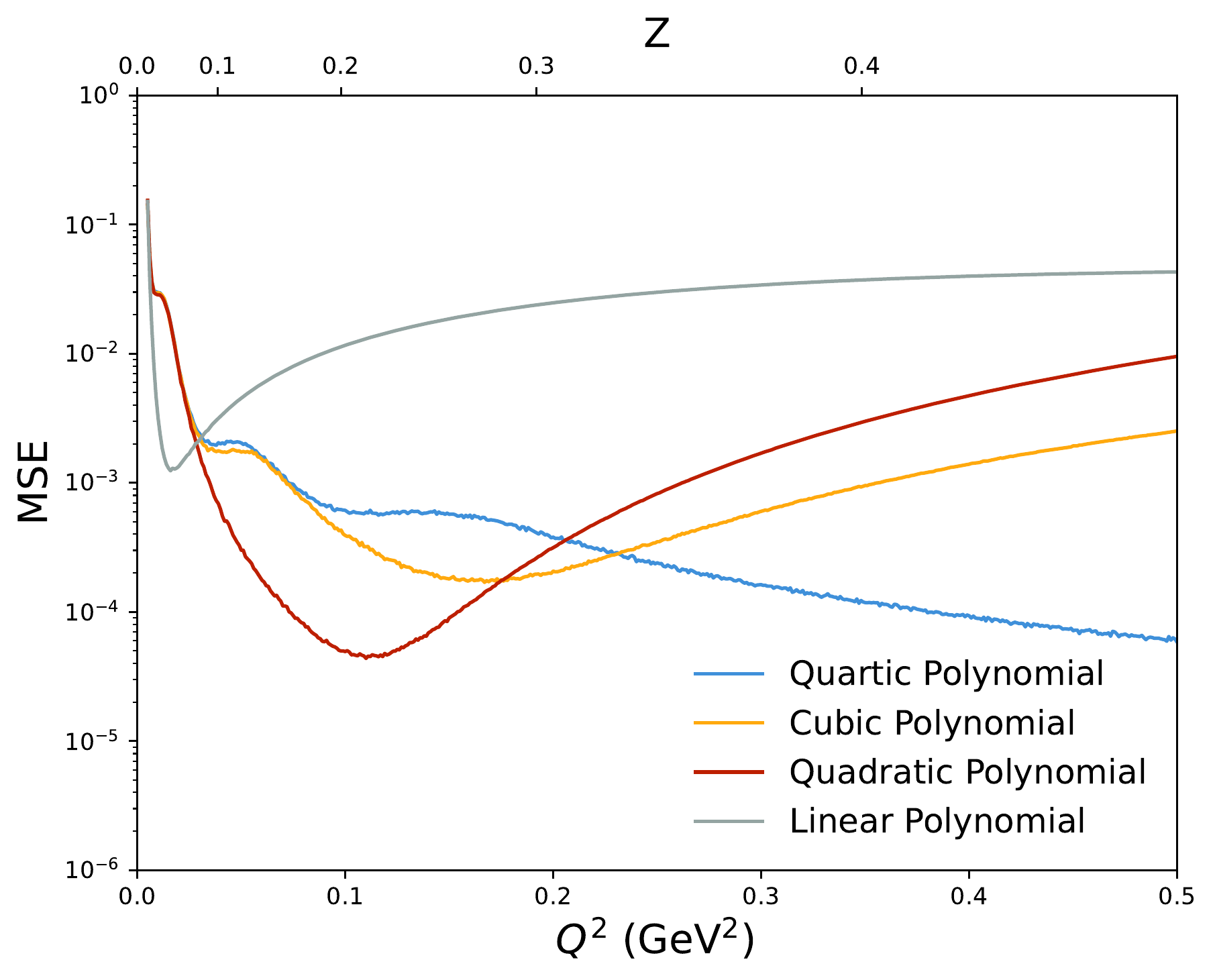}
    \caption{Mean Squared Error plot for bounded polynomial fits in $Z$ of the Arrington \GEp{} parameterization}
\end{figure}

\newpage

\subsubsection{Figure Descriptions}

Three plots, each with “$Q^2$ (GeV$^2$)” from 0 to 0.5 on the x axis. A second x axis on top shows the mapping of $Q^2$ to $Z$, which spreads out low $Q^2$ points and bunches up high $Q^2$ points. Each plot has four sets of data shown, one for each power of polynomial used, labeled as “Linear Polynomial”, “Quadratic Polynomial”, “Cubic Polynomial”, and “Quartic Polynomial”.

The top plot has “Extracted Radius - Input Radius (fm)” from -0.1 to 0.1 on the y axis. Two horizontal lines drawn at $\pm$0.04 fm show the magnitude of the proton radius puzzle. Each data set is shown as a band. The low $Q^2$ variance is significantly tempered from the unbound fits in $Z$. However, as $Q^2$ increases the curves are not qualitatively different from the unbound fits.

The middle plot has $\chi^2$ from 0 to 2 on the y axis. Each data set is shown as a band. The data is not qualitatively different than the unbound fits.

The bottom plot has MSE, Mean Squared Error as described in Eq.~\ref{eq:MSE}, from $10^{-6}$ to $10^0$ with a logarithmic scale on the y axis. Each set of data is represented as a curve. The data approaches their minima more rapidly than the unbound fits, but as $Q^2$ increases they are not qualitatively different from the unbound fits.

\newpage

\subsection{Rational(N,M) Fits in $Q^2$}

\begin{figure}[hb]
    \centering
    \includegraphics[height=0.2\textheight]{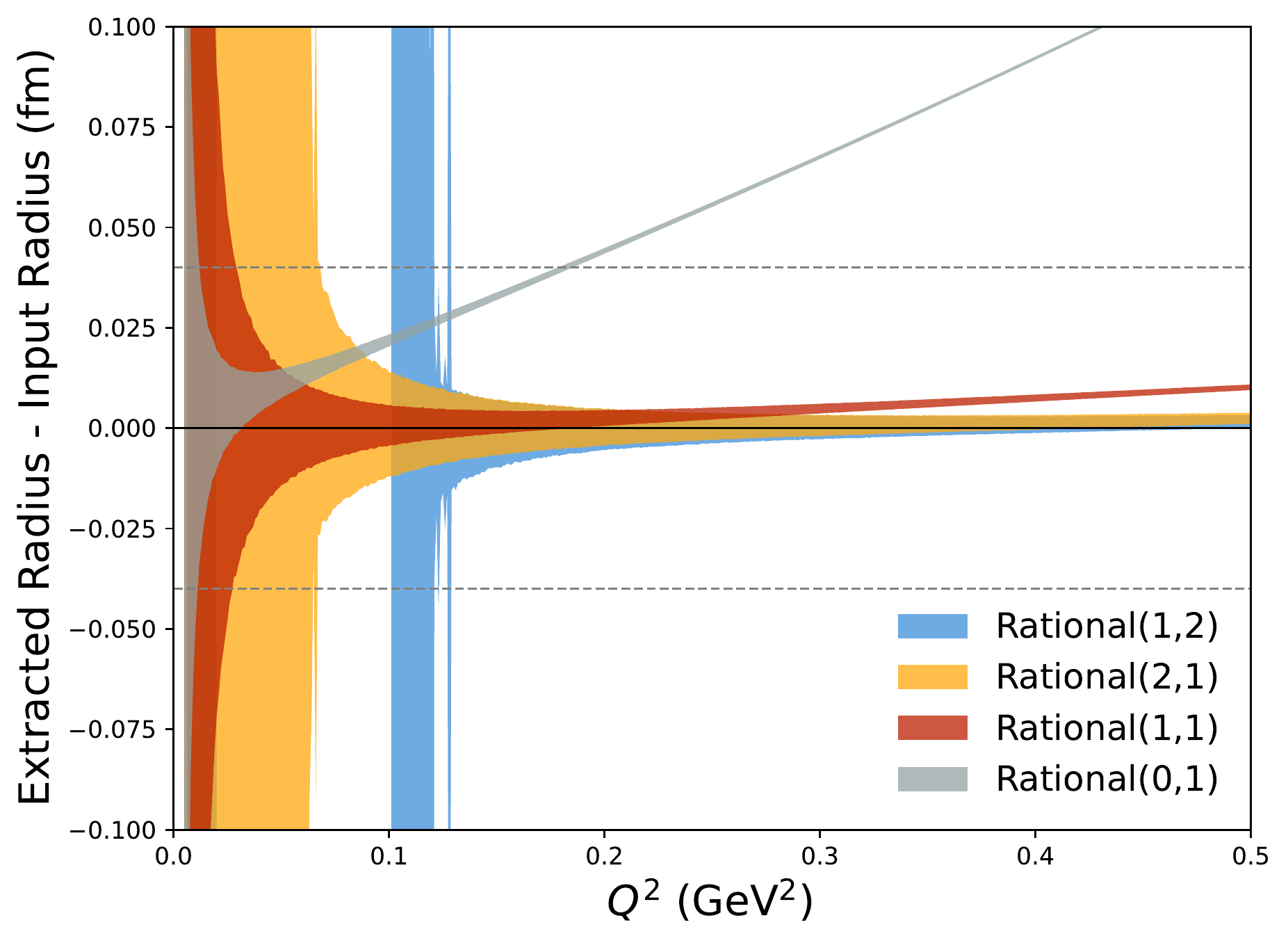}
    \caption{Bias-Variance plot for Rational(N,M) fits in $Q^2$ of the Arrington \GEp{} parameterization}
\end{figure}

\begin{figure}[hb]
    \centering
    \includegraphics[height=0.2\textheight]{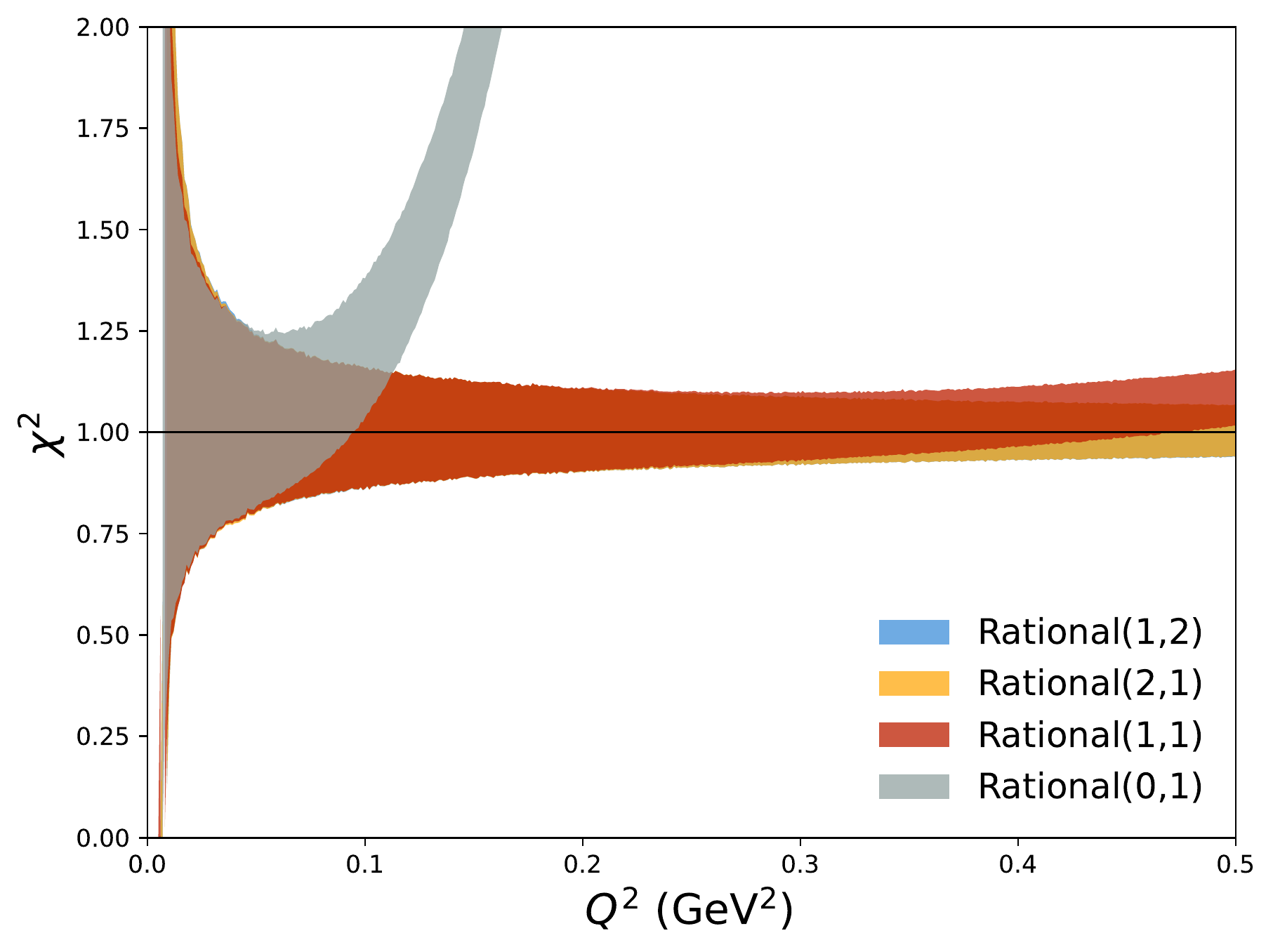}
    \caption{$\chi^2$ plot for Rational(N,M) fits in $Q^2$ of the Arrington \GEp{} parameterization}
\end{figure}

\begin{figure}[hb]
    \centering
    \includegraphics[height=0.2\textheight]{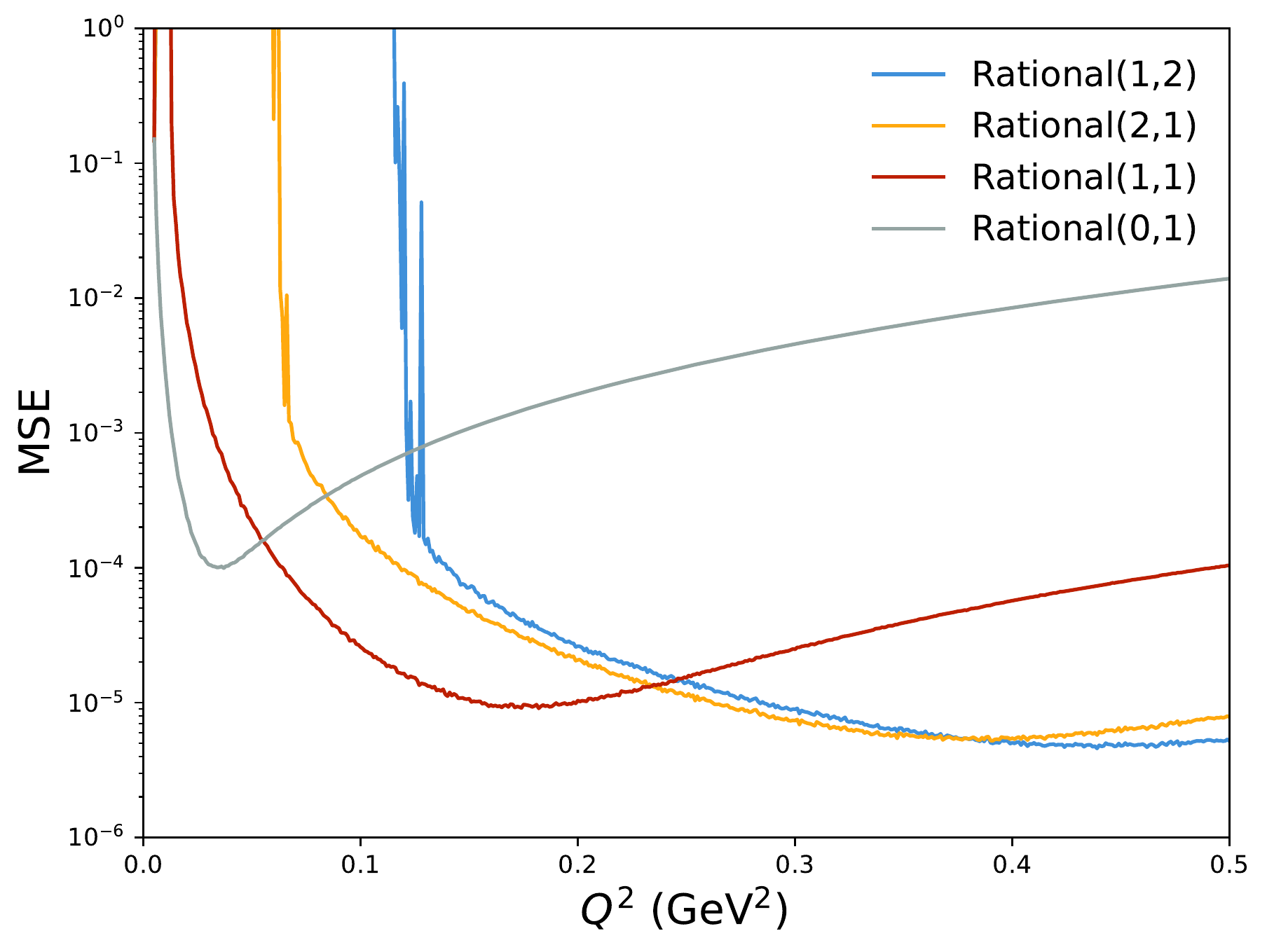}
    \caption{Mean Squared Error plot for Rational(N,M) fits in $Q^2$ of the Arrington \GEp{} parameterization}
\end{figure}

\newpage

\subsubsection{Figure Descriptions}

Three plots, each with “$Q^2$ (GeV$^2$)” from 0 to 0.5 on the x axis. Each plot has four sets of data shown, one for each power of Rational(N,M) function used, labeled as “Rational(0,1)”, “Rational(1,1)”, “Rational(2,1)”, and “Rational(1,2)”.

The top plot has “Extracted Radius - Input Radius (fm)” from -0.1 to 0.1 on the y axis. Two horizontal lines drawn at $\pm$0.04 fm show the magnitude of the proton radius puzzle. Each data set is shown as a band. This data shows larger variance at low $Q^2$ than any other fit and the variance increases with increasing N and M, with M having a larger influence on the variance. The Rational(0,1) curve rapidly diverges from the input radius. The Rational(1,2) curve shows no data below $Q^2\approx0.1$ GeV$^2$ as the mean radius was unphysical. Curves with both N and M of at least 1 stay very close to the input radius, but trend towards a large radius with increasing $Q^2$. The variances of all Rational(N,M) fits are less than half that of any of the other fits.

The middle plot has $\chi^2$ from 0 to 2 on the y axis. Each data set is shown as a band. All begin centered at 1, with Rational(0,1) having a rapid upturn. The others stay centered at 1 for the full range, with Rational(1,1) slightly increasing at the highest $Q^2$.

The bottom plot has MSE, Mean Squared Error as described in Eq.~\ref{eq:MSE}, from $10^{-6}$ to $10^0$ with a logarithmic scale on the y axis. Each set of data is represented as a curve. The Rational(0,1) fit reaches a minima of $1\cdot10^{-5}$ and then quickly rises. The others approach their minima and then remain near there for the full range. These are the lowest MSE values of all fit types.

\newpage

\section{Bernauer}

\subsection{Polynomial Fits in $Q^2$}

\begin{figure}[hb]
    \centering
    \includegraphics[height=0.2\textheight]{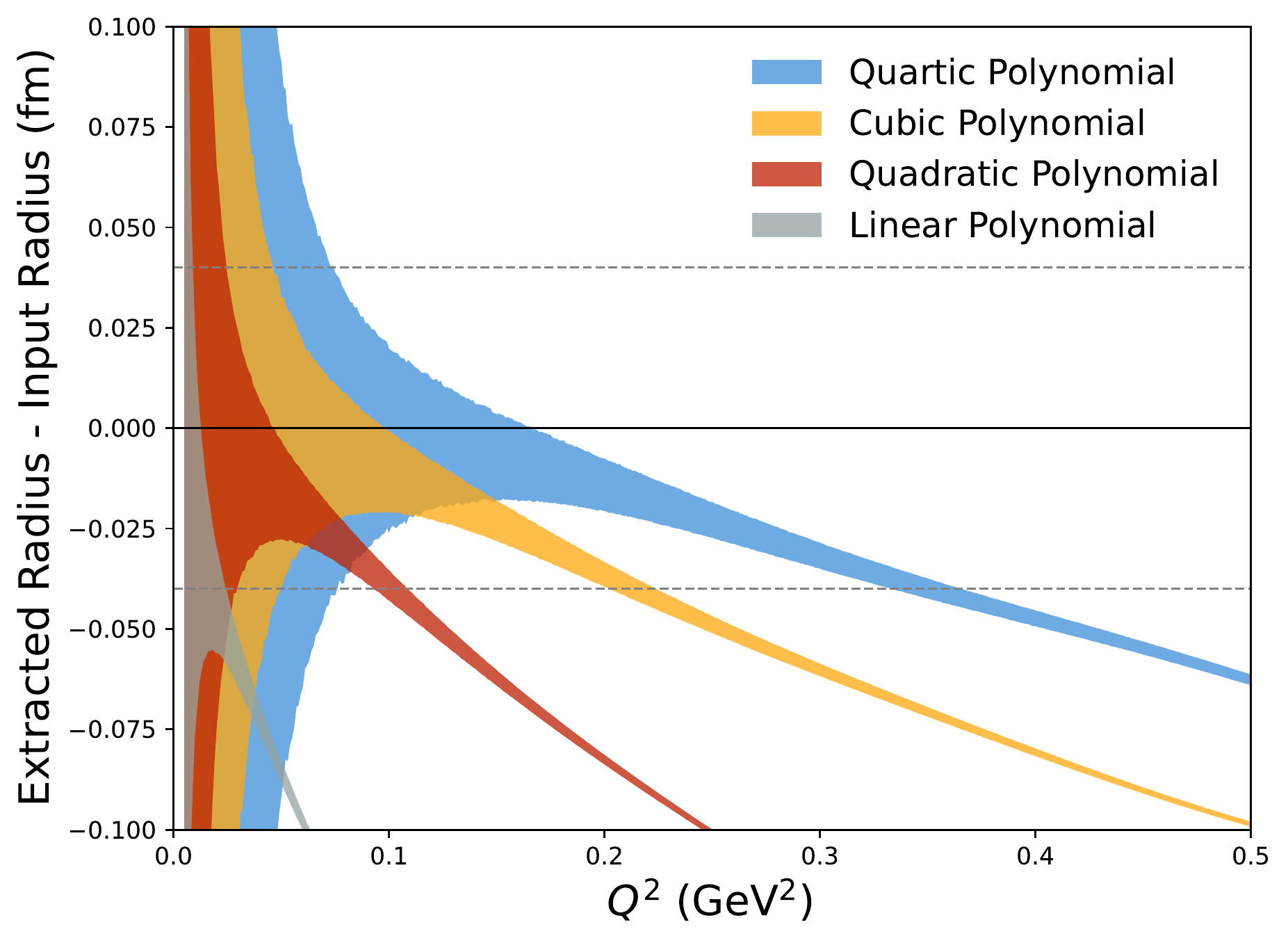}
    \caption{Bias-Variance plot for polynomial fits in $Q^2$ of the Bernauer \GEp{} parameterization}
\end{figure}

\begin{figure}[hb]
    \centering
    \includegraphics[height=0.2\textheight]{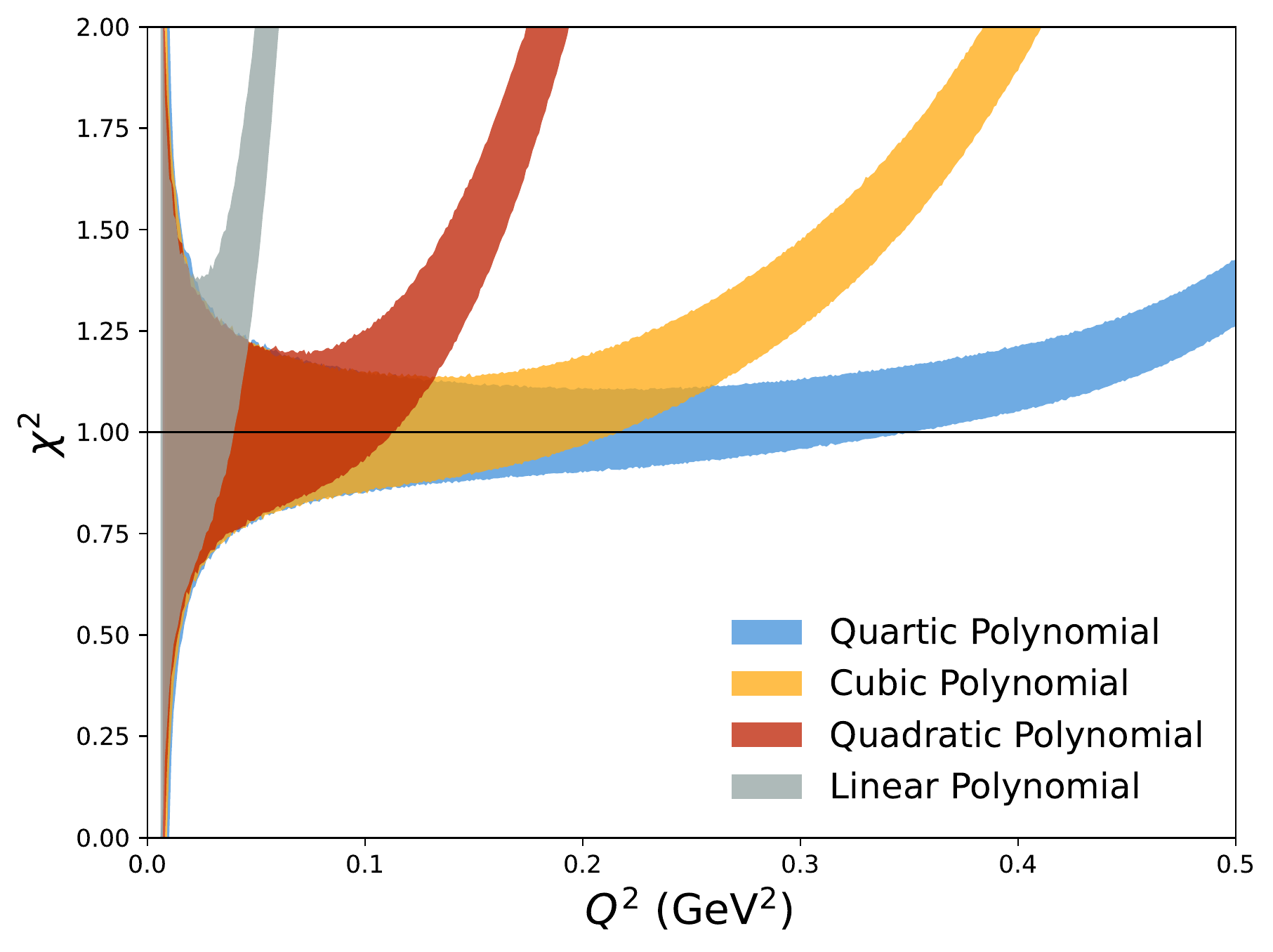}
    \caption{$\chi^2$ plot for polynomial fits in $Q^2$ of the Bernauer \GEp{} parameterization}
\end{figure}

\begin{figure}[hb]
    \centering
    \includegraphics[height=0.2\textheight]{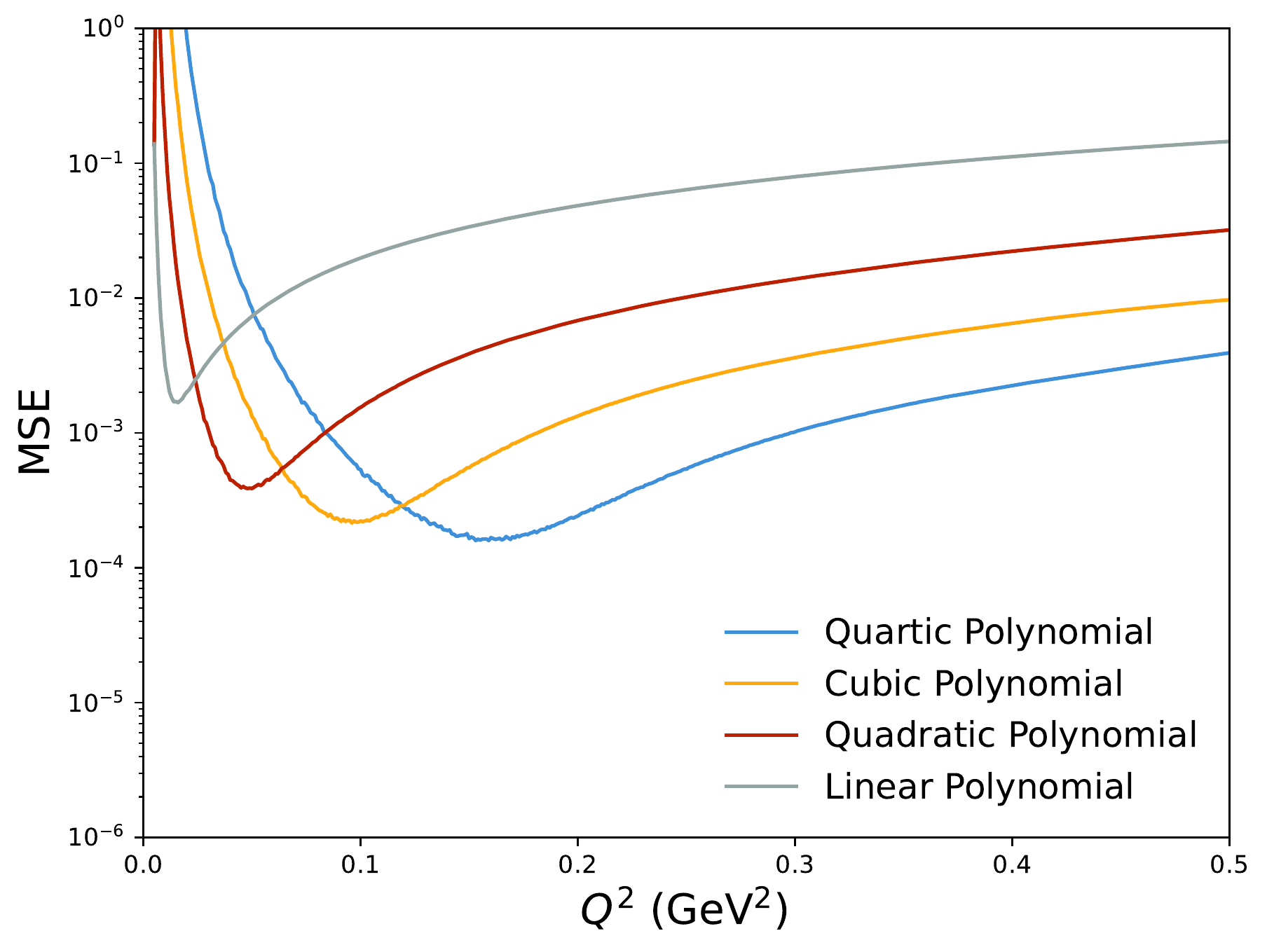}
    \caption{Mean Squared Error plot for polynomial fits in $Q^2$ of the Bernauer \GEp{} parameterization}
\end{figure}

\newpage

\subsubsection{Figure Descriptions}

Three plots, each with “$Q^2$ (GeV$^2$)” from 0 to 0.5 on the x axis. Each plot has four sets of data shown, one for each power of polynomial used, labeled as “Linear Polynomial”, “Quadratic Polynomial”, “Cubic Polynomial”, and “Quartic Polynomial”.

The top plot has “Extracted Radius - Input Radius (fm)” from -0.1 to 0.1 on the y axis. Two horizontal lines drawn at $\pm$0.04 fm show the magnitude of the proton radius puzzle. Each data set is shown as a band. All bands have variance higher than the axis scale at small $Q^2$ and systematically return a small radius that decreases with increasing $Q^2$ range. The bands deviate from the input radius much faster than with other parameterizations.

The middle plot has $\chi^2$ from 0 to 2 on the y axis. Each data set is shown as a band. All bands begin centered at 1 and then at some $Q^2$ have a rapid upturn. Higher order polynomials delay the upturn until higher $Q^2$. The deviation from $\chi^2\approx1$ occurs at significantly higher $Q^2$ than when the extracted radius deviates from the input radius.

The bottom plot has MSE, Mean Squared Error as described in Eq.~\ref{eq:MSE}, from $10^{-6}$ to $10^0$ with a logarithmic scale on the y axis. Each set of data is represented as a curve. Each curve rapidly drops to the minimum MSE for the fit and then begins to increase with $Q^2$ and start to level off about two orders of magnitude higher than the minima. Higher order polynomials have lower minima that occur at higher $Q^2$ and then increase slower. The lowest minima is about $2\cdot10^{-4}$.

\newpage

\subsection{Bounded Polynomial Fits in $Q^2$}

\begin{figure}[hb]
    \centering
    \includegraphics[height=0.2\textheight]{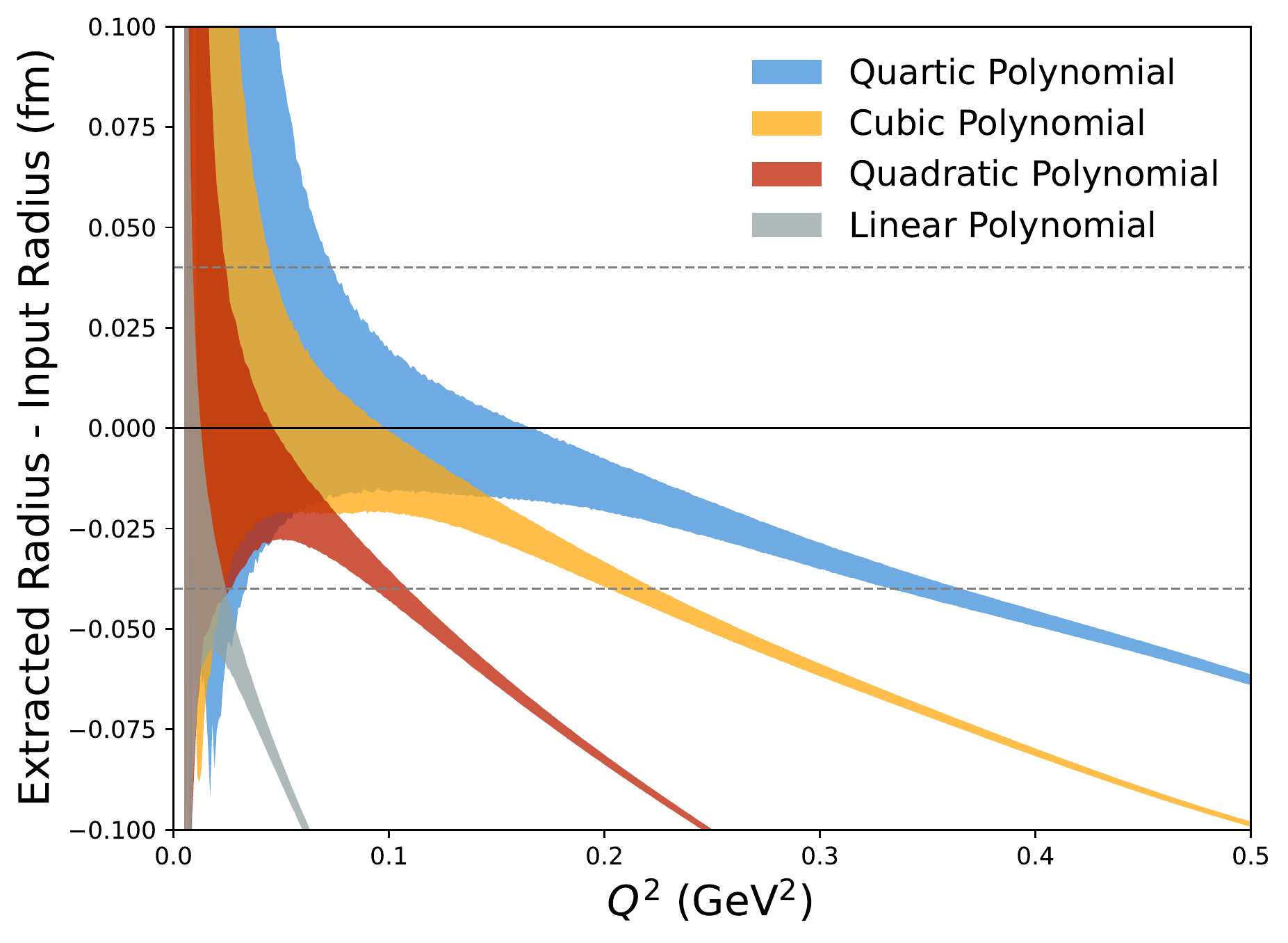}
    \caption{Bias-Variance plot for bounded polynomial fits in $Q^2$ of the Bernauer \GEp{} parameterization}
\end{figure}

\begin{figure}[hb]
    \centering
    \includegraphics[height=0.2\textheight]{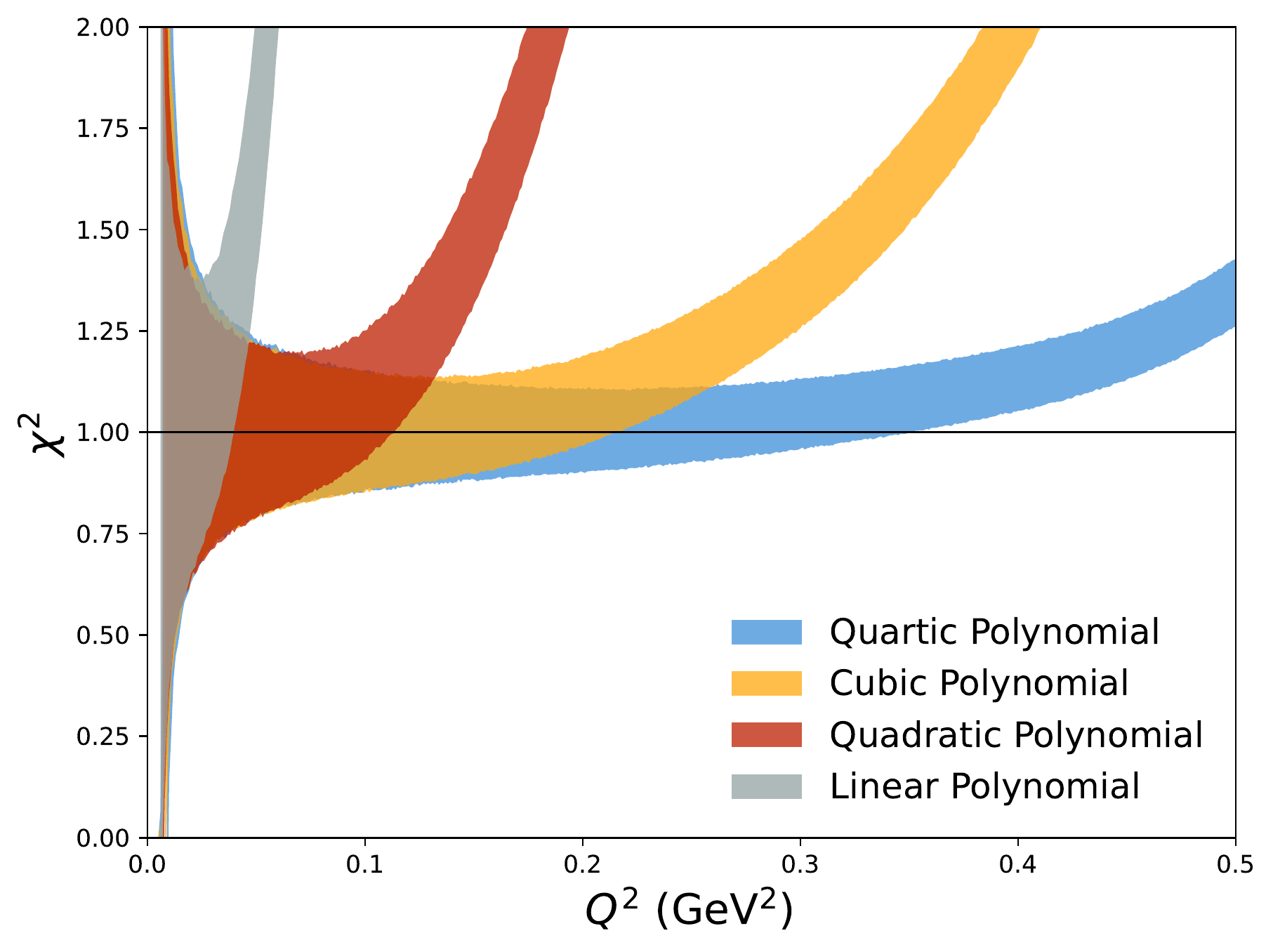}
    \caption{$\chi^2$ plot for bounded polynomial fits in $Q^2$ of the Bernauer \GEp{} parameterization}
\end{figure}

\begin{figure}[hb]
    \centering
    \includegraphics[height=0.2\textheight]{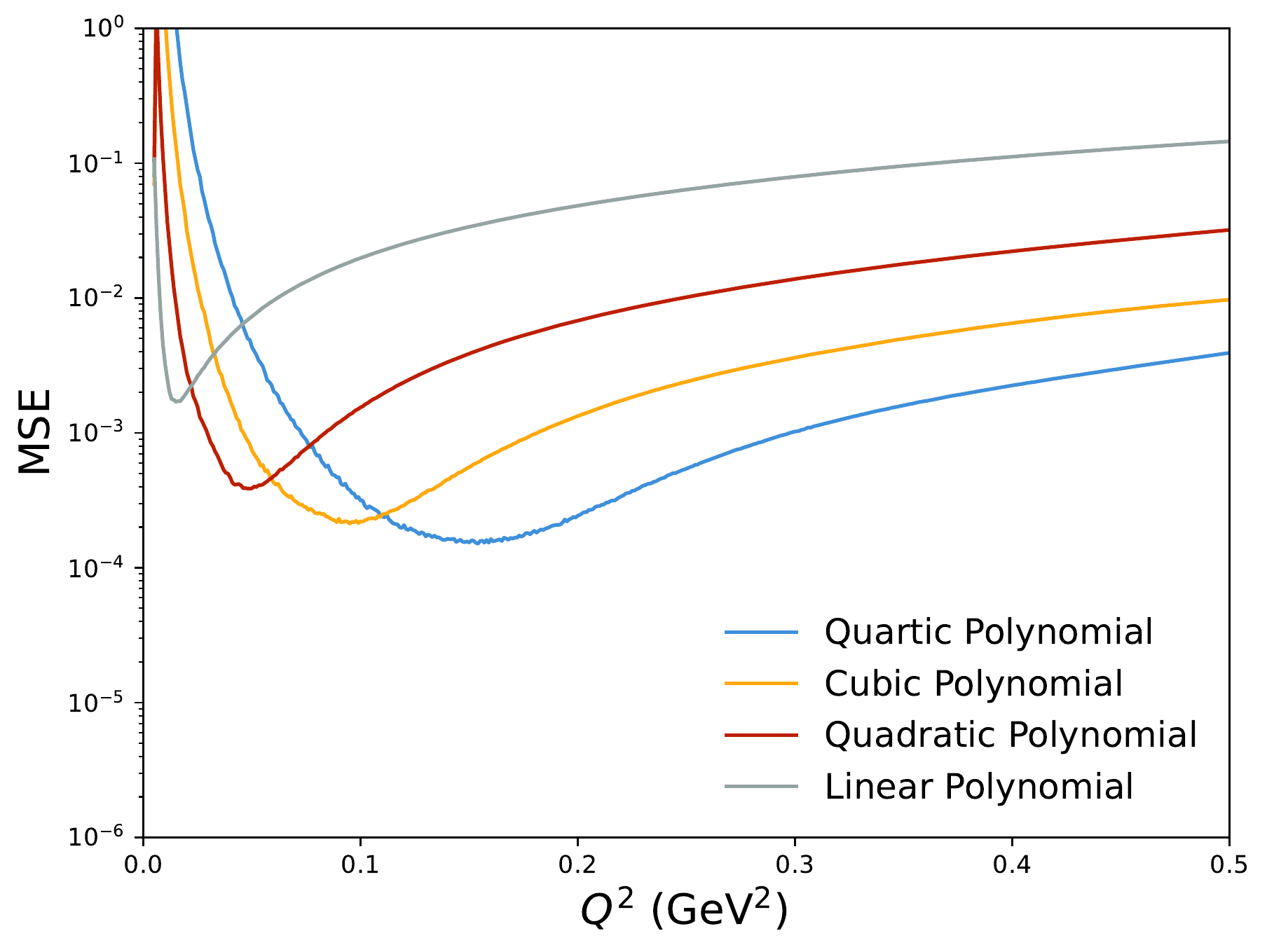}
    \caption{Mean Squared Error plot for bounded polynomial fits in $Q^2$ of the Bernauer \GEp{} parameterization}
\end{figure}

\newpage

\subsubsection{Figure Descriptions}

Three plots, each with “$Q^2$ (GeV$^2$)” from 0 to 0.5 on the x axis. Each plot has four sets of data shown, one for each power of polynomial used, labeled as “Linear Polynomial”, “Quadratic Polynomial”, “Cubic Polynomial”, and “Quartic Polynomial”.

The top plot has “Extracted Radius - Input Radius (fm)” from -0.1 to 0.1 on the y axis. Two horizontal lines drawn at $\pm$0.04 fm show the magnitude of the proton radius puzzle. Each data set is shown as a band. The low $Q^2$ variance is smaller than the unbound fits, but is not overall qualitatively different.

The middle plot has $\chi^2$ from 0 to 2 on the y axis. Each data set is shown as a band. The data is not qualitatively different than the unbound fits.

The bottom plot has MSE, Mean Squared Error as described in Eq.~\ref{eq:MSE}, from $10^{-6}$ to $10^0$ with a logarithmic scale on the y axis. Each set of data is represented as a curve. The data is not qualitatively different than the unbound fits.

\newpage

\subsection{Polynomial Fits in $Z$}

\begin{figure}[hb]
    \centering
    \includegraphics[height=0.2\textheight]{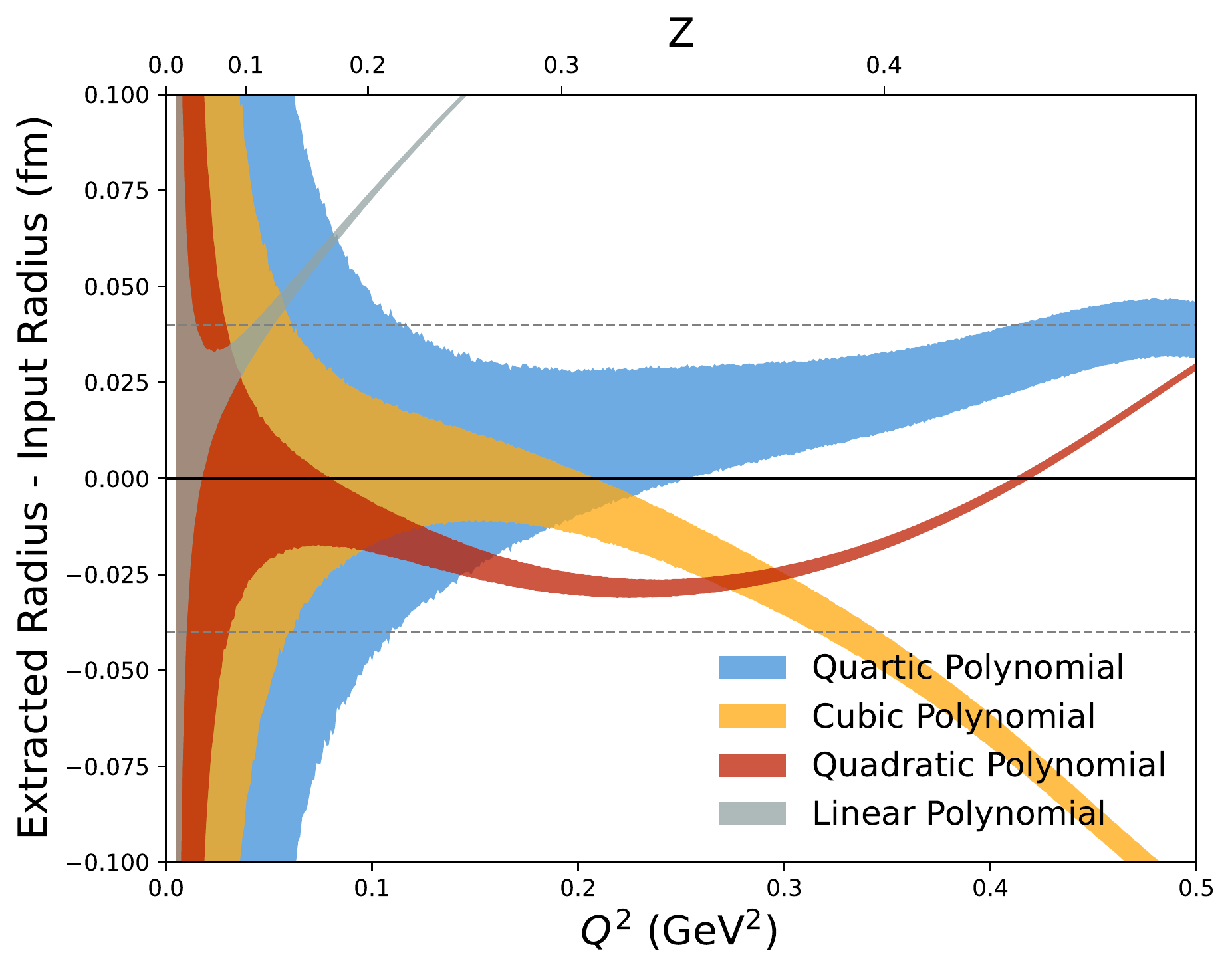}
    \caption{Bias-Variance plot for polynomial fits in $Z$ of the Bernauer \GEp{} parameterization}
\end{figure}

\begin{figure}[hb]
    \centering
    \includegraphics[height=0.2\textheight]{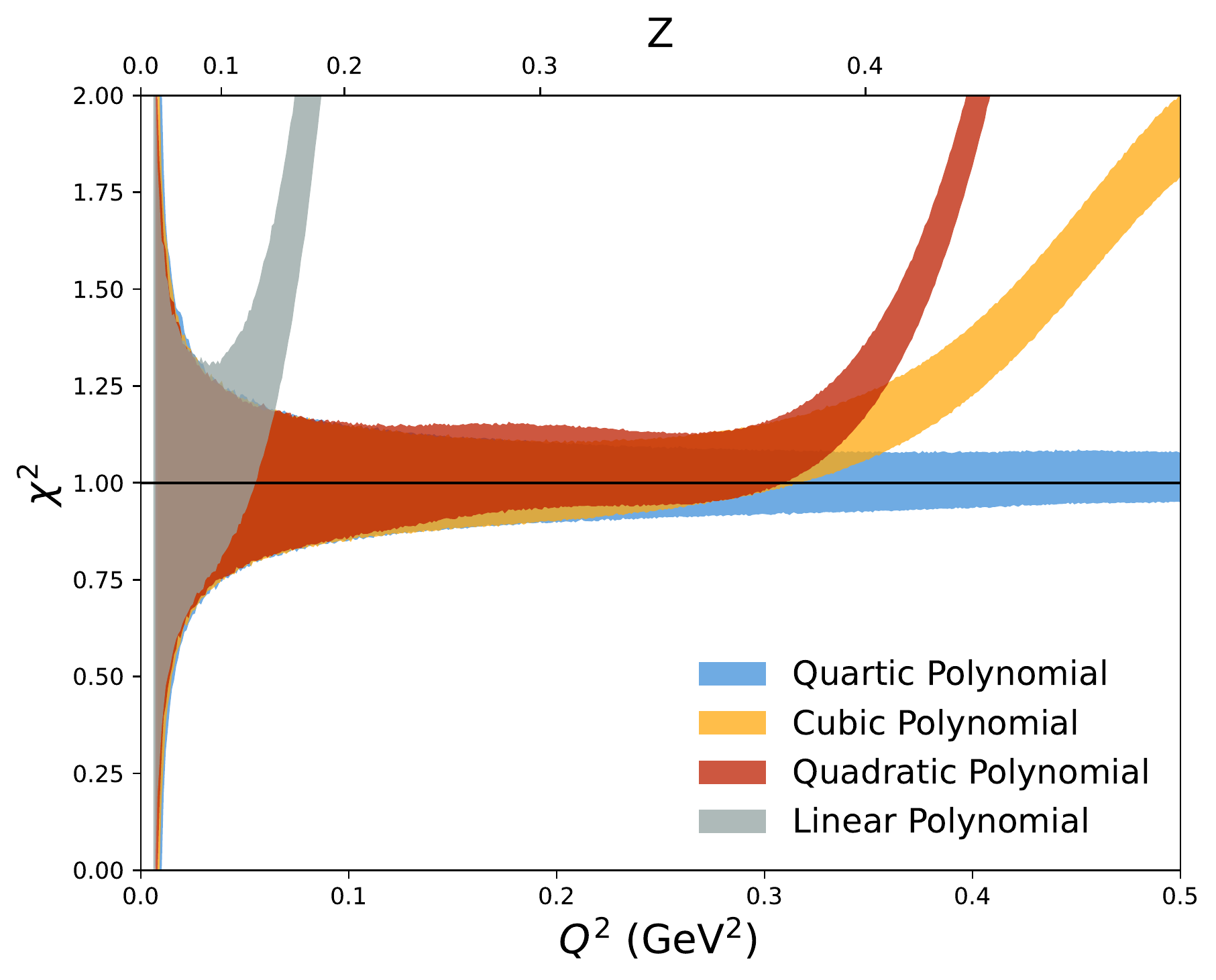}
    \caption{$\chi^2$ plot for polynomial fits in $Z$ of the Bernauer \GEp{} parameterization}
\end{figure}

\begin{figure}[hb]
    \centering
    \includegraphics[height=0.2\textheight]{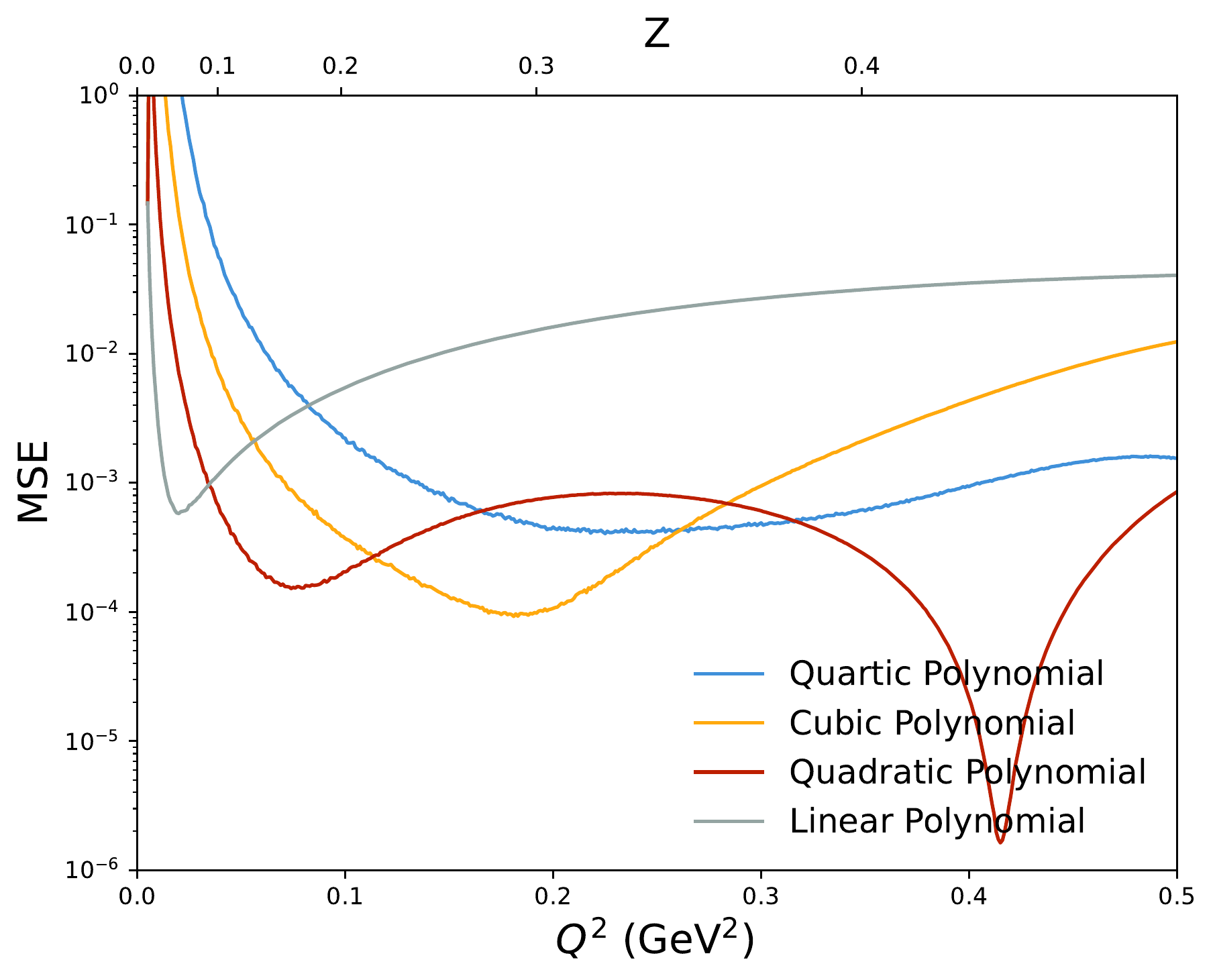}
    \caption{Mean Squared Error plot for polynomial fits in $Z$ of the Bernauer \GEp{} parameterization}
\end{figure}

\newpage

\subsubsection{Figure Descriptions}

Three plots, each with “$Q^2$ (GeV$^2$)” from 0 to 0.5 on the x axis. A second x axis on top shows the mapping of $Q^2$ to $Z$, which spreads out low $Q^2$ points and bunches up high $Q^2$ points. Each plot has four sets of data shown, one for each power of polynomial used, labeled as “Linear Polynomial”, “Quadratic Polynomial”, “Cubic Polynomial”, and “Quartic Polynomial”.

The top plot has “Extracted Radius - Input Radius (fm)” from -0.1 to 0.1 on the y axis. Two horizontal lines drawn at $\pm$0.04 fm show the magnitude of the proton radius puzzle. Each data set is shown as a band. The high variance region extends to higher $Q^2$ than fits in $Q^2$ and the variance is higher than the $Q^2$ fits for the full range of the plot. The curves rapidly diverge from the input radius at approximately the same $Q^2$ as the fits in $Q^2$. The linear polynomial diverges to a large radius. The other curves exhibit inflections that change the direction of the deviations.

The middle plot has $\chi^2$ from 0 to 2 on the y axis. Each data set is shown as a band. All bands begin centered at 1 and then at some $Q^2$ have a rapid upturn. Higher order polynomials delay the upturn until higher $Q^2$. Much like the fits in $Q^2$, the deviation from $\chi^2\approx1$ occurs at significantly higher $Q^2$ than when the extracted radius deviates from the input radius.

The bottom plot has MSE, Mean Squared Error as described in Eq.~\ref{eq:MSE}, from $10^{-6}$ to $10^0$ with a logarithmic scale on the y axis. Each set of data is represented as a curve. Each curve rapidly drops to the minimum MSE for the fit and then begins to increase with $Q^2$ and start to level off about two orders of magnitude higher than the minima. The quadratic polynomial has the lowest minimum at about $2\cdot10^{-6}$. The minima are spread over a larger $Q^2$ range and the range grows with polynomial order. Of note is that the lowest minima corresponds to the point after the inflection when the extracted radius briefly cross the input radius.

\newpage

\subsection{Bounded Polynomial Fits in $Z$}

\begin{figure}[hb]
    \centering
    \includegraphics[height=0.2\textheight]{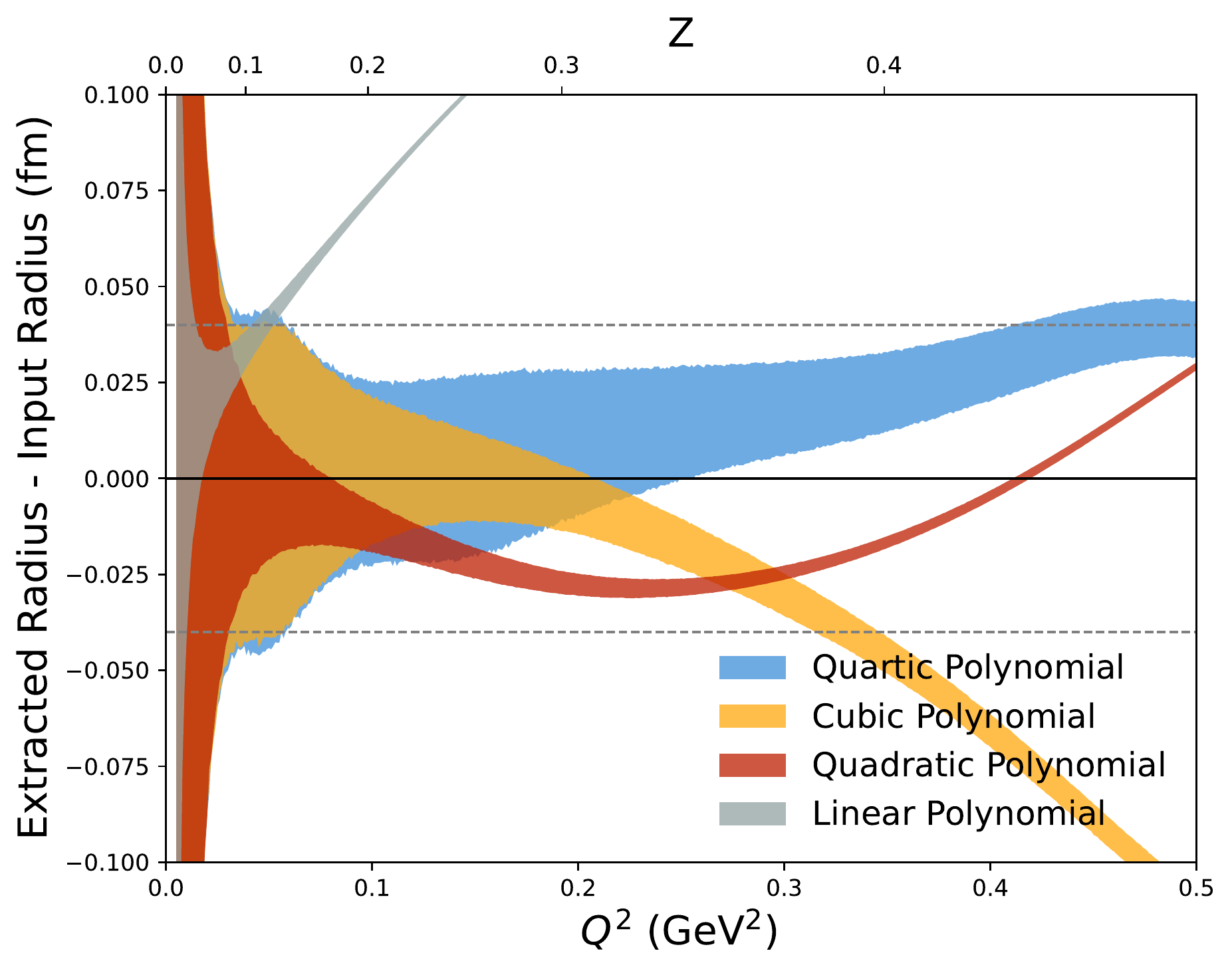}
    \caption{Bias-Variance plot for bounded polynomial fits in $Z$ of the Bernauer \GEp{} parameterization}
\end{figure}

\begin{figure}[hb]
    \centering
    \includegraphics[height=0.2\textheight]{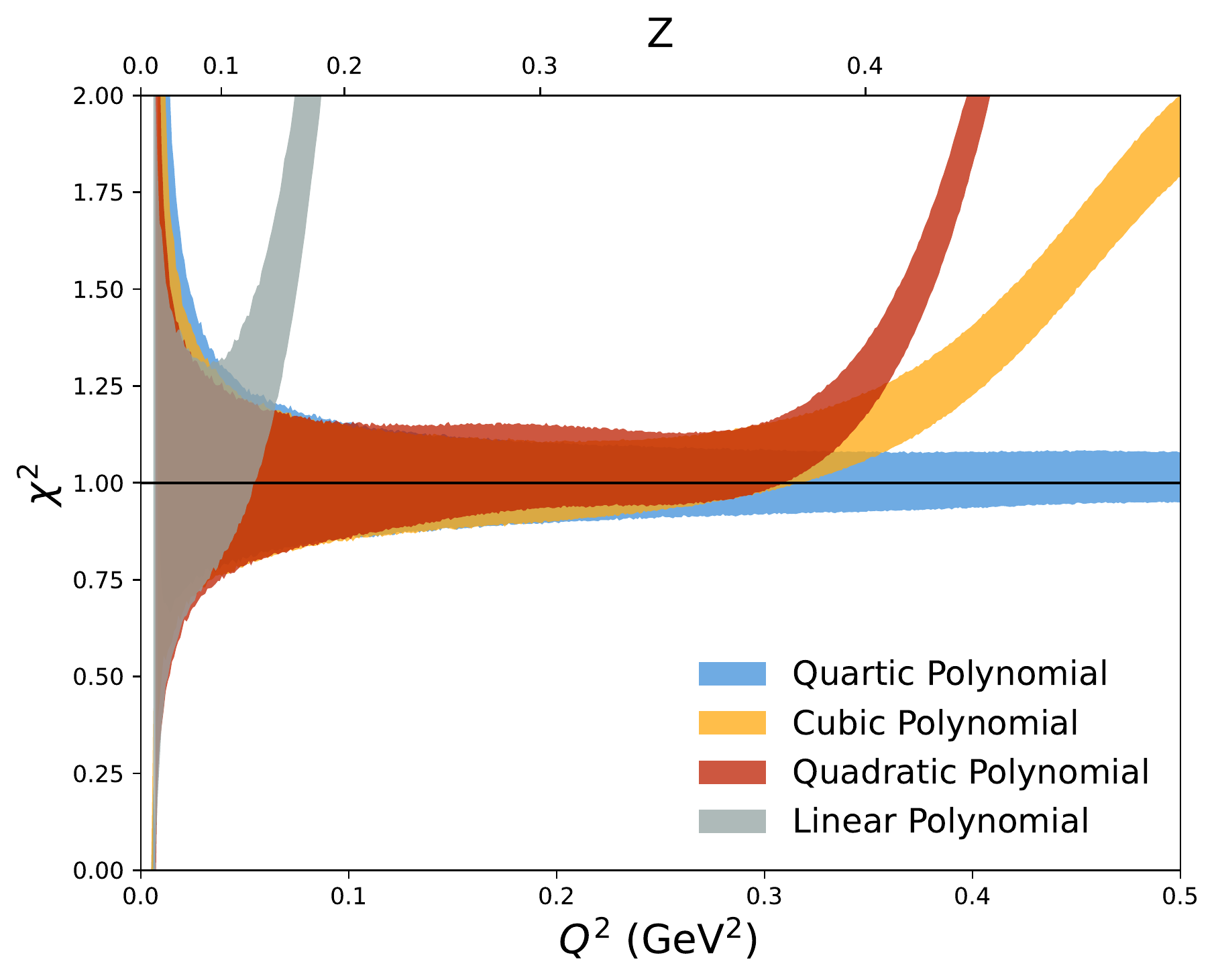}
    \caption{$\chi^2$ plot for bounded polynomial fits in $Z$ of the Bernauer \GEp{} parameterization}
\end{figure}

\begin{figure}[hb]
    \centering
    \includegraphics[height=0.2\textheight]{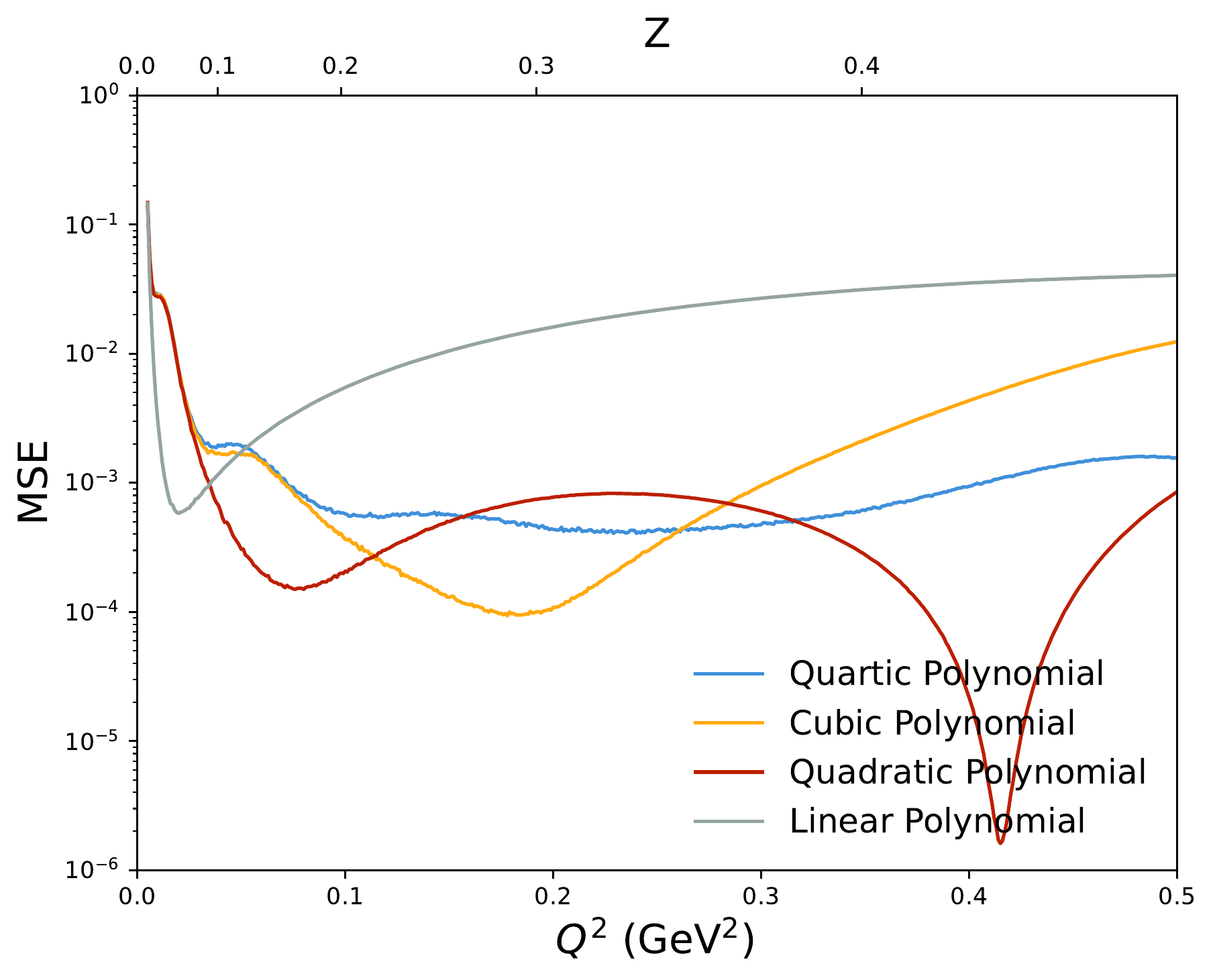}
    \caption{Mean Squared Error plot for bounded polynomial fits in $Z$ of the Bernauer \GEp{} parameterization}
\end{figure}

\newpage

\subsubsection{Figure Descriptions}

Three plots, each with “$Q^2$ (GeV$^2$)” from 0 to 0.5 on the x axis. A second x axis on top shows the mapping of $Q^2$ to $Z$, which spreads out low $Q^2$ points and bunches up high $Q^2$ points. Each plot has four sets of data shown, one for each power of polynomial used, labeled as “Linear Polynomial”, “Quadratic Polynomial”, “Cubic Polynomial”, and “Quartic Polynomial”.

The top plot has “Extracted Radius - Input Radius (fm)” from -0.1 to 0.1 on the y axis. Two horizontal lines drawn at $\pm$0.04 fm show the magnitude of the proton radius puzzle. Each data set is shown as a band. The low $Q^2$ variance is significantly tempered from the unbound fits in $Z$. However, as $Q^2$ increases the curves are not qualitatively different from the unbound fits.

The middle plot has $\chi^2$ from 0 to 2 on the y axis. Each data set is shown as a band. The data is not qualitatively different than the unbound fits.

The bottom plot has MSE, Mean Squared Error as described in Eq.~\ref{eq:MSE}, from $10^{-6}$ to $10^0$ with a logarithmic scale on the y axis. Each set of data is represented as a curve. The data approaches their minima more rapidly than the unbound fits, but as $Q^2$ increases they are not qualitatively different from the unbound fits.

\newpage

\subsection{Rational(N,M) Fits in $Q^2$}

\begin{figure}[hb]
    \centering
    \includegraphics[height=0.2\textheight]{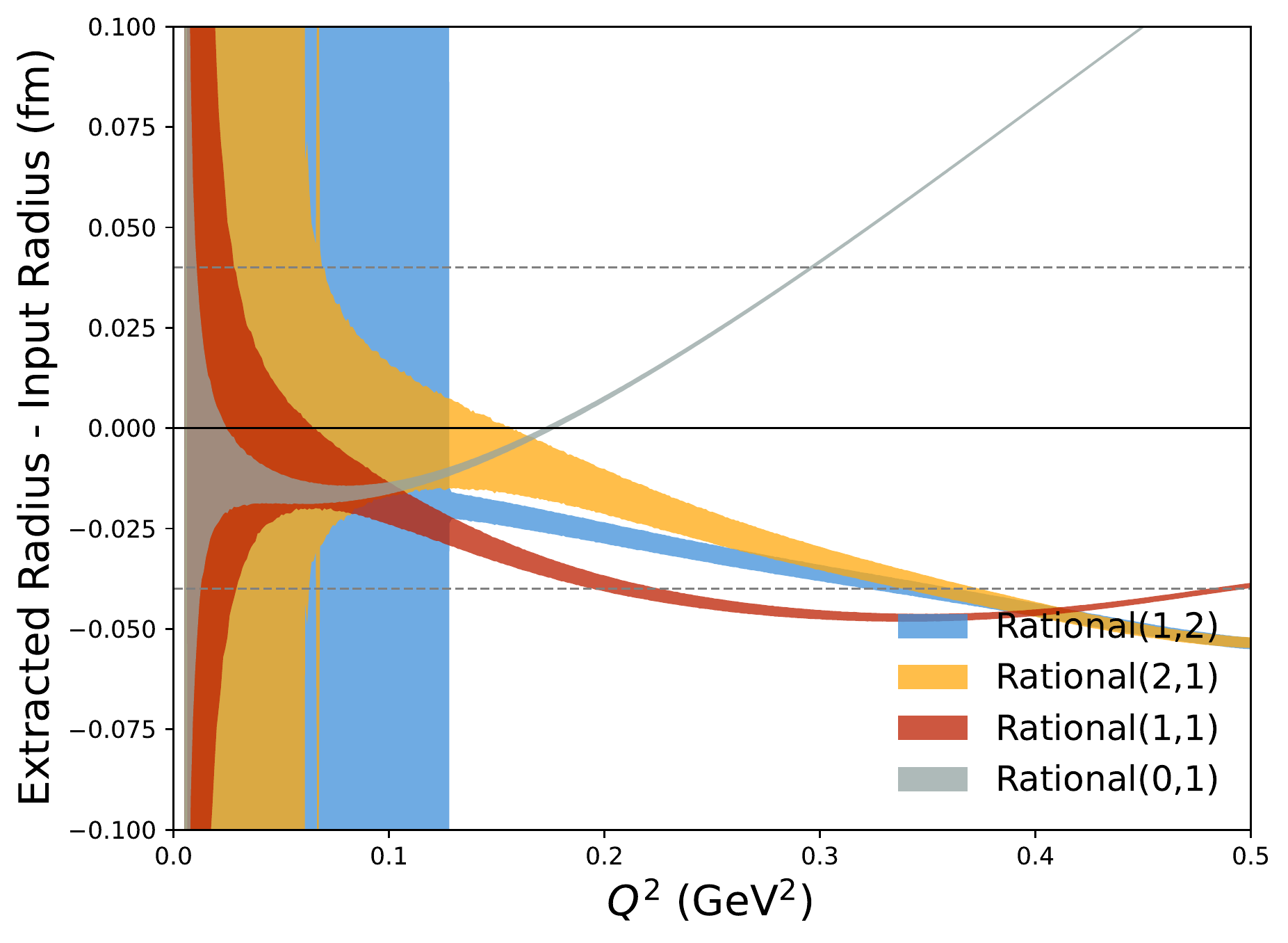}
    \caption{Bias-Variance plot for Rational(N,M) fits in $Q^2$ of the Bernauer \GEp{} parameterization}
\end{figure}

\begin{figure}[hb]
    \centering
    \includegraphics[height=0.2\textheight]{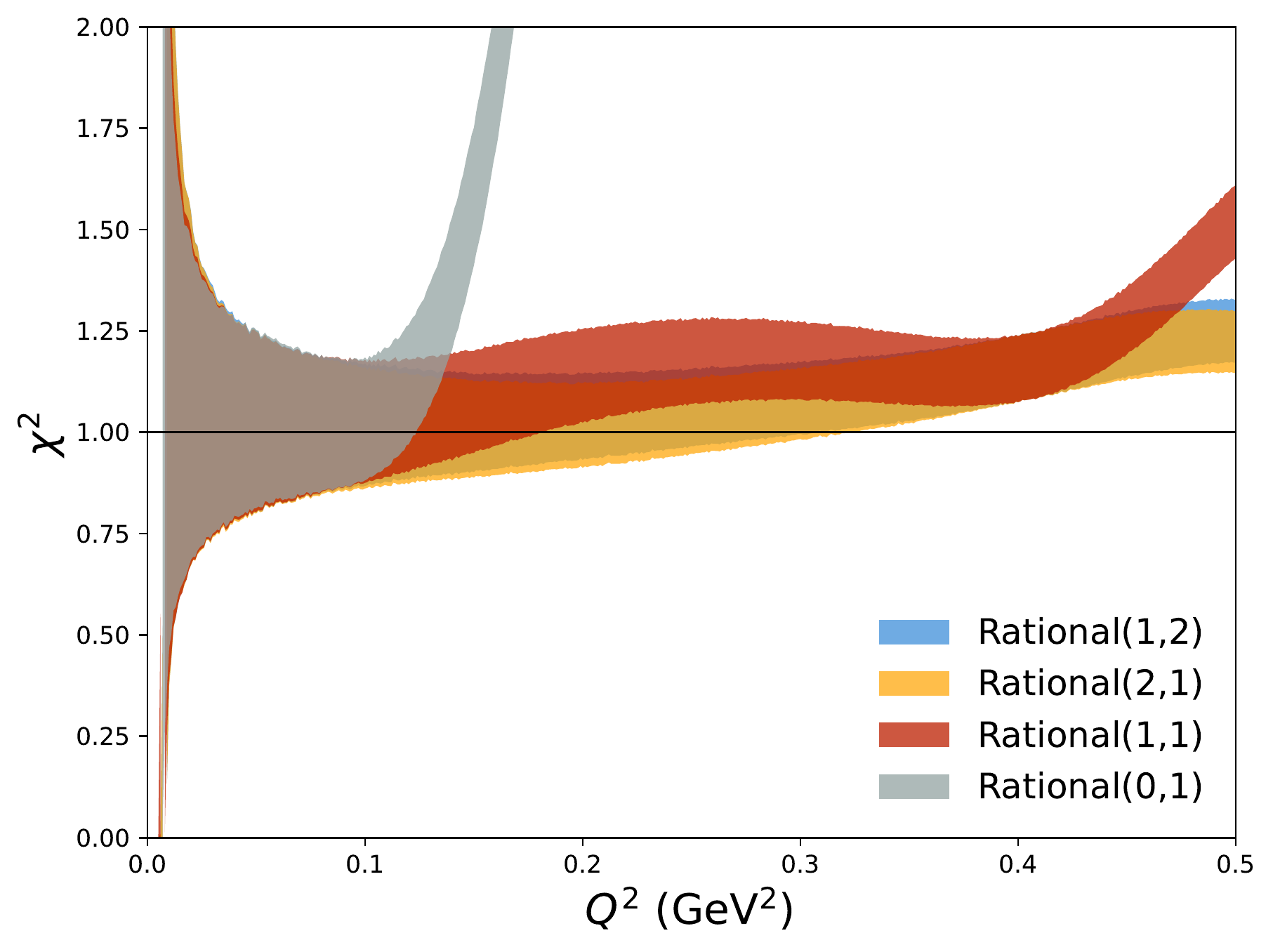}
    \caption{$\chi^2$ plot for Rational(N,M) fits in $Q^2$ of the Bernauer \GEp{} parameterization}
\end{figure}

\begin{figure}[hb]
    \centering
    \includegraphics[height=0.2\textheight]{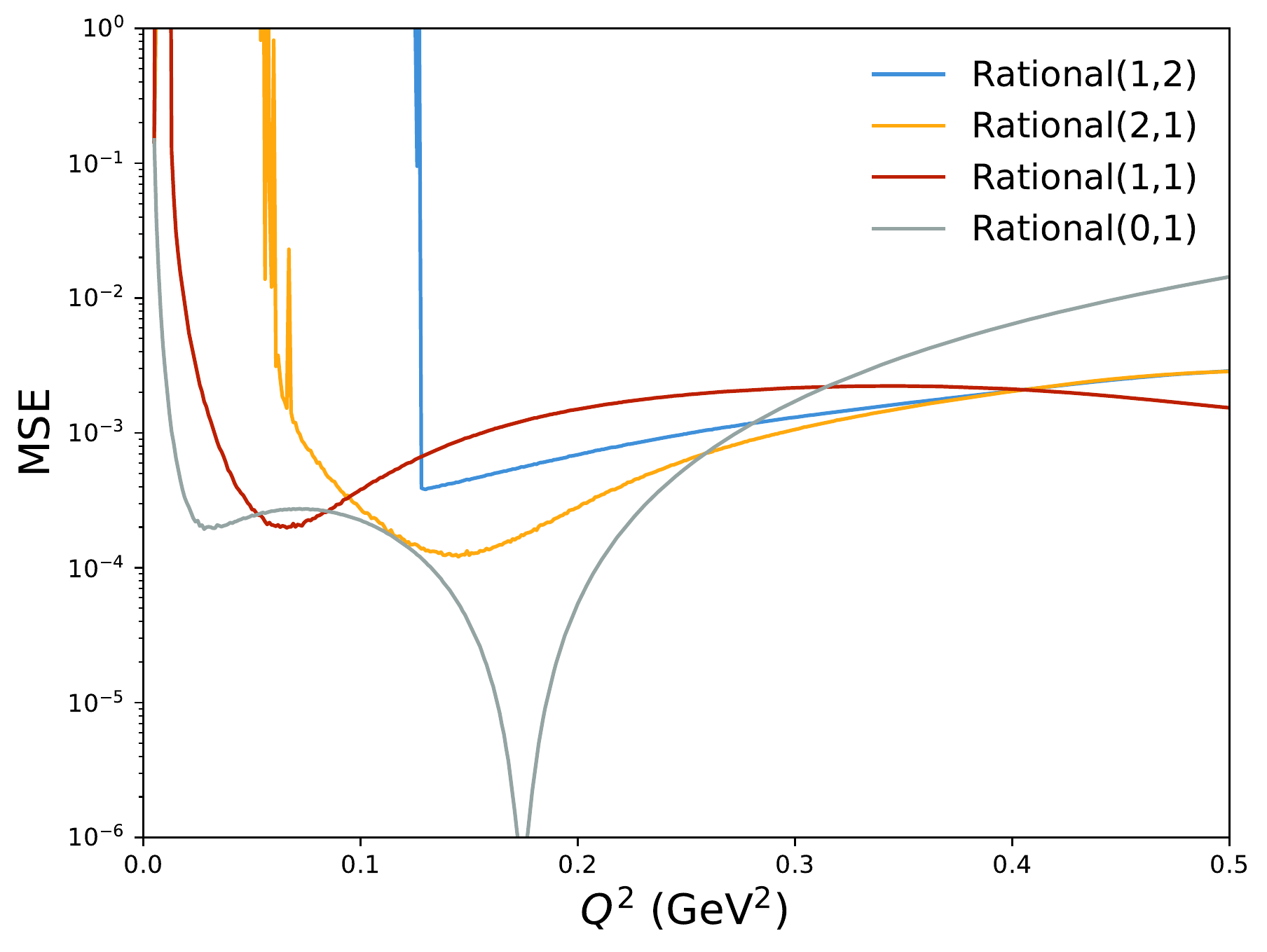}
    \caption{Mean Squared Error plot for Rational(N,M) fits in $Q^2$ of the Bernauer \GEp{} parameterization}
\end{figure}

\newpage

\subsubsection{Figure Descriptions}

Three plots, each with “$Q^2$ (GeV$^2$)” from 0 to 0.5 on the x axis. Each plot has four sets of data shown, one for each power of Rational(N,M) function used, labeled as “Rational(0,1)”, “Rational(1,1)”, “Rational(2,1)”, and “Rational(1,2)”.

The top plot has “Extracted Radius - Input Radius (fm)” from -0.1 to 0.1 on the y axis. Two horizontal lines drawn at $\pm$0.04 fm show the magnitude of the proton radius puzzle. Each data set is shown as a band. This data shows larger variance at low $Q^2$ than any other fit and the variance increases with increasing N and M, with M having a larger influence on the variance. All fits rapidly diverge from the input radius. Rational(0,1) diverges to a large radius, while the others diverge to a small radius. The variances of all Rational(N,M) fits are less than half that of any of the other fits.

The middle plot has $\chi^2$ from 0 to 2 on the y axis. Each data set is shown as a band. All begin centered at 1, with Rational(0,1) having a rapid upturn. The others show a steady increase with $Q^2$ over the full range.

The bottom plot has MSE, Mean Squared Error as described in Eq.~\ref{eq:MSE}, from $10^{-6}$ to $10^0$ with a logarithmic scale on the y axis. Each set of data is represented as a curve. All curves quickly approach a minimum of about $2\cdot10^{-4}$ and then begin to rise. The Rational(0,1) curve has a brief dip to below the scale of the y axis when the increasing extracted radius crosses the input radius value.

\newpage

\section{Dipole}

\subsection{Polynomial Fits in $Q^2$}

\begin{figure}[hb]
    \centering
    \includegraphics[height=0.2\textheight]{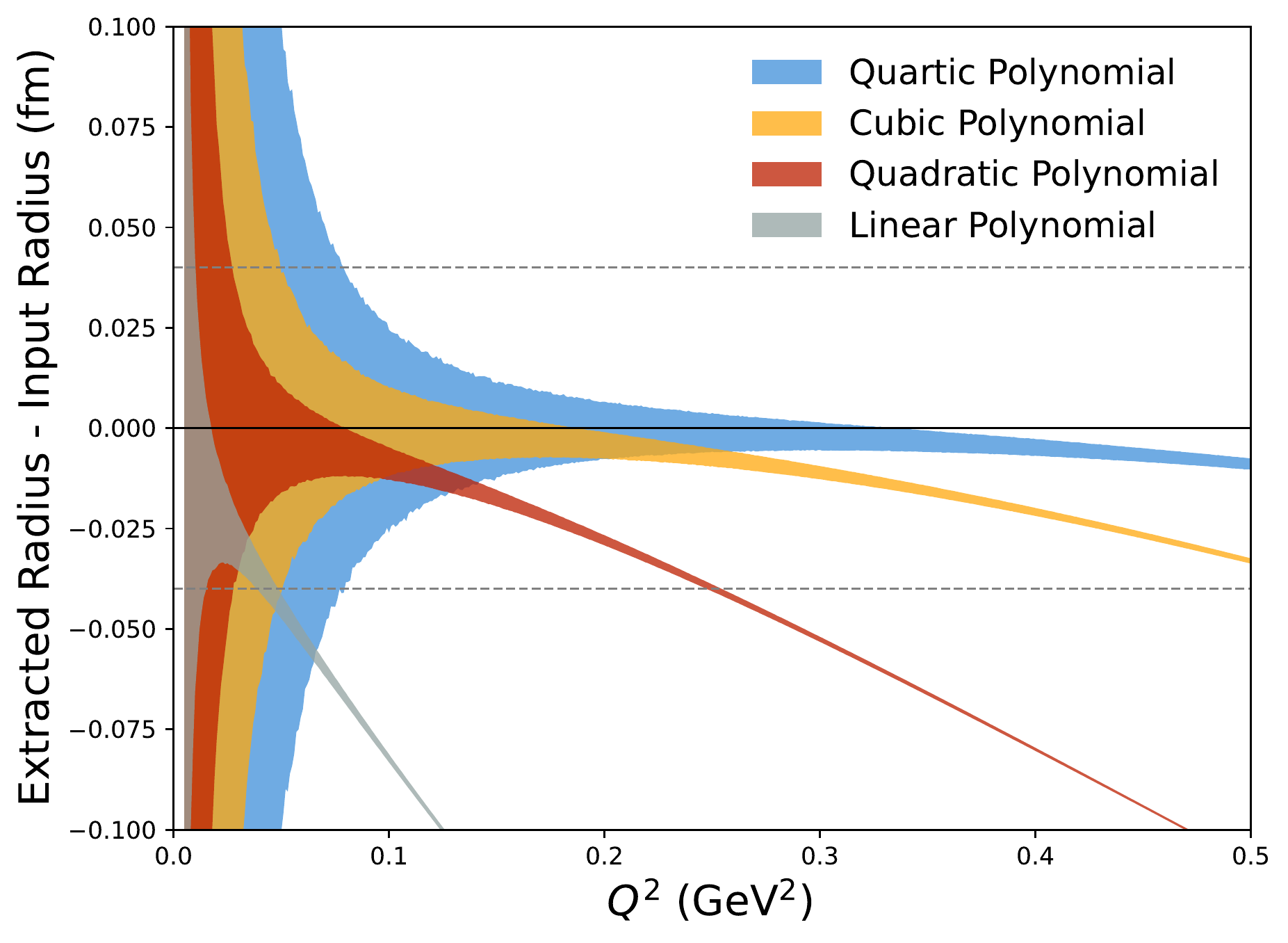}
    \caption{Bias-Variance plot for polynomial fits in $Q^2$ of the Dipole \GEp{} parameterization}
\end{figure}

\begin{figure}[hb]
    \centering
    \includegraphics[height=0.2\textheight]{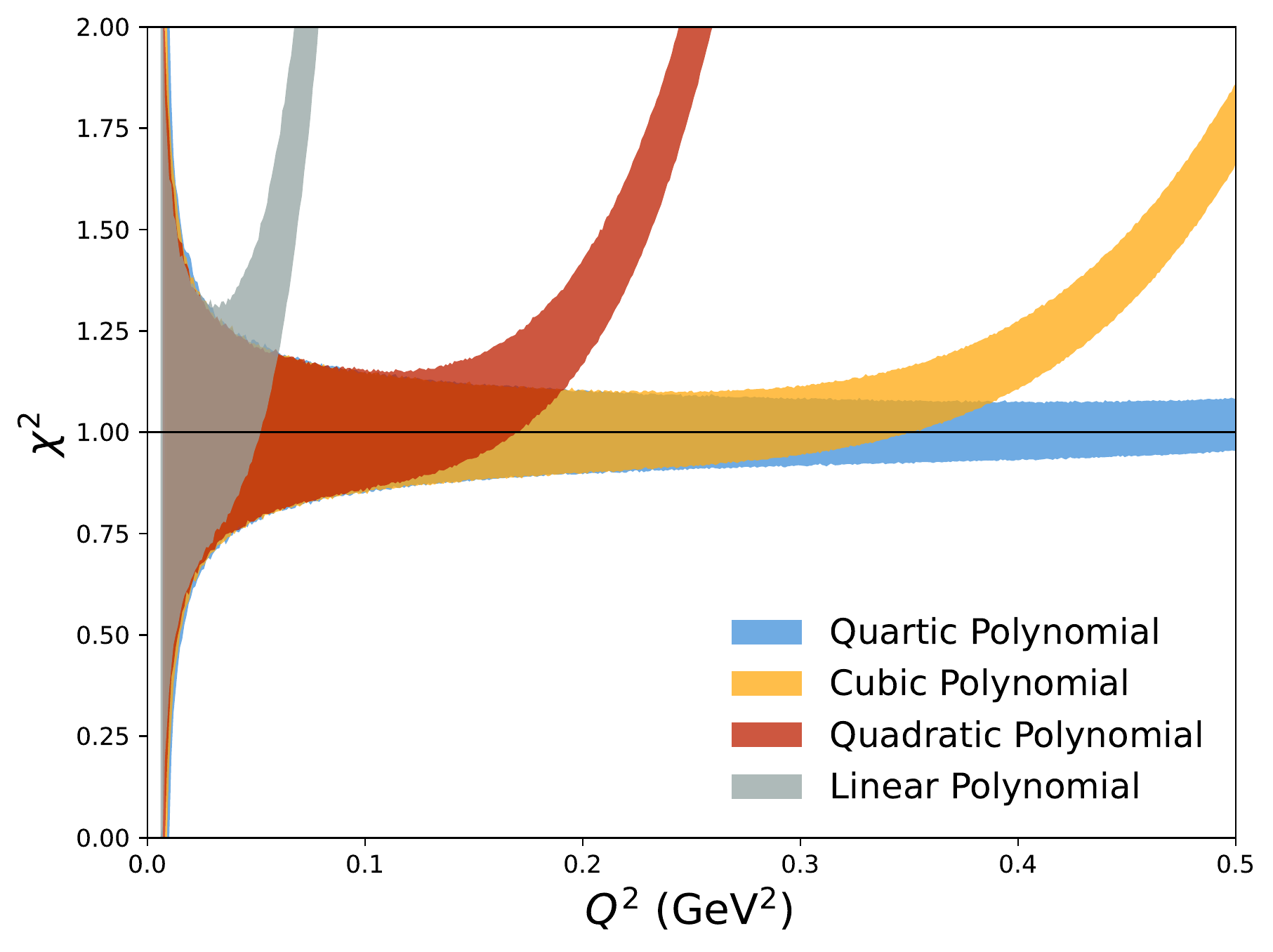}
    \caption{$\chi^2$ plot for polynomial fits in $Q^2$ of the Dipole \GEp{} parameterization}
\end{figure}

\begin{figure}[hb]
    \centering
    \includegraphics[height=0.2\textheight]{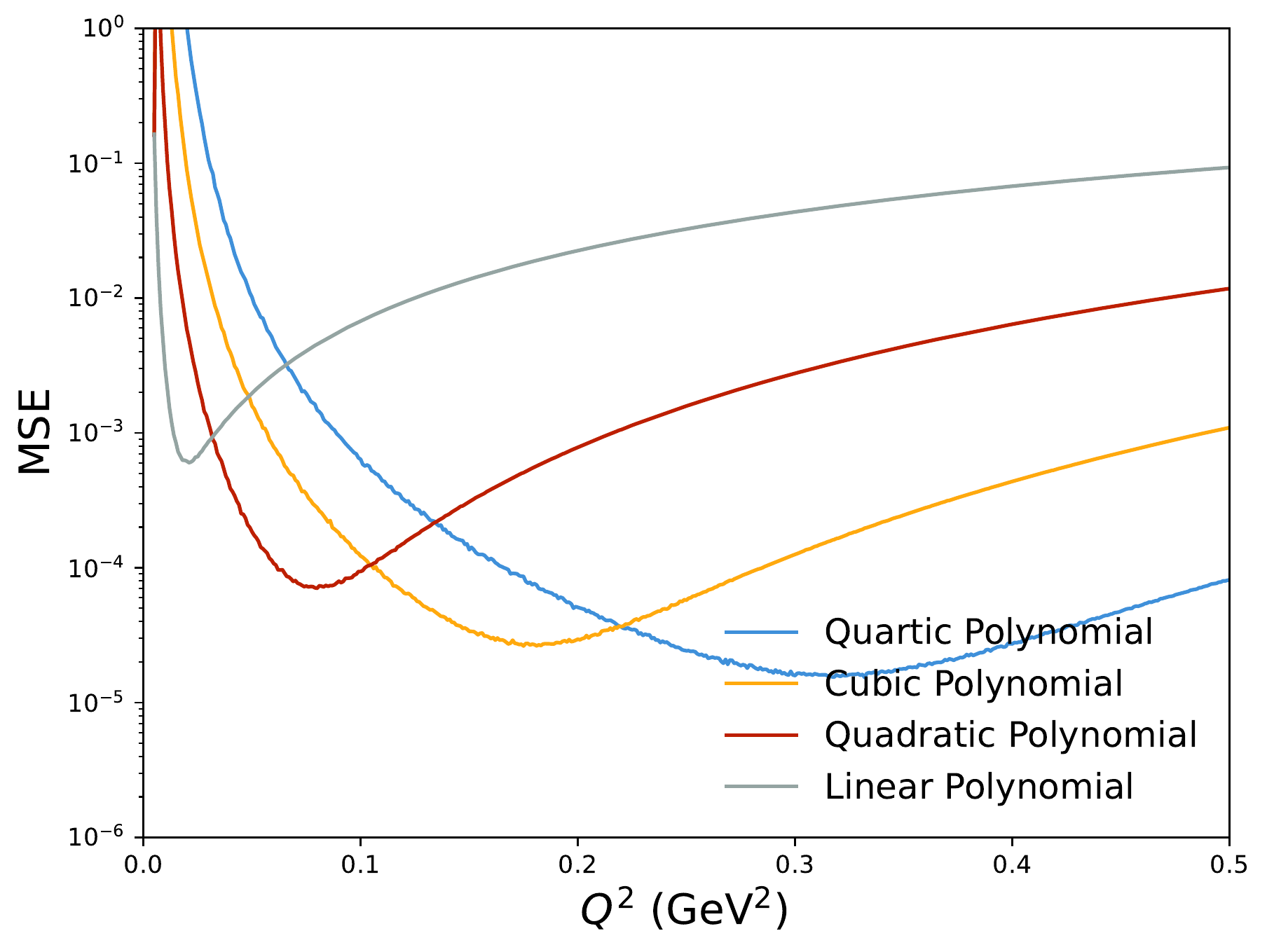}
    \caption{Mean Squared Error plot for polynomial fits in $Q^2$ of the Dipole \GEp{} parameterization}
\end{figure}

\newpage

\subsubsection{Figure Descriptions}

Three plots, each with “$Q^2$ (GeV$^2$)” from 0 to 0.5 on the x axis. Each plot has four sets of data shown, one for each power of polynomial used, labeled as “Linear Polynomial”, “Quadratic Polynomial”, “Cubic Polynomial”, and “Quartic Polynomial”.

The top plot has “Extracted Radius - Input Radius (fm)” from -0.1 to 0.1 on the y axis. Two horizontal lines drawn at $\pm$0.04 fm show the magnitude of the proton radius puzzle. Each data set is shown as a band. All bands have variance higher than the axis scale at small $Q^2$ and systematically return a small radius that decreases with increasing $Q^2$ range. Higher order polynomials decrease slower.

The middle plot has $\chi^2$ from 0 to 2 on the y axis. Each data set is shown as a band. All bands begin centered at 1 and then at some $Q^2$ have a rapid upturn. Higher order polynomials delay the upturn until higher $Q^2$. The deviation from $\chi^2\approx1$ occurs at significantly higher $Q^2$ than when the extracted radius deviates from the input radius.

The bottom plot has MSE, Mean Squared Error as described in Eq.~\ref{eq:MSE}, from $10^{-6}$ to $10^0$ with a logarithmic scale on the y axis. Each set of data is represented as a curve. Each curve rapidly drops to the minimum MSE for the fit and then begins to increase with $Q^2$ and start to level off about two orders of magnitude higher than the minima. Higher order polynomials have lower minima that occur at higher $Q^2$ and then increase slower. The lowest minima is about $2\cdot10^{-5}$.

\newpage

\subsection{Bounded Polynomial Fits in $Q^2$}

\begin{figure}[hb]
    \centering
    \includegraphics[height=0.2\textheight]{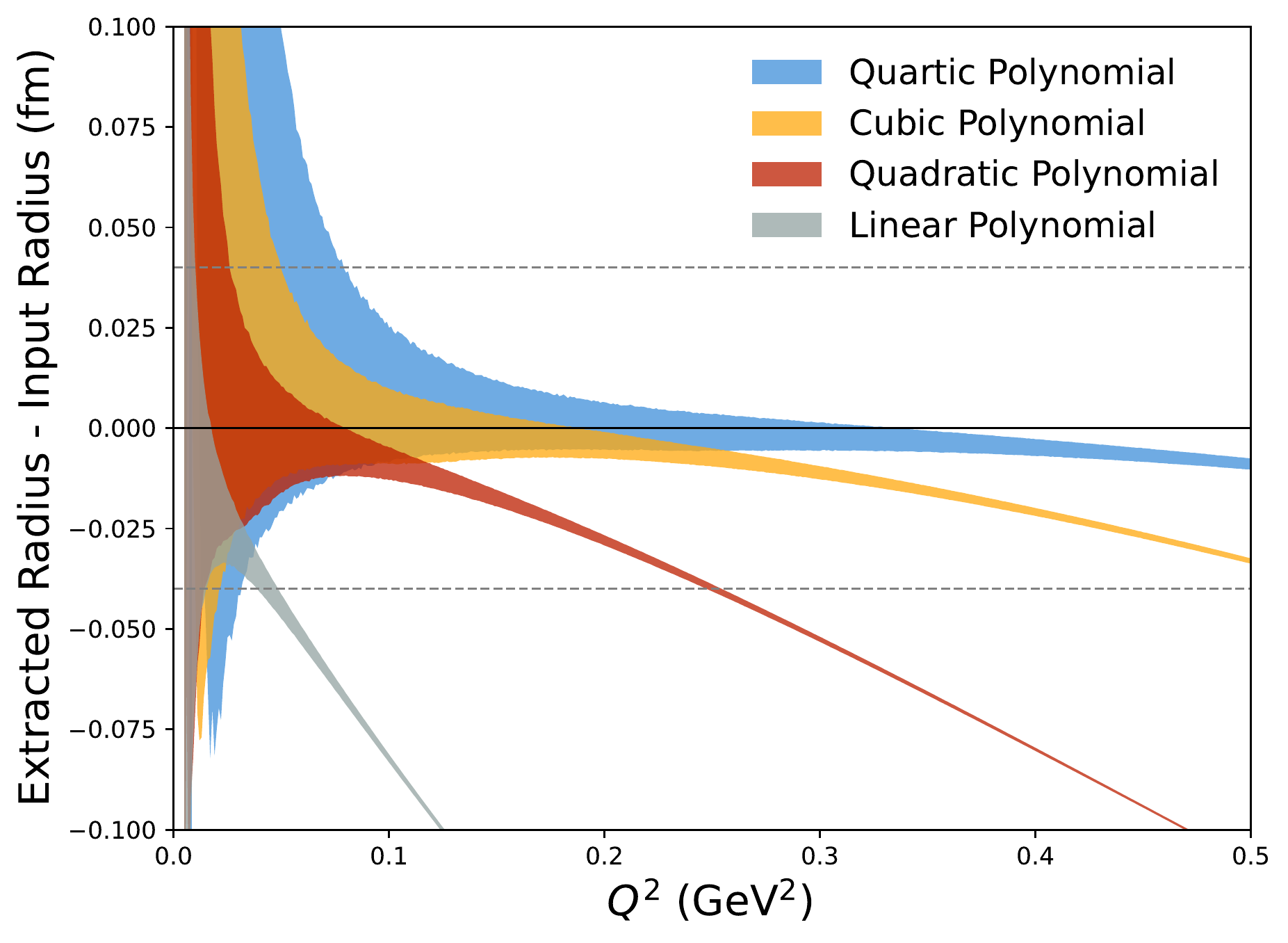}
    \caption{Bias-Variance plot for bounded polynomial fits in $Q^2$ of the Dipole \GEp{} parameterization}
\end{figure}

\begin{figure}[hb]
    \centering
    \includegraphics[height=0.2\textheight]{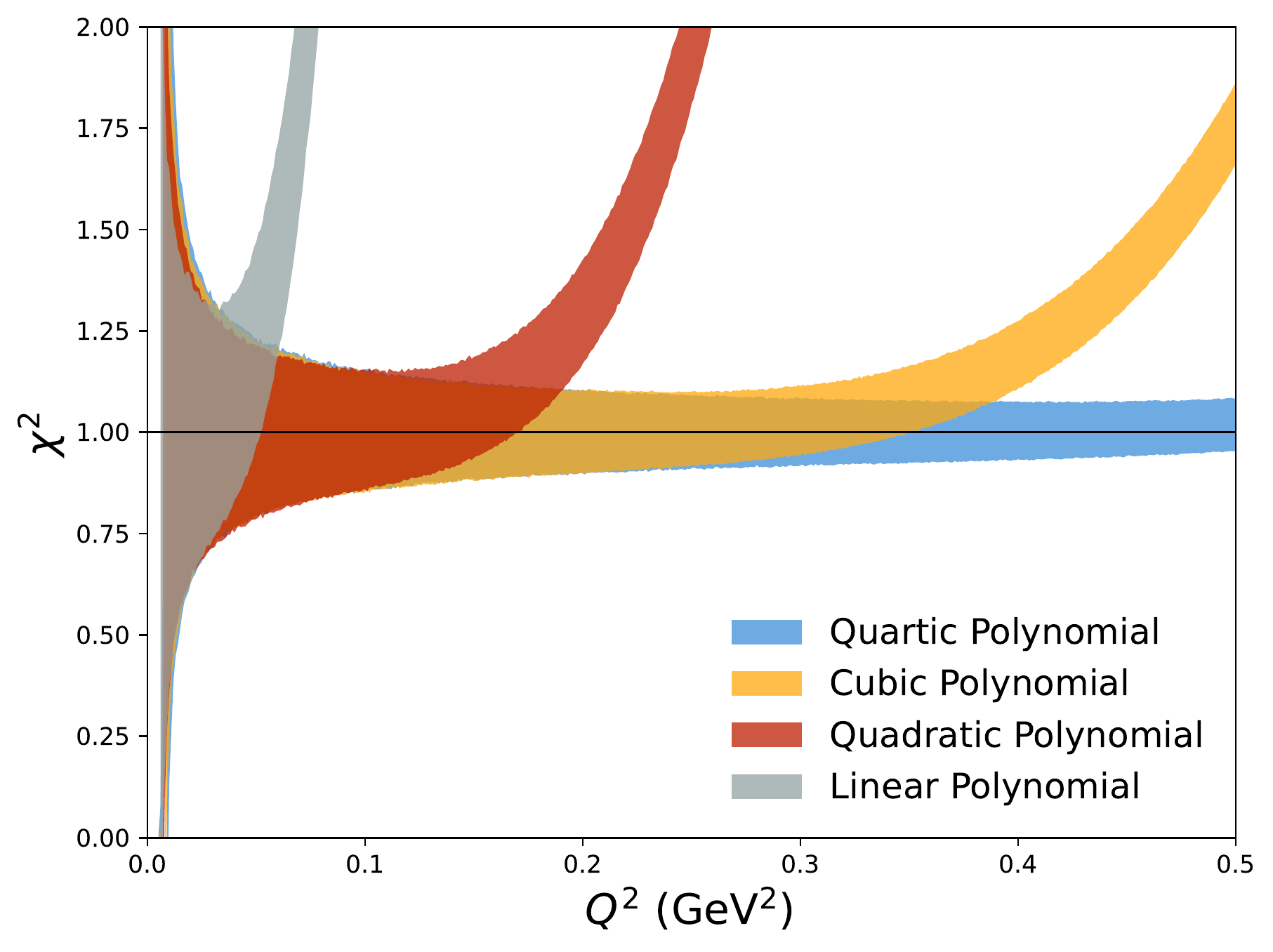}
    \caption{$\chi^2$ plot for bounded polynomial fits in $Q^2$ of the Dipole \GEp{} parameterization}
\end{figure}

\begin{figure}[hb]
    \centering
    \includegraphics[height=0.2\textheight]{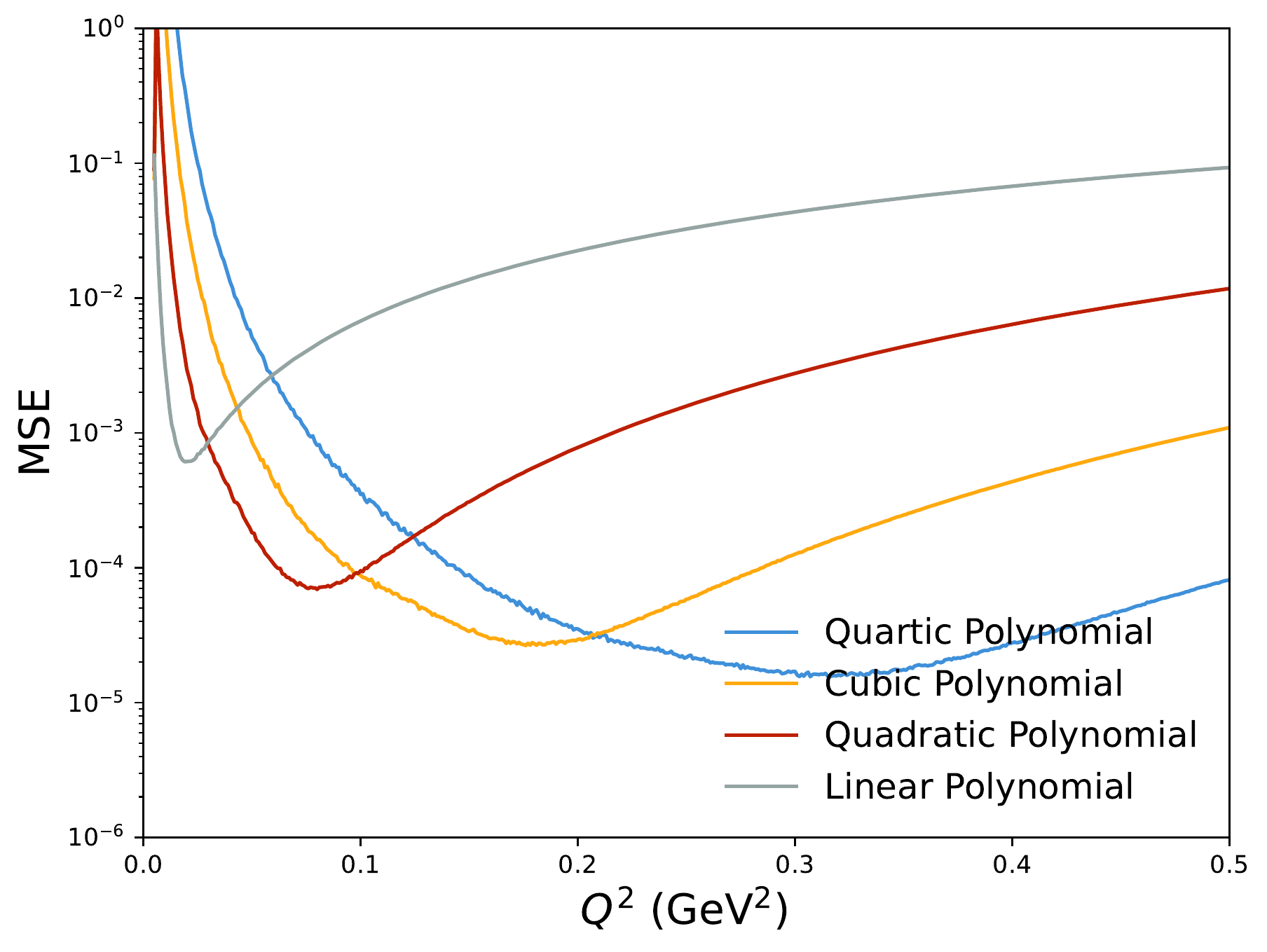}
    \caption{Mean Squared Error plot for bounded polynomial fits in $Q^2$ of the Dipole \GEp{} parameterization}
\end{figure}

\newpage

\subsubsection{Figure Descriptions}

Three plots, each with “$Q^2$ (GeV$^2$)” from 0 to 0.5 on the x axis. Each plot has four sets of data shown, one for each power of polynomial used, labeled as “Linear Polynomial”, “Quadratic Polynomial”, “Cubic Polynomial”, and “Quartic Polynomial”.

The top plot has “Extracted Radius - Input Radius (fm)” from -0.1 to 0.1 on the y axis. Two horizontal lines drawn at $\pm$0.04 fm show the magnitude of the proton radius puzzle. Each data set is shown as a band. The low $Q^2$ variance is smaller than the unbound fits, but is not overall qualitatively different.

The middle plot has $\chi^2$ from 0 to 2 on the y axis. Each data set is shown as a band. The data is not qualitatively different than the unbound fits.

The bottom plot has MSE, Mean Squared Error as described in Eq.~\ref{eq:MSE}, from $10^{-6}$ to $10^0$ with a logarithmic scale on the y axis. Each set of data is represented as a curve. The data is not qualitatively different than the unbound fits.

\newpage

\subsection{Polynomial Fits in $Z$}

\begin{figure}[hb]
    \centering
    \includegraphics[height=0.2\textheight]{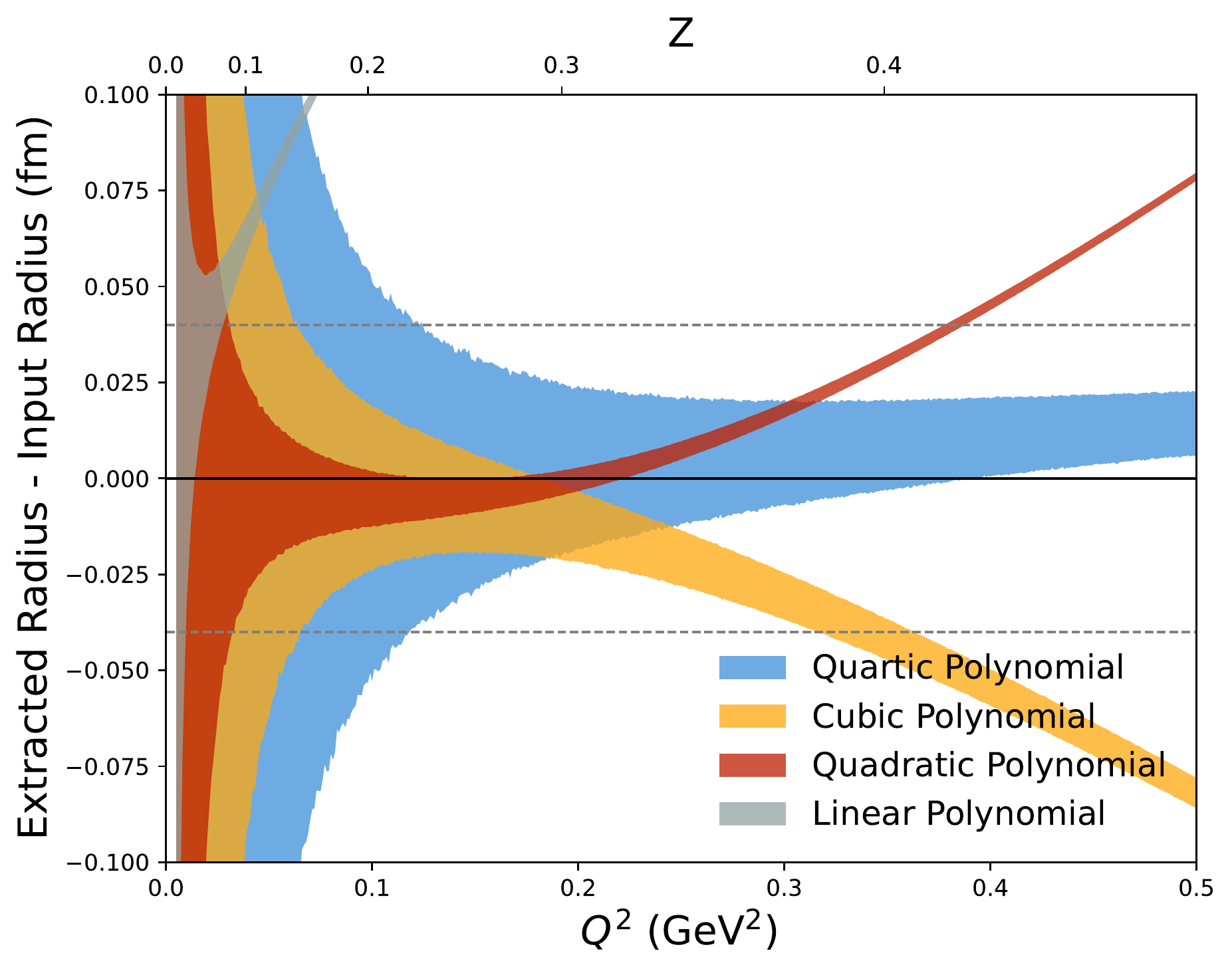}
    \caption{Bias-Variance plot for polynomial fits in $Z$ of the Dipole \GEp{} parameterization}
\end{figure}

\begin{figure}[hb]
    \centering
    \includegraphics[height=0.2\textheight]{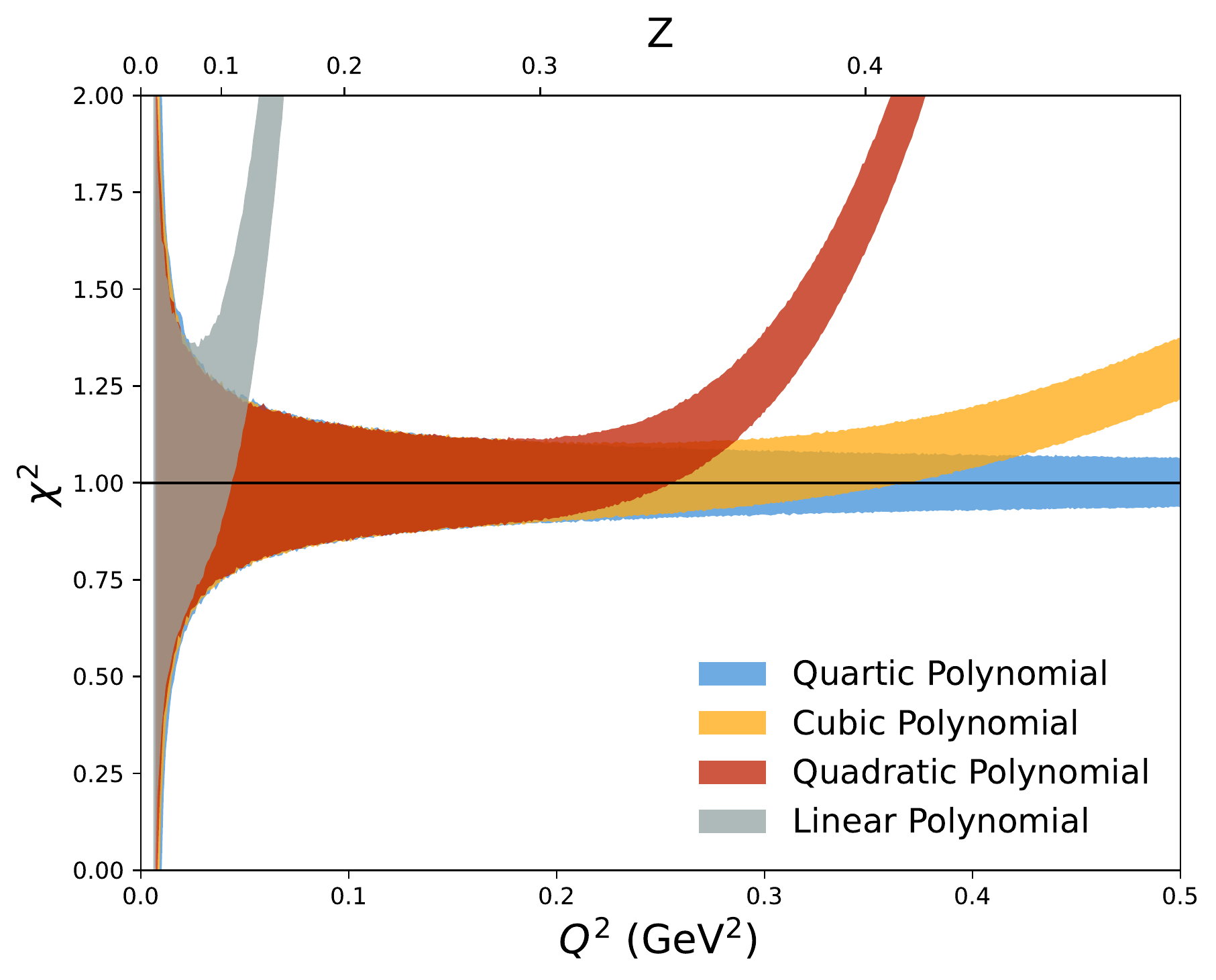}
    \caption{$\chi^2$ plot for polynomial fits in $Z$ of the Dipole \GEp{} parameterization}
\end{figure}

\begin{figure}[hb]
    \centering
    \includegraphics[height=0.2\textheight]{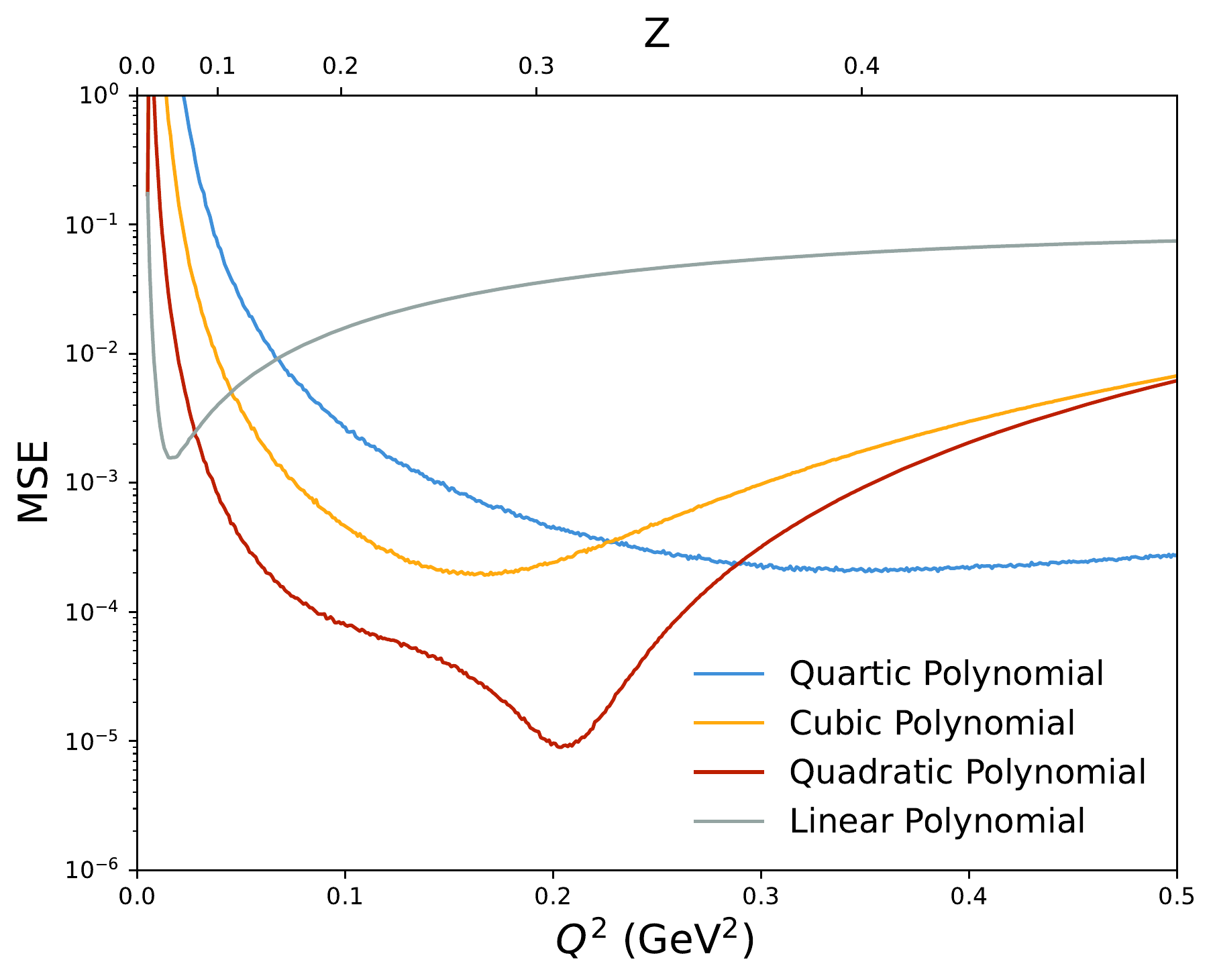}
    \caption{Mean Squared Error plot for polynomial fits in $Z$ of the Dipole \GEp{} parameterization}
\end{figure}

\newpage

\subsubsection{Figure Descriptions}

Three plots, each with “$Q^2$ (GeV$^2$)” from 0 to 0.5 on the x axis. A second x axis on top shows the mapping of $Q^2$ to $Z$, which spreads out low $Q^2$ points and bunches up high $Q^2$ points. Each plot has four sets of data shown, one for each power of polynomial used, labeled as “Linear Polynomial”, “Quadratic Polynomial”, “Cubic Polynomial”, and “Quartic Polynomial”.

The top plot has “Extracted Radius - Input Radius (fm)” from -0.1 to 0.1 on the y axis. Two horizontal lines drawn at $\pm$0.04 fm show the magnitude of the proton radius puzzle. Each data set is shown as a band. The high variance region extends to higher $Q^2$ than fits in $Q^2$ and the variance is higher than the $Q^2$ fits for the full range of the plot. The curves rapidly diverge from the input radius at approximately the same $Q^2$ as the fits in $Q^2$, but are better centered at the correct radius at lower $Q^2$. The linear, quadratic, and quartic polynomials diverge to large radii and the cubic polynomial diverges to a small radius.

The middle plot has $\chi^2$ from 0 to 2 on the y axis. Each data set is shown as a band. All bands begin centered at 1 and then at some $Q^2$ have a rapid upturn. Higher order polynomials delay the upturn until higher $Q^2$. Much like the fits in $Q^2$, the deviation from $\chi^2\approx1$ occurs at significantly higher $Q^2$ than when the extracted radius deviates from the input radius. The quartic polynomial remains well centered at 1 for the full range.

The bottom plot has MSE, Mean Squared Error as described in Eq.~\ref{eq:MSE}, from $10^{-6}$ to $10^0$ with a logarithmic scale on the y axis. Each set of data is represented as a curve. Each curve rapidly drops to the minimum MSE for the fit and then begins to increase with $Q^2$ and start to level off about two orders of magnitude higher than the minima. The quadratic polynomial has the lowest minimum at about $1\cdot10^{-5}$. The minima are spread over a larger $Q^2$ range and the range grows with polynomial order.

\newpage

\subsection{Bounded Polynomial Fits in $Z$}

\begin{figure}[hb]
    \centering
    \includegraphics[height=0.2\textheight]{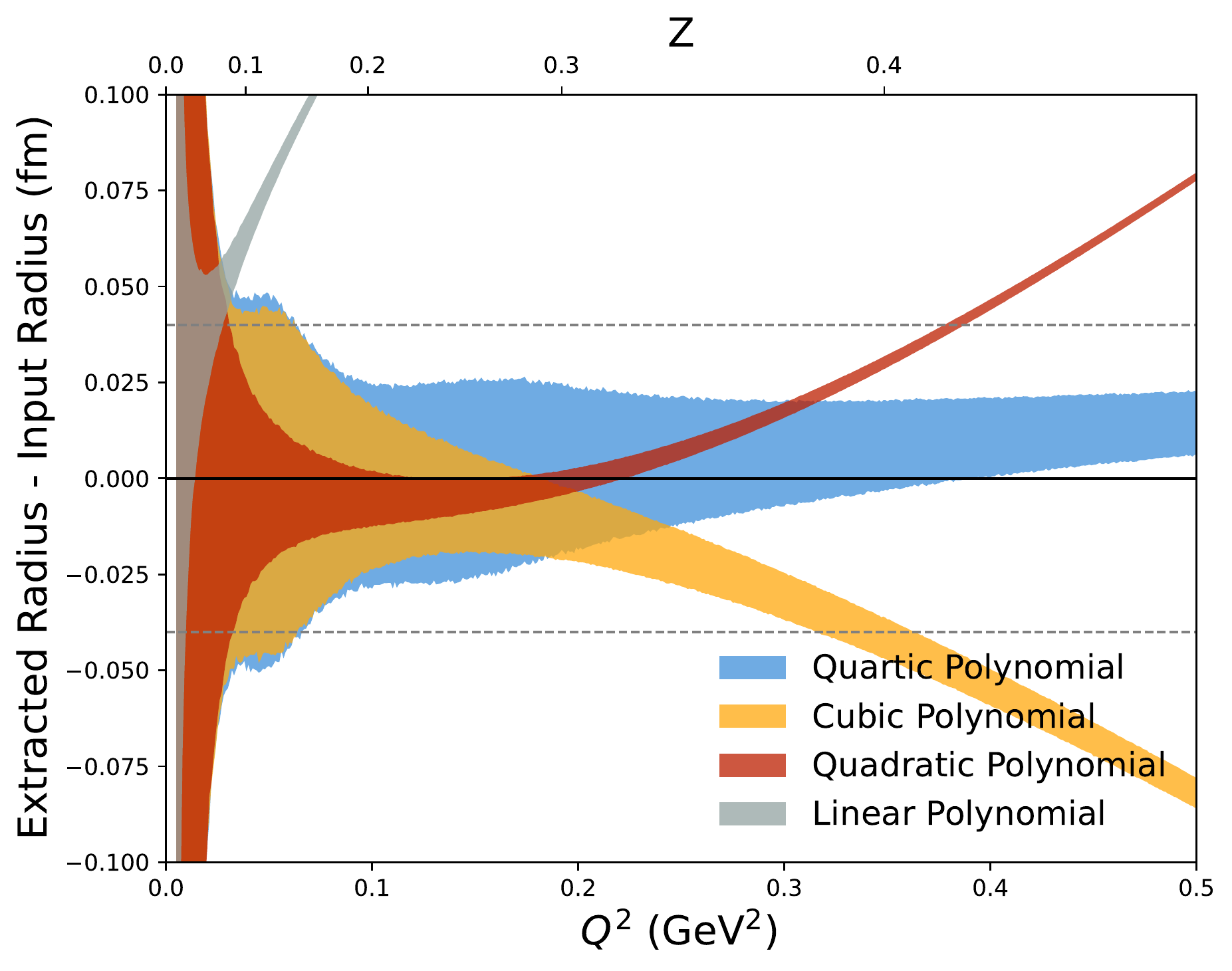}
    \caption{Bias-Variance plot for bounded polynomial fits in $Z$ of the Dipole \GEp{} parameterization}
\end{figure}

\begin{figure}[hb]
    \centering
    \includegraphics[height=0.2\textheight]{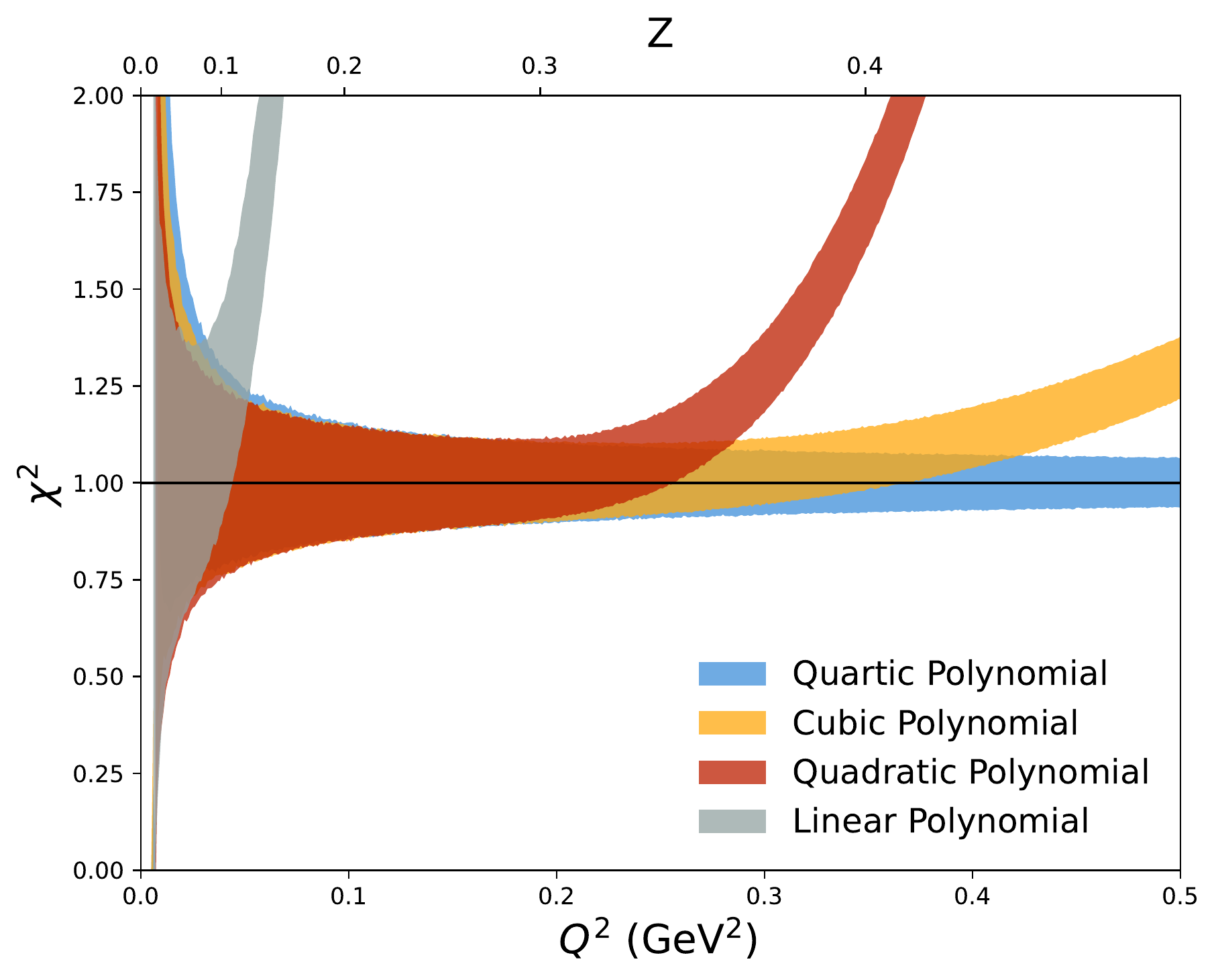}
    \caption{$\chi^2$ plot for bounded polynomial fits in $Z$ of the Dipole \GEp{} parameterization}
\end{figure}

\begin{figure}[hb]
    \centering
    \includegraphics[height=0.2\textheight]{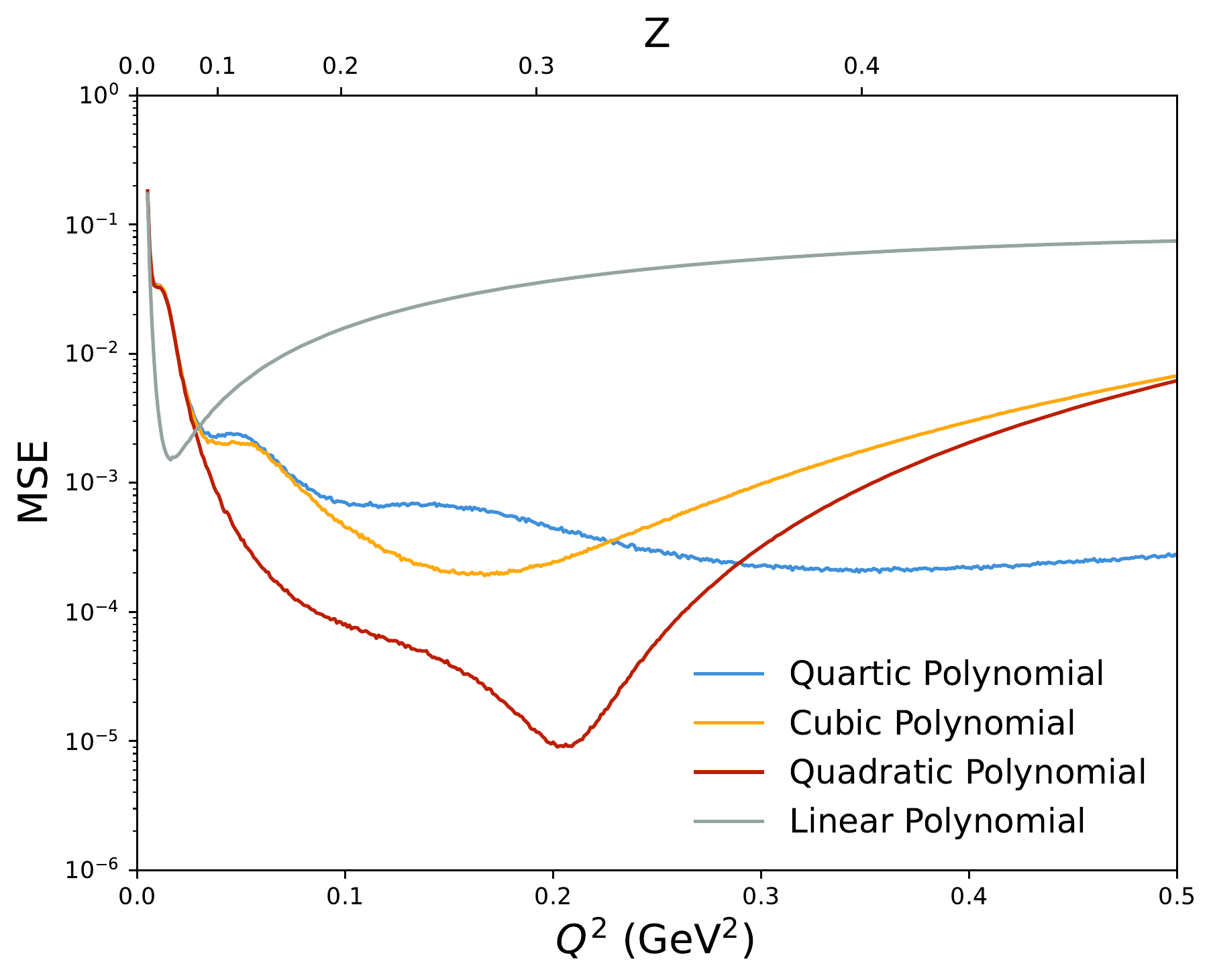}
    \caption{Mean Squared Error plot for bounded polynomial fits in $Z$ of the Dipole \GEp{} parameterization}
\end{figure}

\newpage

\subsubsection{Figure Descriptions}

Three plots, each with “$Q^2$ (GeV$^2$)” from 0 to 0.5 on the x axis. A second x axis on top shows the mapping of $Q^2$ to $Z$, which spreads out low $Q^2$ points and bunches up high $Q^2$ points. Each plot has four sets of data shown, one for each power of polynomial used, labeled as “Linear Polynomial”, “Quadratic Polynomial”, “Cubic Polynomial”, and “Quartic Polynomial”.

The top plot has “Extracted Radius - Input Radius (fm)” from -0.1 to 0.1 on the y axis. Two horizontal lines drawn at $\pm$0.04 fm show the magnitude of the proton radius puzzle. Each data set is shown as a band. The low $Q^2$ variance is significantly tempered from the unbound fits in $Z$. However, as $Q^2$ increases the curves are not qualitatively different from the unbound fits.

The middle plot has $\chi^2$ from 0 to 2 on the y axis. Each data set is shown as a band. The data is not qualitatively different than the unbound fits.

The bottom plot has MSE, Mean Squared Error as described in Eq.~\ref{eq:MSE}, from $10^{-6}$ to $10^0$ with a logarithmic scale on the y axis. Each set of data is represented as a curve. The data approaches their minima more rapidly than the unbound fits, but as $Q^2$ increases they are not qualitatively different from the unbound fits.

\newpage

\subsection{Rational(N,M) Fits in $Q^2$}

\begin{figure}[hb]
    \centering
    \includegraphics[height=0.2\textheight]{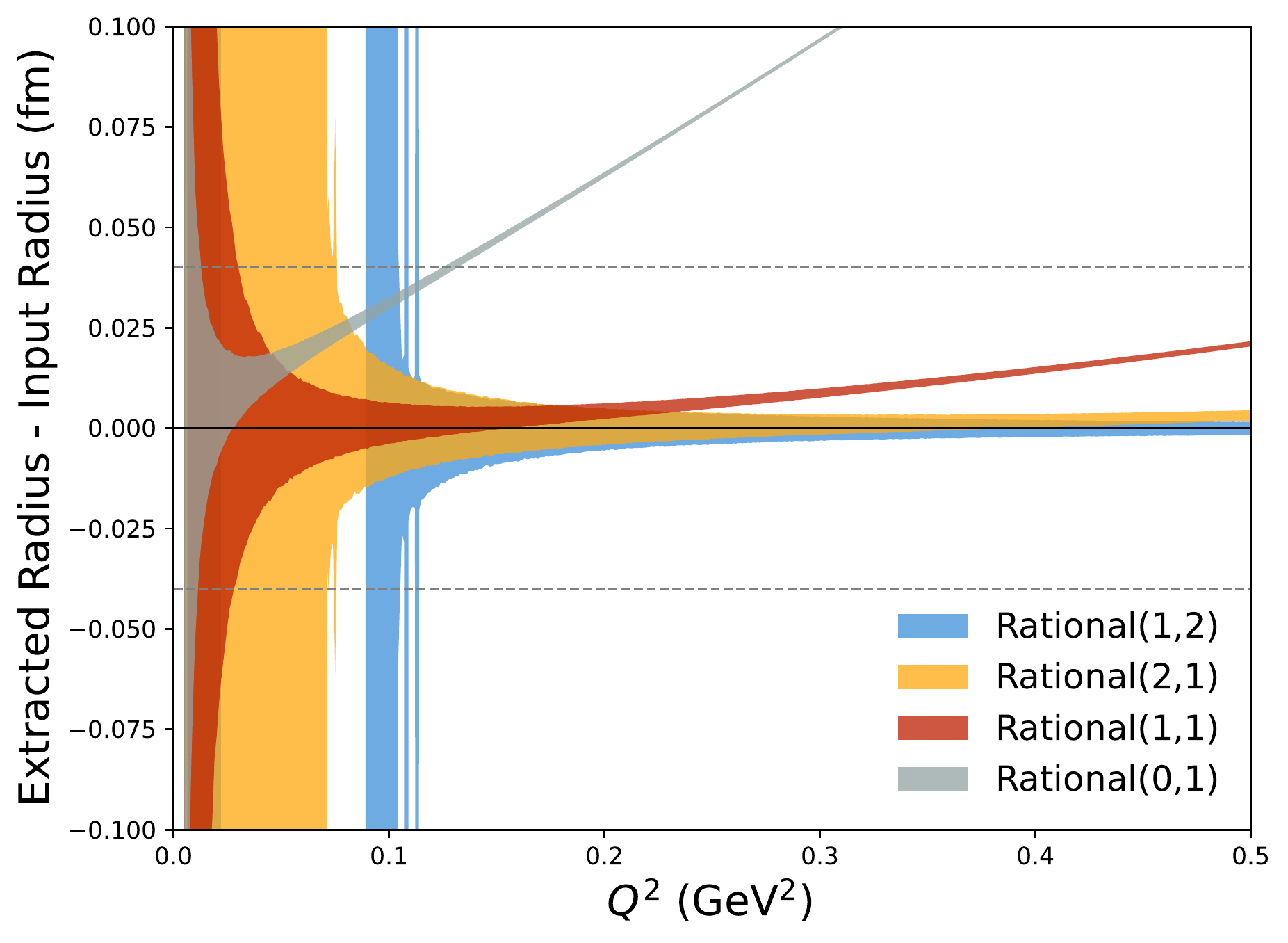}
    \caption{Bias-Variance plot for Rational(N,M) fits in $Q^2$ of the Dipole \GEp{} parameterization}
\end{figure}

\begin{figure}[hb]
    \centering
    \includegraphics[height=0.2\textheight]{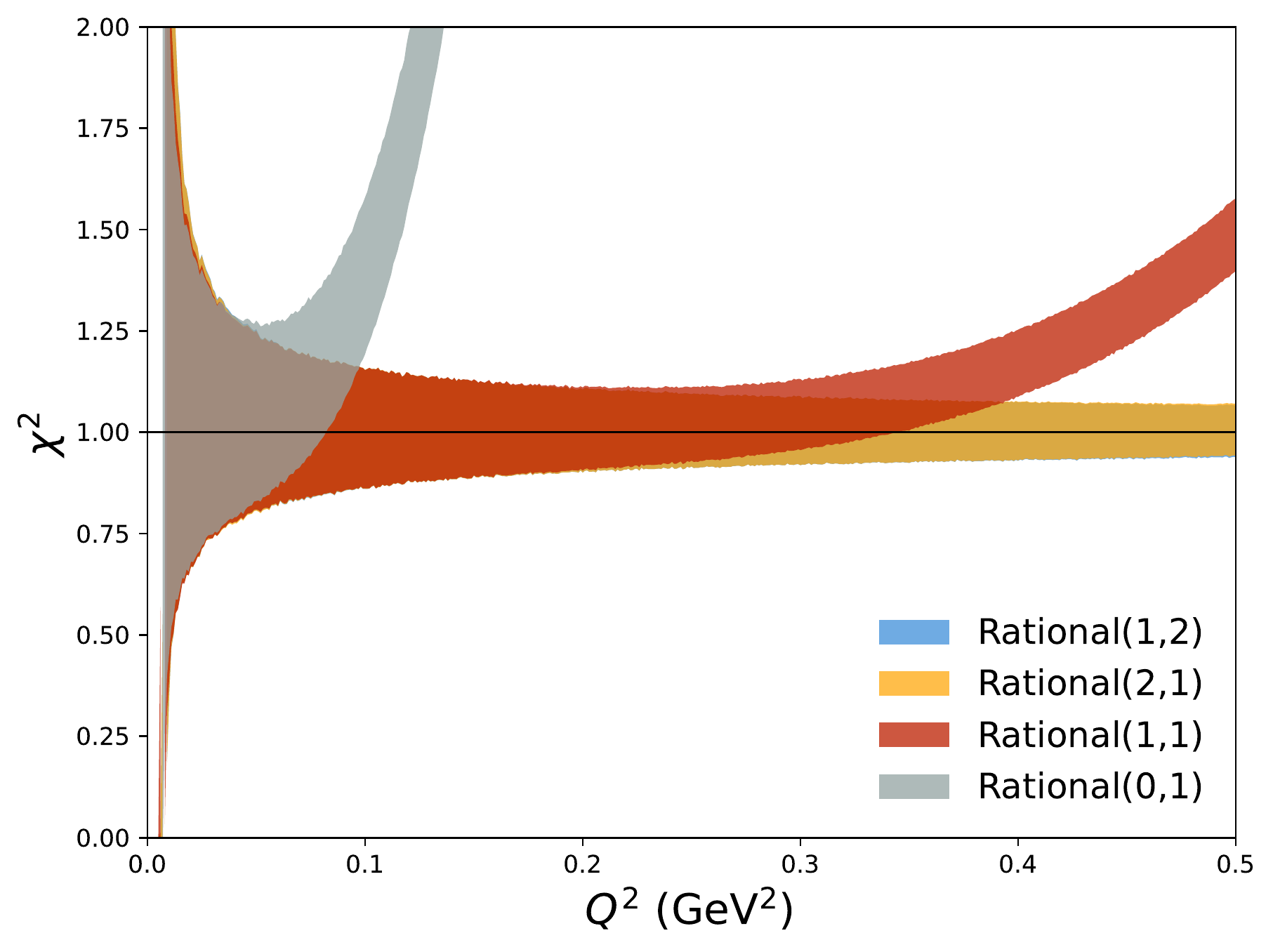}
    \caption{$\chi^2$ plot for Rational(N,M) fits in $Q^2$ of the Dipole \GEp{} parameterization}
\end{figure}

\begin{figure}[hb]
    \centering
    \includegraphics[height=0.2\textheight]{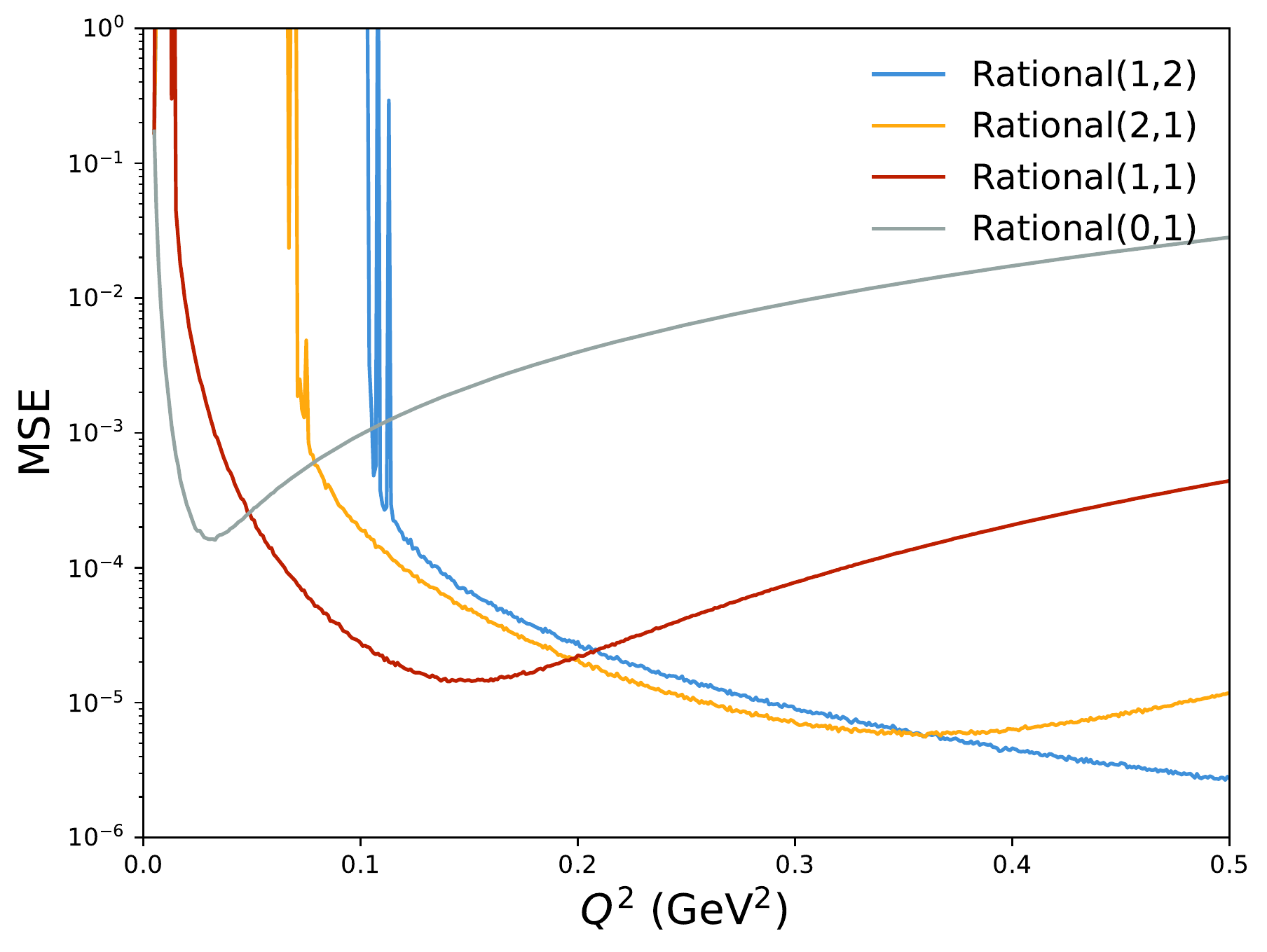}
    \caption{Mean Squared Error plot for Rational(N,M) fits in $Q^2$ of the Dipole \GEp{} parameterization}
\end{figure}

\newpage

\subsubsection{Figure Descriptions}

Three plots, each with “$Q^2$ (GeV$^2$)” from 0 to 0.5 on the x axis. Each plot has four sets of data shown, one for each power of Rational(N,M) function used, labeled as “Rational(0,1)”, “Rational(1,1)”, “Rational(2,1)”, and “Rational(1,2)”.

The top plot has “Extracted Radius - Input Radius (fm)” from -0.1 to 0.1 on the y axis. Two horizontal lines drawn at $\pm$0.04 fm show the magnitude of the proton radius puzzle. Each data set is shown as a band. This data shows larger variance at low $Q^2$ than any other fit and the variance increases with increasing N and M, with M having a larger influence on the variance. The Rational(0,1) curve rapidly diverges from the input radius. The Rational(1,2) curve shows no data below $Q^2\approx0.1$ GeV$^2$ as the mean radius was unphysical. The Rational(1,1) curve slowly diverges to a large radius with increasing $Q^2$. The Rational(1,2) and Rational(2,1) curves stay well centered on the input radius with the Rational(2,1) curve slightly deviating to a large radius at high $Q^2$. The variances of all Rational(N,M) fits are less than half that of any of the other fits.

The middle plot has $\chi^2$ from 0 to 2 on the y axis. Each data set is shown as a band. All begin centered at 1, with Rational(0,1) having a rapid upturn and Rational(1,1) having a slow upturn. The others stay centered at 1 for the full range.

The bottom plot has MSE, Mean Squared Error as described in Eq.~\ref{eq:MSE}, from $10^{-6}$ to $10^0$ with a logarithmic scale on the y axis. Each set of data is represented as a curve. With the exception of Rational(1,2), all approach their minima and then begin to increase. Rational(1,2) is steadily decreasing over the full range and is approximately $3\cdot10^{-6}$ at the highest $Q^2$. These are the lowest MSE values of all fit types.

\newpage

\section{Kelly}

\subsection{Polynomial Fits in $Q^2$}

\begin{figure}[hb]
    \centering
    \includegraphics[height=0.2\textheight]{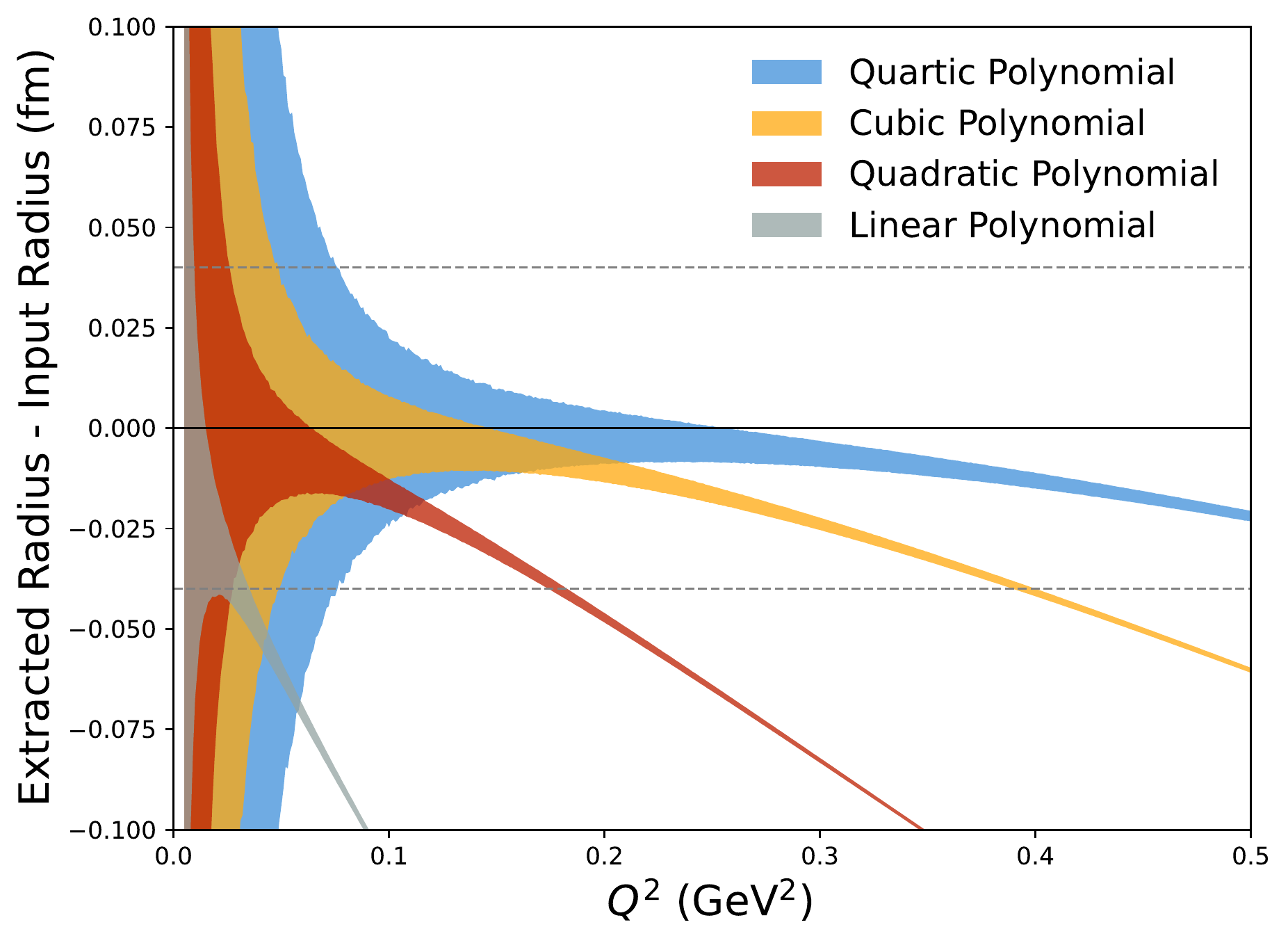}
    \caption{Bias-Variance plot for polynomial fits in $Q^2$ of the Kelly \GEp{} parameterization}
\end{figure}

\begin{figure}[hb]
    \centering
    \includegraphics[height=0.2\textheight]{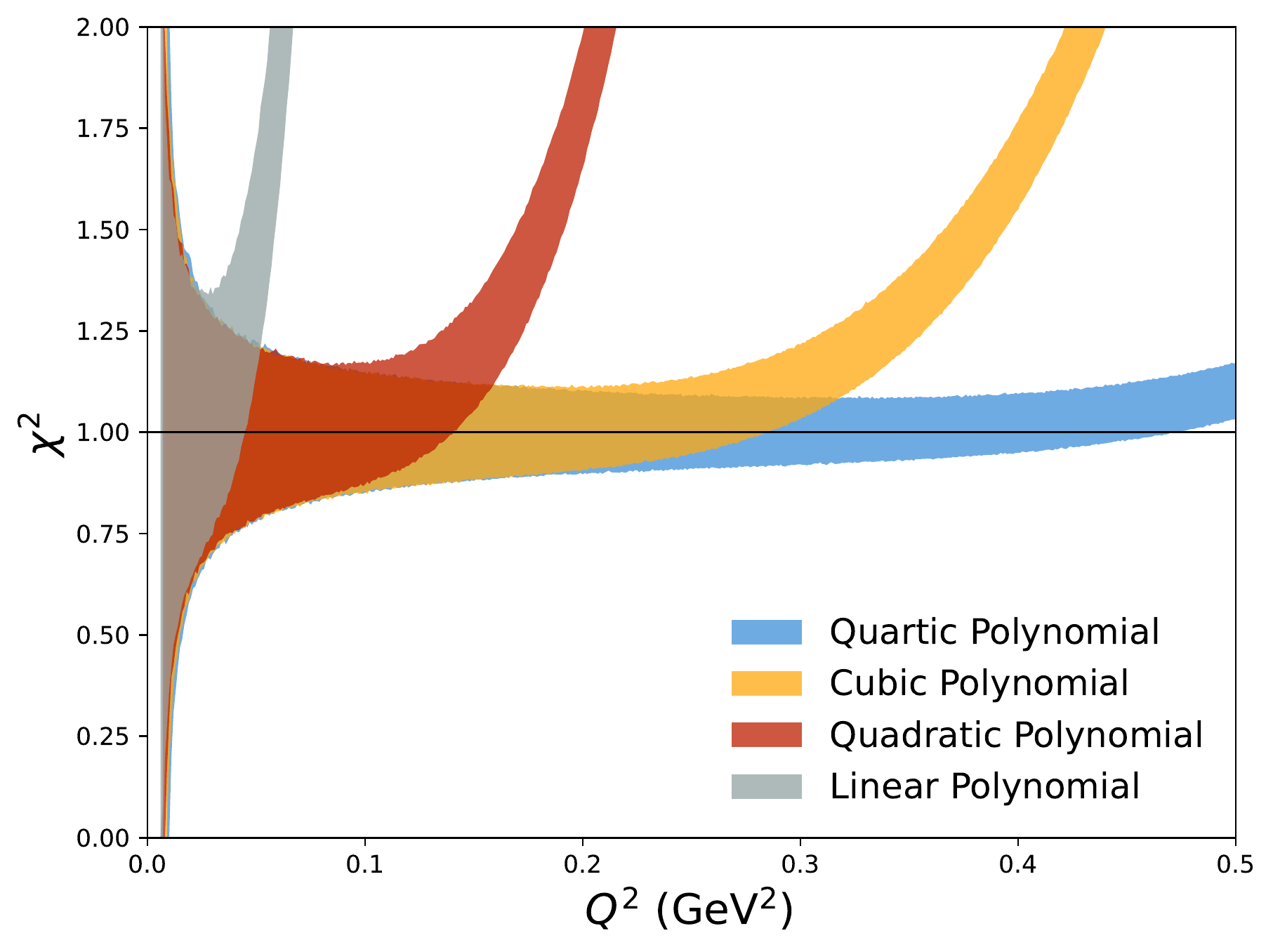}
    \caption{$\chi^2$ plot for polynomial fits in $Q^2$ of the Kelly \GEp{} parameterization}
\end{figure}

\begin{figure}[hb]
    \centering
    \includegraphics[height=0.2\textheight]{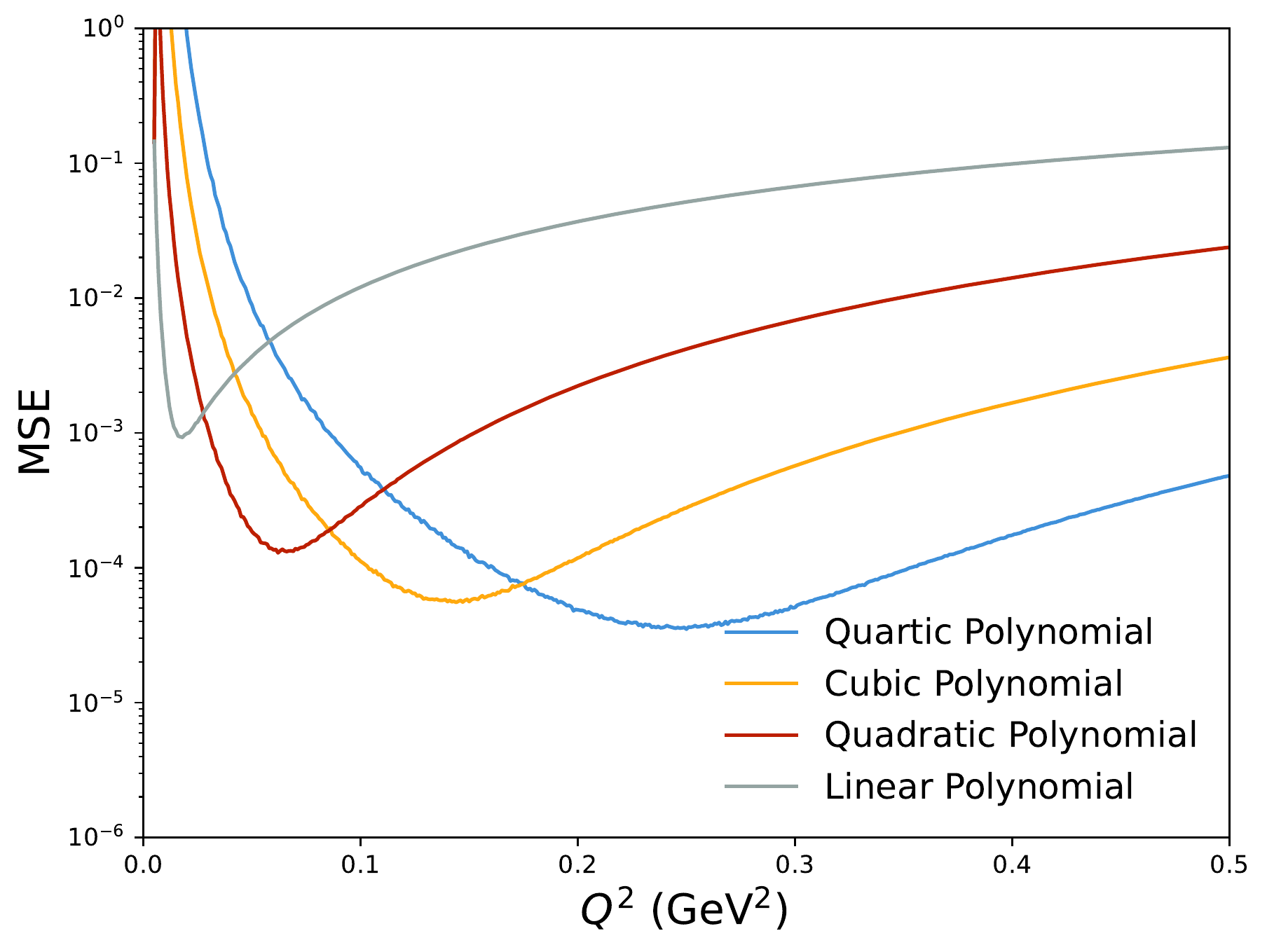}
    \caption{Mean Squared Error plot for polynomial fits in $Q^2$ of the Kelly \GEp{} parameterization}
\end{figure}

\newpage

\subsubsection{Figure Descriptions}

Three plots, each with “$Q^2$ (GeV$^2$)” from 0 to 0.5 on the x axis. Each plot has four sets of data shown, one for each power of polynomial used, labeled as “Linear Polynomial”, “Quadratic Polynomial”, “Cubic Polynomial”, and “Quartic Polynomial”.

The top plot has “Extracted Radius - Input Radius (fm)” from -0.1 to 0.1 on the y axis. Two horizontal lines drawn at $\pm$0.04 fm show the magnitude of the proton radius puzzle. Each data set is shown as a band. All bands have variance higher than the axis scale at small $Q^2$ and systematically return a small radius that decreases with increasing $Q^2$ range. Higher order polynomials decrease slower.

The middle plot has $\chi^2$ from 0 to 2 on the y axis. Each data set is shown as a band. All bands begin centered at 1 and then at some $Q^2$ have a rapid upturn. Higher order polynomials delay the upturn until higher $Q^2$. The deviation from $\chi^2\approx1$ occurs at significantly higher $Q^2$ than when the extracted radius deviates from the input radius.

The bottom plot has MSE, Mean Squared Error as described in Eq.~\ref{eq:MSE}, from $10^{-6}$ to $10^0$ with a logarithmic scale on the y axis. Each set of data is represented as a curve. Each curve rapidly drops to the minimum MSE for the fit and then begins to increase with $Q^2$ and start to level off about two orders of magnitude higher than the minima. Higher order polynomials have lower minima that occur at higher $Q^2$ and then increase slower. The lowest minima is about $4\cdot10^{-5}$.

\newpage

\subsection{Bounded Polynomial Fits in $Q^2$}

\begin{figure}[hb]
    \centering
    \includegraphics[height=0.2\textheight]{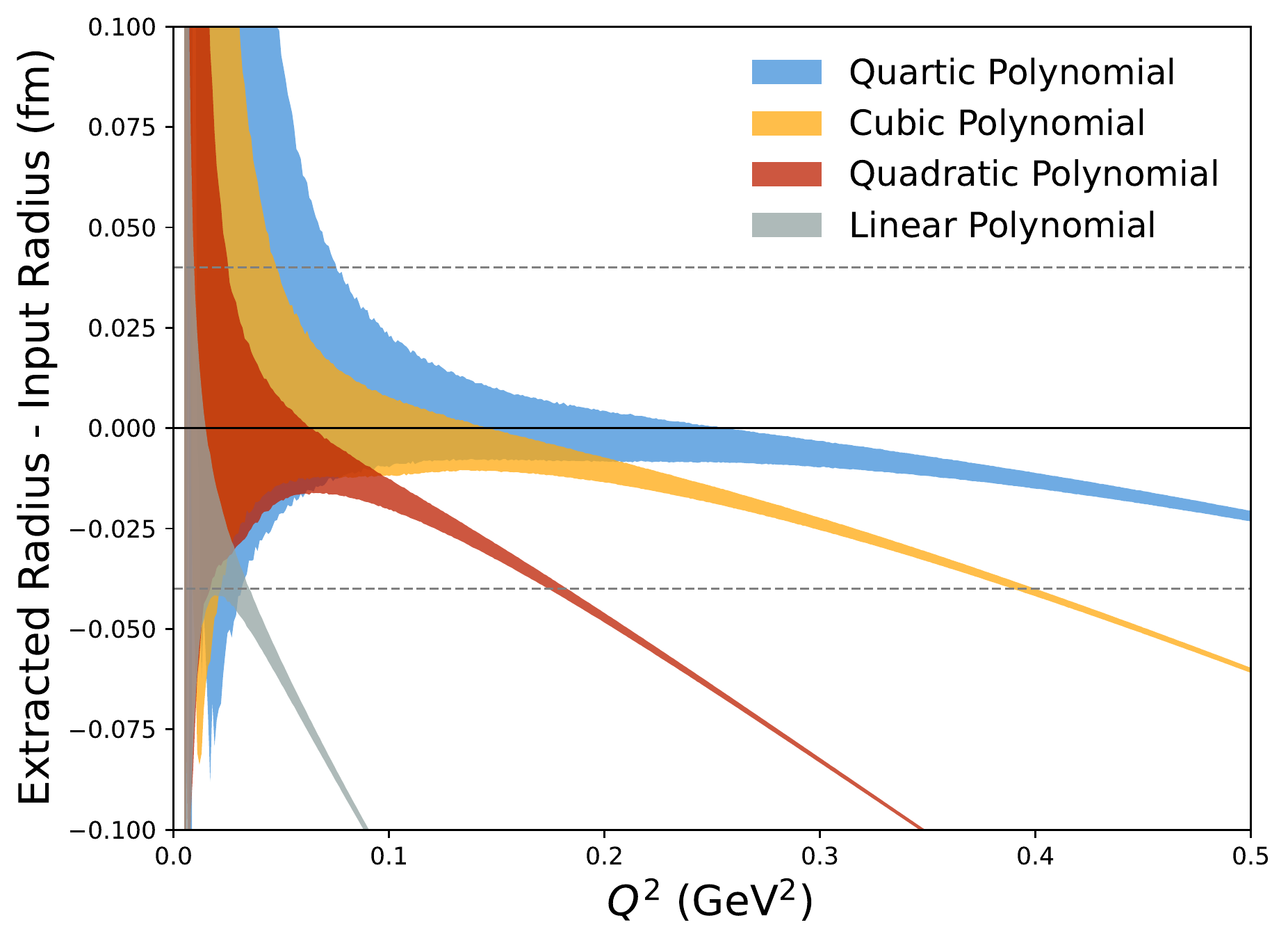}
    \caption{Bias-Variance plot for bounded polynomial fits in $Q^2$ of the Kelly \GEp{} parameterization}
\end{figure}

\begin{figure}[hb]
    \centering
    \includegraphics[height=0.2\textheight]{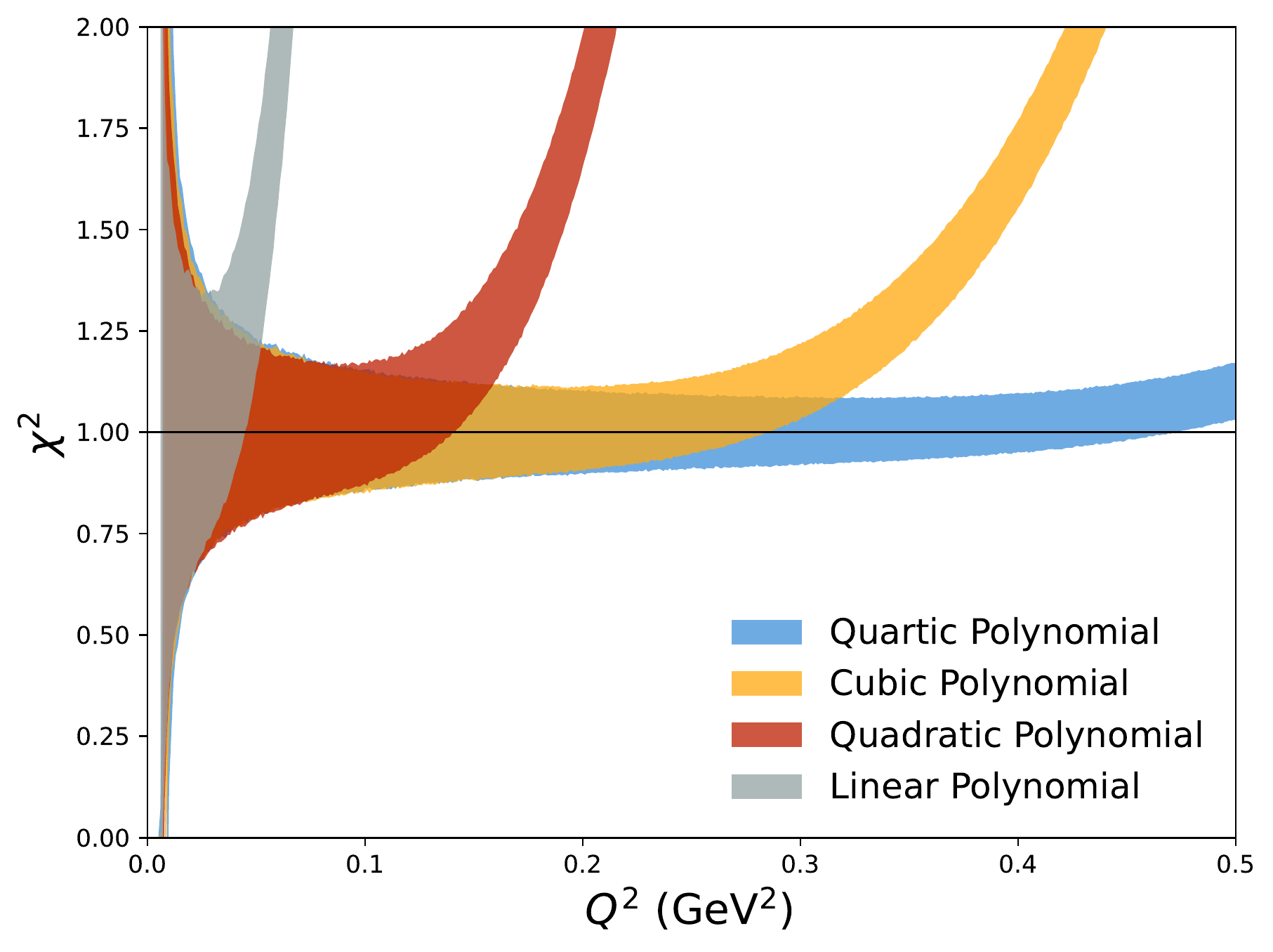}
    \caption{$\chi^2$ plot for bounded polynomial fits in $Q^2$ of the Kelly \GEp{} parameterization}
\end{figure}

\begin{figure}[hb]
    \centering
    \includegraphics[height=0.2\textheight]{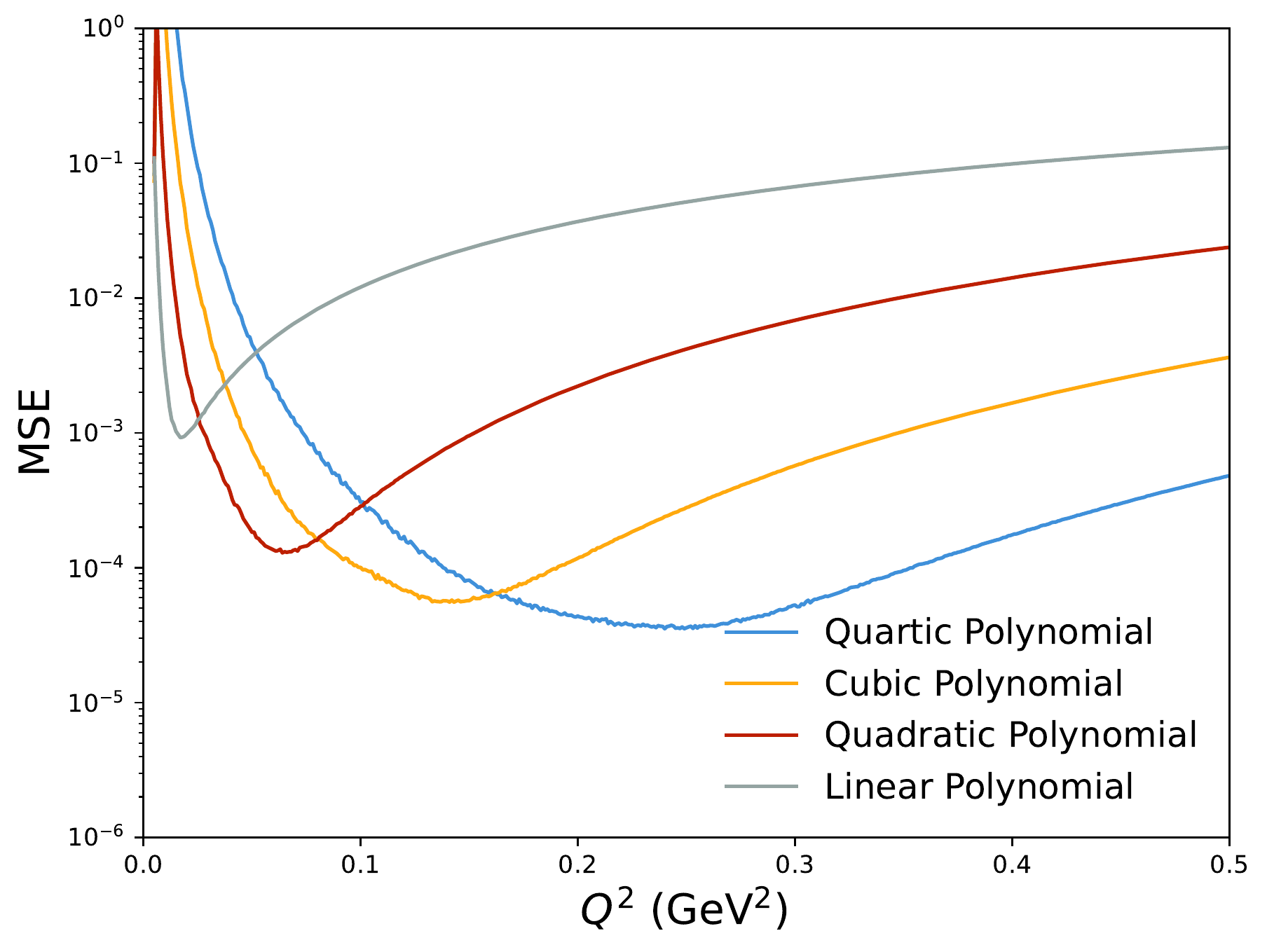}
    \caption{Mean Squared Error plot for bounded polynomial fits in $Q^2$ of the Kelly \GEp{} parameterization}
\end{figure}

\newpage

\subsubsection{Figure Descriptions}

Three plots, each with “$Q^2$ (GeV$^2$)” from 0 to 0.5 on the x axis. Each plot has four sets of data shown, one for each power of polynomial used, labeled as “Linear Polynomial”, “Quadratic Polynomial”, “Cubic Polynomial”, and “Quartic Polynomial”.

The top plot has “Extracted Radius - Input Radius (fm)” from -0.1 to 0.1 on the y axis. Two horizontal lines drawn at $\pm$0.04 fm show the magnitude of the proton radius puzzle. Each data set is shown as a band. The low $Q^2$ variance is smaller than the unbound fits, but is not overall qualitatively different.

The middle plot has $\chi^2$ from 0 to 2 on the y axis. Each data set is shown as a band. The data is not qualitatively different than the unbound fits.

The bottom plot has MSE, Mean Squared Error as described in Eq.~\ref{eq:MSE}, from $10^{-6}$ to $10^0$ with a logarithmic scale on the y axis. Each set of data is represented as a curve. The data is not qualitatively different than the unbound fits.

\newpage

\subsection{Polynomial Fits in $Z$}

\begin{figure}[hb]
    \centering
    \includegraphics[height=0.2\textheight]{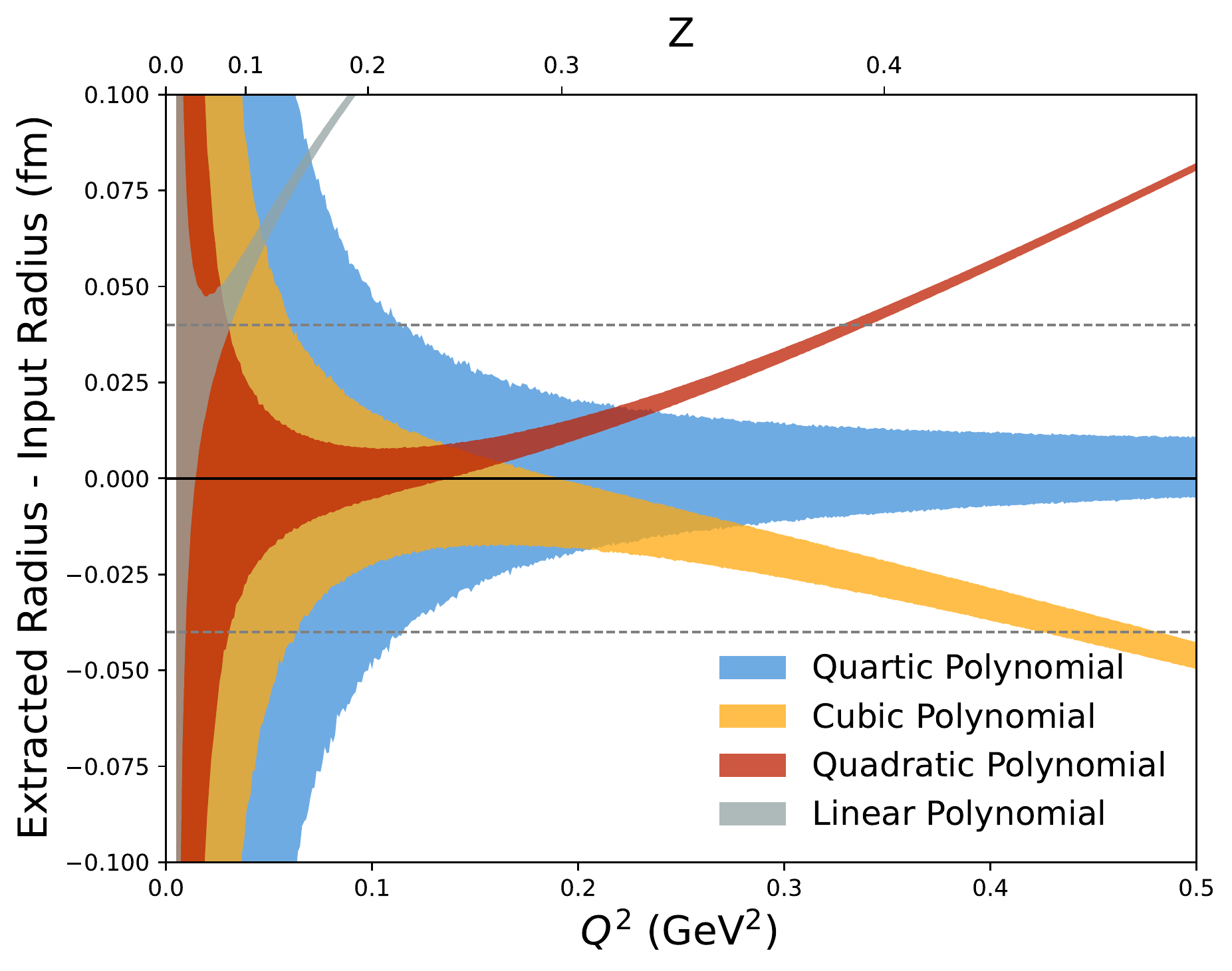}
    \caption{Bias-Variance plot for polynomial fits in $Z$ of the Kelly \GEp{} parameterization}
\end{figure}

\begin{figure}[hb]
    \centering
    \includegraphics[height=0.2\textheight]{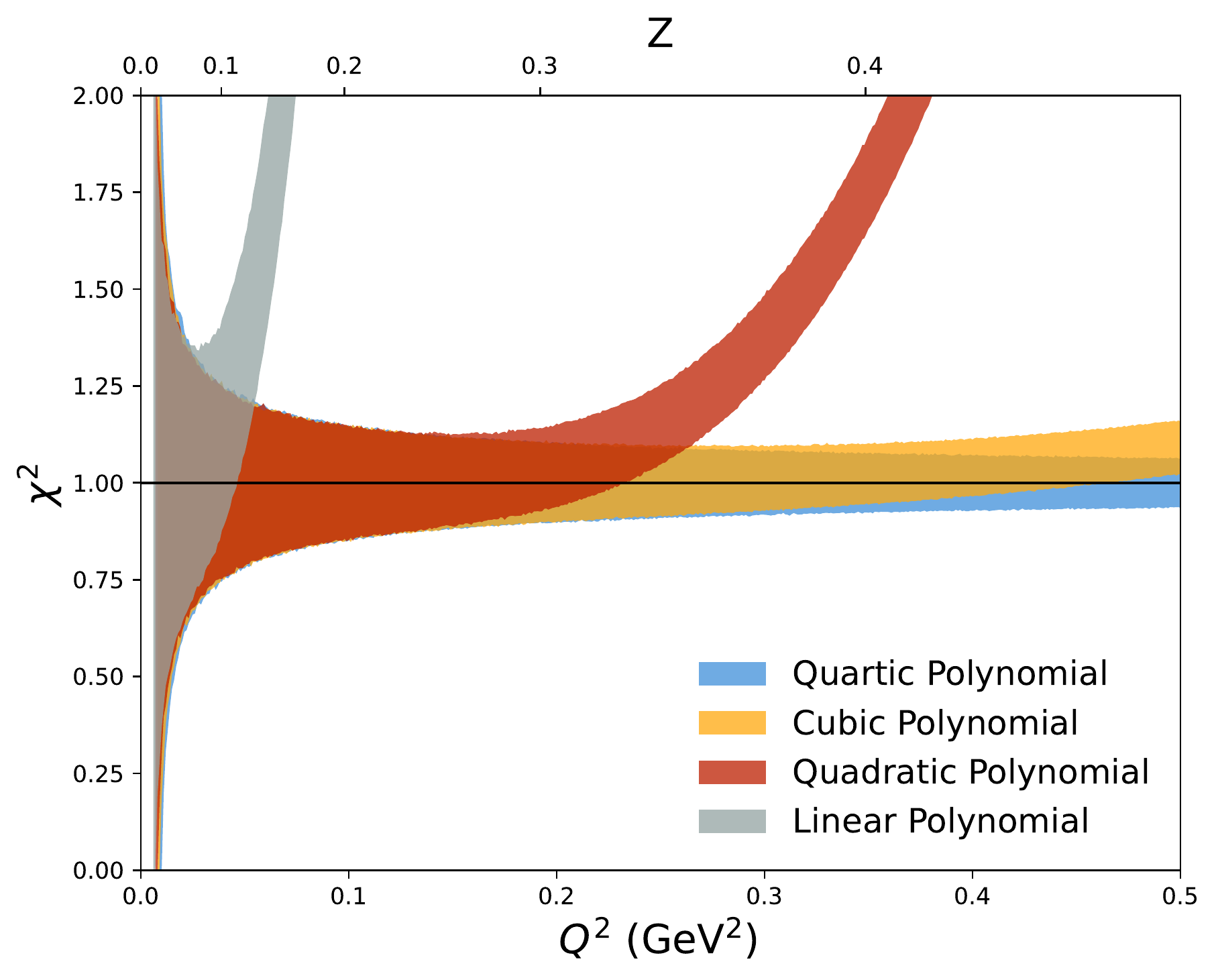}
    \caption{$\chi^2$ plot for polynomial fits in $Z$ of the Kelly \GEp{} parameterization}
\end{figure}

\begin{figure}[hb]
    \centering
    \includegraphics[height=0.2\textheight]{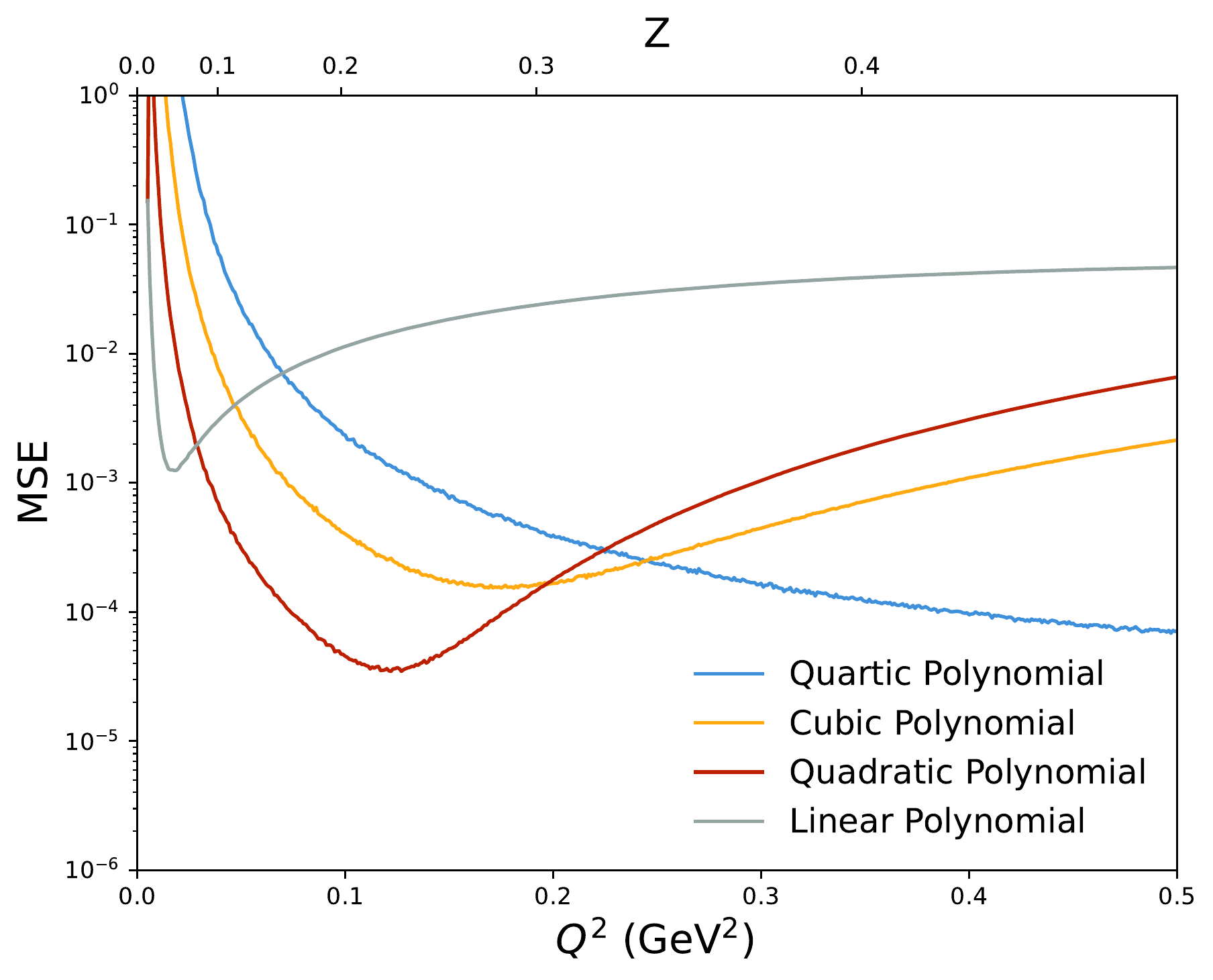}
    \caption{Mean Squared Error plot for polynomial fits in $Z$ of the Kelly \GEp{} parameterization}
\end{figure}

\newpage

\subsubsection{Figure Descriptions}

Three plots, each with “$Q^2$ (GeV$^2$)” from 0 to 0.5 on the x axis. A second x axis on top shows the mapping of $Q^2$ to $Z$, which spreads out low $Q^2$ points and bunches up high $Q^2$ points. Each plot has four sets of data shown, one for each power of polynomial used, labeled as “Linear Polynomial”, “Quadratic Polynomial”, “Cubic Polynomial”, and “Quartic Polynomial”.

The top plot has “Extracted Radius - Input Radius (fm)” from -0.1 to 0.1 on the y axis. Two horizontal lines drawn at $\pm$0.04 fm show the magnitude of the proton radius puzzle. Each data set is shown as a band. The high variance region extends to higher $Q^2$ than fits in $Q^2$ and the variance is higher than the $Q^2$ fits for the full range of the plot. With the exception of the quartic polynomial, the curves rapidly diverge from the input radius at approximately the same $Q^2$ as the fits in $Q^2$, but are better centered at the correct radius at lower $Q^2$. The quartic polynomial remains well centered with a slight indication of trending towards a large radius at high $Q^2$. The linear and quadratic polynomials diverge to large radii and the cubic polynomial diverges to a small radius.

The middle plot has $\chi^2$ from 0 to 2 on the y axis. Each data set is shown as a band. All bands begin centered at 1 and then at some $Q^2$ have a rapid upturn. Higher order polynomials delay the upturn until higher $Q^2$. Much like the fits in $Q^2$, the deviation from $\chi^2\approx1$ occurs at significantly higher $Q^2$ than when the extracted radius deviates from the input radius. The quartic polynomial remains well centered at 1 for the full range.

The bottom plot has MSE, Mean Squared Error as described in Eq.~\ref{eq:MSE}, from $10^{-6}$ to $10^0$ with a logarithmic scale on the y axis. Each set of data is represented as a curve. Each curve rapidly drops to the minimum MSE for the fit and then begins to increase with $Q^2$ and start to level off about two orders of magnitude higher than the minima. The quadratic polynomial has the lowest minimum at about $3\cdot10^{-5}$. The minima are spread over a larger $Q^2$ range and the range grows with polynomial order. The quartic polynomial curve is decreasing over the full range.

\newpage

\subsection{Bounded Polynomial Fits in $Z$}

\begin{figure}[hb]
    \centering
    \includegraphics[height=0.2\textheight]{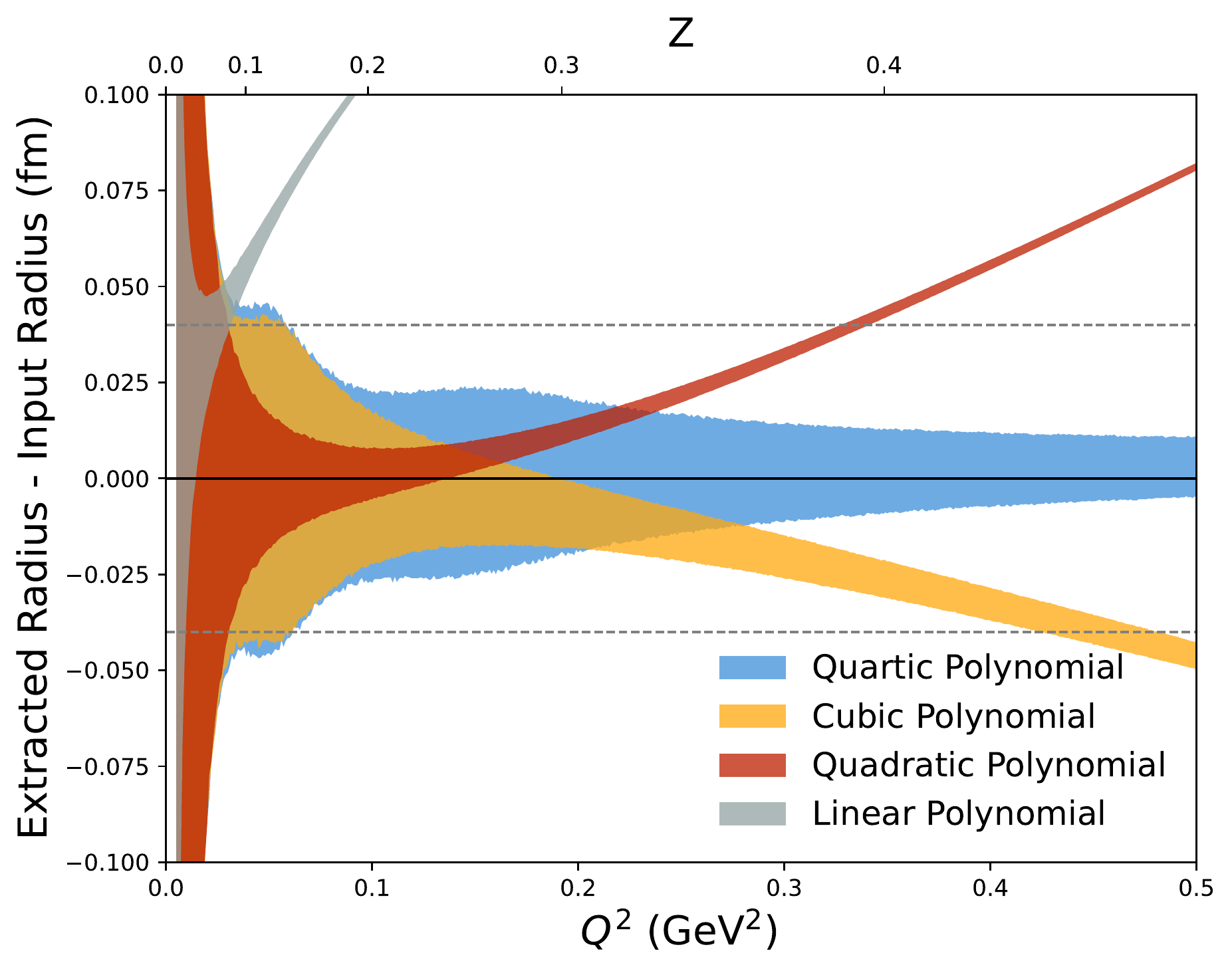}
    \caption{Bias-Variance plot for bounded polynomial fits in $Z$ of the Kelly \GEp{} parameterization}
\end{figure}

\begin{figure}[hb]
    \centering
    \includegraphics[height=0.2\textheight]{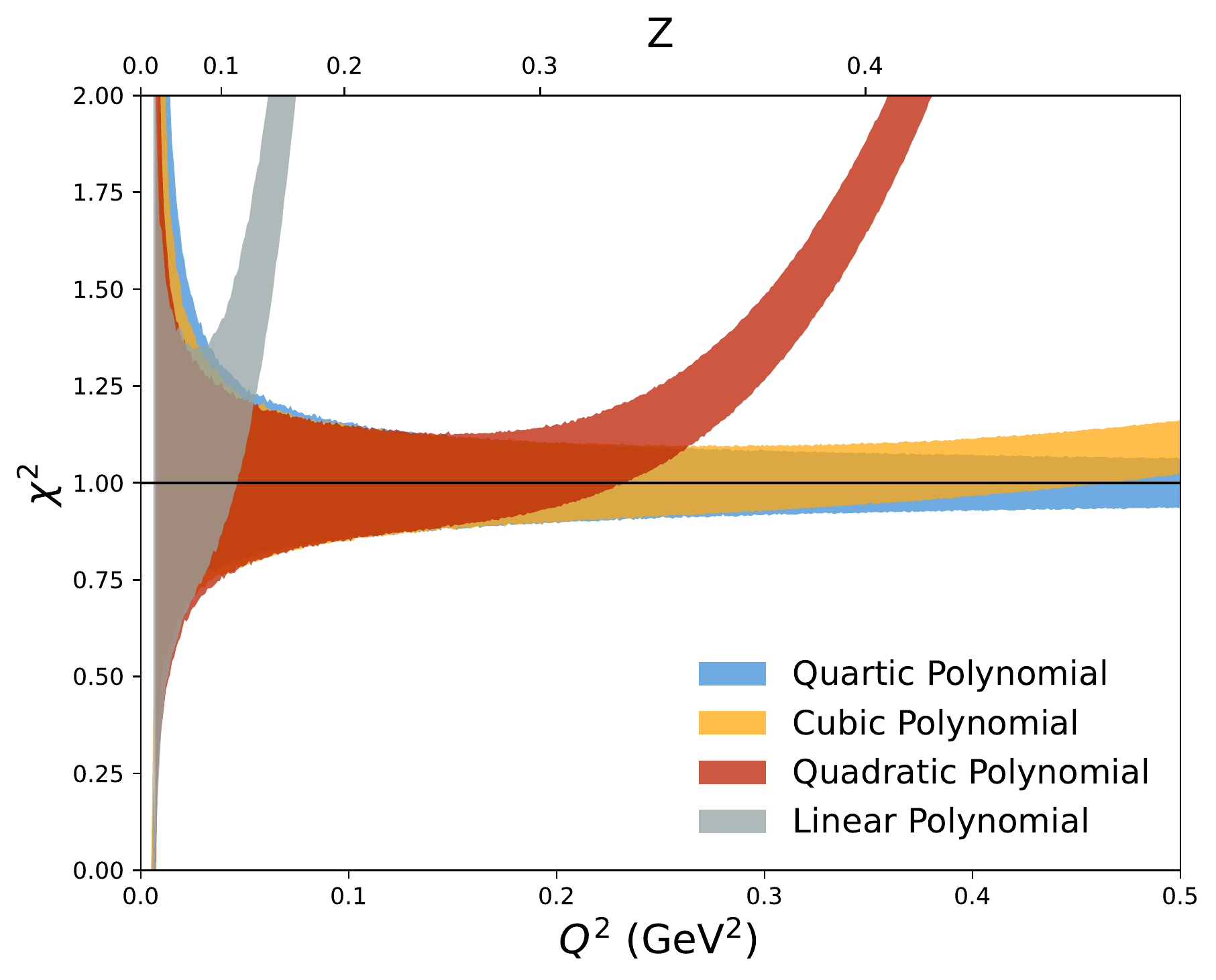}
    \caption{$\chi^2$ plot for bounded polynomial fits in $Z$ of the Kelly \GEp{} parameterization}
\end{figure}

\begin{figure}[hb]
    \centering
    \includegraphics[height=0.2\textheight]{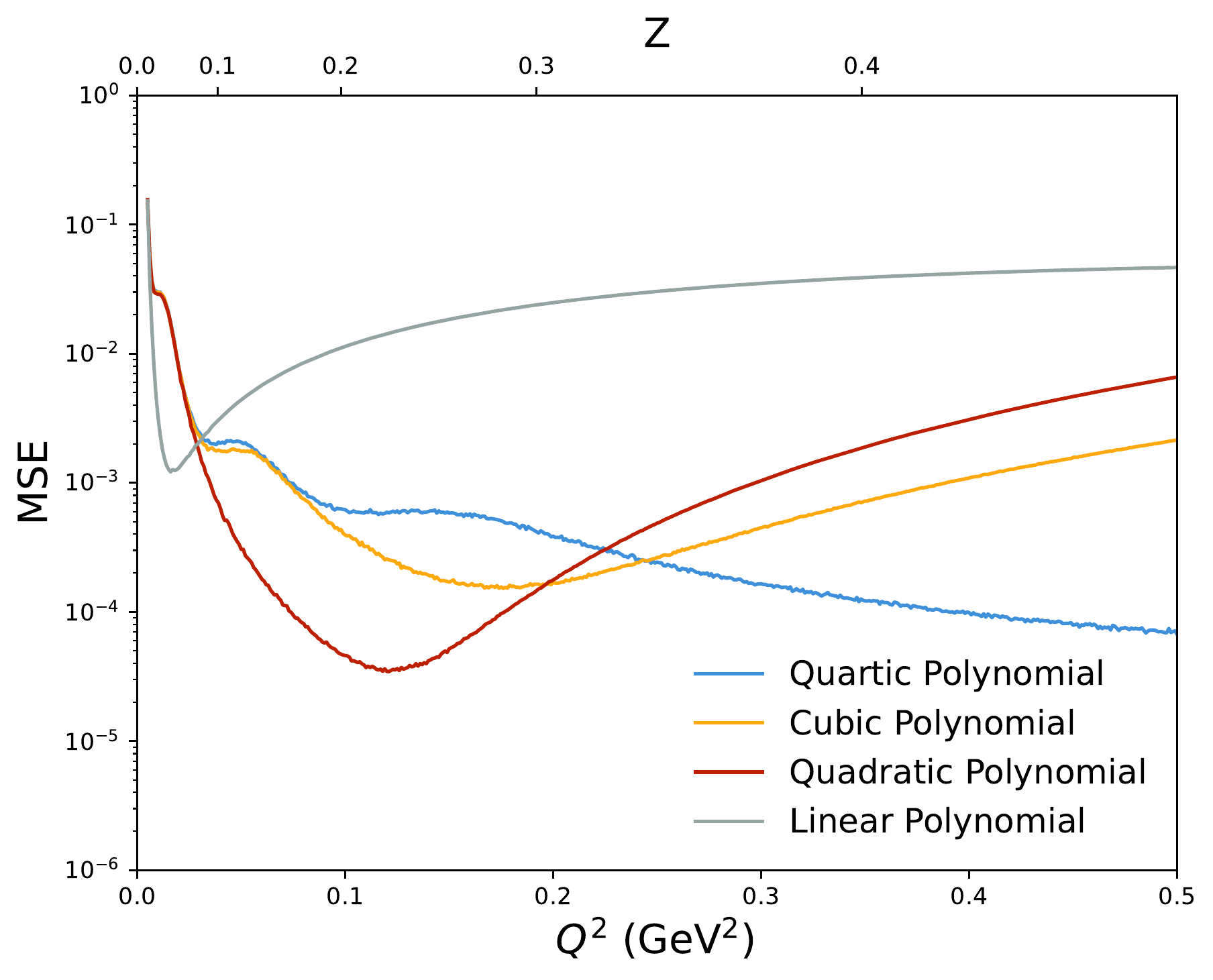}
    \caption{Mean Squared Error plot for bounded polynomial fits in $Z$ of the Kelly \GEp{} parameterization}
\end{figure}

\newpage

\subsubsection{Figure Descriptions}

Three plots, each with “$Q^2$ (GeV$^2$)” from 0 to 0.5 on the x axis. A second x axis on top shows the mapping of $Q^2$ to $Z$, which spreads out low $Q^2$ points and bunches up high $Q^2$ points. Each plot has four sets of data shown, one for each power of polynomial used, labeled as “Linear Polynomial”, “Quadratic Polynomial”, “Cubic Polynomial”, and “Quartic Polynomial”.

The top plot has “Extracted Radius - Input Radius (fm)” from -0.1 to 0.1 on the y axis. Two horizontal lines drawn at $\pm$0.04 fm show the magnitude of the proton radius puzzle. Each data set is shown as a band. The low $Q^2$ variance is significantly tempered from the unbound fits in $Z$. However, as $Q^2$ increases the curves are not qualitatively different from the unbound fits.

The middle plot has $\chi^2$ from 0 to 2 on the y axis. Each data set is shown as a band. The data is not qualitatively different than the unbound fits.

The bottom plot has MSE, Mean Squared Error as described in Eq.~\ref{eq:MSE}, from $10^{-6}$ to $10^0$ with a logarithmic scale on the y axis. Each set of data is represented as a curve. The data approaches their minima more rapidly than the unbound fits, but as $Q^2$ increases they are not qualitatively different from the unbound fits.

\newpage

\subsection{Rational(N,M) Fits in $Q^2$}

\begin{figure}[hb]
    \centering
    \includegraphics[height=0.2\textheight]{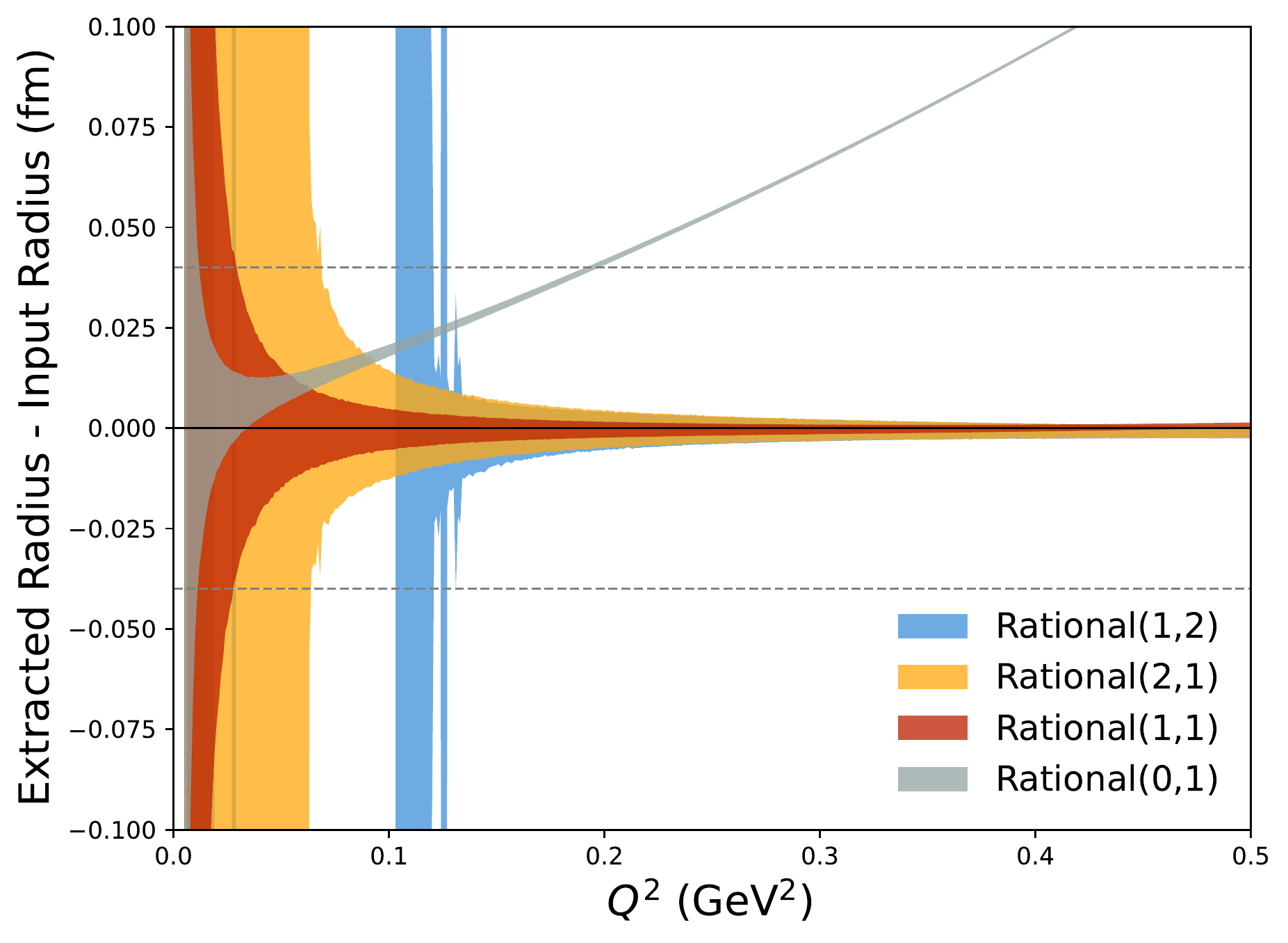}
    \caption{Bias-Variance plot for Rational(N,M) fits in $Q^2$ of the Kelly \GEp{} parameterization}
\end{figure}

\begin{figure}[hb]
    \centering
    \includegraphics[height=0.2\textheight]{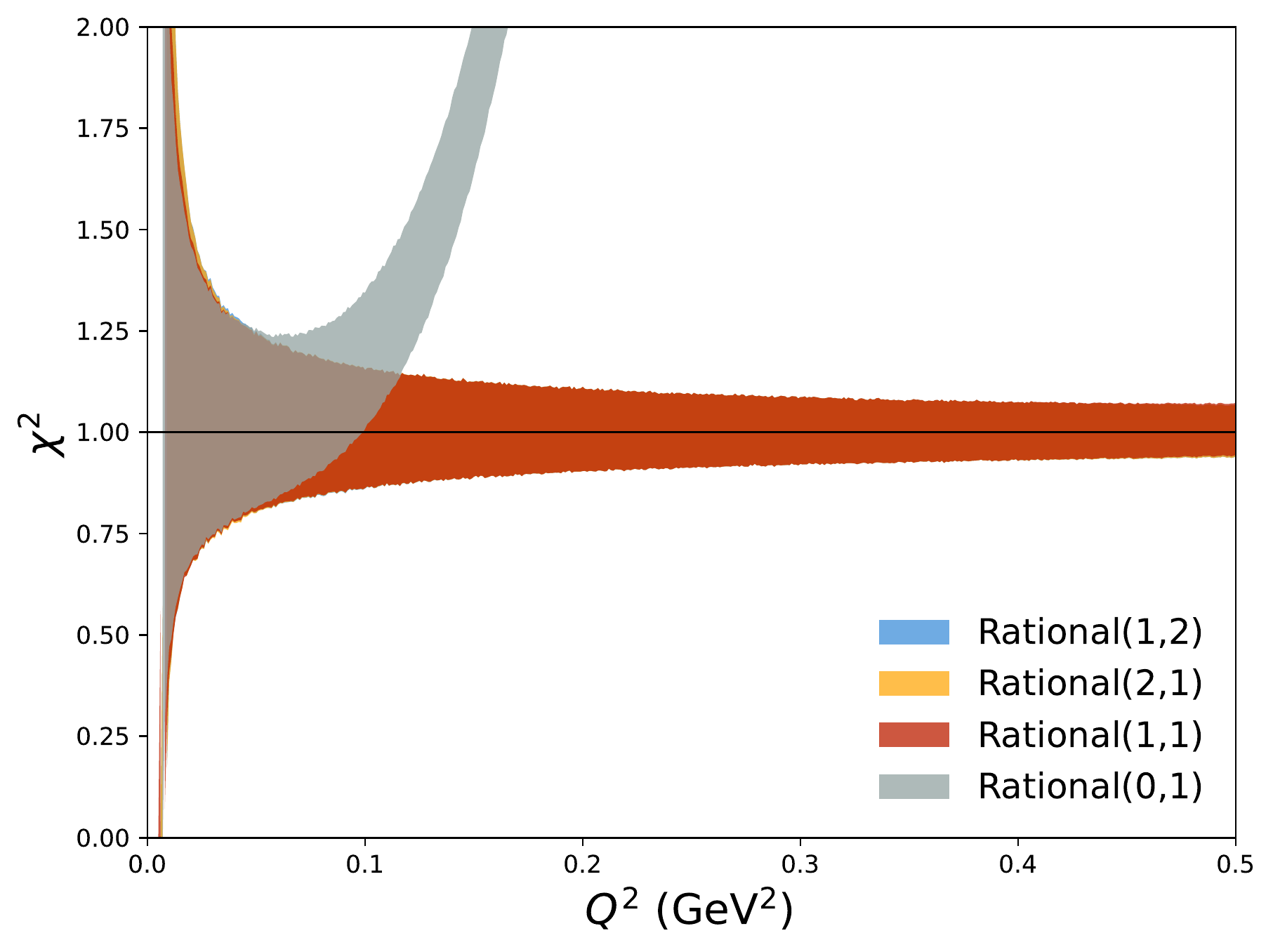}
    \caption{$\chi^2$ plot for Rational(N,M) fits in $Q^2$ of the Kelly \GEp{} parameterization}
\end{figure}

\begin{figure}[hb]
    \centering
    \includegraphics[height=0.2\textheight]{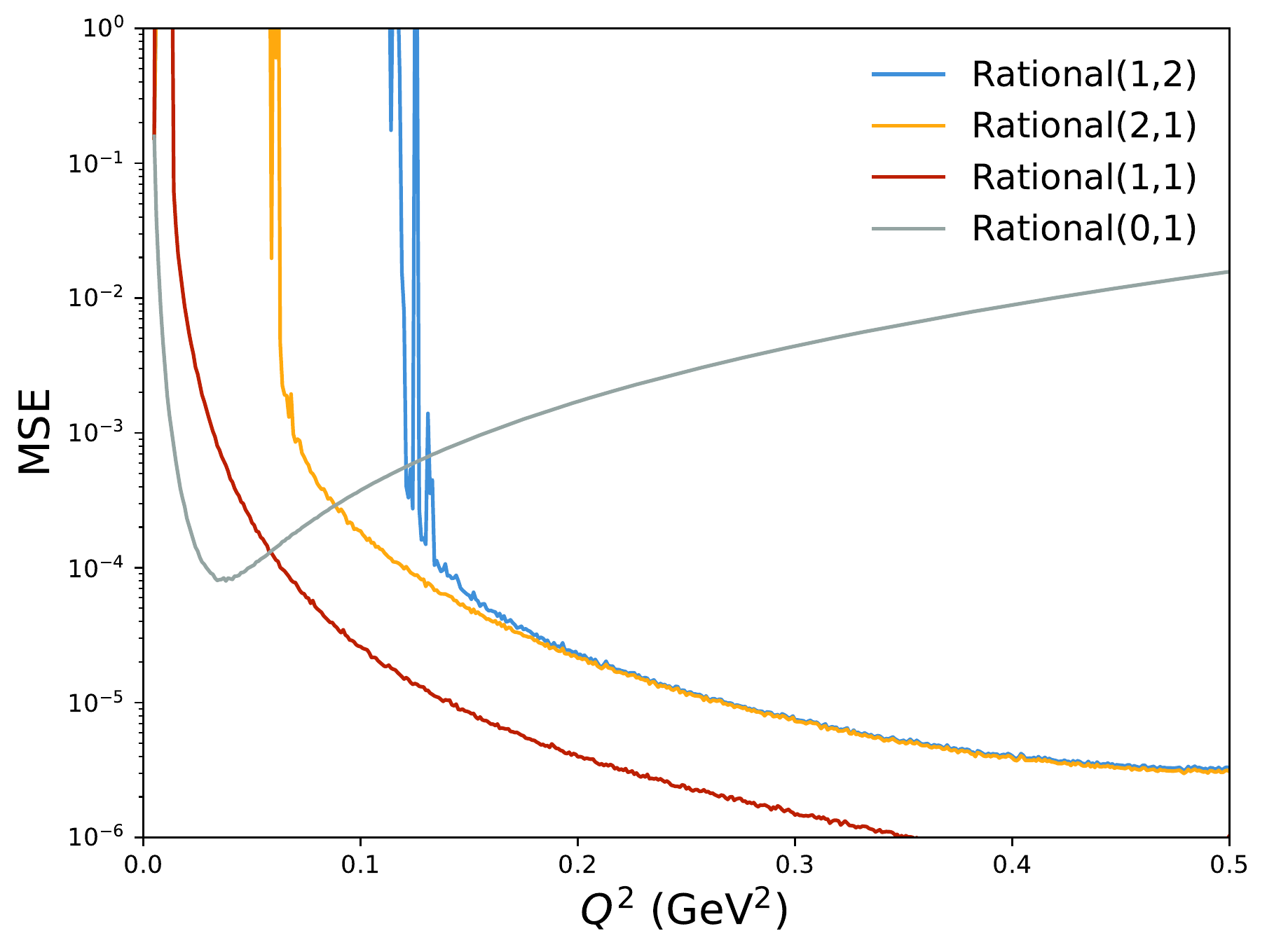}
    \caption{Mean Squared Error plot for Rational(N,M) fits in $Q^2$ of the Kelly \GEp{} parameterization}
\end{figure}

\newpage

\subsubsection{Figure Descriptions}

Three plots, each with “$Q^2$ (GeV$^2$)” from 0 to 0.5 on the x axis. Each plot has four sets of data shown, one for each power of Rational(N,M) function used, labeled as “Rational(0,1)”, “Rational(1,1)”, “Rational(2,1)”, and “Rational(1,2)”.

The top plot has “Extracted Radius - Input Radius (fm)” from -0.1 to 0.1 on the y axis. Two horizontal lines drawn at $\pm$0.04 fm show the magnitude of the proton radius puzzle. Each data set is shown as a band. This data shows larger variance at low $Q^2$ than any other fit and the variance increases with increasing N and M, with M having a larger influence on the variance. The Rational(0,1) curve rapidly diverges from the input radius. The Rational(1,2) curve shows no data below $Q^2\approx0.1$ GeV$^2$ as the mean radius was unphysical. Curves with both N and M of at least 1 stay very well centered on the input radius for the full range. The variances of all Rational(N,M) fits are less than half that of any of the other fits.

The middle plot has $\chi^2$ from 0 to 2 on the y axis. Each data set is shown as a band. All begin centered at 1, with Rational(0,1) having a rapid upturn. The others stay centered at 1 for the full range.

The bottom plot has MSE, Mean Squared Error as described in Eq.~\ref{eq:MSE}, from $10^{-6}$ to $10^0$ with a logarithmic scale on the y axis. Each set of data is represented as a curve. The Rational(0,1) curve rapidly approaches a minimum of $8\cdot10^{-5}$ and then begins to increase. The other curves continue to decrease over the full range. The Rational(1,1) curve drops below the y axis minimum at about $Q^2=0.35$ GeV$^2$.

\newpage

\bibliography{ZTransform.bib}